\documentclass{aa}  
\usepackage{natbib}
\usepackage{CJK}

\usepackage{tikz,xcolor}
\definecolor{lime}{HTML}{A6CE39}
\DeclareRobustCommand{\orcidicon}{
        \begin{tikzpicture}
        \draw[lime, fill=lime] (0,0) 
        circle [radius=0.16] 
        node[white] {{\fontfamily{qag}\selectfont \tiny ID}};
        \draw[white, fill=white] (-0.0625,0.095) 
        circle [radius=0.007];
        \end{tikzpicture}
        \hspace{-2mm}
}
\foreach \x in {A, ..., Z}{\expandafter\xdef\csname orcid\x\endcsname{\noexpand\href{https://orcid.org/\csname orcidauthor\x\endcsname}
                        {\noexpand\orcidicon}}
}

\usepackage{graphicx}
\usepackage{txfonts}
\usepackage[colorlinks=true, allcolors=blue]{hyperref}

\usepackage[normalem]{ulem}

\begin{document}

   \title{Direct measurements of carbon and sulfur isotope ratios\\ in the Milky Way}


   \author{{\protect\begin{CJK*}{UTF8}{gkai}Y. T. Yan (闫耀庭)\protect\end{CJK*}} \inst{\ref{inst.mpifr}}\fnmsep\thanks{Member of the International Max Planck Research School (IMPRS) for Astronomy and Astrophysics at the universities of Bonn and Cologne.}\orcidB{}
         \and C. Henkel\inst{\ref{inst.mpifr},\ref{inst.KingAbdulazizU},\ref{inst.xao}}\orcidA{}
         \and C. Kobayashi\inst{\ref{inst.UH}}\orcidJ{}
         \and K. M. Menten\inst{\ref{inst.mpifr}}\orcidC{}
         \and {\protect\begin{CJK*}{UTF8}{gkai}Y. Gong (龚\protect\end{CJK*}\protect\begin{CJK*}{UTF8}{bkai}龑\protect\end{CJK*})} \inst{\ref{inst.mpifr}}\orcidD{}
        \and {\protect\begin{CJK*}{UTF8}{gkai}J. S. Zhang (张江水)\protect\end{CJK*}}\inst{\ref{inst.gzhu}}\orcidE{}
        \and {\protect\begin{CJK*}{UTF8}{gkai}H. Z. Yu (余鸿智)\protect\end{CJK*}}\inst{\ref{inst.ufu},\ref{inst.gzhu}}
         \and {\protect\begin{CJK*}{UTF8}{gkai}K. Yang (杨楷)\protect\end{CJK*}}\inst{\ref{inst.nju},\ref{inst.keylaboratorynj}}\orcidG{}
        \and {\protect\begin{CJK*}{UTF8}{gkai}J. J. Xie (谢津津)\protect\end{CJK*}}\inst{\ref{inst.shao}}\orcidH{}
        \and {\protect\begin{CJK*}{UTF8}{gkai}Y. X. Wang (汪友鑫)\protect\end{CJK*}}\inst{\ref{inst.gzhu}}\orcidI{}
}

   \institute{
\label{inst.mpifr}Max-Planck-Institut f\"{u}r Radioastronomie, Auf dem H\"{u}gel 69, 53121 Bonn, Germany\\  \email{yyan@mpifr-bonn.mpg.de, astrotingyan@gmail.com}
\and\label{inst.KingAbdulazizU}Astronomy Department, Faculty of Science, King Abdulaziz University, P.~O.~Box 80203, Jeddah 21589, Saudi Arabia
\and\label{inst.xao}Xinjiang Astronomical Observatory, Chinese Academy of Sciences, 830011 Urumqi, PR China
\and\label{inst.UH}Centre for Astrophysics Research, Department of Physics, Astronomy and Mathematics, University of Hertfordshire, Hatfield, AL10 9AB, UK
\and\label{inst.gzhu}Center for Astrophysics, Guangzhou University, 510006 Guangzhou, People's Republic of China
\and\label{inst.ufu}Ural Federal University, 19 Mira Street, 620002 Ekaterinburg, Russia
\and\label{inst.nju}School of Astronomy and Space Science, Nanjing University, 163 Xianlin Avenue, Nanjing 210023, People's Republic of China
\and\label{inst.keylaboratorynj}Key Laboratory of Modern Astronomy and Astrophysics (Nanjing University), Ministry of Education, Nanjing 210023, People's Republic of China
\and\label{inst.shao}Shanghai Astronomical Observatory, Shanghai 200030, People's Republic of China
}

   \date{Received XXX; accepted YYY}

 
  \abstract
   {Isotope abundance ratios provide a powerful tool for tracing stellar nucleosynthesis, evaluating the composition of stellar ejecta, and constraining the chemical evolution of the Milky Way. }
   { We aim to measure the $^{12}$C/$^{13}$C, $^{32}$S/$^{34}$S, $^{32}$S/$^{33}$S, $^{32}$S/$^{36}$S, $^{34}$S/$^{33}$S, $^{34}$S/$^{36}$S, and $^{33}$S/$^{36}$S isotope ratios across the Milky Way.}
   {With the IRAM 30 meter telescope, we performed observations of the $J$ = 2-1 transitions of CS, C$^{33}$S, C$^{34}$S, C$^{36}$S, $^{13}$CS, $^{13}$C$^{33}$S, and $^{13}$C$^{34}$S as well as the $J$ = 3-2 transitions of C$^{33}$S, C$^{34}$S, C$^{36}$S, and $^{13}$CS toward a large sample of 110 high-mass star-forming regions.}
   {We measured the $^{12}$C/$^{13}$C, $^{32}$S/$^{34}$S, $^{32}$S/$^{33}$S, $^{32}$S/$^{36}$S, $^{34}$S/$^{33}$S, $^{34}$S/$^{36}$S, and $^{33}$S/$^{36}$S abundance ratios with rare isotopologs of CS, thus avoiding significant saturation effects. With accurate distances obtained from parallax data, we confirm previously identified $^{12}$C/$^{13}$C and $^{32}$S/$^{34}$S gradients as a function of galactocentric distance. In the central molecular zone, $^{12}$C/$^{13}$C ratios are higher than suggested by a linear fit to the disk values as a function of galactocentric radius. While $^{32}$S/$^{34}$S ratios near the Galactic center and in the inner disk are similar, this is not the case for $^{12}$C/$^{13}$C, when comparing central values with those near galactocentric radii of 5 kpc. As was already known, there is no $^{34}$S/$^{33}$S gradient but the average ratio of 4.35~$\pm$~0.44 derived from the $J$ = 2-1 transition lines of C$^{34}$S and C$^{33}$S is well below previously reported values. A comparison between solar and local interstellar $^{32}$S/$^{34}$S and $^{34}$S/$^{33}$S ratios suggests that the Solar System may have been formed from gas with a particularly high $^{34}$S abundance. For the first time, we report positive gradients of $^{32}$S/$^{33}$S, $^{34}$S/$^{36}$S, $^{33}$S/$^{36}$S, and $^{32}{\rm S}/^{36}{\rm S}$ in our Galaxy. The predicted $^{12}$C/$^{13}$C ratios from the latest Galactic chemical-evolution models \citep[e.g.,][]{2020ApJ...900..179K,2021A&A...653A..72R,2022arXiv220910620C} are in good agreement with our results. While $^{32}$S/$^{34}$S and $^{32}$S/$^{36}$S ratios show larger differences at larger galactocentric distances, $^{32}$S/$^{33}$S ratios show an offset across the entire inner 12 kpc of the Milky Way.}
  {}

   \keywords{nucleosynthesis --
                Galaxy: evolution --
                Galaxy: formation --
                ISM: abundances --
                \ion{H}{II} regions --
                ISM: molecules
               }
\maketitle
%
\section{Introduction}
\label{sect_introduction}

Isotope abundance ratios provide a powerful tool for tracing stellar nucleosynthesis, evaluating the composition of stellar ejecta, and constraining the chemical evolution of the Milky Way \citep{1994ARA&A..32..191W}. In particular, the $^{12}$C/$^{13}$C ratio is one of the most useful tracers of the relative degree of primary to secondary processing. $^{12}$C is a predominantly primary nucleus formed by He burning in massive stars on short timescales \citep[e.g.,][]{1995ApJS...98..617T}. $^{13}$C is produced on a longer timescale via CNO processing of $^{12}$C seeds from earlier stellar generations during the red giant phase in low- and intermediate-mass stars or novae \citep[e.g.,][]{1994LNP...439...72H,1994ARA&A..32..153M,1994ARA&A..32..191W}. $^{12}$C/$^{13}$C ratios are expected to be low in the central molecular zone (CMZ) and high in the Galactic outskirts, because the Galaxy formed from the inside out (e.g., \citealt{2001ApJ...554.1044C}; \citealt{2012A&A...540A..56P}).

Observations indeed indicate a gradient of $^{12}$C/$^{13}$C ratios across the Galaxy. \citet{1976A&A....51..303W} and \citet{1979MNRAS.188..445W} measured the $J_{Ka,Kc}=1_{1,0}-1_{1,1}$ lines of H$_2^{12}$CO and H$_2^{13}$CO near 5 GHz toward 11 and 24 Galactic continuum sources, respectively. While ignoring effects of photon trapping, the results suggested that the $^{12}$C/$^{13}$C ratios may vary with galactocentric distance ($R_{\rm GC}$). With the additional measurement of the $J_{Ka,Kc}=2_{1,1}-2_{1,2}$ line of H$_2$CO at 14.5 GHz, \citet{1980A&A....82...41H, 1982A&A...109..344H, 1983A&A...127..388H, 1985A&A...143..148H} also reported a gradient after correcting for effects of optical depth and photon trapping. \citet{1990ApJ...357..477L} used the optically thin lines of C$^{18}$O and $^{13}$C$^{18}$O to trace the carbon isotope ratios. They also found a systematic $^{12}$C/$^{13}$C gradient across the Galaxy, ranging from about 20--25 near the Galactic center, to 30--50 in the inner Galactic disk, to $\sim$70 in the local interstellar medium (ISM). \citet{1996A&AS..119..439W}, complementing these investigations by also including the far outer Galaxy, encountered ratios in excess of 100 and demonstrated that the gradient found in the inner Galaxy continues farther out. \citet{2005ApJ...634.1126M} obtained $^{12}$C/$^{13}$C = 6.01$R_{\rm GC}$ + 12.28 based on the CN measurements of \citet{2002ApJ...578..211S}. Here and elsewhere, $R_{\rm GC}$ denotes the galactocentric distance in units of kiloparsecs (kpc). By combining previously obtained H$_2$CO and C$^{18}$O results with these CN data, \citet{2005ApJ...634.1126M} obtained $^{12}$C/$^{13}$C = 6.21$R_{\rm GC}$ + 18.71.

More recently, \citet{2017ApJ...845..158H} reported observations of a variety of molecules (e.g., H$_2$CS, CH$_3$CCH, NH$_2$CHO, CH$_2$CHCN, and CH$_3$CH$_2$CN) and their $^{13}$C-substituted species toward Sgr B2(N). These authors obtained an average $^{12}$C/$^{13}$C value of 24 $\pm$ 7 in the Galactic center region, which is close to results using $^{12}$CH/$^{13}$CH (15.8 $\pm$ 2.4, Sgr B2(M)) by \citet{2020A&A...640A.125J} and the particularly solid $^{12}$C$^{34}$S/$^{13}$C$^{34}$S ratio (22.1$^{+3.3}_{-2.4}$, $+$50 km s$^{-1}$ Cloud) from \citet{2020A&A...642A.222H} who use a variety of CS isotopologs and rotational transitions. \citet{2019ApJ...877..154Y} proposed a linear fit of $^{12}$C/$^{13}$C = (5.08 $\pm$ 1.10)$R_{\rm GC}$ + (11.86 $\pm$ 6.60) based on a large survey of H$_2$CO. The latter includes data from the center to the outskirts of the Milky Way well beyond the Perseus Arm. However, data from the CMZ are not similar to those thoroughly traced by \citet{2020A&A...642A.222H}, not many sources from the innermost Galactic disk could be included in this survey, and also the number of sources beyond the Perseus arm was small, meaning that there is still space for improvement.

While the carbon isotope ratio has drawn much attention in the past, it is not the only isotope ratio that can be studied at radio wavelengths and that has a significant impact on our understanding of the chemical evolution of the Galaxy. The isotope ratios of sulfur are providing complementary information on stellar nucleosynthesis that is not traced by the carbon isotope ratio. Sulfur is special in that it provides a total of four stable isotopes, $^{32}$S, $^{34}$S, $^{33}$S, and $^{36}$S. In the Solar System, abundance ratios are 95.02 : 4.21 : 0.75 : 0.021, respectively \citep{1989GeCoA..53..197A}. $^{32}$S and $^{34}$S are synthesized during stages of hydrostatic oxygen-burning preceding a type II supernova event or during stages of explosive oxygen-burning in a supernova of type Ia; $^{33}$S is synthesized in explosive oxygen- and neon-burning, which is also related to massive stars; and $^{36}$S may be an s-process nucleus. The comprehensive calculations of \citet{1995ApJS..101..181W} indicate that $^{32}$S and $^{33}$S are primary (in the sense that the stellar yields do not strongly depend on the initial metallicity of the stellar model), while $^{34}$S is not a clean primary isotope; its yield decreases with decreasing metallicity. According to \citet{1985ApJ...295..604T} and \citet{1989A&A...210...93L}, $^{36}$S is produced as a purely secondary isotope in massive stars, with a possible (also secondary) contribution from asymptotic giant branch (AGB) stars. Only a small fraction of $^{36}$S is destroyed during supernova explosions (Woosley, priv. comm.). Comparing ``primary'' and ``secondary'' nuclei, we might therefore expect the presence of weak $^{32}$S/$^{34}$S and $^{34}$S/$^{33}$S gradients and a stronger $^{32}$S/$^{36}$S gradient as a function of galactocentric radius.

There is a strong and widespread molecular species that allows us to measure carbon and sulfur isotope ratios simultaneously, namely carbon monosulfide (CS). CS is unique in that it is a simple diatomic molecule exhibiting strong line emission and possessing eight stable isotopologs, which allows us to determine the above-mentioned carbon and sulfur isotope ratios. Six isotopologs have been detected so far in the ISM (e.g., \citealt{1996A&A...305..960C}; \citealt{1996A&A...313L...1M}; \citealt{2020A&A...642A.222H}; \citealt{2020ApJ...899..145Y}). 

Making use of the CS species, \citet{1996A&A...305..960C} and \citet{1996A&A...313L...1M} obtained average abundance ratios of 24.4 $\pm$ 5.0, 6.3 $\pm$ 1.0, and 115 $\pm$ 17 for $^{32}$S/$^{34}$S, $^{34}$S/$^{33}$S, and $^{34}$S/$^{36}$S for the ISM, respectively. The latter is approximately half the solar value, but similar to the value found in IRC+10216 \citep{2004A&A...426..219M}. Recently, \citet{2020A&A...642A.222H} published $^{32}$S/$^{34}$S ratios of 16.3$^{+2.1}_{-1.7}$ and 17.9 $\pm$ 5.0 for the $+$50 km s$^{-1}$ cloud and Sgr B2(N) near the Galactic center, respectively. These are only slightly lower than the value of 22 in the Solar System. There is an obvious and confirmed $^{32}$S/$^{34}$S gradient \citep{1996A&A...305..960C, 2020ApJ...899..145Y} from the inner Galaxy out to a galactocentric distance of 12.0 kpc. Nevertheless, there is a lack of data at small and large galactocentric distances.

We are performing systematic observational studies on isotope ratios in the Milky Way, including $^{12}$C/$^{13}$C \citep{2019ApJ...877..154Y}, $^{14}$N/$^{15}$N \citep{2021ApJS..257...39C},$^{18}$O/$^{17}$O (\citealt{2015ApJS..219...28Z}, \citeyear{2020ApJS..249....6Z}, \citeyear{2020IAUGA..30..278Z}; \citealt{2016RAA....16...47L}), and $^{32}$S/$^{34}$S \citep{2020ApJ...899..145Y}. We have thus performed a more systematic study on CS and its isotopologs toward 110 high-mass star-forming regions (HMSFRs). $^{12}$C/$^{13}$C and $^{32}$S/$^{34}$S ratios can be directly derived from integrated $^{12}$C$^{34}$S/$^{13}$C$^{34}$S (hereafter C$^{34}$S/$^{13}$C$^{34}$S) and $^{13}$C$^{32}$S/$^{13}$C$^{34}$S (hereafter $^{13}$CS/$^{13}$C$^{34}$S, see Section \ref{ratios_13c34s}) intensities, respectively. Also, $^{34}$S/$^{33}$S and $^{34}$S/$^{36}$S values could be obtained with measurements of C$^{34}$S, $^{12}$C$^{33}$S (hereafter C$^{33}$S), and $^{12}$C$^{36}$S (hereafter C$^{36}$S). Furthermore, $^{32}$S/$^{33}$S and $^{32}$S/$^{36}$S ratios can then be derived with the resulting $^{34}$S/$^{33}$S and $^{34}$S/$^{36}$S values combined with the $^{32}$S/$^{34}$S ratios (see Sections \ref{section_34s33s} to \ref{section_32s36s}). In Section \ref{sou_selection}, we describe the source selection and observations for our large sample. Section \ref{results} presents our results on $^{12}$C/$^{13}$C, $^{32}$S/$^{34}$S, $^{34}$S/$^{33}$S, $^{32}$S/$^{33}$S, $^{34}$S/$^{36}$S, and $^{32}$S/$^{36}$S ratios. Section \ref{discussion} discusses potential processes that could contaminate and affect the isotope ratios derived in the previous section and provides a detailed comparison with results from earlier studies. Our main results are summarized in Section \ref{summary}.

\section{Source selection and observations}
\label{sou_selection}

\subsection{Sample selection and distance}
\label{section_distance}

In 2019, we selected 18 HMSFRs from the Galactic center region to the outer Galaxy beyond the Perseus arm. To enlarge this sample, we chose 92 sources from the Bar and Spiral Structure Legacy (BeSSeL) Survey\footnote{http://bessel.vlbi-astrometry.org} in 2020. These 92 targets were recently released by the BeSSeL project \citep{2019ApJ...885..131R} and not observed by \citet{2020ApJ...899..145Y}. In total, 110 objects in the Galaxy are part of our survey. The coordinates of our sample sources are listed in Table \ref{table_sources}. Determining trigonometric parallaxes is a very direct and accurate method to measure the distance of sources from the Sun \citep{2009ApJ...700..137R, 2014ApJ...783..130R, 2019ApJ...885..131R}. Over the past decade, mainly thanks to the BeSSeL project, the trigonometric parallaxes of approximately 200 HMSFRs have been determined across the Milky Way through dedicated high-resolution observations of molecular maser lines. Therefore, this is a good opportunity to investigate carbon and sulfur isotope ratios with well-determined distances across the Galaxy. The galactocentric distance ($R_{\rm GC}$) can be obtained with the heliocentric distance $d$ from the trigonometric parallax data base of the BeSSeL project using \begin{equation}
R_{\rm GC} = \sqrt{( R_0 cos(l) - d )^2 + R_0^2 sin^2(l)}
,\end{equation}
\citep{2009ApJ...699.1153R}. $R_0$ = 8.178 $\pm$ 0.013$\rm _{stat.}$ $\pm$ 0.022$\rm _{sys.}$ kpc \citep{2019A&A...625L..10G} describes the distance from the Sun to the Galactic center, $l$ is the Galactic longitude of the source, and $d$ is the distance either directly derived from the trigonometric parallax data based on the BeSSeL project or a kinematic distance in cases where no such distance is yet available. Because the uncertainty in $R_0$ is very small, it will be neglected in the following analysis. For 12 of our targets without trigonometric parallax data, we estimated their kinematic distances from the Revised Kinematic Distance calculator\footnote{http://bessel.vlbi-astrometry.org/revised\_kd\_2014} \citep{2014ApJ...783..130R}. The resulting distances indicate that 6 of these 12 sources are located in the CMZ, namely SgrC, the $+$20 km~s$^{-1}$ cloud, the $+$50 km~s$^{-1}$ cloud, G0.25, G1.28$+$0.07, and SgrD. Four targets belong to the inner Galactic disk, namely PointC1, CloudD, Clump2, and PointD1. Two sources, W89-380 and WB89-391, are in the outer regions beyond the Perseus arm. The heliocentric distances ($d$) and the galactocentric distances ($R_{\rm GC}$) for our sample are listed in Columns 6 and 7 of Table~\ref{table_sources}.

\subsection{Observations}
We observed the $J$ = 2-1 transitions of CS, C$^{33}$S, C$^{34}$S, C$^{36}$S, $^{13}$CS, $^{13}$C$^{33}$S, and $^{13}$C$^{34}$S as well as the $J$ = 3-2 transitions of C$^{33}$S, C$^{34}$S, C$^{36}$S, and $^{13}$CS toward 110 HMSFRs with the IRAM 30 meter telescope\footnote{IRAM is supported by INSU/CNRS (France), MPG (Germany) and IGN (Spain).} in 2019 June, July, and October under project 045-19 (PI, Christian Henkel) as well as in 2020 August within project 022-20 (PI, Hongzhi Yu). The on$+$off source integration times for our sources range from 4.0 minutes to 10.8 hours. These values are given in Appendix~\ref{appendix_table} (Table \ref{fitting_all}). The EMIR receiver with two bands, E090 and E150, was used to cover a bandwidth of $\sim$16 GHz (from 90.8 to 98.2 GHz and 138.4 to 146.0 GHz) simultaneously in dual polarisation. We used the wide-mode FTS backend with a resolution of 195 kHz, corresponding to $\sim$0.6 km s$^{-1}$ and $\sim$0.4 km s$^{-1}$ at 96 GHz and 145 GHz, respectively. The observations were performed in total-power position-switching mode and the off position was set at 30$\arcmin$ in azimuth. Pointing was checked every 2 hours using nearby quasars. Focus calibrations were done at the beginning of the observations and during sunset and sunrise toward strong quasars. The main beam brightness temperature, $T_{\rm MB}$, was obtained from the antenna temperature $T_{\rm A}^*$ via the relation $T_{\rm MB}$ = $T_{\rm A}^*\times F_{\rm eff}$/$B_{\rm eff}$ ($F_{\rm eff}$: forward hemisphere efficiency; $B_{\rm eff}$: main beam efficiency) with corresponding telescope efficiencies\footnote{https://publicwiki.iram.es/Iram30mEfficiencies}: $F_{\rm eff}$/$B_{\rm eff}$ are 0.95/0.81 and 0.93/0.73 in the frequency ranges of 90.8-98.2 GHz and 138.4-146.0 GHz, respectively. The system temperatures were 100-160 K and 170-300 K on a $T_{\rm A}^*$ scale for the E090 and E150 band observations. The half power beam width (HPBW) for each transition was calculated as HPBW($\arcsec$)=2460/$\nu$(GHz). Rest frequencies, excitations of the upper levels above the ground state, Einstein coefficients for spontaneous emission, and respective beam sizes are listed in Table~\ref{table_linelist}.

\begin{table*}[h]
\caption{Observed spectral line parameters$^a$.}
\centering
\begin{tabular}{cccccc}
\hline\hline
Isotopolog &  Transition  &  $\nu_0$\tablefootmark{b}  &  $E_{\rm up}$\tablefootmark{c} & $A_{\rm u,l}$\tablefootmark{d} & HPBW\tablefootmark{e}  \\
  & \    & (MHz)   &  (K) &  (s$^{-1}$) &  ($\arcsec$) \\
\hline
\label{table_linelist}
CS               & 2-1 & 97980.953   &7.1  & 1.68 $\times$ 10$^{-5}$ & 25.1 \\
C$^{33}$S        & 2-1 & 97172.064   &7.0  & 1.64 $\times$ 10$^{-5}$ & 25.3 \\
C$^{34}$S        & 2-1 & 96412.95    &6.9  & 1.60 $\times$ 10$^{-5}$ & 25.5 \\
C$^{36}$S        & 2-1 & 95016.722   &6.8  & 1.53 $\times$ 10$^{-5}$ & 25.9 \\
$^{13}$CS        & 2-1 & 92494.308   &6.7  & 1.41 $\times$ 10$^{-5}$ & 26.6 \\
$^{13}$C$^{33}$S & 2-1 & 91685.241   &6.6  & 1.38 $\times$ 10$^{-5}$ & 26.8 \\
$^{13}$C$^{34}$S & 2-1 & 90926.026   &6.5  & 1.34 $\times$ 10$^{-5}$ & 27.1 \\
C$^{33}$S        & 3-2 & 145755.732  &14.0 & 5.92 $\times$ 10$^{-5}$ & 16.9 \\
C$^{34}$S        & 3-2 & 144617.101  &13.9 & 5.78 $\times$ 10$^{-5}$ & 17.0 \\
C$^{36}$S        & 3-2 & 142522.785  &13.7 & 5.54 $\times$ 10$^{-5}$ & 17.3 \\
$^{13}$CS        & 3-2 & 138739.335  &13.3 & 5.11 $\times$ 10$^{-5}$ & 17.7 \\
\hline
\end{tabular}
\tablefoot{
\tablefoottext{a}{From the Cologne Database for Molecular Spectroscopy \citep[CDMS,][]{2005JMoSt.742..215M,2016JMoSp.327...95E}.}
\tablefoottext{b}{Rest frequency.}
\tablefoottext{c}{Upper energy level.}
\tablefoottext{d}{Einstein coefficient for spontaneous emission from upper $u$ to lower $l$ level.}
\tablefoottext{e}{Half power beam width.}}
\end{table*}

\subsection{Data reduction}
\label{datareduction}

\begin{figure}[h]
\centering
   \includegraphics[width=0.3\textwidth]{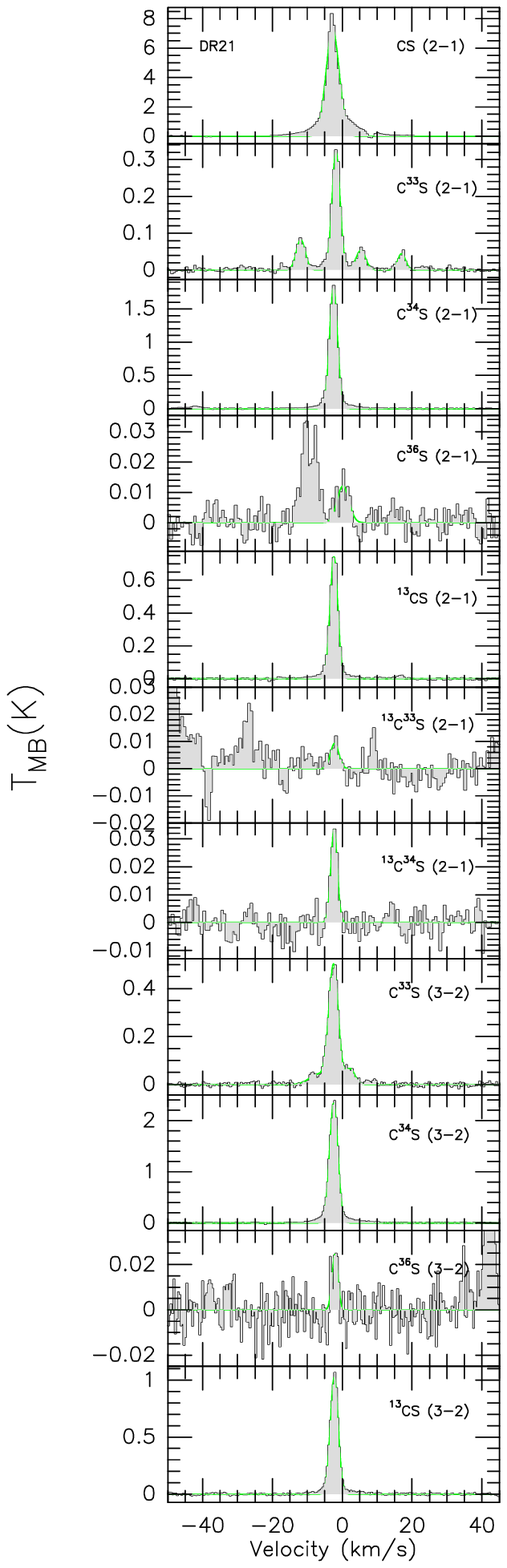}
\caption{Line profiles of the $J$ = 2-1 transitions of CS, C$^{33}$S, C$^{34}$S, C$^{36}$S, $^{13}$CS, $^{13}$C$^{33}$S, and $^{13}$C$^{34}$S as well as the $J$ = 3-2 transitions of C$^{33}$S, C$^{34}$S, C$^{36}$S, and $^{13}$CS toward one typical target (DR21) of our large sample of 110 sources, after subtracting first-order polynomial baselines. The main beam brightness temperature scales are presented on the left hand side of the profiles. The spectra of all 110 objects in our sample are shown in Appendix~\ref{appendix_spectra} (Fig.~\ref{spectra_all}).}
  \label{fig_dr21}
\end{figure}

We used the GILDAS/CLASS\footnote{https://www.iram.fr/IRAMFR/GILDAS/} package to analyze the spectral line data. The spectra of the $J$ = 2-1 transitions of CS, C$^{33}$S, C$^{34}$S, C$^{36}$S, $^{13}$CS, $^{13}$C$^{33}$S, and $^{13}$C$^{34}$S as well as the $J$ = 3-2 transitions of C$^{33}$S, C$^{34}$S, C$^{36}$S, and $^{13}$CS toward one of our targets, DR21, are shown in Fig.~\ref{fig_dr21}, after subtracting first-order polynomial baselines and applying Hanning smoothing. The spectra of all 110 targets, also after first-order polynomial-baseline removal and Hanning smoothing, are presented in Appendix~\ref{appendix_spectra} (Fig.~\ref{spectra_all}).

Among our sample of 110 targets, we detected the $J$ = 2-1 line of CS toward 106 sources, which yields a detection rate of 96\%. The $J$ = 2-1 transitions of C$^{34}$S, $^{13}$CS, C$^{33}$S, and $^{13}$C$^{34}$S were successfully detected in 90, 82, 46, and  17 of our sources with signal-to-noise (S/N) ratios of greater than 3, respectively. The $J$ = 3-2 lines of C$^{34}$S, $^{13}$CS, and C$^{33}$S were detected in 87, 71, and 42 objects  with S/Ns  of $\ge$3.0. Line parameters from Gaussian fitting are listed in Appendix~\ref{appendix_table} (Table \ref{fitting_all}). Relevant for the evaluation of isotope ratios is the fact that  for 17 sources with 19 velocity components, the S/Ns of the $J$ = 2-1 transition of $^{13}$C$^{34}$S are greater than 3, which allows us to determine the $^{12}$C/$^{13}$C and $^{32}$S/$^{34}$S ratios directly with the $J$ = 2-1 lines of C$^{34}$S, $^{13}$CS, and $^{13}$C$^{34}$S.  Toward 82 targets with 90 radial velocity components, the $J$ = 2-1 transitions of C$^{34}$S and $^{13}$CS were both detected with S/Ns of $\ge$3.0. The $J$ = 3-2 lines of C$^{34}$S and $^{13}$CS were both found in 71 objects with 73 radial velocity components and S/Ns of $\ge$3.0. Furthermore, the $J$ = 2-1 and $J$ = 3-2 transitions of C$^{34}$S and C$^{33}$S were detected  with S/Ns of $\ge$3.0 toward 46 and 42 sources, respectively.

The C$^{36}$S $J$ = 2-1 line was successfully detected with S/Ns of $\ge$3.0 toward three targets, namely W3OH, the $+$50 km~s$^{-1}$ cloud near the Galactic center, and DR21. As C$^{36}$S and $^{13}$C$^{33}$S are the least abundant among the CS isotopologs, tentative detection with S/Ns of $\sim$2.0 are also presented here but not included in further analyses. In another five objects, the C$^{36}$S $J$ = 2-1 line was tentatively detected. For the C$^{36}$S $J$ = 3-2 transitions, we report one detection with an S/N larger than 3.0 toward Orion-KL and five tentative detections. The $J$ = 2-1 lines of $^{13}$C$^{33}$S were tentatively detected toward three sources, namely Orion-KL, W51-IRS2, and DR21.
Integration times and 1$\sigma$ noise levels of the observed transitions are listed in Columns 3 and 4 of Table~\ref{fitting_all} for each target.

\section{Results}
\label{results}
In the following, we first estimate the optical depths of the various lines to avoid problems with line saturation that might affect our results. We then present the carbon and sulfur isotope ratios derived from different detected CS isotopologs. 

\subsection{Optical depth}
\label{section_opacities}

The main isotopolog, CS, is usually optically thick in massive star-forming regions (e.g. \citealt{1980ApJ...235..437L}; \citealt{2020ApJ...899..145Y}). Therefore, the $^{12}$C/$^{13}$C and $^{32}$S/$^{34}$S ratios cannot be determined from the line intensity ratios of $I$(CS)/$I$($^{13}$CS) and $I$(CS)/$I$(C$^{34}$S). However, assuming that the $J$ = 2-1 transitions of CS, C$^{34}$S, and $^{13}$CS share the same beam filling factor and excitation temperature, we can estimate the maximum optical depth of the $^{13}$CS $J$ = 2-1 line from:
\begin{equation}
\frac{T_{\rm {mb}}(^{12}\rm C\rm S)}{T_{\rm mb}(^{13}\rm C\rm S)} \sim \frac{1 - e^{-\tau(^{13}\rm C\rm S)R_C}}{1 - e^{-\tau(^{13}\rm C\rm S)}},\,R_C = \frac{^{12}\rm C}{^{13}\rm C},
\end{equation}
where $T_{\rm {mb}}$ is the peak main beam brightness temperature derived from the best Gaussian-fitting result and listed in Column 8 of Table~\ref{fitting_all}. In this case, the $^{12}$C/$^{13}$C ratios can be derived from the integrated line intensities of C$^{34}$S and $^{13}$C$^{34}$S with the assumption of $\tau(\rm C^{34}\rm S) \textless 1.0$, which then also implies $\tau(^{13}\rm C^{34}\rm S) \textless 1.0$ (see details in Section \ref{ratios_13c34s}). Multiplying $\tau(^{13}\rm CS)$ by $R_C$ = $^{12}$C/$^{13}$C, we can get the peak opacity $\tau(\rm CS)$ = $\tau(^{13}{\rm CS})R_C$. The maximum optical depth of C$^{34}$S can be obtained from:
\begin{equation}
\frac{T_{\rm mb}(\rm C\rm ^{34}S)}{T_{\rm mb}(^{13}\rm C\rm S)} = \frac{1 - e^{-\tau(\rm C\rm ^{34}S)}}{1 - e^{-\tau(^{13}\rm C\rm S)}},
\end{equation}
where $T_{\rm {mb}}$ is the peak main beam brightness temperature derived from the best Gaussian-fitting result and listed in Column 8 of Table~\ref{fitting_all}. As shown in Table \ref{table_13c34sresults}, the peak optical depths of the $J$ = 2-1 lines of CS, C$^{34}$S, and $^{13}$CS for our 17 targets with detections of $^{13}$C$^{34}$S range from 1.29 to 8.79, 0.12 to 0.55, and 0.05 to 0.34, respectively. Therefore, C$^{34}$S and $^{13}$CS in these 17 objects are optically thin, even though they belong, on average, to the more opaque ones,  being successfully detected
in $^{13}$C$^{34}$S (see below). Nevertheless, the corrections for optical depth are applied to C$^{34}$S and $^{13}$CS with factors of $f_1$ and $f_2$, respectively.
\begin{equation}
f_1=\frac{\tau(\rm C\rm ^{34}S)}{1-e^{-\tau(\rm C\rm ^{34}S)}} {\quad\rm and}
\end{equation}
\begin{equation}
f_2=\frac{\tau(^{13}\rm C\rm S)}{1-e^{-\tau(^{13}\rm C\rm S)}}
\end{equation}
are listed in Columns 7 and 8 of Table~\ref{table_13c34sresults}, respectively.

For those 82 sources with detections of $J$ = 2-1 CS, C$^{34}$S, and $^{13}$CS, the optical depths were calculated based on the $^{12}$C/$^{13}$C gradient that we derived from our C$^{34}$S and $^{13}$C$^{34}$S measurements (for details, see Section \ref{ratios_13c34s}). In Table \ref{table_doubleisotope}, the peak opacities of the $J$ = 2-1 lines of CS, C$^{34}$S, and $^{13}$CS for these 82 targets range from 0.34 to 14.48, 0.02 to 0.74, and 0.01 to 0.39, respectively. The CS $J$ = 2-1 lines are optically thick with $\tau(\rm CS) \textgreater 1.0$ in most sources (89\%) of our sample, while they tend to be optically thin in seven objects, namely Point C1 ($\tau(\rm CS) \leq 0.59$), Sgr C ($\tau(\rm CS) \leq 0.54$), Cloud D ($\tau(\rm CS) \leq 0.64$), G1.28$+$0.07 ($\tau(\rm CS) \leq 0.82$), Sgr D ($\tau(\rm CS) \leq 0.45$), and Point D1 ($\tau(\rm CS) \leq 0.34$). In contrast, the transitions from rare isotopologs, the C$^{34}$S and $^{13}$CS $J$ = 2-1 lines in our sample, are all optically thin, as their maximum optical depths are less than 0.8 and 0.4, respectively. In the following, we are therefore motivated to consider all CS isotopologs as optically thin, except CS itself. This allows us to use ratios of integrated intensity of all the rare CS isotopologs ---but not CS itself--- to derive the carbon and sulfur isotope ratios we intend to study. Small corrections accounting for the optical depths are applied to the C$^{34}$S and $^{13}$CS $J$ = 2-1 lines with factors of $f_1$ and $f_2$, respectively, and are listed in Columns 7 and 8 of Table~\ref{table_doubleisotope}. The optical depths of the $J$ = 3-2 transitions cannot be estimated, as the CS $J$ = 3-2 line was not covered by our observations because of bandwidth limitations. However, the $^{32}$S/$^{34}$S ratios for a given source obtained through the double isotope method from the $J$ = 2-1 and $J$ = 3-2 transitions are in good agreement, indicating that the C$^{34}$S and $^{13}$CS $J$ = 3-2 lines in our sample are also optically thin (see details in Section~\ref{section_double_32s34s}). 

\begin{figure}[h]
\centering
   \includegraphics[width=0.5\textwidth]{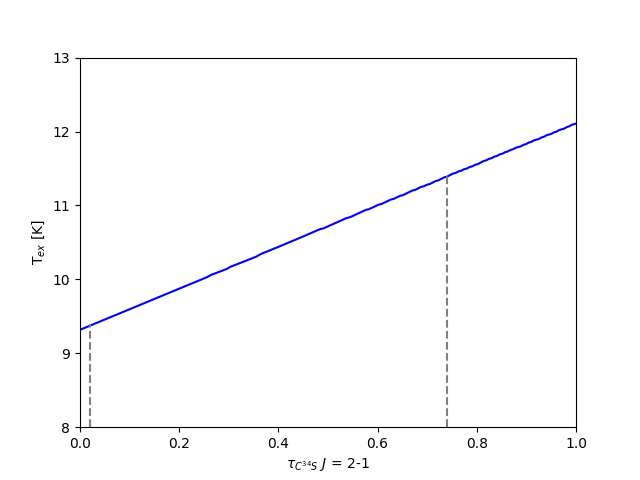}
\caption{Excitation temperature, $T_{ex}$, as a function of optical depth for the $J$ = 2-1 transition of C$^{34}$S. The gray dashed lines indicate the range of opacities for our sample sources.}
  \label{fig_c34stex}
\end{figure}

The RADEX non Local Thermodynamic Equilibrium (LTE)  model \citep{2007A&A...468..627V} was used to calculate the variation of excitation temperature, $T_{ex}$, with optical depth. Frequencies, energy levels, and Einstein A coefficients for spontaneous emission were taken from the Cologne Database for Molecular Spectroscopy \citep[CDMS;][]{2005JMoSt.742..215M,2016JMoSp.327...95E}. Recent collision rates for CS with para- and ortho-H$_2$ \citep{2018MNRAS.478.1811D} were used. Figure~\ref{fig_c34stex} shows the excitation temperatures and opacities of C$^{34}$S $J$ = 2-1 for a kinetic temperature of 30 K and a molecular hydrogen density of 10$^5$ cm$^{-3}$. Variations of $T_{ex}$ within about 2 K for our sample targets with optical depths of 0.02 $\leq \tau(\rm C^{34}S) \leq 0.74$ can barely affect our results.

\begin{figure*}[h]
\centering
   \includegraphics[width=460pt]{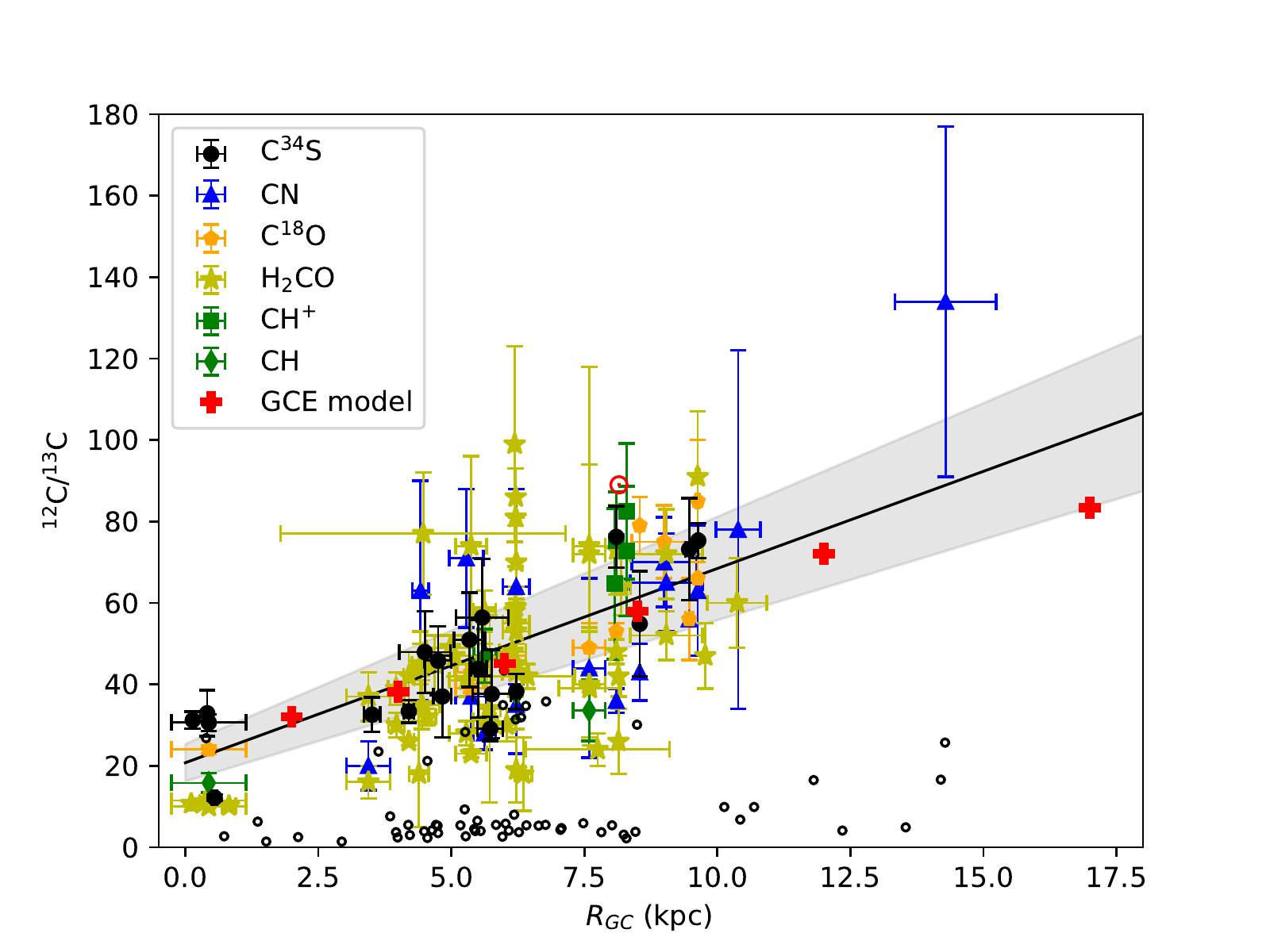}
     \caption{$^{12}$C/$^{13}$C isotope ratios from C$^{34}$S/$^{13}$C$^{34}$S, CN/$^{13}$CN, C$^{18}$O/$^{13}$C$^{18}$O, H$_2$CO/H$_2^{13}$CO, CH$^+$/$^{13}$CH$^+$, and CH/$^{13}$CH are plotted as functions of the distance from the Galactic center. The red symbol $\odot$ indicates the $^{12}$C/$^{13}$C isotope ratio of the Sun. The filled black circles are the results obtained from C$^{34}$S with corrections of opacity in the current work, and the resulting first-order polynomial fit is plotted as a solid line, with the gray-shaded area showing the 1$\sigma$ interval of the fit. The open black circles are the 3$\sigma$ lower limits obtained from nondetections of $^{13}$C$^{34}$S in the current work. The blue triangles, orange pentagons, yellow stars, green squares, and green diamonds are values determined from CN \citep{2002ApJ...578..211S,2005ApJ...634.1126M}, C$^{18}$O \citep{1990ApJ...357..477L,1996A&AS..119..439W,1998ApJ...494L.107K}, H$_2$CO \citep{1980A&A....82...41H,1982A&A...109..344H,1983A&A...127..388H,1985A&A...143..148H,2019ApJ...877..154Y}, CH$^+$ \citep{2011ApJ...728...36R}, and CH \citep{2020A&A...640A.125J}, respectively, using the most up-to-date distances. The red crosses visualize the results from the GCE model of \citet[][see also Section~\ref{section_discussion_model}]{2011MNRAS.414.3231K,2020ApJ...900..179K}. }
  \label{fig_gradient_12C13C}
\end{figure*}

\begin{table*}[h]
\caption{Isotope ratios derived with the $J$ = 2-1 transitions of C$^{34}$S, $^{13}$CS, and $^{13}$C$^{34}$S.}
\centering
\small
\begin{tabular}{lcc|ccc|cc|cc}
\hline\hline
Source &  $V_{\rm LSR}$ & $R_{GC}$  &   \multicolumn{3}{c}{Optical depth}  &   \multicolumn{2}{c}{Corrections for} &  $^{12}$C/$^{13}$C  &  $^{32}$S/$^{34}$S  \\
  &                &         &     &           &           &  \multicolumn{2}{c}{optical depth} &   &    \\
  &  (km s$^{-1}$) & (kpc)   &  CS & C$^{34}$S & $^{13}$CS &  $f_1$  &  $f_2$ &   &    \\
\hline
\label{table_13c34sresults}
W3OH                    & -46.96 &  9.64 $\pm$ 0.03   & 8.289 $\pm$ 0.083 & 0.292 $\pm$ 0.003 & 0.127 $\pm$ 0.001 &  1.2  &  1.1  &  75.30 $\pm$ 4.27  &    29.80 $\pm$ 1.68 \\
Orion-KL                &   8.17 &  8.54 $\pm$ 0.00   & 3.722 $\pm$ 0.037 & 0.180 $\pm$ 0.002 & 0.074 $\pm$ 0.001 &  1.1  &  1.0  &  54.89 $\pm$ 12.90 &    21.24 $\pm$ 4.67 \\
G359.61$-$00.24         &  19.33 &  5.51 $\pm$ 0.15   & 3.531 $\pm$ 0.035 & 0.204 $\pm$ 0.002 & 0.089 $\pm$ 0.001 &  1.1  &  1.0  &  43.85 $\pm$ 11.95 &    17.31 $\pm$ 4.79 \\
$+$50~km~s$^{-1}$~cloud &  46.79 &  0.02 $\pm$ 0.04   & 4.278 $\pm$ 0.043 & 0.192 $\pm$ 0.002 & 0.151 $\pm$ 0.002 &  1.1  &  1.1  &  31.22 $\pm$ 2.06  &    21.71 $\pm$ 1.40 \\
SgrB2                   &  53.18 &  0.55 $\pm$ 0.05   & 1.290 $\pm$ 0.013 & 0.140 $\pm$ 0.001 & 0.113 $\pm$ 0.001 &  1.1  &  1.1  &  12.18 $\pm$ 0.73  &    9.77  $\pm$ 0.58 \\
SgrB2                   &  66.54 &  0.44 $\pm$ 0.70   & 8.119 $\pm$ 0.081 & 0.523 $\pm$ 0.005 & 0.341 $\pm$ 0.003 &  1.3  &  1.2  &  30.62 $\pm$ 2.04  &    18.99 $\pm$ 1.31 \\
SgrB2                   &  83.38 &  0.41 $\pm$ 0.02   & 2.417 $\pm$ 0.024 & 0.097 $\pm$ 0.001 & 0.077 $\pm$ 0.001 &  1.0  &  1.0  &  32.90 $\pm$ 5.65  &    26.24 $\pm$ 4.49 \\
G006.79$-$00.25         &  20.87 &  4.75 $\pm$ 0.25   & 8.789 $\pm$ 0.088 & 0.484 $\pm$ 0.005 & 0.242 $\pm$ 0.002 &  1.3  &  1.1  &  45.82 $\pm$ 8.39  &    19.46 $\pm$ 3.63 \\
G010.32$-$00.15         &  11.99 &  5.34 $\pm$ 0.29   & 5.039 $\pm$ 0.050 & 0.237 $\pm$ 0.002 & 0.111 $\pm$ 0.001 &  1.1  &  1.1  &  50.95 $\pm$ 11.58 &    21.55 $\pm$ 5.05 \\
G019.36$-$00.03         &  26.40 &  5.58 $\pm$ 0.49   & 6.789 $\pm$ 0.068 & 0.332 $\pm$ 0.003 & 0.141 $\pm$ 0.001 &  1.2  &  1.1  &  56.41 $\pm$ 14.37 &    21.19 $\pm$ 5.53 \\
G024.78$+$00.08         & 110.72 &  3.51 $\pm$ 0.15   & 8.038 $\pm$ 0.080 & 0.554 $\pm$ 0.006 & 0.322 $\pm$ 0.003 &  1.3  &  1.2  &  32.56 $\pm$ 4.25  &    16.08 $\pm$ 2.12 \\
G028.39$+$00.08         &  78.01 &  4.83 $\pm$ 0.17   & 8.705 $\pm$ 0.087 & 0.443 $\pm$ 0.004 & 0.291 $\pm$ 0.003 &  1.2  &  1.2  &  37.04 $\pm$ 10.05 &    18.74 $\pm$ 5.12 \\
G028.83$-$00.25         &  87.19 &  4.50 $\pm$ 0.48   & 7.283 $\pm$ 0.073 & 0.388 $\pm$ 0.004 & 0.183 $\pm$ 0.002 &  1.2  &  1.1  &  47.93 $\pm$ 10.00 &    19.73 $\pm$ 4.16 \\
G030.70$-$00.06         &  89.94 &  4.20 $\pm$ 0.10   & 7.088 $\pm$ 0.071 & 0.448 $\pm$ 0.004 & 0.263 $\pm$ 0.003 &  1.2  &  1.1  &  33.39 $\pm$ 2.77  &    18.07 $\pm$ 1.52 \\
G030.74$-$00.04         &  91.82 &  5.76 $\pm$ 0.36   & 5.924 $\pm$ 0.059 & 0.404 $\pm$ 0.004 & 0.191 $\pm$ 0.002 &  1.2  &  1.1  &  37.63 $\pm$ 10.87 &    15.61 $\pm$ 4.59 \\
G030.81$-$00.05         &  98.89 &  5.73 $\pm$ 0.24   & 4.691 $\pm$ 0.047 & 0.306 $\pm$ 0.003 & 0.187 $\pm$ 0.002 &  1.2  &  1.1  &  29.07 $\pm$ 2.99  &    14.97 $\pm$ 1.44 \\
W51-IRS2                &  61.07 &  6.22 $\pm$ 0.06   & 1.883 $\pm$ 0.019 & 0.119 $\pm$ 0.001 & 0.052 $\pm$ 0.001 &  1.1  &  1.0  &  38.26 $\pm$ 4.35  &    16.70 $\pm$ 1.92 \\
DR21                    &  -2.49 &  8.10 $\pm$ 0.00   & 7.079 $\pm$ 0.071 & 0.270 $\pm$ 0.003 & 0.106 $\pm$ 0.001 &  1.1  &  1.1  &  76.22 $\pm$ 7.57  &    27.48 $\pm$ 2.84 \\
NGC7538                 & -57.12 &  9.47 $\pm$ 0.07   & 3.433 $\pm$ 0.034 & 0.131 $\pm$ 0.001 & 0.050 $\pm$ 0.001 &  1.1  &  1.0  &  73.19 $\pm$ 12.54 &    26.72 $\pm$ 4.57 \\
\hline
\end{tabular}
\tablefoot{ Velocities were obtained from measurements of C$^{34}$S, see Table \ref{fitting_all} in Appendix \ref{appendix_table}.}
\end{table*}

\subsection{$^{12}$C/$^{13}$C and $^{32}$S/$^{34}$S ratios derived directly from $^{13}$C$^{34}$S}
\label{ratios_13c34s}

\subsubsection{$^{12}$C/$^{13}$C ratios}
\label{section_results_12c13c}

The $^{12}$C/$^{13}$C ratios derived from the integrated intensity ratios of C$^{34}$S and $^{13}$C$^{34}$S with corrections of optical depth are listed in Table~\ref{table_13c34sresults}. Figure~\ref{fig_gradient_12C13C} shows our results as filled black circles. A gradient of $^{12}$C/$^{13}$C is obtained with an unweighted least-squares fit:
\begin{equation}
^{12}{\rm C}/^{13}{\rm C} = (4.77 \pm 0.81)R_{\rm GC}+(20.76 \pm 4.61).
\end{equation}
The correlation coefficient is 0.82. Around the CMZ toward the $+$50 km~s$^{-1}$ cloud and SgrB2, four velocity components of C$^{34}$S and $^{13}$C$^{34}$S were detected and then an average $^{12}$C/$^{13}$C value of 27~$\pm$~3 is derived. The uncertainties given here and below are standard deviations of the mean. Eleven objects within a range of 3.50 kpc < $R_{\rm GC}$ < 6.50 kpc in the inner Galactic disk lead to an average $^{12}$C/$^{13}$C value of 41~$\pm$~9. In the Local arm near the Sun, the $^{13}$C$^{34}$S lines were detected toward two sources, Orion-KL and DR21. These provide an average $^{12}$C/$^{13}$C value of 66~$\pm$~10, which is lower than the Solar System ratio. The other two targets beyond the solar neighborhood belong to the Perseus arm and show a slightly higher value of 74~$\pm$~8.

For sources with detections of C$^{34}$S and nondetections of $^{13}$C$^{34}$S, 3$\sigma$ lower limits of the $^{12}$C/$^{13}$C ratio have been derived and are shown as open black circles in Fig.~\ref{fig_gradient_12C13C}. All these lower limits are below the $^{12}$C/$^{13}$C gradient we describe above.

\subsubsection{$^{32}$S/$^{34}$S ratios}
\label{section_ratios3234_13c34s}

\begin{figure*}[h]
\centering
   \includegraphics[height=600pt]{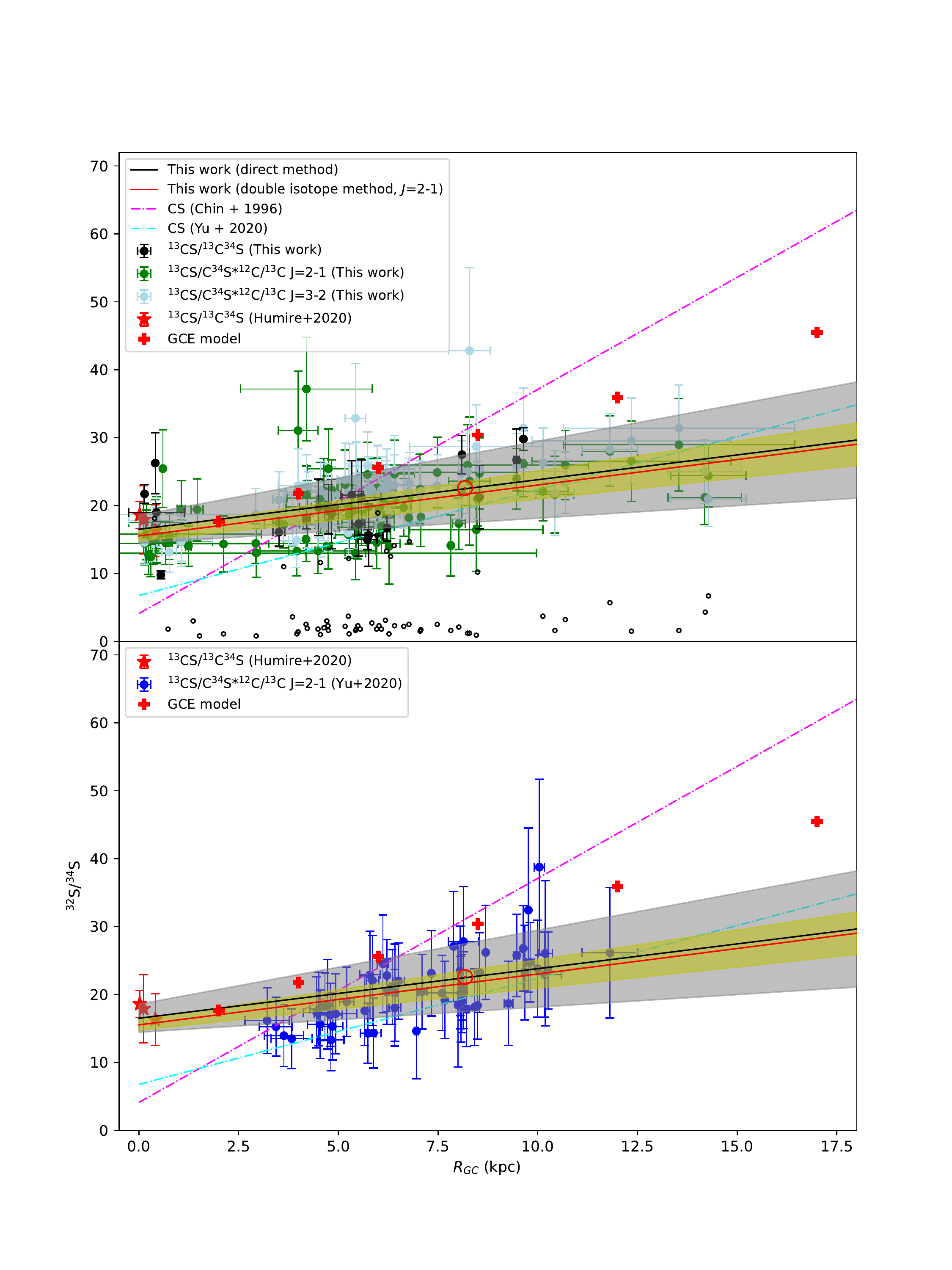}
     \caption{$^{32}$S/$^{34}$S isotope ratios as functions of the distance from the Galactic center. The symbol $\odot$ indicates the $^{32}$S/$^{34}$S isotope ratio in the Solar System. In the upper panel, the $^{32}$S/$^{34}$S ratios directly derived from $^{13}$CS/$^{13}$C$^{34}$S in the $J$ = 2-1 transition and obtained from the double isotope method in the $J$ = 2-1 transition with corrections for optical depth are plotted as black and green dots, respectively. The $^{32}$S/$^{34}$S ratios without corrections of opacity in the $J$ = 3-2 transition are plotted as light blue dots. The 3$\sigma$ lower limits of $^{32}$S/$^{34}$S ratios obtained from nondetections of $^{13}$C$^{34}$S in the current work are shown as open black circles. The $^{32}$S/$^{34}$S ratios in \citet{2020ApJ...899..145Y} derived from the double isotope method in the $J$ = 2-1 transitions are shown as blue dots in the lower panel. The $^{32}$S/$^{34}$S values in the CMZ obtained from $^{13}$CS/$^{13}$C$^{34}$S in \citet{2020A&A...642A.222H} are plotted as red stars in both panels. The resulting first-order polynomial fits to $^{32}$S/$^{34}$S ratios with the direct method and the double isotope method from the $J$ = 2-1 transition in this work are plotted as black and red solid lines in the two panels, respectively, with the gray and yellow shaded areas showing the 1~$\sigma$ intervals of the fits. The magenta and cyan dashed-dotted lines show the $^{32}$S/$^{34}$S gradients from \citet{1996A&A...305..960C} and \citet{2020ApJ...899..145Y}. The red crosses visualize the results from the GCE model of \citet[][see also Section~\ref{section_discussion_model}]{2011MNRAS.414.3231K,2020ApJ...900..179K}. }
  \label{fig_gradient_32S34S}
\end{figure*}

The $^{32}$S/$^{34}$S ratios directly derived from the integrated intensity ratios of $^{13}$CS/$^{13}$C$^{34}$S from the $J$ = 2-1 lines with corrections of optical depth are listed in Table~\ref{table_13c34sresults} and are plotted as a function of galactocentric distance in Fig.~\ref{fig_gradient_32S34S}. With an unweighted least-squares fit, a gradient with a correlation coefficient of 0.47 can be obtained:
\begin{equation}
^{32}{\rm S}/^{34}{\rm S} =(0.73 \pm 0.36)R_{\rm GC}+(16.50 \pm 2.07).
\end{equation}
An average $^{32}$S/$^{34}$S ratio of 19~$\pm$~2 is obtained in the CMZ, which is based on the measurements from two sources, namely the $+$50 km~s$^{-1}$ cloud next to the Galactic center and Sgr B2. In the inner Galactic disk at a range of 3.50 kpc < $R_{\rm GC}$ < 6.50 kpc, $^{13}$C$^{34}$S was detected toward 11 objects, leading to an average $^{32}$S/$^{34}$S value of 18~$\pm$~4. For sources in the Local and the Perseus arm beyond the Sun, the $^{32}$S/$^{34}$S ratios are 24~$\pm$~4 and 28~$\pm$~3, respectively. This reveals a gradient from the inner Galactic disk to the outer Galaxy, but none from the CMZ to the inner disk.

For sources with detections of $^{13}$CS and nondetections of $^{13}$C$^{34}$S, we determined 3$\sigma$ lower limits to the $^{32}$S/$^{34}$S ratio, which are shown as open black circles in Fig.~\ref{fig_gradient_32S34S}. All these lower limits are below the $^{32}$S/$^{34}$S gradient we describe above.

\subsection{$^{32}$S/$^{34}$S ratios obtained through the double isotope method}
\label{section_double_32s34s}

The $^{32}$S/$^{34}$S values can also be derived from measurements of C$^{34}$S and $^{13}$CS  using the carbon gradient obtained from our $^{13}$C$^{34}$S measurements above by applying the following equation:
\begin{equation}
\frac{^{32}{\rm S}}{^{34}{\rm S}} = R_{\rm C} \frac{I(^{13}{\rm CS})}{I(\rm C^{34}{\rm S})},
\end{equation}
where $R_{\rm C}$ is the $^{12}$C/$^{13}$C ratio derived from equation (6). The uncertainty on this latter is also included in our error budget. The $^{32}$S/$^{34}$S ratios in the $J$ = 2-1 transitions were calculated with corrections of optical depth for 83 targets with 90 radial velocity components, in which the C$^{34}$S and $^{13}$CS $J$ = 2-1 lines were both detected, and are listed in Column 7 of Table \ref{table_doubleisotope}. An unweighted least-squares fit to these values yields  
\begin{equation}
^{32}{\rm S}/^{34}{\rm S} (2-1) = (0.75 \pm 0.13)R_{\rm GC}+(15.52 \pm 0.78), 
\end{equation}
with a correlation coefficient of 0.54. The C$^{34}$S and $^{13}$CS $J$ = 3-2 lines were both detected in 71 objects with 73 radial velocity components. The $^{32}$S/$^{34}$S ratios derived with equation (8) from the $J$ = 3-2 transition are shown in Column 8 of Table \ref{table_doubleisotope}. An unweighted least-squares fit to the $J$ = 3-2 transition yields 
\begin{equation}
^{32}{\rm S}/^{34}{\rm S} (3-2) = (0.99 \pm 0.14)R_{\rm GC}+(16.05 \pm 0.95),  
\end{equation}
which is within the errors, and is consistent with the trend obtained from the $J$ = 2-1 transition (equation (9)). However, we note that we do not have CS $J$ = 3-2 data, and therefore no opacity corrections could be applied to our C$^{34}$S $J$ = 3-2 spectra (see also Sect.~\ref{section_34s33s}).

\subsection{$^{34}$S/$^{33}$S ratios}
\label{section_34s33s}

The $^{34}$S/$^{33}$S ratios can be determined directly from the intensity ratios of C$^{34}$S/C$^{33}$S. The $^{34}$S/$^{33}$S ratios from the $J$ = 2-1 lines were then corrected for optical depths derived in Section~\ref{section_opacities}. However, both the $J$ = 2-1 and $J$ = 3-2 transitions of C$^{33}$S are split by hyperfine structure (HFS) interactions \citep{1981CPL....81..256B}, which may affect the deduced values of $^{34}$S/$^{33}$S. 

\begin{figure*}[h]
\centering
   \includegraphics[width=500pt]{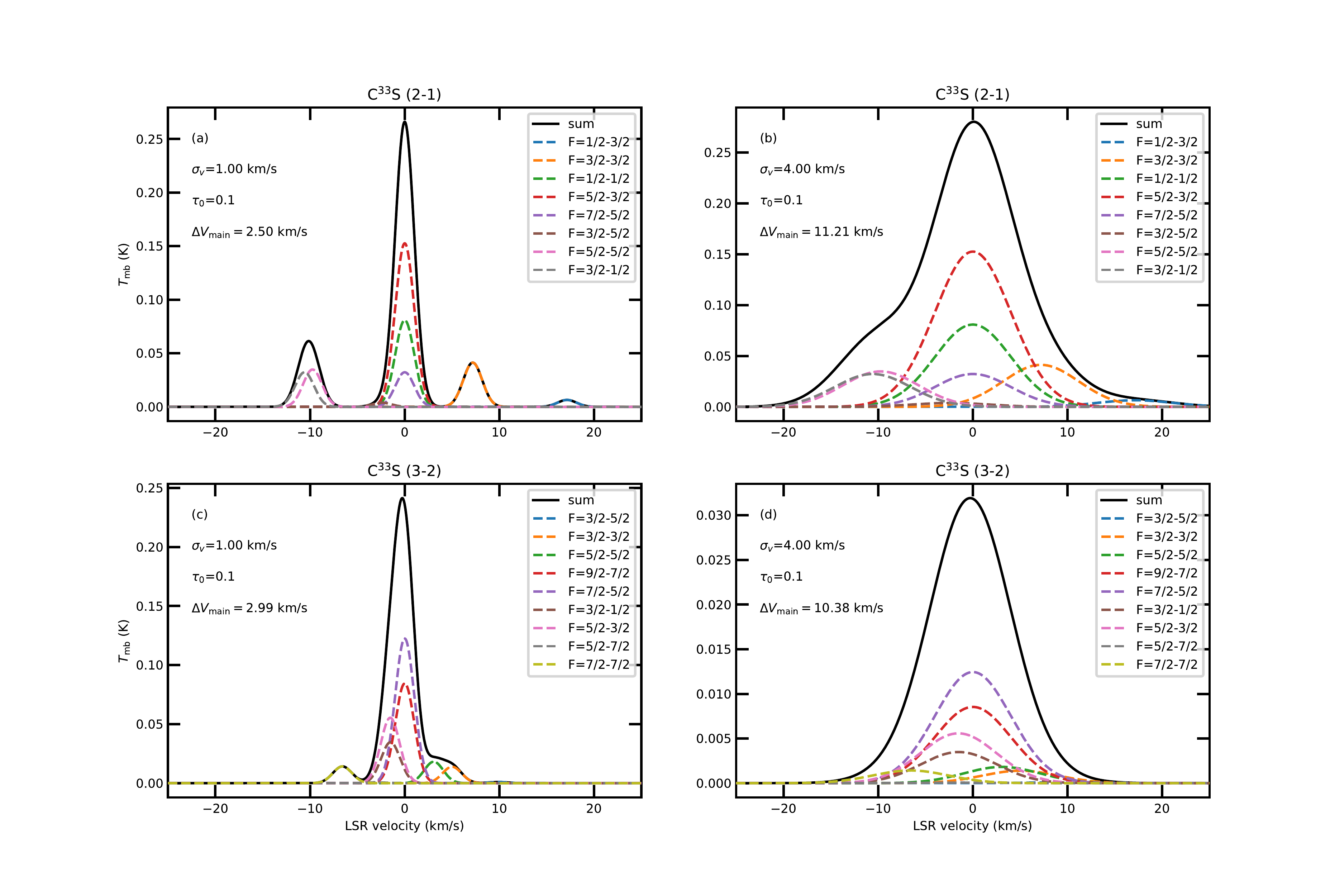}
     \caption{Synthetic C$^{33}$S (2-1) and C$^{33}$S (3-2) spectra for two intrinsic line widths, 1.0~km~s$^{-1}$ (left panels) and 4.0~km~s$^{-1}$ (right panels).}
  \label{fig_c33s_hfs}
\end{figure*}

\begin{figure*}[h]
\centering
   \includegraphics[width=230pt]{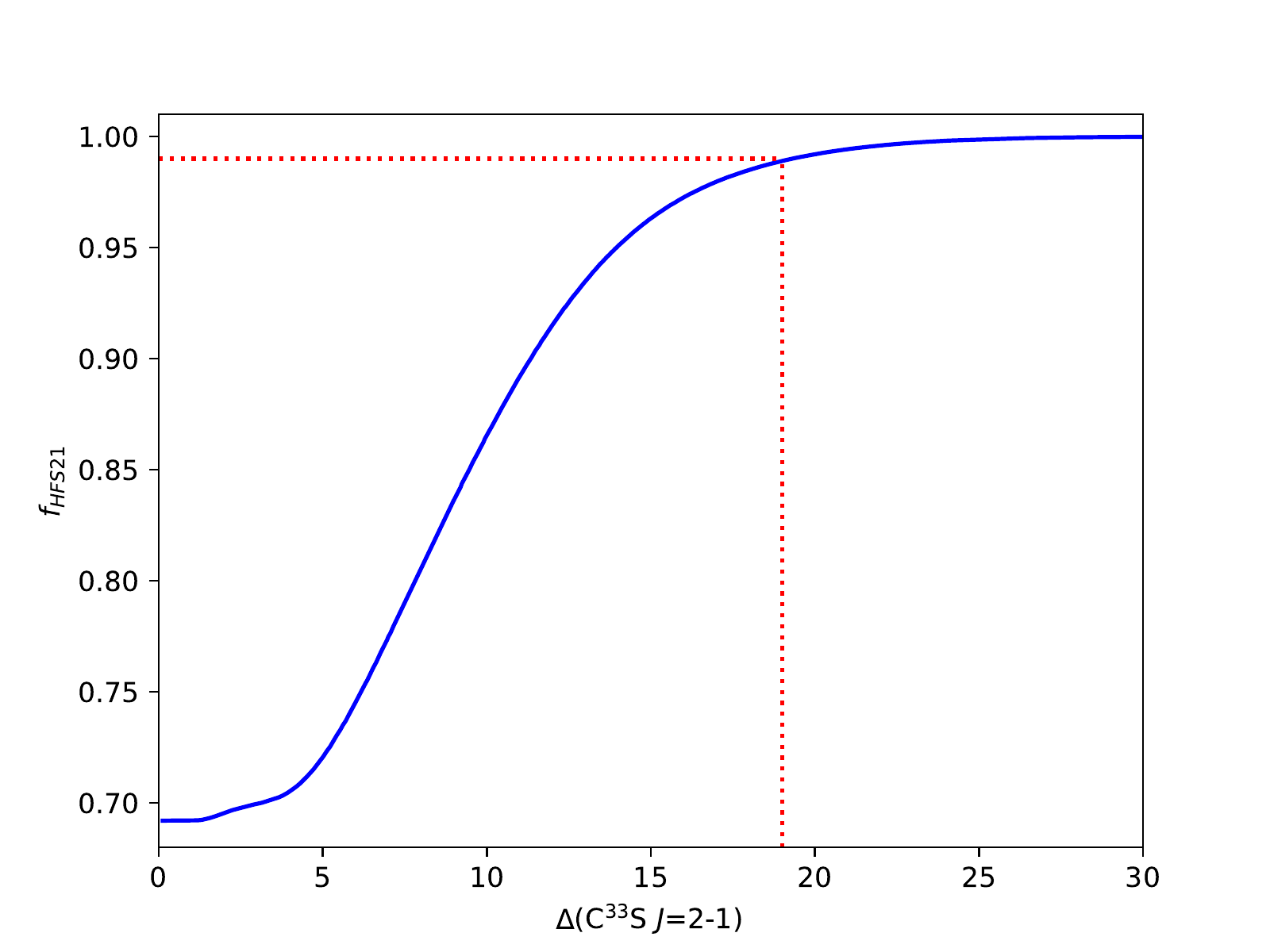}
   \includegraphics[width=230pt]{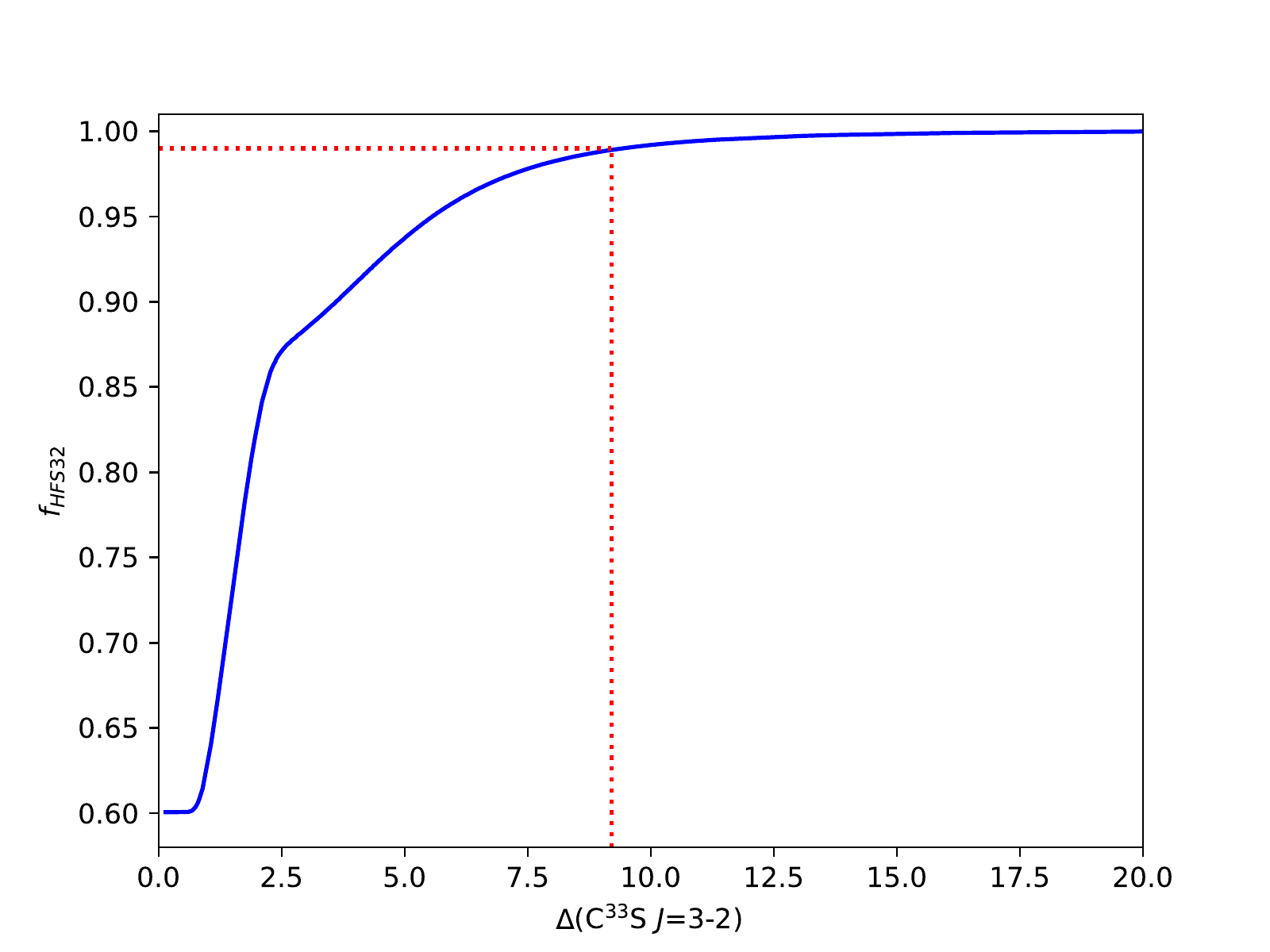}
     \caption{Blue lines are curves showing the theoretical dependencies of the HFS factors, $f_{\rm 21HFS}$ (left) and $f_{\rm 32HFS}$ (right), on the line width of the C$^{33}$S main component for sources in which only the main component was detected. The red dotted vertical and horizontal lines indicate the values of the minimal line widths where the factors are reaching almost 1.0, 19.0~km~s$^{-1}$ for the $J$ = 2-1 transition ($f_{\rm 21HFS} \sim$ 0.99) and 9.2~km~s$^{-1}$ for the $J$ = 3-2 transition ($f_{\rm 32HFS} \sim$ 0.99). }
  \label{fig_model_hfs}
\end{figure*}

The C$^{33}$S $J$ = 2-1 line consists of eight hyperfine components distributed over about 9.0 MHz \citep{2005JMoSt.742..215M,2016JMoSp.327...95E}, which corresponds to a velocity range of about 28~km~s$^{-1}$. Following the method introduced in Appendix D in \citet{2021A&A...646A.170G}, and assuming that the intrinsic width of each HFS line is 1~km~s$^{-1}$, the $J$ = 2-1 line profile can be reproduced by four components (see Fig.~\ref{fig_c33s_hfs}, upper left panel). In this case, the main component ($I_{main}$) consists of four HFS lines ($F$=7/2-5/2, $F$=5/2-3/2, $F$=1/2-1/2, $F$=3/2-5/2), which account for 70\% of the total intensity. Among the 46 sources with detections of the C$^{33}$S $J$ = 2-1 line, all of the four components were detected in 10 targets. Toward 16 objects, only the three components with the lowest velocities were detected, accounting for 98\% of the total intensity. For the remaining 20 sources, only the main component was detected. Based on the above assumptions, 30\% of the total intensity would be missed. The situation is different when the main component becomes broader. If the line width of the main component is larger than 10~km~s$^{-1}$, 87\% of the total intensity is covered by the main spectral feature. When the line width of the main component is larger than 19.0~km~s$^{-1}$, then almost all HFS lines are included. In Fig.~\ref{fig_model_hfs}, we show the dependence of the HFS factor for the $J$ = 2-1 line, $f_{\rm 21HFS}$, on the line width of the main component. Depending on the specific condition of each target, we derived $f_{\rm 21HFS}$ for each source. The values are listed in Table~\ref{table_3433}. 

The C$^{33}$S $J$ = 3-2 line consists of nine hyperfine components covering about 8.0 MHz \citep{2005JMoSt.742..215M,2016JMoSp.327...95E}, corresponding to a velocity range of about 16~km~s$^{-1}$. Assuming that the intrinsic width of each HFS line is 1~km~s$^{-1}$, the $J$ = 3-2 line profile can be characterized by three components (see also Fig.~\ref{fig_c33s_hfs}). All these three components are detected in only two sources, Orion-KL and DR21. The main component consists of four HFS lines ($F$=5/2-3/2, $F$=3/2-1/2, $F$=7/2-5/2, $F$=9/2-7/2), which account for 86\% of the total intensity. When the line width becomes larger than 9.2~km~s$^{-1}$, almost all of the HFS lines overlap. The HFS factors ($f_{\rm 32HFS}$) of the $J$ = 3-2 transition obtained individually for each source are listed in Table~\ref{table_3433}.

We calculated the $^{34}$S/$^{33}$S intensity ratios and present them in Table~\ref{table_3433}. Applying corrections accounting for the effect of hyperfine splitting, the $^{34}$S/$^{33}$S ratios are derived and are 
also listed in Table~\ref{table_3433}. The $^{34}$S/$^{33}$S values obtained from the $J$ = 2-1 transitions are always higher than the ones derived from the $J$ = 3-2 lines toward the same source, with the exception of G097.53$+$03.18. This difference could be caused by the lack of corrections of optical depth in the $J$ = 3-2 transition. The average values of $^{34}$S/$^{33}$S toward our sample are 4.35~$\pm$~0.44 and 3.49~$\pm$~0.26 in the $J$ = 2-1 and $J$ = 3-2 transitions, respectively. The $^{34}$S/$^{33}$S ratios were found to be independent of galactocentric distance (Fig.~\ref{fig_all33S}). 

\begin{figure}[h]
\centering
   \includegraphics[height=270pt]{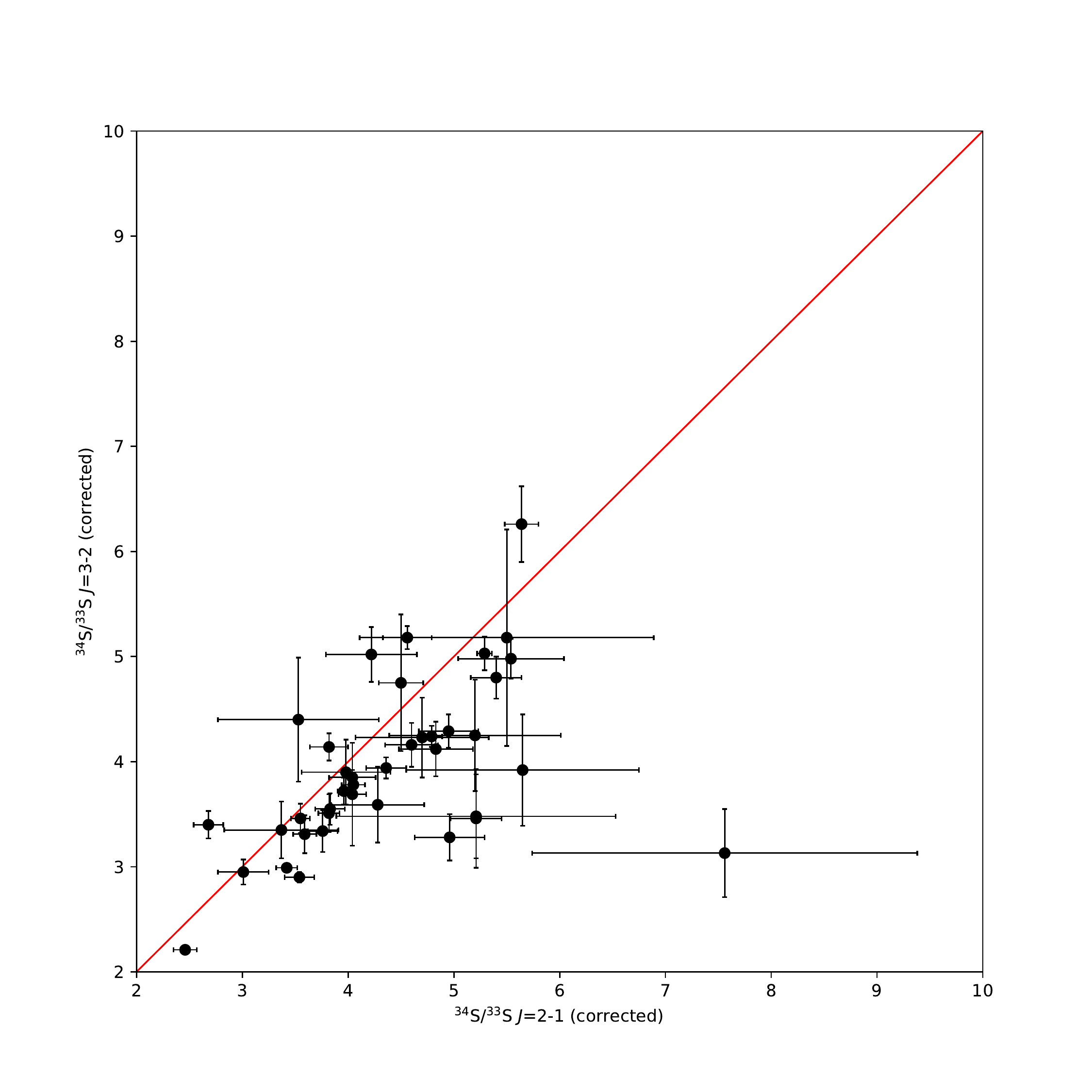}
     \caption{Comparison of $^{34}$S/$^{33}$S ratios between the $J$ = 2-1 and 3-2 data. The $J$ = 2-1 ratios are opacity corrected, while the same correction factors were also applied to the $J$ = 3-2 data. The red solid line indicates that the $^{34}$S/$^{33}$S ratios are equal in these two transitions.}
  \label{fig_33S_compared}
\end{figure}

After applying the opacity correction of the $J$ = 2-1 transition to the $J$ = 3-2 line in the same source, $^{34}$S/$^{33}$S ratios in the $J$ = 3-2 transition are higher than the $^{34}$S/$^{33}$S values in the $J$ = 2-1 transition in seven targets, suggesting that the C$^{34}$S $J$ = 3-2 lines in these seven sources are less opaque than the C$^{34}$S $J$ = 2-1 lines. The seven targets are the $+$20~km~s$^{-1}$~cloud, G023.43$-$00.18, G028.39$+$00.08, G028.83$-$00.25, G073.65$+$00.19, G097.53$+$03.18, and G109.87$+$02.11. A comparison of the corrected $^{34}$S/$^{33}$S ratios in these two transitions is shown in Fig.~\ref{fig_33S_compared}. On the other hand, toward the other 32 objects of the whole sample of 39 sources with detections in these two transitions, the $^{34}$S/$^{33}$S ratios in the $J$ = 3-2 transition are still lower than the $^{34}$S/$^{33}$S values in the $J$ = 2-1 transitions. This suggests that the C$^{34}$S $J$ = 3-2 lines may be more opaque than the C$^{34}$S $J$ = 2-1 lines. The ratios of $^{34}$S/$^{33}$S values without corrections of opacity in the $J$ = 3-2 transition and the corrected $^{34}$S/$^{33}$S values from the $J$ = 2-1 lines in 31 targets of these 32 sources are within the range of 1.02 to 1.71, which suggest that the optical depths of C$^{34}$S in the $J$ = 3-2 transition in these objects range from  0.05 to 1.20 based on equation (4). The maximum optical depth of the C$^{34}$S $J$ = 3-2 line toward the additional source, G024.85$+$00.08, which has not considered until now, is estimated to be 2.6.

\subsection{$^{32}$S/$^{33}$S ratios}
\label{section_32s33s}

The $^{32}$S/$^{33}$S values can also be derived from the $^{34}$S/$^{33}$S ratios in Section~\ref{section_34s33s} using the $^{32}$S/$^{34}$S ratios ---which we directly obtained from $^{13}$CS/$^{13}$C$^{34}$S and present in Section~\ref{section_ratios3234_13c34s}--- by applying the following equation:
\begin{equation}
\frac{^{32}{\rm S}}{^{33}{\rm S}} = \frac{^{34}\rm S}{^{33}\rm S} \times \frac{^{32}\rm S}{^{34}\rm S} .
\end{equation}
For sources where we did not detect $^{13}$C$^{34}$S, the $^{32}$S/$^{34}$S ratios derived from the double isotope method in Section~\ref{section_double_32s34s} are used. Their uncertainty is also included in our error budget. The resulting $^{32}$S/$^{33}$S ratios are listed in Table~\ref{table_3233}. As in the case of the $^{34}$S/$^{33}$S ratios, the $^{32}$S/$^{33}$S values obtained from the $J$ = 2-1 transitions with corrections for optical depth are slightly larger than the ones from the $J$ = 3-2 transitions without opacity corrections toward the same source. 

In the CMZ, $^{32}$S/$^{33}$S $J$ = 2-1 ratios toward four targets, namely the $+$20 km~s$^{-1}$ cloud, the $+$50 km~s$^{-1}$ cloud, Sgr B2, and Sgr D, lead to an average value of 70 $\pm$ 16. In the inner Galaxy, in a galactocentric distance range of 2.0~kpc~$\le R_{\rm GC} \le$~6.0~kpc, an average $^{32}$S/$^{33}$S ratio of 82 $\pm$ 19 was derived from values in 20 sources. Near the solar neighborhood, in a galactocentric distance range of 7.5~kpc~$\le R_{\rm GC} \le$~8.5~kpc, $^{32}$S/$^{33}$S ratios were obtained toward four objects, Orion-KL, G071.31$+$00.82, DR21, and G109.87$+$02.11, resulting in an average value of 88 $\pm$ 21. For the outer Galaxy, beyond the local arm, we were able to deduce an average $^{32}$S/$^{33}$S ratio of 105 $\pm$ 19  from $^{32}$S/$^{33}$S values in four sources. All these average values provide us with an indication of the existence of a $^{32}$S/$^{33}$S gradient in the Galaxy. An unweighted least-squares fit to the $J$ = 2-1 transition data from 46 sources yields
\begin{equation}
^{32}{\rm S}/^{33}{\rm S} (2-1) = (2.64 \pm 0.77)R_{\rm GC}+(70.80 \pm 5.57).  
\end{equation}
In the $J$ = 3-2 transition with data from 42 targets, an unweighted least-squares fit can be obtained with
\begin{equation}
^{32}{\rm S}/^{33}{\rm S} (3-2) = (2.80 \pm 0.59)R_{\rm GC}+(59.30 \pm 4.22).  
\end{equation}
The $^{32}$S/$^{33}$S gradients derived from these two transitions have similar slopes but obviously different intercepts. The lower intercept in the $J$ = 3-2 transition could be due to the fact that we could not correct the values for optical-depth effects. If this difference is caused only by the opacity, then an average optical depth in the C$^{34}$S $J$ = 3-2 transition of about 0.4 can be derived with equation (4). 

\setcounter{table}{3}

\begin{table*}[h]
\centering
\caption{\label{table_3233}  $^{32}$S/$^{33}$S isotope ratios}
\begin{tabular}{lccc}
\hline\hline
Source &  $R_{GC}$  &   \multicolumn{2}{c}{$^{32}$S/$^{33}$S}  \\
       &  (kpc)     &  $J$ = 2-1    &     $J$ = 3-2       \\
\hline
WB89-380 & 14.19 $\pm$ 0.92                & 90.74  $\pm$ 19.53  &   83.93  $\pm$ 17.90  \\
WB89-391 & 14.28 $\pm$ 0.94                & 127.07 $\pm$ 40.26  &   65.46  $\pm$ 14.69  \\
W3OH & 9.64 $\pm$ 0.03                     & 105.49 $\pm$ 7.26   &   74.20  $\pm$ 4.35   \\
Orion-KL & 8.54 $\pm$ 0.00                 & 74.50  $\pm$ 16.85  &   69.12  $\pm$ 13.51  \\
$+$20 km~s$^{-1}$ cloud & 0.03 $\pm$ 0.03  & 82.63  $\pm$ 18.20  &   78.98  $\pm$ 17.86  \\
G359.61$-$00.24 & 5.51 $\pm$ 0.15          & 75.75  $\pm$ 14.99  &   71.69  $\pm$ 16.86  \\
$+$50 km~s$^{-1}$ cloud & 0.02 $\pm$ 0.04  & 76.49  $\pm$ 16.96  &   66.68  $\pm$ 14.92  \\
G000.31$-$00.20 & 5.25 $\pm$ 0.36          & 55.61  $\pm$ 14.84  &   $\cdots$           \\
SgrB2 & 0.44 $\pm$ 0.70                    & 42.98  $\pm$ 9.62   &   35.66  $\pm$ 7.84   \\
SgrD & 0.45 $\pm$ 0.07                     & 77.48  $\pm$ 20.88  &   50.61  $\pm$ 13.66  \\
G006.79$-$00.25 & 4.75 $\pm$ 0.25          & 86.65  $\pm$ 17.01  &   70.08  $\pm$ 13.79  \\
G007.47$+$00.05 & 12.35 $\pm$ 2.49         & 107.98 $\pm$ 30.91  &   $\cdots$           \\
G010.32$-$00.15 & 5.34 $\pm$ 0.29          & 88.27  $\pm$ 17.82  &   79.56  $\pm$ 15.99  \\
G010.62$-$00.33 & 4.26 $\pm$ 0.21          & 106.56 $\pm$ 33.99  &   $\cdots$           \\
G011.10$-$00.11 & 5.49 $\pm$ 0.56          & 72.13  $\pm$ 29.12  &   $\cdots$           \\
G016.86$-$02.15 & 5.97 $\pm$ 0.47          & 86.63  $\pm$ 17.00  &   76.23  $\pm$ 15.01  \\
G017.02$-$02.40 & 6.40 $\pm$ 0.36          & 100.03 $\pm$ 20.54  &   86.36  $\pm$ 17.47  \\
G017.63$+$00.15 & 6.77 $\pm$ 0.04          & 73.12  $\pm$ 23.04  &   $\cdots$           \\
G018.34$+$01.76 & 6.31 $\pm$ 0.07          & 92.66  $\pm$ 18.57  &   86.44  $\pm$ 16.87  \\
G019.36$-$00.03 & 5.58 $\pm$ 0.49          & 94.55  $\pm$ 18.78  &   80.16  $\pm$ 15.89  \\
G023.43$-$00.18 & 3.63 $\pm$ 0.49          & 72.63  $\pm$ 16.26  &   80.58  $\pm$ 16.63  \\
G024.78$+$00.08 & 3.51 $\pm$ 0.15          & 70.88  $\pm$ 14.22  &   55.64  $\pm$ 11.12  \\
G024.85$+$00.08 & 3.85 $\pm$ 0.23          & 134.87 $\pm$ 43.15  &   48.25  $\pm$ 11.88  \\
G028.30$-$00.38 & 4.71 $\pm$ 0.26          & $\cdots$            &   76.82  $\pm$ 20.06  \\
G028.39$+$00.08 & 4.83 $\pm$ 0.17          & 85.30  $\pm$ 17.17  &   76.11  $\pm$ 15.02  \\
G028.83$-$00.25 & 4.50 $\pm$ 0.48          & 78.57  $\pm$ 15.87  &   80.37  $\pm$ 15.83  \\
G030.70$-$00.06 & 4.20 $\pm$ 0.10          & 76.54  $\pm$ 15.18  &   62.68  $\pm$ 12.57  \\
G030.74$-$00.04 & 5.76 $\pm$ 0.36          & 86.28  $\pm$ 17.07  &   72.73  $\pm$ 14.12  \\
G030.78$+$00.20 & 4.19 $\pm$ 0.05          & $\cdots$            &   72.50  $\pm$ 17.46  \\
G030.81$-$00.05 & 5.73 $\pm$ 0.24          & 88.21  $\pm$ 17.23  &   72.72  $\pm$ 14.52  \\
G031.24$-$00.11 & 7.48 $\pm$ 2.00          & $\cdots$            &   90.57  $\pm$ 21.01  \\
G032.74$-$00.07 & 4.55 $\pm$ 0.23          & 75.67  $\pm$ 14.99  &   68.63  $\pm$ 13.88  \\
G032.79$+$00.19 & 5.26 $\pm$ 1.57          & 79.87  $\pm$ 16.07  &   76.72  $\pm$ 14.93  \\
G034.41$+$00.23 & 5.99 $\pm$ 0.06          & 80.54  $\pm$ 15.82  &   72.63  $\pm$ 14.63  \\
G034.79$-$01.38 & 6.21 $\pm$ 0.09          & 103.60 $\pm$ 22.03  &   82.58  $\pm$ 16.21  \\
G036.11$+$00.55 & 5.45 $\pm$ 0.43          & 39.43  $\pm$ 13.64  &   $\cdots$           \\
G037.42$+$01.51 & 6.78 $\pm$ 0.05          & 110.92 $\pm$ 21.76  &   76.06  $\pm$ 17.77  \\
G040.28$-$00.21 & 6.02 $\pm$ 0.10          & 110.40 $\pm$ 31.44  &   71.86  $\pm$ 17.32  \\
G045.45$+$00.06 & 6.41 $\pm$ 0.50          & 83.04  $\pm$ 21.44  &   76.44  $\pm$ 16.33  \\
G048.99$-$00.29 & 6.18 $\pm$ 0.02          & 107.90 $\pm$ 27.60  &   87.21  $\pm$ 20.50  \\
W51-IRS2 & 6.22 $\pm$ 0.06                 & 74.69  $\pm$ 14.46  &   66.43  $\pm$ 12.72  \\
G071.31$+$00.82 & 8.02 $\pm$ 0.16          & 95.37  $\pm$ 31.90  &   106.87 $\pm$ 30.60  \\
G073.65$+$00.19 & 13.54 $\pm$ 2.90         & 102.16 $\pm$ 32.61  &   128.66 $\pm$ 31.17  \\
G075.29$+$01.32 & 10.69 $\pm$ 0.58         & 122.06 $\pm$ 29.02  &   92.08  $\pm$ 19.53  \\
DR21 & 8.10 $\pm$ 0.00                     & 86.80  $\pm$ 16.37  &   80.51  $\pm$ 15.38  \\
G090.92$+$01.48 & 10.13 $\pm$ 0.63         & 87.88  $\pm$ 19.68  &   95.73  $\pm$ 20.14  \\
G097.53$+$03.18 & 11.81 $\pm$ 0.70         & 74.99  $\pm$ 14.50  &   87.92  $\pm$ 16.65  \\
G109.87$+$02.11 & 8.49 $\pm$ 0.01          & 94.41  $\pm$ 18.40  &   98.52  $\pm$ 23.17  \\
NGC7538 & 9.47 $\pm$ 0.07                  & 104.53 $\pm$ 19.54  &   $\cdots$           \\
\hline
\end{tabular}
\tablefoot{The $^{32}$S/$^{33}$S isotope ratios from the $J$ = 2-1 transition are corrected for the line saturation effects, while the ones in the 3-2 line are not corrected because of the missing main isotopic species.}
\end{table*}

\subsection{$^{34}$S/$^{36}$S ratios}
\label{section_34s36s}

The detection of C$^{36}$S in the $J$ = 2-1 and $J$ = 3-2 transitions allows us to also calculate $^{34}$S/$^{36}$S ratios. Around the CMZ, the C$^{36}$S $J$ = 2-1 line in the $+$50 km~s$^{-1}$ cloud was detected, resulting in a $^{34}$S/$^{36}$S ratio of 41~$\pm$~4. In the local arm toward DR21 and Orion-KL \citep{2007A&A...474..515M,2013ApJ...769...15X}, the C$^{36}$S $J$ = 2-1 and $J$ = 3-2 transitions were detected, respectively, leading to $^{34}$S/$^{36}$S values of 117~$\pm$~24 and 83~$\pm$~7. This yields an average $^{34}$S/$^{36}$S ratio of 100~$\pm$~16 in the ISM near the Sun. In the Perseus arm beyond the Solar System \citep{2006Sci...311...54X}, we detected the C$^{36}$S $J$ = 2-1 line toward W3OH and obtained a $^{34}$S/$^{36}$S value of 140~$\pm$~16. These results reveal the possibility of the existence of a $^{34}$S/$^{36}$S gradient from the Galactic center region to the outer Galaxy. Five tentative detections in the C$^{36}$S $J$ = 2-1 line and also five tentative detections in the $J$ = 3-2 line provide additional $^{34}$S/$^{36}$S ratios but with large uncertainties. All of the $^{34}$S/$^{36}$S values are listed in Table~\ref{table_all36s} and are plotted as a function of the distance to the Galactic center in Fig.~\ref{fig_all36S}.

\subsection{$^{32}$S/$^{36}$S ratios}
\label{section_32s36s}

\begin{figure*}[h]
\centering
   \includegraphics[height=550pt]{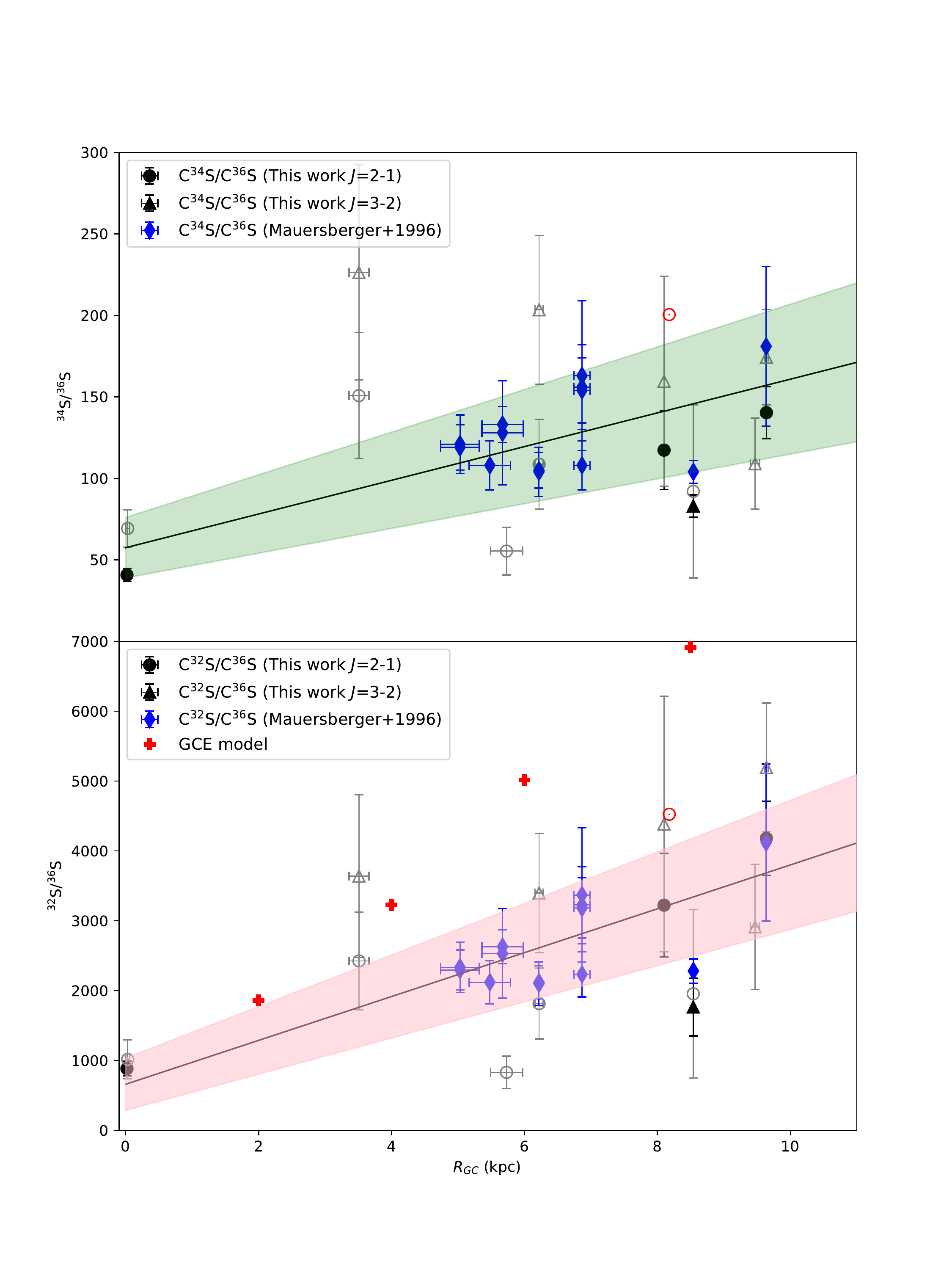}
     \caption{ $J$ = 2-1 (opacity corrected) and $J$ = 3-2 (no opacity corrections) $^{34}$S/$^{36}$S and $^{32}$S/$^{36}$S isotope ratios (see equations (18) and (19)) plotted as functions of the distance from the Galactic center. \textbf{Top:} Filled black circles and filled black triangle present the $^{34}$S/$^{36}$S ratios in the $J$ = 2-1 and $J$ = 3-2 transitions derived from C$^{34}$S/C$^{36}$S in this work with detections of C$^{36}$S, respectively. The open gray circles and open gray triangles present the $^{34}$S/$^{36}$S ratios in the $J$ = 2-1 and $J$ = 3-2 transitions derived from C$^{34}$S/C$^{36}$S in this work with tentative detections of C$^{36}$S, respectively. The blue diamonds show the $^{34}$S/$^{36}$S ratios in \citet{1996A&A...313L...1M}. The red symbol $\odot$ indicates the $^{34}$S/$^{36}$S isotope ratio in the Solar System. The resulting first-order polynomial fit to the $^{34}$S/$^{36}$S ratios in \citet{1996A&A...313L...1M} and this work, excluding the values from tentative detections, is plotted as a black solid line, with the green shaded area showing the 1~$\sigma$ interval of the fit. \textbf{Bottom:}  $^{32}$S/$^{36}$S ratios obtained from $^{34}$S/$^{36}$S ratios combined with the $^{32}$S/$^{34}$S ratios derived in this work. The filled black circles and filled black triangle present the values in the $J$ = 2-1 and $J$ = 3-2 transitions from this work with detections of C$^{36}$S, respectively. The open gray circles and open gray triangles present ratios in the $J$ = 2-1 and $J$ = 3-2 transitions from this work with tentative detections of C$^{36}$S, respectively. The $^{32}$S/$^{36}$S ratios derived with $^{34}$S/$^{36}$S values from \citet{1996A&A...313L...1M} are plotted as blue diamonds. The red symbol $\odot$ indicates the $^{32}$S/$^{36}$S isotope ratio in the Solar System. The $^{32}$S/$^{36}$S gradient, excluding tentative detections, is plotted as a black solid line, with the pink shaded area showing the 1~$\sigma$ interval of the fit. The red crosses visualize results from the GCE model of \citet[][see Section~\ref{section_discussion_model}]{2011MNRAS.414.3231K,2020ApJ...900..179K}. }
  \label{fig_all36S}
\end{figure*}

As in the case of the $^{32}$S/$^{33}$S ratios, the $^{32}$S/$^{36}$S values could also be obtained from the resulting $^{34}$S/$^{36}$S ratios in Section~\ref{section_34s36s} and the $^{32}$S/$^{34}$S ratios in Section~\ref{section_ratios3234_13c34s}  using the following equation:
\begin{equation}
\frac{^{32}{\rm S}}{^{36}{\rm S}} = \frac{^{34}\rm S}{^{36}\rm S} \times \frac{^{32}\rm S}{^{34}\rm S}.
\end{equation}
The uncertainties of both isotope ratios in the product on the right-hand side of the equation are included in our error budget. The resulting $^{32}$S/$^{36}$S ratios are listed in Table~\ref{table_all36s}. In the CMZ, a $^{32}$S/$^{36}$S ratio of 884~$\pm$~104 is obtained toward the $+$50 km~s$^{-1}$ cloud. $^{32}$S/$^{36}$S values of 1765~$\pm$~414 and 3223~$\pm$~742 are derived toward Orion-KL and DR21, leading to an average $^{32}$S/$^{36}$S value of 2494 $\pm$ 578 in the local regions near the Solar System. In the Perseus arm beyond the solar neighborhood toward W3OH, we obtain the highest $^{32}$S/$^{36}$S value, 4181~$\pm$~531. All these results indicate that there could be a positive $^{32}$S/$^{36}$S gradient from the Galactic center region to the outer Galaxy. Figure~\ref{fig_all36S} shows the $^{32}$S/$^{36}$S ratios plotted as a function of the distance to the Galactic center.

\begin{figure*}[h]
\centering
   \includegraphics[height=620pt]{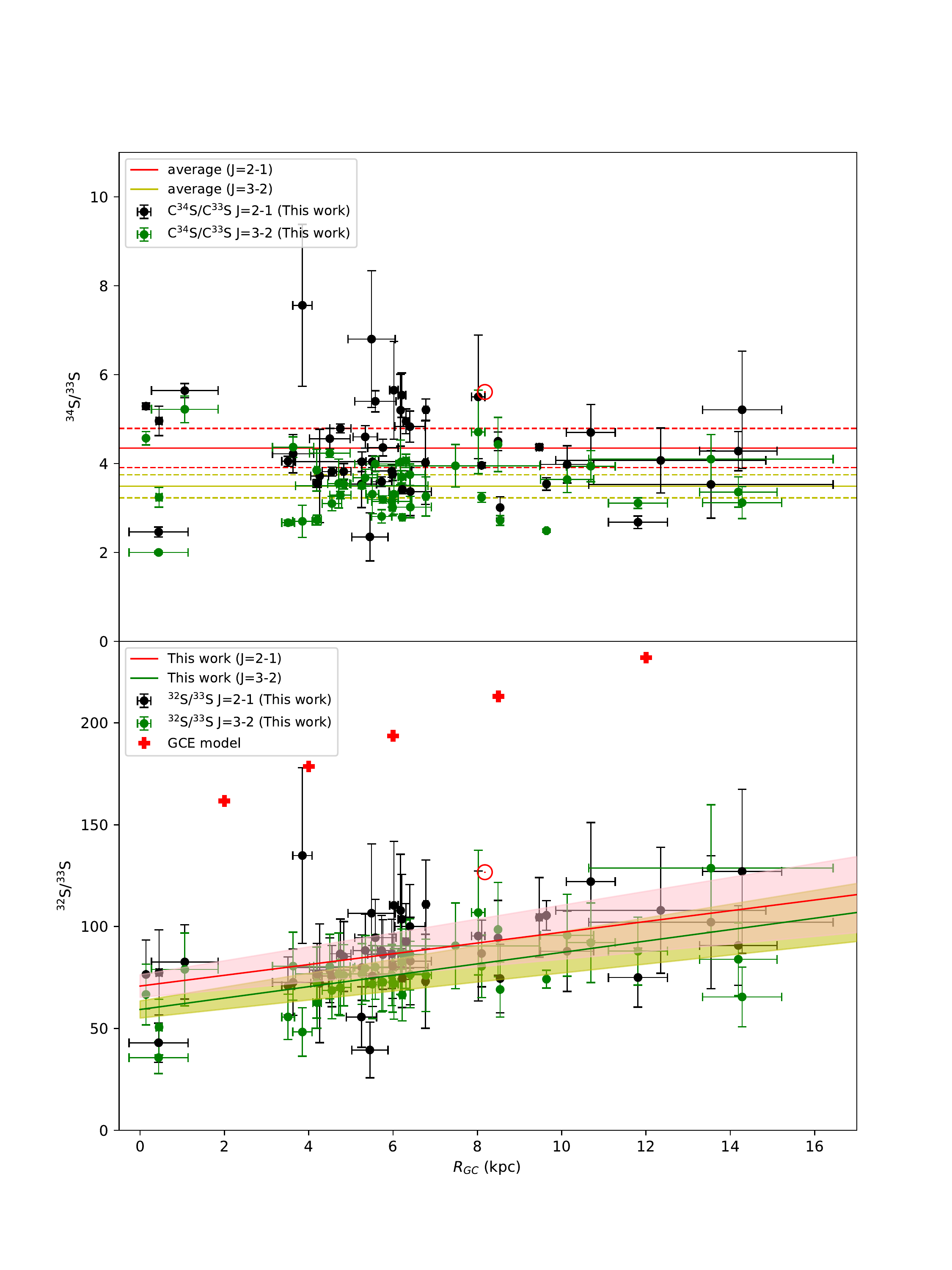}
     \caption{ $^{34}$S/$^{33}$S and $^{32}$S/$^{33}$S isotope ratios (for the latter, see equations (12) and (13)) plotted as functions of the distance from the Galactic center. \textbf{Top:}  $^{34}$S/$^{33}$S ratios derived from C$^{34}$S/C$^{33}$S in the $J$ = 2-1 and $J$ = 3-2 transitions plotted as black and green dots, respectively. The red solid and the two dashed lines show the average value and its standard deviation, 4.35~$\pm$~0.44, of $^{34}$S/$^{33}$S with corrections of optical depth toward our sample in the $J$ = 2-1 transition. The yellow solid and the two dashed lines show the average value and its standard deviation, 3.49~$\pm$~0.26, of $^{34}$S/$^{33}$S toward our sample in the $J$ = 3-2 transition. The red symbol $\odot$ indicates the $^{34}$S/$^{33}$S isotope ratio in the Solar System. \textbf{Bottom:} Black and green dots show the $^{32}$S/$^{33}$S ratios in the $J$ = 2-1 and $J$ = 3-2 transitions, respectively. The red symbol $\odot$ indicates the $^{32}$S/$^{33}$S value in the Solar System. The resulting first-order polynomial fits to the $^{32}$S/$^{33}$S ratios in the $J$ = 2-1 and $J$ = 3-2 transitions in this work are plotted as red and green solid lines, respectively, with the pink and yellow shaded area showing the 1~$\sigma$ standard deviations. The red crosses are the results from the GCE model of \citet[][see also Section~\ref{section_discussion_model}]{2011MNRAS.414.3231K,2020ApJ...900..179K}. }
  \label{fig_all33S}
\end{figure*}

\begin{table*}[h]
\centering
\caption{\label{table_all36s}Isotope ratios of $^{34}$S/$^{36}$S and $^{32}$S/$^{36}$S}
\begin{tabular}{lc|cc|cc}
\hline\hline
Source  &   $R_{\rm GC}$   &  \multicolumn{2}{c}{$^{34}$S/$^{36}$S}   &  \multicolumn{2}{c}{$^{32}$S/$^{36}$S}    \\
        &    (kpc)           &    $J$ = 2-1    &   $J$ = 3-2          &    $J$ = 2-1    &   $J$ = 3-2            \\
\hline
W3OH            & 9.64 $\pm$ 0.03   &  140 $\pm$ 16  &  174 $\pm$ 29\tablefootmark{*}                   &  4181 $\pm$ 531  &  5195 $\pm$ 919\tablefootmark{*} \\
Orion-KL        & 8.54 $\pm$ 0.00   &  92  $\pm$ 53\tablefootmark{*}  &  83  $\pm$ 7                    &  1954 $\pm$ 1207\tablefootmark{*}  &  1765 $\pm$ 414  \\
$+$20 km~s$^{-1}$ cloud         & 0.03 $\pm$ 0.03   &  69  $\pm$ 11\tablefootmark{*}   &        $\cdots$                       &  1015 $\pm$ 278\tablefootmark{*}   &        $\cdots$     \\
$+$50 km~s$^{-1}$ cloud     & 0.02 $\pm$ 0.04   &  41  $\pm$ 4   &        $\cdots$                                        &  884 $\pm$ 104   &        $\cdots$     \\
G024.78$+$00.08 & 3.51 $\pm$ 0.15   &  151 $\pm$ 39\tablefootmark{*}  &  226 $\pm$ 66\tablefootmark{*}  &  2424 $\pm$ 699\tablefootmark{*}  &  3640 $\pm$ 1164\tablefootmark{*} \\
G030.81$-$00.05 & 5.73 $\pm$ 0.24   &  55  $\pm$ 15\tablefootmark{*}  &       $\cdots$                        &  829 $\pm$ 233\tablefootmark{*}  &       $\cdots$      \\
W51-IRS2        & 6.22 $\pm$ 0.06   &  109 $\pm$ 27\tablefootmark{*}  &  203 $\pm$ 45\tablefootmark{*}  &  1814 $\pm$ 506\tablefootmark{*}  &  3397 $\pm$ 855\tablefootmark{*} \\
DR21            & 8.10  $\pm$ 0.00   &  117 $\pm$ 24  &  159 $\pm$ 64\tablefootmark{*}                  &  3223 $\pm$ 742  &  4384 $\pm$ 1829\tablefootmark{*} \\
NGC7538         & 9.47 $\pm$ 0.07   &       $\cdots$       &  109 $\pm$ 28\tablefootmark{*}                   &       $\cdots$       &  2912 $\pm$ 897\tablefootmark{*} \\
\hline
\multicolumn{6}{c}{Below are the isotope ratios of $^{34}$S/$^{36}$S and $^{32}$S/$^{36}$S from \citet{1996A&A...313L...1M}} \\
\hline
W3OH                         & 9.64 $\pm$ 0.03   &   $\cdots$       &  181 $\pm$ 49  &  $\cdots$       &  4118 $\pm$ 1124  \\
Orion-KL                     & 8.54 $\pm$ 0.00   &   $\cdots$       &  104 $\pm$ 7   &  $\cdots$       &  2281 $\pm$ 174 \\
IRAS15491\tablefootmark{**}  & 5.48 $\pm$ 0.31   &   $\cdots$       &  108 $\pm$ 15  &  $\cdots$       &  2120 $\pm$ 307 \\
IRAS15520\tablefootmark{**}  & 5.67 $\pm$ 0.31   &   128 $\pm$ 32   &  133 $\pm$ 11  & 2531 $\pm$ 641  &  2629 $\pm$ 243 \\
IRAS16172\tablefootmark{**}  & 5.03 $\pm$ 0.29   &   121 $\pm$ 18   &  119 $\pm$ 14  & 2334 $\pm$ 361  &  2296 $\pm$ 287 \\
NGC6334A                     & 6.87 $\pm$ 0.12   &   108 $\pm$ 15   &  154 $\pm$ 20  & 2232 $\pm$ 322  &  3183 $\pm$ 431 \\
NGC6334B                     & 6.87 $\pm$ 0.12   &   163 $\pm$ 46   &  156 $\pm$ 26  & 3369 $\pm$ 960  &  3225 $\pm$ 552 \\
W51(M)                       & 6.22 $\pm$ 0.004  &   105 $\pm$ 11   &  104 $\pm$ 15  & 2119 $\pm$ 237  &  2099 $\pm$ 313 \\
\hline
\end{tabular}
\tablefoot{\tablefoottext{*}{Values with large uncertainties are derived from the tentative detection of C$^{36}$S lines.} \tablefoottext{**}{For these three sources without parallax data, their kinematic distances were estimated (for details, see Section~\ref{section_distance}).} The $^{34}$S/$^{36}$S and $^{32}$S/$^{36}$S isotope ratios in this work from the $J$ = 2-1 transition are corrected for the optical depth effects, while the ones for the 3-2 line could not be corrected.}
\end{table*}

\section{Discussion}
\label{discussion}

With the measurements of optically thin lines of the rare CS isotopologs, C$^{34}$S, $^{13}$CS, C$^{33}$S, $^{13}$C$^{34}$S, and C$^{36}$S, we derived $^{12}$C/$^{13}$C, $^{32}$S/$^{34}$S, $^{34}$S/$^{33}$S, $^{32}$S/$^{33}$S, $^{34}$S/$^{36}$S, and $^{32}$S/$^{36}$S isotope ratios. Combined with accurate galactocentric distances, we established a $^{32}$S/$^{33}$S gradient for the first time and confirmed the existing gradients of $^{12}$C/$^{13}$C and $^{32}$S/$^{34}$S, as well as uniform $^{34}$S/$^{33}$S ratios across the Milky Way, which are lower than previously reported. Furthermore, we may have detected  $^{34}$S/$^{36}$S and $^{32}$S/$^{36}$S gradients  for the first time. In Section~\ref{section_comparisons_12c13c}, we compare the $^{12}$C/$^{13}$C gradient obtained in this work with previous published ones derived from a variety of molecular species. A comparison between the $^{32}$S/$^{34}$S gradients we obtained and previously published ones is presented in Section~\ref{section_comparisons_32s34s}. The condition of LTE for C$^{33}$S with its HFS line ratios is discussed in Section~\ref{section_hfs_c33s}. We then also compare our results on $^{34}$S/$^{33}$S ratios with previously published values and discuss the $^{32}$S/$^{33}$S gradient. In Section~\ref{section_discussion_all36} we evaluate whether or not $^{34}$S/$^{36}$S, $^{33}$S/$^{36}$S, and $^{32}$S/$^{36}$S ratios may show gradients with galactocentric distance. Observational bias due to distance effects, beam size effects, and chemical fractionation are discussed. A comparison of several isotopes with respect to primary or secondary synthesis is provided in Section~\ref{section_discussion_all}. Results from a Galactic chemical evolution (GCE) model, trying to simulate the observational data, are presented in Section~\ref{section_discussion_model}.

\subsection{Comparisons of $^{12}$C/$^{13}$C ratios determined with different species}
\label{section_comparisons_12c13c}

$^{12}$C/$^{13}$C ratios have been well studied in the CMZ where the value is about 20--25 \citep[e.g.,][]{1983A&A...127..388H,1985A&A...149..195G,1990ApJ...357..477L,2005ApJ...634.1126M,2010A&A...523A..51R,2013A&A...559A..47B,2017ApJ...845..158H,2020A&A...642A.222H}, similar to the results that we obtained from C$^{34}$S in the current work. In the inner Galaxy, the $^{12}$C/$^{13}$C ratios are higher than the values in the CMZ, which were $\sim$50 as derived from H$_2^{12}$CO/H$_2^{13}$CO in \citet{1985A&A...143..148H} and 41~$\pm$~9 from C$^{34}$S/$^{13}$C$^{34}$S in this work. As can be inferred from Fig.~\ref{fig_gradient_12C13C}, $^{12}$C/$^{13}$C ratios at 3.0 kpc $\le R_{\rm GC} \le$ 4.0 kpc might be as low as in the CMZ, but this is so far tentative and uncertain, as we only have one detection (G024.78$+$00.08) in this region and data from other groups referring to this small galactocentric interval are also relatively few. In the local regions near the Solar System, an average $^{12}$C/$^{13}$C ratio of 66~$\pm$~10 was obtained from C$^{34}$S/$^{13}$C$^{34}$S in this work, which is consistent with $^{12}$C/$^{13}$C values derived from other molecular species and their $^{13}$C isotopes, that are 75~$\pm$~9 from C$^{18}$O \citep{1998ApJ...494L.107K}, 60~$\pm$~19 from CN \citep{2005ApJ...634.1126M}, 74~$\pm$~8 from CH$^+$ \citep{2011ApJ...728...36R}, and 53~$\pm$~16 from H$_2$CO \citep{2019ApJ...877..154Y}. All these $^{12}$C/$^{13}$C values for the solar neighborhood are well below the value for the Sun \citep[89,][]{1989GeCoA..53..197A,2007ApJ...656L..33M}. This indicates that $^{13}$C has been enriched in the local ISM during the last 4.6 billion years following the  formation of the Solar System. Beyond the Sun, at a galactocentric distance of about 10~kpc, our results from C$^{34}$S/$^{13}$C$^{34}$S show a slightly higher value of 67~$\pm$~8, which is similar to $^{12}$C/$^{13}$C ratios from CN (66~$\pm$~20), C$^{18}$O (69~$\pm$~10), and H$_2$CO (64~$\pm$~10). These values are still below the value for the Sun. In the far outer Galaxy at 13.8~kpc toward WB89~437, \citet{1996A&AS..119..439W} found a 3$\sigma$ lower limit of 201~$\pm$~15 from C$^{18}$O/$^{13}$C$^{18}$O. This suggests that the $^{12}$C/$^{13}$C gradient  extends well beyond the solar neighborhood to the outer Galaxy. Additional sources with large galactocentric distances have to be measured to further improve the statistical significance of this result.

\begin{figure*}[h]
\centering
   \includegraphics[width=520pt]{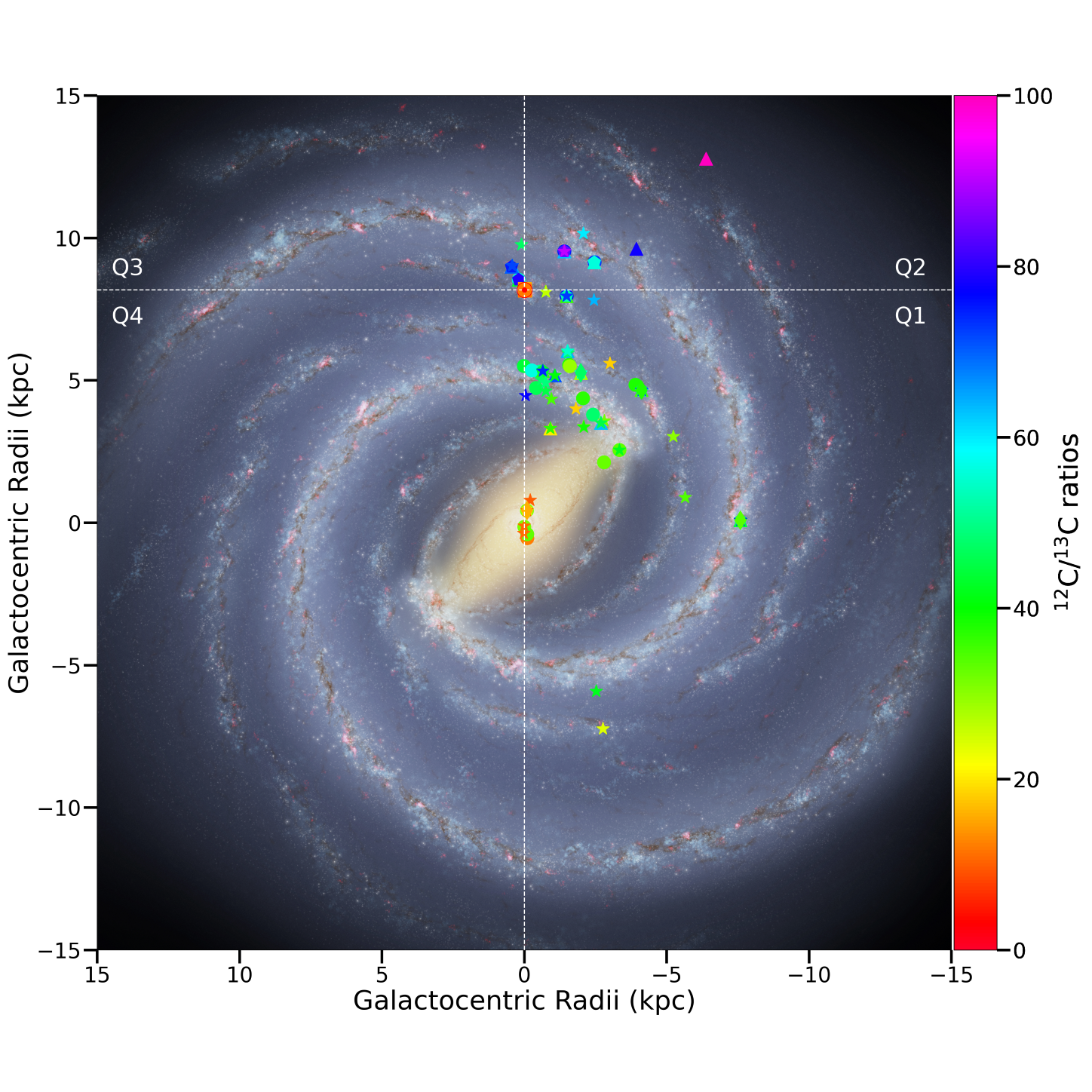}
     \caption{Distribution of $^{12}$C/$^{13}$C ratios from 93 sources in the Milky Way. The background image is the structure of the Milky Way from the artist's impression [Credit:
NASA/JPL-Caltech/ESO/R. Hurt]. The $^{12}$C/$^{13}$C isotope ratios with corrections for optical depth from C$^{34}$S/$^{13}$C$^{34}$S in this work are plotted as circles. The triangles, pentagons, stars, squares, and diamonds indicate the $^{12}$C/$^{13}$C ratios derived from CN/$^{13}$CN \citep{2002ApJ...578..211S,2005ApJ...634.1126M}, C$^{18}$O/$^{13}$C$^{18}$O \citep{1990ApJ...357..477L,1996A&AS..119..439W,1998ApJ...494L.107K}, H$_2$CO/H$_2^{13}$CO \citep{1980A&A....82...41H,1982A&A...109..344H,1983A&A...127..388H,1985A&A...143..148H,2019ApJ...877..154Y}, CH$^+$/$^{13}$CH$^+$ \citep{2011ApJ...728...36R}, and CH/$^{13}$CH \citep{2020A&A...640A.125J}, respectively. The red symbol $\odot$ indicates the position of the Sun. The color bar on the right-hand side indicates the range of the $^{12}$C/$^{13}$C ratios. }
  \label{distribution_12C13C}
\end{figure*}

Previously published $^{12}$C/$^{13}$C ratios derived from CN \citep{2002ApJ...578..211S,2005ApJ...634.1126M}, C$^{18}$O \citep{1990ApJ...357..477L,1996A&AS..119..439W,1998ApJ...494L.107K}, H$_2$CO \citep{1980A&A....82...41H,1982A&A...109..344H,1983A&A...127..388H,1985A&A...143..148H,2019ApJ...877..154Y}, CH$^+$ \citep{2011ApJ...728...36R}, and CH \citep{2020A&A...640A.125J} are shown in Fig.~\ref{fig_gradient_12C13C}, but with respect to the new distance values (see details in Section~\ref{section_distance}). In Fig.~\ref{distribution_12C13C}, the $^{12}$C/$^{13}$C values from different molecular species are projected onto the Galactic plane. This also  visualizes the gradient from the Galactic center to the Galactic outer regions beyond the Solar System. In Table~\ref{table_all12C13Cgradients}, the fitting results for the old and new distances are presented. The comparison shows that the adoption of the new distances has indeed an effect on the fitting results, such as for the $^{12}$CN/$^{13}$CN gradient. In \citet{2002ApJ...578..211S} and \citet{2005ApJ...634.1126M}, the slope and intercept become (6.75~$\pm$~1.44) and (5.77~$\pm$~11.29) from (6.01~$\pm$~1.19) and (12.28~$\pm$~9.33), respectively. The fitting for H$_2^{12}$CO/H$_2^{13}$CO from  \citet{1980A&A....82...41H,1982A&A...109..344H,1983A&A...127..388H,1985A&A...143..148H} and \citet{2019ApJ...877..154Y} becomes (5.43~$\pm$~1.04) and (13.87~$\pm$~6.38) from (5.08~$\pm$~1.10) and (11.86~$\pm$~6.60), respectively. The Galactic $^{12}$C/$^{13}$C gradients derived based on measurements of CN, C$^{18}$O, and  H$_2$CO are in agreement with our results from C$^{34}$S and therefore indicate that chemical fractionation has little effect on the $^{12}$C/$^{13}$C ratios. It is noteworthy that all these fits show a significant discrepancy with observations from the Galactic center. Indeed, they suggest values of 5--17 at $R_{\rm GC}$=0, substantially below observed values of 20--25 (see also Tables~\ref{table_all12C13Cgradients}, \ref{table_allratios}). While the values in the CMZ are clearly lower than in the inner Galactic disk (with the potential exception at galactocentric distances of 2.0--4.0 kpc), they are larger than suggested by a linear fit encompassing the entire inner 12.0 kpc of the Galaxy.

\subsection{The $^{32}$S/$^{34}$S gradient across the Milky Way}
\label{section_comparisons_32s34s}

The existence of a $^{32}$S/$^{34}$S gradient was first proposed by \citet{1996A&A...305..960C} based on observations of $^{13}$CS and C$^{34}$S $J$ = 2-1 lines toward 20 mostly southern HMSFRs with galactocentric distances of between 3.0 and 9.0 kpc. Very recently, \citet{2020ApJ...899..145Y} confirmed the existence of this $^{32}$S/$^{34}$S gradient and enlarged the sample of measurements of $^{13}$CS and C$^{34}$S $J$ = 2-1 lines to a total of 61 HMSFRs from the inner Galaxy out to a galactocentric distance of 12.0 kpc. In the CMZ, \citet{2020A&A...642A.222H} found $^{32}$S/$^{34}$S ratios of 16.3$^{+2.1}_{-1.7}$ and 17.9~$\pm$~5.0 for the $+$50 km s$^{-1}$ cloud and Sgr B2(N), which is consistent with our values derived from $^{13}$C$^{34}$S and also with our results using the double isotope method. In the inner disk at 2.0~kpc $\le R_{\rm GC} \le$ 6.0~kpc, a similar $^{32}$S/$^{34}$S value of 18~$\pm$~4 was derived based on our results in Sections~\ref{section_ratios3234_13c34s} and \ref{section_double_32s34s}. While $^{12}$C/$^{13}$C ratios in the inner disk at $R_{\rm GC} \ge$ 4.0~kpc are clearly higher than in the CMZ (see details in Section~\ref{section_comparisons_12c13c}), this is not the case for $^{32}$S/$^{34}$S. On the contrary, $^{32}$S/$^{34}$S ratios in the CMZ and inner disk are similar, as already suggested for the first time by \citet{2020A&A...642A.222H}. In the local ISM, our results lead to an average $^{32}$S/$^{34}$S ratio of 24~$\pm$~4, which is close to the value in the Solar System \citep[22.57,][]{1989GeCoA..53..197A}. That is also differing from the $^{12}$C/$^{13}$C ratio and its clearly subsolar value in the local ISM. A more detailed discussion is given in Section~\ref{section_discussion_all}.

For the first time, we established a $^{32}$S/$^{34}$S gradient directly from measurements of $^{13}$CS and $^{13}$C$^{34}$S (see Section~\ref{section_ratios3234_13c34s} for details). Similar $^{32}$S/$^{34}$S gradients were also found in the $^{32}$S/$^{34}$S values derived by the double isotope method with the $J$ = 2-1 and $J$ = 3-2 transitions (for details, see Section~\ref{section_double_32s34s}). A gradient of $^{32}$S/$^{34}$S = (0.75 $\pm$ 0.13)$R_{\rm GC}$+(15.52 $\pm$ 0.78) was obtained based on a large dataset of 90 values from our detections of $^{13}$CS and C$^{34}$S $J$ = 2-1 lines with corrections of opacity, which is flatter than previous ones presented by \citet{1996A&A...305..960C} and \citet{2020ApJ...899..145Y}. Following \citet{2020ApJ...899..145Y}, the gradient does not significantly change when the ratios in the CMZ or in the outer regions are excluded, indicating that the $^{32}$S/$^{34}$S gradient is robust.

\begin{table*}[h]
\caption{Measurements of the $^{12}$C/$^{13}$C gradient.}
\centering
\begin{tabular}{c|cc|cc}
\hline\hline
  &\multicolumn{2}{c}{Previous fitting results} & \multicolumn{2}{c}{This work} \\
  &  slope  & intercept  &  slope &  intercept \\
\hline
\label{table_all12C13Cgradients}
CN\tablefootmark{a}        & 6.01 $\pm$ 1.19 & 12.28 $\pm$ 9.33 & 6.75 $\pm$ 1.44 & 5.57 $\pm$ 11.29 \\  
C$^{18}$O\tablefootmark{b} & 5.41 $\pm$ 1.07 & 19.03 $\pm$ 7.90 & 5.72 $\pm$ 1.20 & 14.56 $\pm$ 9.25 \\  
H$_2$CO\tablefootmark{c}   & 5.08 $\pm$ 1.10 & 11.86 $\pm$ 6.60 & 5.43 $\pm$ 1.04 & 13.87 $\pm$ 6.38 \\  
C$^{34}$S (this work)      &     $\cdots$    &     $\cdots$     & 4.77 $\pm$ 0.81 & 20.76 $\pm$ 4.61 \\  
\hline
\end{tabular}
\tablefoot{Fitting results for the old (left) and new (right) distances, respectively. 
\tablefoottext{a}{From \citet{2002ApJ...578..211S} and \citet{2005ApJ...634.1126M}.}
\tablefoottext{b}{From \citet{1990ApJ...357..477L}, \citet{1996A&AS..119..439W}, and \citet{1998ApJ...494L.107K}.}
\tablefoottext{c}{From \citet{1980A&A....82...41H,1982A&A...109..344H,1983A&A...127..388H,1985A&A...143..148H} and \citet{2019ApJ...877..154Y}.}}
\end{table*}

\begin{figure*}[h]
\centering
   \includegraphics[width=520pt]{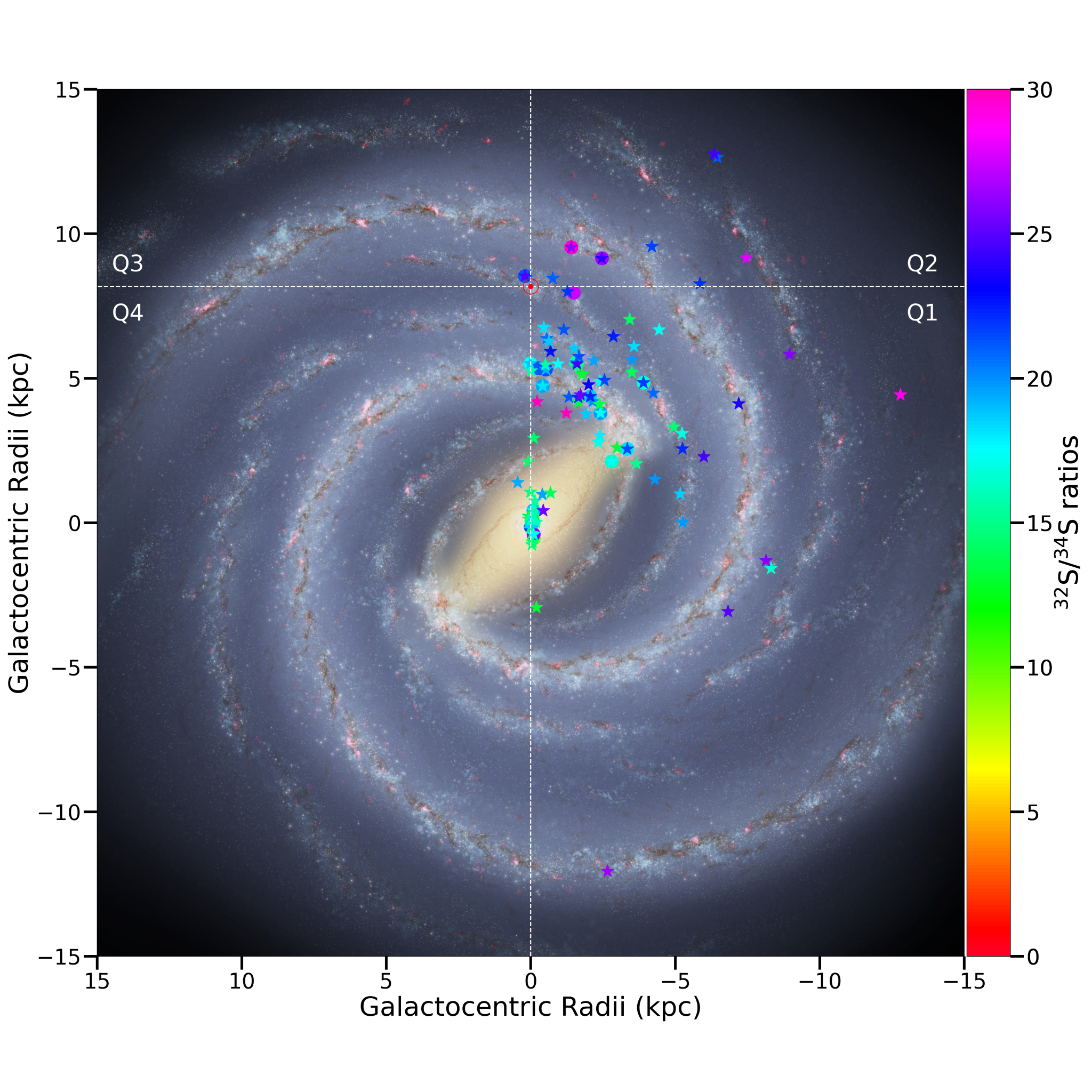}
     \caption{Distribution of $^{32}$S/$^{34}$S ratios from 112 sources in the Milky Way. The background image is the structure of the Milky Way from the artist's impression [Credit:
NASA/JPL-Caltech/ESO/R. Hurt]. The $^{32}$S/$^{34}$S isotope ratios with corrections of opacity derived in this work from $^{13}$CS/$^{13}$C$^{34}$S and the double isotope method in the $J$ = 2-1 transition are plotted as circles and stars, respectively. The triangles indicate the results from the CMZ obtained by \citet{2020A&A...642A.222H}. The red symbol $\odot$ indicates the position of the Sun. The color bar on the right-hand side indicates the range of the $^{32}$S/$^{34}$S ratios.}
  \label{distribution_32S34S}
\end{figure*}

\subsection{C$^{33}$S}
\label{section_hfs_c33s}

\begin{figure}[h]
\centering
   \includegraphics[width=220pt]{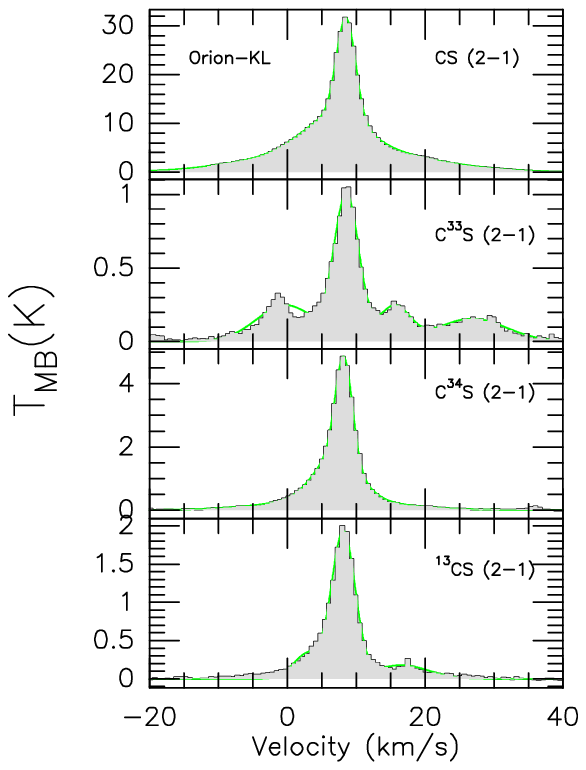}
     \caption{Line profiles of the $J$ = 2-1 transitions of CS, C$^{33}$S, C$^{34}$S, and $^{13}$CS toward Orion-KL.}
  \label{fig_orion}
\end{figure}

 We detected at least three components of the C$^{33}$S $J$ = 2-1 line toward 26 sources. This will guide us to obtain information with respect to LTE or nonLTE conditions. As mentioned in Section~\ref{section_34s33s}, the main component ($I_{main}$) consists of four HFS lines ($F$=7/2-5/2, $F$=5/2-3/2, $F$=1/2-1/2, $F$=3/2-5/2). Under LTE conditions in the optically thin case, the ratio between the other identified components (not belonging to the main one) and the main component can be obtained:
\begin{equation}
\begin{aligned}
R_{211} &=  \frac{I(F=3/2-1/2 + F=5/2-5/2)}{I_{main}}\\ &= 0.25,
\end{aligned}
\end{equation}
\begin{equation}
\begin{aligned}
R_{212} &=  \frac{I(F=3/2-3/2)}{I_{main}}\\ &= 0.15,
\end{aligned}
\end{equation}
\begin{equation}
\begin{aligned}
R_{213} &=  \frac{I(F=1/2-3/2)}{I_{main}}\\ &= 0.02.
\end{aligned}
\end{equation}
A detailed examination of our 26 objects is presented in Table~\ref{table_c33s_hfs}. Except for Orion-KL there is no evidence for nonLTE effects. All the remaining sources are found to be compatible with LTE conditions. The spectra of CS, C$^{34}$S, and $^{13}$CS toward Orion-KL show broad wings (see Fig.~\ref{fig_orion}), which might lead to a highly complex C$^{33}$S $J$ = 2-1 line shape, which may be what is causing apparent LTE deviations in Orion-KL.

No systematic dependence of the $^{34}$S/$^{33}$S ratios on galactocentric distance was found in previous studies. The average $^{34}$S/$^{33}$S values were 6.3~$\pm$~1.0 and 5.9~$\pm$~1.5 in \citet{1996A&A...305..960C} and \citet{2020ApJ...899..145Y}, respectively. A particularly low $^{34}$S/$^{33}$S ratio of 4.3~$\pm$~0.2 was found in the Galactic center region by \citet{2020A&A...642A.222H}. These authors speculated about the possible presence of a gradient with low values at the center. However, in view of the correction for HFS in Section~\ref{section_34s33s},  it appears that this is simply an effect of different HFS correction factors, with the one at the Galactic center with its wide spectral lines being 1.0, thus (exceptionally) not requiring a downward correction. \citet{2020ApJ...899..145Y} considered the effect of HFS, while they overestimated the ratio of the main component to the total intensity of the C$^{33}$S $J$ = 2-1 line (for details, see their Section~4.2). This resulted in higher $^{34}$S/$^{33}$S values. However, the $^{34}$S/$^{33}$S ratios appear  to be independent of galactocentric distance based on our results (see details in Section~\ref{section_34s33s}). The average $^{34}$S/$^{33}$S value with corrections of optical depth toward our sample in the $J$ = 2-1 transition is 4.35~$\pm$~0.44, which is similar to the value in the Galactic center region derived by \citet{2020A&A...642A.222H} and lower than the value of 5.61 in the Solar System \citep{1989GeCoA..53..197A}. This indicates that no systematic variation exists in the $^{34}$S/$^{33}$S ratios in our Galaxy, that $^{33}$S is (with respect to stellar nucleosynthesis) similar to $^{34}$S, and that the Solar System (see Fig.~\ref{fig_all33S}) must be peculiar. An approximately constant $^{34}$S/$^{33}$S ratio with opacity correction from the $J$ = 2-1 transition across the Galactic plane leads to a $^{32}$S/$^{33}$S gradient in our Galaxy as we already mentioned in Section~\ref{section_32s33s}: $^{32}{\rm S}/^{33}{\rm S}$ = $(2.64 \pm 0.77)R_{\rm GC}+(70.80 \pm 5.57)$, with a correlation coefficient of 0.46.

\begin{table*}[h]
\caption{Line intensity ratios of C$^{33}$S $J$ = 2-1 hyperfine components.}
\centering
\begin{tabular}{lcccc}
\hline\hline
Source  &     FWHM  & $R_{211}$ & $R_{212}$ & $R_{213}$     \\
        &    (km s$^{-1}$)           &           &           &                \\
\hline
\label{table_c33s_hfs}
W3OH            &  4.41  & 0.29 $\pm$ 0.04  &  0.15 $\pm$ 0.03  &  0.05 $\pm$ 0.03   \\ 
Orion-KL        &  4.25  & 0.60 $\pm$ 0.02  &  0.30 $\pm$ 0.01  &  0.44 $\pm$ 0.02   \\ 
G359.61$-$00.24 &  3.17  & 0.40 $\pm$ 0.05  &  0.62 $\pm$ 0.05  &  0.14 $\pm$ 0.04   \\ 
G006.79$-$00.25 &  2.73  & 0.28 $\pm$ 0.02  &  0.17 $\pm$ 0.02  &     $\cdots$       \\ 
G010.32$-$00.15 &  2.66  & 0.27 $\pm$ 0.04  &  0.23 $\pm$ 0.06  &     $\cdots$       \\ 
G016.86$-$02.15 &  3.32  & 0.30 $\pm$ 0.05  &  0.18 $\pm$ 0.03  &     $\cdots$       \\ 
G017.02$-$02.40 &  3.74  & 0.17 $\pm$ 0.03  &  0.17 $\pm$ 0.08  &     $\cdots$       \\ 
G018.34$+$01.76 &  2.30  & 0.24 $\pm$ 0.07  &  0.13 $\pm$ 0.06  &     $\cdots$       \\ 
G019.36$-$00.03 &  3.02  & 0.29 $\pm$ 0.06  &  0.14 $\pm$ 0.04  &     $\cdots$       \\ 
G023.43$-$00.18 &  3.81  & 0.40 $\pm$ 0.14  &  1.15 $\pm$ 0.34  &     $\cdots$       \\ 
G024.78$+$00.08 &  4.29  & 0.34 $\pm$ 0.03  &  0.21 $\pm$ 0.03  &  0.08 $\pm$ 0.03   \\ 
G028.39$+$00.08 &  3.04  & 0.26 $\pm$ 0.04  &  0.24 $\pm$ 0.06  &     $\cdots$       \\ 
G028.83$-$00.25 &  2.27  & 0.25 $\pm$ 0.04  &  0.18 $\pm$ 0.03  &     $\cdots$       \\ 
G030.70$-$00.06 &  4.85  & 0.33 $\pm$ 0.02  &  0.23 $\pm$ 0.03  &  0.11 $\pm$ 0.02   \\ 
G030.74$-$00.04 &  3.55  & 0.19 $\pm$ 0.03  &  0.25 $\pm$ 0.06  &     $\cdots$       \\ 
G030.81$-$00.05 &  6.15  & 0.27 $\pm$ 0.03  &  0.16 $\pm$ 0.02  &  0.14 $\pm$ 0.02   \\ 
G032.74$-$00.07 &  4.54  & 0.36 $\pm$ 0.04  &  0.16 $\pm$ 0.03  &     $\cdots$       \\ 
G032.79$+$00.19 &  4.78  & 0.09 $\pm$ 0.05  &  1.72 $\pm$ 0.28  &     $\cdots$       \\ 
G034.41$+$00.23 &  4.56  & 0.32 $\pm$ 0.04  &  0.28 $\pm$ 0.04  &  0.12 $\pm$ 0.03   \\ 
G034.79$-$01.38 &  2.40  & 0.44 $\pm$ 0.15  &  0.14 $\pm$ 0.06  &     $\cdots$       \\ 
G037.42$+$01.51 &  3.18  & 0.20 $\pm$ 0.05  &  0.10 $\pm$ 0.02  &     $\cdots$       \\ 
W51-IRS2        &  8.55  & 0.13 $\pm$ 0.09  &  3.63 $\pm$ 0.36  &  0.71 $\pm$ 0.13   \\ 
DR21            &  2.71  & 0.28 $\pm$ 0.01  &  0.20 $\pm$ 0.01  &  0.14 $\pm$ 0.01   \\ 
G097.53$+$03.18 &  5.22  & 0.30 $\pm$ 0.05  &  0.13 $\pm$ 0.04  &  0.14 $\pm$ 0.04   \\ 
G109.87$+$02.11 &  3.19  & 0.23 $\pm$ 0.06  &  0.38 $\pm$ 0.07  &     $\cdots$       \\ 
NGC7538         &  3.71  & 0.22 $\pm$ 0.01  &  0.21 $\pm$ 0.01  &     $\cdots$       \\ 
\hline
\end{tabular}
\tablefoot{ Full width at half maximum values of the main component were obtained from measurements of C$^{33}$S; see Table \ref{fitting_all}. The errors provided here are 1$\sigma$.
}
\end{table*}

\subsection{C$^{36}$S}
\label{section_discussion_all36}

As mentioned in Section~\ref{section_34s36s}, we find novel potential indications for a positive $^{34}$S/$^{36}$S gradient with galactocentric radius. The $^{36}$S-bearing molecule C$^{36}$S was first detected by \citet{1996A&A...313L...1M}. These authors observed the $J$ = 2-1 and 3-2 transitions toward eight Galactic molecular hot cores at galactocentric distances of between 5.0 kpc and 10.0 kpc.  \citet{1996A&A...313L...1M} reported an average $^{34}$S/$^{36}$S ratio of 115~$\pm$~17, which is smaller than the value in the Solar System \citep[200.5,][]{1989GeCoA..53..197A}. This is consistent with this nucleus being of  a purely secondary nature. Combining the ratios of \citet{1996A&A...313L...1M} --- after applying new distances (see details in Table~\ref{table_all36s})--- with our results in the $J$ = 2-1 transition, the following fit could be achieved:
\begin{equation}
^{34}{\rm S}/^{36}{\rm S} = (10.34 \pm 2.74)R_{\rm GC}+(57.45 \pm 18.59), 
\end{equation}
with a correlation coefficient of 0.71. As the $^{34}$S/$^{33}$S ratios show a uniform distribution across our Galaxy (see details in Section~\ref{section_hfs_c33s}), a $^{33}$S/$^{36}$S gradient is also expected. We obtain (2.38~$\pm$~0.67)$R_{\rm GC}$+(13.21~$\pm$~4.48). After applying our $^{32}$S/$^{34}$S gradient to the $^{34}$S/$^{36}$S ratios in \citet{1996A&A...313L...1M} with equation (9), $^{32}$S/$^{36}$S ratios were then derived and listed in Table~\ref{table_all36s}. Combined with our results in the $J$ = 2-1 transition, a linear fit to the $^{32}$S/$^{36}$S ratios is obtained:
\begin{equation}
 ^{32}{\rm S}/^{36}{\rm S} = (314 \pm 55)R_{\rm GC}+(659 \pm 374), 
 \end{equation}
with a correlation coefficient of 0.84. The $^{34}$S/$^{36}$S and $^{32}$S/$^{36}$S ratios are plotted as functions of galactocentric distances in Fig.~\ref{fig_all36S}. Measurements of $^{34}$S/$^{36}$S, $^{33}$S/$^{36}$S, and $^{32}$S/$^{36}$S are still not numerous. More sources with detected C$^{36}$S lines would be highly
desirable, especially in the CMZ and the inner disk within $R_{\rm GC}$~=~5.0~kpc.

\subsection{Observational bias due to distance effects}
\label{section_discussion_distance}

While we have so far analyzed isotope ratios as a function of galactocentric distances, there might be a bias in the sense that the ratios could at least in part also depend on the distance from Earth, a bias caused by different linear resolutions. In Appendix~\ref{appendix_spectra}, the $^{12}$C/$^{13}$C, $^{32}$S/$^{34}$S, $^{34}$S/$^{33}$S, and $^{32}$S/$^{33}$S, as well as the $^{34}$S/$^{36}$S and $^{32}$S/$^{36}$S isotope ratios are plotted as functions of the distance from the Sun and shown in Figs.~\ref{fig_12C13C_2Sun} to \ref{fig_all36S_2Sun}, respectively. No apparent gradients can be found, which indicates that any observational bias on account of distance-dependent effects is not significant for $^{12}$C/$^{13}$C, $^{32}$S/$^{34}$S, $^{32}$S/$^{33}$S, $^{34}$S/$^{36}$S, and $^{32}$S/$^{36}$S. This agrees with the findings of \citet[][see their Section~4.5]{2020ApJ...899..145Y}.

\subsection{Beam size effects}
\label{section_beam}

A good way to check whether the different beam sizes for different lines could affect our results is to compare the isotope ratios derived from different transitions at different frequencies observed with different beam sizes. As shown in Section~\ref{section_double_32s34s}, $^{32}$S/$^{34}$S ratios obtained from the double isotope method in the $J$ = 2-1 and 3-2 transitions are in good agreement, suggesting that the effect of beam size is negligible. Furthermore, \citet{2020A&A...642A.222H} found an average $^{32}$S/$^{34}$S ratio of 17.9~$\pm$~5.0 in the envelope of Sgr B2(N) with the Atacama Large Millimetre/submillimetre Array (ALMA) with 1$\farcs$6 beam size, which is consistent with our results in the CMZ from the IRAM 30 meter telescope with beam sizes of about 27$\arcsec$. \citet{2020ApJ...899..145Y} derived similar $^{32}$S/$^{34}$S and $^{34}$S/$^{33}$S ratios from different telescopes ---that is, the IRAM 30 meter and the ARO 12 meter--- toward six HMSFRs and concluded that beam-size effects are insignificant (see details in their Section~4.1). All this suggests that beam-size effects could not obviously affect our results.

\subsection{Chemical fractionation}
\label{section_fractionation}

Isotopic fractionation could possibly affect the isotope ratios derived from the molecules in the interstellar medium. \citet{1976ApJ...205L.165W} firstly proposed that gas-phase CO should have a tendency to be enriched in $^{13}$CO because of the charge-exchange reaction of CO with $^{13}$C$^+$. Several theoretical studies also support this mechanism \citep[e.g.,][]{1984ApJ...277..581L,2020MNRAS.497.4333V}, which was then extended to other carbon bearing species \citep{2020MNRAS.498.4663L}. Formaldehyde forming in the gas phase is suggested to be depleted in the $^{13}$C bearing isotopolog \citep[e.g.,][]{1984ApJ...277..581L}. However, if H$_2$CO originates from dust grain mantles, then the $^{13}$C bearing isotopolog might be enhanced relative to species like methanol and CO \citep{2012LPI....43.1611W,2019ApJ...877..154Y}. Recently, \citet{2020A&A...640A..51C} performed a new gas-grain chemical model and proposed that molecules formed starting from atomic carbon could also show $^{13}$C enhancements through the reaction $^{13}$C + C$_{3}$ $\rightarrow$ $^{12}$C + $^{13}$CC$_{2}$. As already mentioned in Section~\ref{section_comparisons_12c13c}, the Galactic $^{12}$C/$^{13}$C gradient derived from C$^{34}$S in this work is in good agreement with previous results based on measurements of CN, C$^{18}$O, and H$_2$CO. Therefore, chemical fractionation cannot greatly affect the carbon isotope ratios. 

To date, little is known about sulfur fractionation. \citet{2019MNRAS.485.5777L} proposed a low $^{34}$S enrichment through the reaction of $^{34}$S$^+$ + CS $\rightarrow$ S$^+$ + C$^{34}$S in dense clouds. A slight enrichment in $^{13}$C was predicted for CS with the $^{13}$C$^{+}$ + CS $\rightarrow$ C$^+$ + $^{13}$CS reaction \citep{2020MNRAS.498.4663L}. $^{32}$S/$^{34}$S ratios derived directly from $^{13}$CS and $^{13}$C$^{34}$S and the double isotope method involving $^{12}$C/$^{13}$C ratios (equation 8) turn out to agree very well, suggesting that sulfur fractionation is negligible as previously suggested by \citet{2020A&A...642A.222H} and in this work. 

\subsection{Interstellar C, N, O, and S isotope ratios}
\label{section_discussion_all}

The data collected so far allow us to evaluate the status of several isotopes with respect to primary or secondary synthesis in stellar objects. From the data presented here in Table~\ref{table_allratios}, we choose the $^{12}$C/$^{13}$C, $^{32}$S/$^{34}$S, $^{32}$S/$^{33}$S, and $^{32}$S/$^{36}$S ratios because $^{12}$C is mostly primary \citep{1995ApJS...98..617T} while $^{32}$S is definitely a primary nucleus \citep{1995ApJS..101..181W}, against which the other isotopes can be evaluated. A question arises as to whether all these ratios, as well as those from nitrogen and oxygen, can be part of the same scheme.

 Comparing the CMZ with the ratios in the inner disk, the CMZ with the outer Galaxy, the inner disk with the ratios in the outer Galaxy, and the local ISM values with those of the Solar System, we obtain increases in values for $^{12}$C/$^{13}$C, $^{32}$S/$^{34}$S, $^{32}$S/$^{33}$S, and $^{32}$S/$^{36}$S ratios. All of these values are listed in Table~\ref{table_allratios_com}. Percentages are clearly highest between $^{32}$S and $^{36}$S. These indicate that $^{36}$S is, as opposed to $^{32}$S, definitely secondary. Percentages between $^{12}$C and $^{13}$C are also high but not as extreme, presumably because $^{12}$C is also synthesized in longer lived stars of intermediate mass (e.g., \citealt{2020ApJ...900..179K}). More difficult to interpret are the $^{32}$S/$^{33}$S and $^{32}$S/$^{34}$S ratios, where percentages are smaller, indicating that $^{33}$S and $^{34}$S are, as already mentioned in Section~\ref{section_hfs_c33s}, neither fully primary nor secondary. However, percentages in the case of the $^{32}$S/$^{33}$S ratio systematically surpass those of the $^{32}$S/$^{34}$S ratio, suggesting a more secondary origin of $^{33}$S with respect to $^{34}$S, even though $^{34}$S/$^{33}$S appears to be constant across the Galaxy. Finally, local interstellar $^{32}$S/$^{34}$S and $^{32}$S/$^{33}$S ratios behave strikingly differently with respect to solar values. While $^{32}$S/$^{34}$S is (almost) solar, $^{32}$S/$^{33}$S is far below the solar value. Peculiar Solar System abundance ratios may be the easiest way to explain this puzzling situation. Most likely there is an overabundance of $^{34}$S in the gas and dust that formed the Solar System.

Another clearly primary isotope is $^{16}$O, which allows us to look for $^{16}$O/$^{18}$O and $^{16}$O/$^{17}$O ratios \citep{1993A&A...274..730H,1994LNP...439...72H,1999RPPh...62..143W,2008A&A...487..237W,2020ApJS..249....6Z}. The high percentages in the $^{16}$O/$^{17}$O ratios show that $^{17}$O is more secondary than $^{18}$O, which is consistent with models of stellar nucleosynthesis, because $^{17}$O is a main product of the CNO cycle while $^{18}$O can also be synthesized by helium burning in massive stars.

$^{14}$N/$^{15}$N can also be measured; both nuclei can be synthesized in rotating massive stars and AGB stars as primary products \citep[e.g.,][]{2002A&A...390..561M,2011MNRAS.414.3231K,2020ApJ...900..179K,2018ApJS..237...13L}. However, most of the $^{14}$N is produced through CNO cycling, and is therefore secondary \citep[e.g.,][]{2011MNRAS.414.3231K,2014PASA...31...30K}. The production of $^{15}$N remains to be understood and may be related to novae \citep[e.g.,][]{2020ApJ...900..179K,2022arXiv221004350R}. None of the stable nitrogen isotopes are purely primary. While $^{14}$N appears to be less secondary than $^{15}$N \citep{1975A&A....43...71A,1994LNP...439...72H,1999RPPh...62..143W,2012ApJ...744..194A,2015ApJ...804L...3R,2018A&A...609A.129C,2021ApJS..257...39C}, in this case we do not have a clear calibration against an isotope that can be considered to be mainly primary. Remarkably, \citet{2022arXiv220910620C} reported a rising $^{14}$N/$^{15}$N gradient that peaks at $R_{\rm GC}$ = 11 kpc and then decreases, and suggested that $^{15}$N could be mainly produced by novae on long timescales.

\begin{table*}[h]
\caption{Interstellar C, N, O, and S isotope ratios.}
\centering
\begin{tabular}{ccccccc}
\hline\hline
  & Molecule & CMZ & Inner disk &Local ISM & Outer Galaxy & Solar System\tablefootmark{*} \\
\hline
\label{table_allratios}
$^{12}$C/$^{13}$C & C$^{34}$S\tablefootmark{a}&  27 $\pm$ 3     &   41 $\pm$ 9  &   66 $\pm$ 10  &   74 $\pm$ 8   & 89     \\ 
          &        CN\tablefootmark{b}        & $\cdots$        &  44 $\pm$ 12  &   41 $\pm$ 11  &   66 $\pm$ 19  &      \\ 
          &      C$^{18}$O\tablefootmark{c}   &  24 $\pm$ 1     &   41 $\pm$ 2  &   60 $\pm$ 5   &   70 $\pm$ 10  &      \\ 
          &      H$_2$CO\tablefootmark{d}     & $\cdots$        &   40 $\pm$ 7  &   50 $\pm$ 13  &   64 $\pm$ 10  &      \\ 
          &            average                &   25 $\pm$ 2    &   42 $\pm$ 9  &   54 $\pm$ 10  &   69 $\pm$ 12  &       \\
\hline
$^{14}$N/$^{15}$N  & CN\tablefootmark{e}      & $\cdots$        &   269 $\pm$ 59   &   314 $\pm$ 104  &   289 $\pm$ 85  &  270 \\ 
          &   HCN\tablefootmark{f}            & $\cdots$        &    284 $\pm$ 63  &   398 $\pm$ 48   &   388 $\pm$ 32  &      \\ 
          &   HNC\tablefootmark{f}            & $\cdots$        &   363 $\pm$ 100  &   378 $\pm$ 79   &   395 $\pm$ 74  &      \\ 
          &    N$_2$H$^{+}$\tablefootmark{g}  & $\cdots$        &   900 $\pm$ 250  &   496 $\pm$ 65   &   581 $\pm$ 140 &      \\ 
          &   NH$_3$\tablefootmark{h}         &  40 $\pm$ 13    &   175 $\pm$ 46   &   297 $\pm$ 99   &   96 $\pm$ 44   &      \\
\hline
$^{16}$O/$^{18}$O & H$_2$CO\tablefootmark{i}  &   263 $\pm$ 45  &  327 $\pm$ 32    &  560 $\pm$ 25    &  625 $\pm$ 144  &  490 \\
$^{18}$O/$^{17}$O & CO\tablefootmark{j}       &  3.4 $\pm$ 0.1  &   3.6 $\pm$ 0.2  &   3.9 $\pm$ 0.4  &   4.8 $\pm$ 0.6 &  5.5 \\
$^{16}$O/$^{17}$O\tablefootmark{**}  &        &  894 $\pm$ 155  &  1177 $\pm$ 132  &  2184 $\pm$ 244  & 3000 $\pm$ 786  &  2625 \\
\hline
$^{32}$S/$^{34}$S\tablefootmark{a}  &         &  19 $\pm$ 2     &   18 $\pm$ 4      &   24 $\pm$ 4     &   28 $\pm$ 3     & 23   \\
$^{34}$S/$^{33}$S\tablefootmark{a}  &         &  4.2 $\pm$ 0.2  &   4.3 $\pm$ 0.4   &   4.2 $\pm$ 0.5  &   4.1 $\pm$ 0.3  & 5.6  \\
$^{32}$S/$^{33}$S\tablefootmark{a}  &         &   70 $\pm$ 16   &   82 $\pm$ 19     &   88 $\pm$ 21    &   105 $\pm$ 19   & 127  \\
$^{34}$S/$^{36}$S\tablefootmark{a}  &         &  41 $\pm$ 4     &   122 $\pm$ 18    &   111 $\pm$ 16   &   161 $\pm$ 32   & 200  \\
$^{32}$S/$^{36}$S\tablefootmark{a}  &         &  884 $\pm$ 104  &   2382 $\pm$ 368  &   2752 $\pm$ 458 &   4150 $\pm$ 828 & 4525 \\
\hline
\end{tabular}
\tablefoot{The inner disk values refer to the mean values at galactocentric distances of 2.0 kpc~$\le R_{\rm GC} \le$~6.0 kpc. The local ISM values refer to 7.5 kpc~$\le R_{\rm GC} \le$~8.5 kpc. The outer Galaxy values point to 9.0 kpc~$\le R_{\rm GC} \le$~11.0 kpc. 
\tablefoottext{*}{From \citet{1989GeCoA..53..197A}.} 
\tablefoottext{a}{This work.}
\tablefoottext{b}{From \citet{2002ApJ...578..211S} and \citet{2005ApJ...634.1126M}.}
\tablefoottext{c}{From \citet{1990ApJ...357..477L,1993ApJ...408..539L}, \citet{1996A&AS..119..439W}, and \citet{1998ApJ...494L.107K}.}
\tablefoottext{d}{From \citet{1980A&A....82...41H,1982A&A...109..344H,1983A&A...127..388H,1985A&A...143..148H} and \citet{2019ApJ...877..154Y}.}  
\tablefoottext{e}{From \citet{2012ApJ...744..194A}, \citet{2015ApJ...804L...3R}, and \citet{2015ApJ...808L..46F}. }  
\tablefoottext{f}{From \citet{2012ApJ...744..194A} and \citet{2018A&A...609A.129C}}  
\tablefoottext{g}{From \citet{2015ApJ...804L...3R}. }  
\tablefoottext{h}{From \citet{2021ApJS..257...39C}. }  
\tablefoottext{i}{From \citet{1981MNRAS.194P..37G} and \citet{1994ARA&A..32..191W}. }  
\tablefoottext{j}{From \citet{2020ApJS..249....6Z}. } 
\tablefoottext{**}{The $^{16}$O/$^{17}$O ratios are derived with values of $^{16}$O/$^{18}$O and $^{18}$O/$^{17}$O.}    }
\end{table*}

\begin{table*}[h]
\caption{Comparison of isotope ratios at different galactocentric distances. Given are percentage enhancements.}
\centering
\begin{tabular}{ccccc}
\hline\hline
   & CMZ          &     CMZ      & Inner disk   &  Local ISM  \\
   & $\downarrow$ & $\downarrow$ & $\downarrow$ & $\downarrow$ \\
   &  Inner disk  & Outer Galaxy & Outer Galaxy & Solar System \\
\hline
\label{table_allratios_com}
$^{12}$C/$^{13}$C & 68 $\pm$ 33  &  176 $\pm$ 54  &  64 $\pm$ 21  &  65 $\pm$ 31  \\
\hline
$^{16}$O/$^{18}$O  & 24 $\pm$ 9  &  138 $\pm$ 61  &  91 $\pm$ 43   &  -12 $\pm$ 5   \\
$^{16}$O/$^{17}$O  & 32 $\pm$ 8  &  236 $\pm$ 111  &  155 $\pm$ 73  &  20 $\pm$ 13    \\
\hline
$^{32}$S/$^{34}$S  & -5 $\pm$ 11  &  47 $\pm$ 10  &  56 $\pm$ 18  &  -4 $\pm$ 17   \\
$^{32}$S/$^{33}$S  & 17 $\pm$ 8   &  50 $\pm$ 16  &  28 $\pm$ 6  &  44 $\pm$ 34 \\
$^{32}$S/$^{36}$S  & 169 $\pm$ 50  &  369 $\pm$ 125  &  74 $\pm$ 31  &  64 $\pm$ 27   \\
\hline
\end{tabular}
\end{table*}

\subsection{Galactic chemical environment}
\label{section_discussion_model}

\citet{2020ApJ...900..179K} established a Galactic chemical evolution (GCE) model based on the GCE model in \citet{2011MNRAS.414.3231K} with updates with respect to new solar abundances and also accounting for failed supernovae, super-AGB stars, the s-process from AGB stars, and various r-process sites. Based on this GCE model, the predicted $^{12}$C/$^{13}$C, $^{32}$S/$^{34}$S, $^{32}$S/$^{33}$S, and $^{32}$S/$^{36}$S ratios at $R_{\rm GC}$~=~2.0, 4.0, 6.0, 8.5, 12.0, and 17.0~kpc are obtained and plotted in Figs.~\ref{fig_gradient_12C13C}, \ref{fig_gradient_32S34S}, \ref{fig_all33S}, and \ref{fig_all36S}. The initial mass function and nucleosynthesis yields are the same for different galactic radii but star formation and inflow timescales ($\tau_{\rm s}$ and $\tau_{\rm i}$) depend on the Galactic radius (see \citealt{2000ApJ...539...26K} for the definition of the timescales). Adopted values are $\tau_{\rm s}$~=~1.0, 2.0, 3.0, 4.6, 6.5, and 8.8 Gyr as well as $\tau_{\rm i}$~=~4.0, 5.0, 5.0, 5.0, 7.0, and 50.0 Gyr for $R_{\rm GC}$~=~2.0, 4.0, 6.0, 8.5, 12.0, and 17.0~kpc, respectively. The predicted $^{12}$C/$^{13}$C ratios are in good agreement with our results, while $^{32}$S/$^{34}$S and $^{32}$S/$^{36}$S ratios show significant deviations at larger galactocentric distances. $^{32}$S/$^{33}$S ratios show an offset along the entire inner 12 kpc of the Milky Way. This indicates that current models of Galactic chemical evolution are still far from perfect. In this context, our data will serve as a useful guideline for further even more refined GCE models.

\begin{figure}[h]
\centering
   \includegraphics[width=280pt]{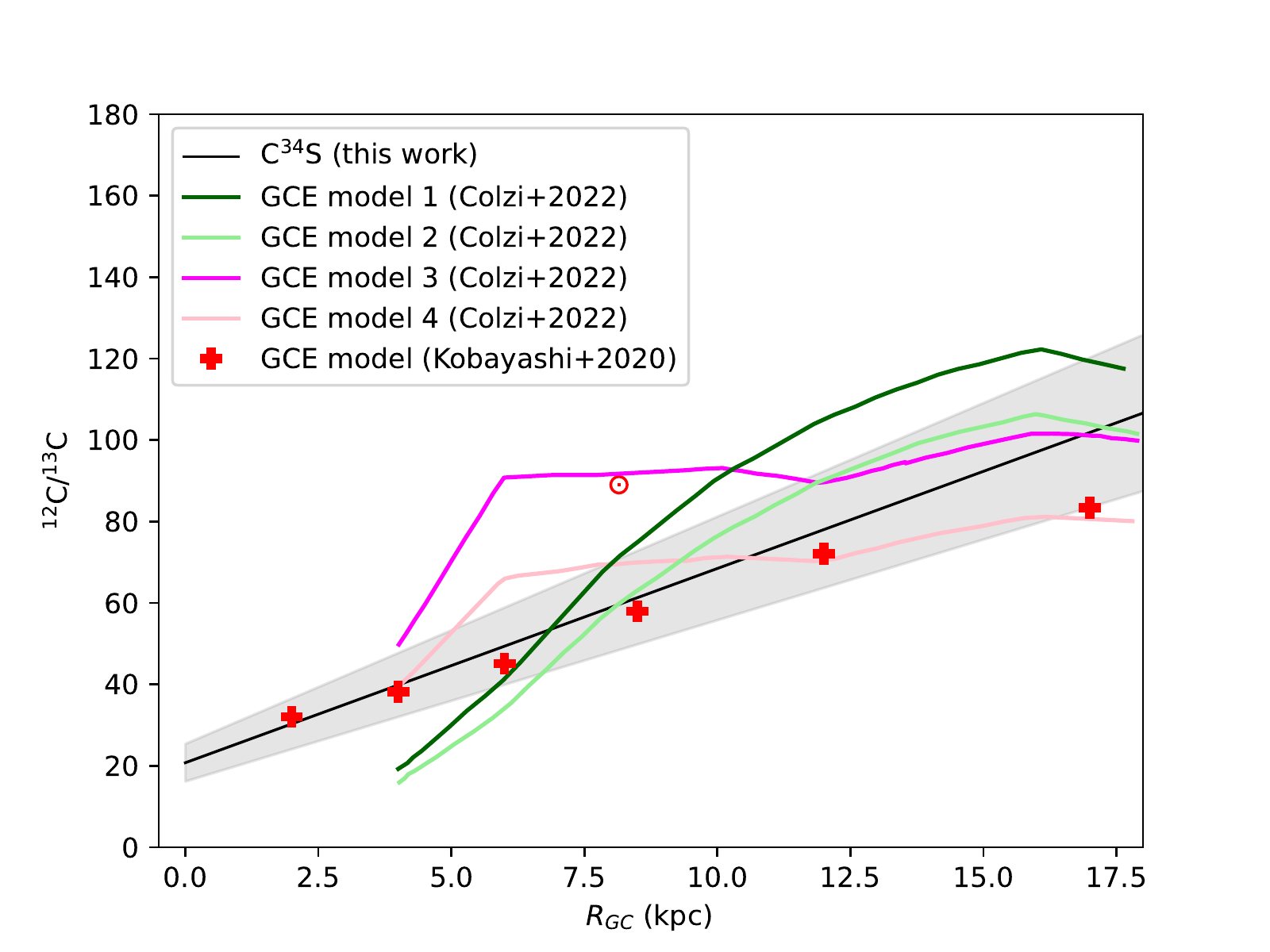}
     \caption{$^{12}$C/$^{13}$C isotope ratios from observations in this work and GCE models. The red symbol $\odot$ indicates the $^{12}$C/$^{13}$C isotope ratio of the Sun. The $^{12}$C/$^{13}$C gradient obtained from C$^{34}$S in the current work is plotted as a black solid line, with the gray shaded area showing the 1$\sigma$ interval of the fit. The red crosses visualize the results from the GCE model of \citet[][see also Section~\ref{section_discussion_model}]{2011MNRAS.414.3231K,2020ApJ...900..179K}. The dark green, light green, magenta, and pink lines refer to the predicted gradients from models in Table 2 from \citet{2022arXiv220910620C}.  }
  \label{fig_12C13C_GCEmodels}
\end{figure}

Very recently, \citet{2022arXiv220910620C} predicted $^{12}$C/$^{13}$C gradients with four different models addressing nova systems (see details in their Table~2 and Sect.~4), following \citet{2017MNRAS.470..401R,2019MNRAS.490.2838R,2021A&A...653A..72R}. The gradients from these four models are shown in Fig.~\ref{fig_12C13C_GCEmodels}. The results from model 1 show a large deviation with respect to the observed values. The other three models could reproduce the ratios within the dispersion at galactocentric radii beyond the solar neighborhood, while the inner Galaxy is not as well reproduced.

\section{Summary}
\label{summary}

We used the IRAM 30 meter telescope to perform observations of the $J$ = 2-1 transitions of CS, C$^{33}$S, C$^{34}$S, C$^{36}$S, $^{13}$CS, $^{13}$C$^{33}$S, and $^{13}$C$^{34}$S as well as the $J$ = 3-2 transitions of C$^{33}$S, C$^{34}$S, C$^{36}$S, and $^{13}$CS toward a large sample of 110 HMSFRs. The CS $J$ = 2-1 line was detected toward 106 sources, with a detection rate of 96\%. The $J$ = 2-1 transitions of C$^{34}$S, $^{13}$CS, C$^{33}$S, $^{13}$C$^{34}$S, and C$^{36}$S were successfully detected in 90, 82, 46, 17, and 3 of our sources, respectively. The $J$ = 3-2 lines of C$^{34}$S, $^{13}$CS, C$^{33}$S, and C$^{36}$S were detected in 87, 71, 42, and 1 object(s). All the detected rare CS isotopologs exhibit optically thin lines and allow us to measure the isotope ratios of $^{12}$C/$^{13}$C, $^{32}$S/$^{34}$S, $^{32}$S/$^{33}$S, $^{32}$S/$^{36}$S, $^{34}$S/$^{33}$S, $^{34}$S/$^{36}$S, and $^{33}$S/$^{36}$S with only minor saturation corrections. Our main results are as follows:
\begin{itemize}
    \item Based on the measurements of C$^{34}$S and $^{13}$C$^{34}$S $J$ = 2-1 transitions, we directly measured the $^{12}$C/$^{13}$C ratios with corrections of opacity. With accurate distances obtained from parallax data \citep{2009ApJ...700..137R, 2014ApJ...783..130R, 2019ApJ...885..131R}, we confirm the previously determined $^{12}$C/$^{13}$C gradient. A least-squares fit to our data results in $^{12}\rm C$/$^{13}\rm C$ = (4.77~$\pm$~0.81)$R_{\rm GC}$+(20.76~$\pm$~4.61), with a correlation coefficient of 0.82. 
    \item The Galactic $^{12}$C/$^{13}$C gradients derived based on measurements of CN \citep{2002ApJ...578..211S,2005ApJ...634.1126M}, C$^{18}$O \citep{1990ApJ...357..477L,1996A&AS..119..439W,1998ApJ...494L.107K}, and H$_2$CO \citep{1980A&A....82...41H,1982A&A...109..344H,1983A&A...127..388H,1985A&A...143..148H,2019ApJ...877..154Y} are in agreement with our results from C$^{34}$S and emphasize that chemical fractionation has little effect on $^{12}$C/$^{13}$C ratios.
    \item While previously it had been assumed that a linear fit would provide a good simulation of carbon isotope ratios as a function of galactocentric distance, our analysis reveals that this does not hold for the Galactic center region. While $^{12}$C/$^{13}$C ratios are lowest in this part of the Milky Way, they clearly surpass values expected from a linear fit to the Galactic disk sources. This indicates that there is no strict linear correlation of carbon isotope ratios across the Galaxy.
    \item We confirm the previously determined $^{32}$S/$^{34}$S gradients \citep{1996A&A...305..960C,2020ApJ...899..145Y,2020A&A...642A.222H} with the direct method from $^{13}$CS and $^{13}$C$^{34}$S, as well as the double isotope method also using $^{12}$C/$^{13}$C ratios in the $J$ = 2-1 and $J$ = 3-2 transitions. Opacity corrections could be applied to the $J$ = 2-1 transitions, but not to the $J$ = 3-2 lines that may show, on average, slightly higher opacities. A $^{32}$S/$^{34}$S gradient of (0.75 $\pm$ 0.13)$R_{\rm GC}$+(15.52 $\pm$ 0.78) was obtained based on a large dataset of 90 values from our double isotope method in the $J$ = 2-1 transition. The 19 sources permitting the direct determination of this ratio with $^{13}$CS/$^{13}$C$^{34}$S yield $^{32}$S/$^{34}$S=(0.73 $\pm$ 0.36)$R_{\rm GC}$+(16.50 $\pm$ 2.07). 
    \item Differences between the behavior of the $^{12}$C/$^{13}$C and $^{32}$S/$^{34}$S ratios as a function of galactocentric distance are reported and should be used as input for further chemical models: (a) In the inner disk the $^{12}$C/$^{13}$C ratios at $R_{\rm GC} \ge$ 4.0 kpc are clearly higher than the value in the CMZ, while the $^{32}$S/$^{34}$S ratios in the CMZ and inner disk are similar, as already suggested for the first time by \citet{2020A&A...642A.222H}. (b) In the local ISM, the $^{12}$C/$^{13}$C ratio is well below the Solar System value but $^{32}$S/$^{34}$S is still quite close to it. All of this indicates that, unlike $^{13}$C, $^{34}$S is not a clean secondary isotope.
    \item There is no notable $^{34}$S/$^{33}$S gradient across the Galaxy. Ratios are well below the values commonly reported in earlier publications. This is a consequence of accounting for the full hyperfine structure splitting of the C$^{33}$S lines. The average value of $^{34}$S/$^{33}$S derived from the $J$ = 2-1 transition lines after corrections for opacity toward our sample is 4.35~$\pm$~0.44.
    \item While there is no $^{34}$S/$^{33}$S gradient with galactocentric radius, interstellar
    $^{34}$S/$^{33}$S values near the solar neighborhood are well below the Solar System ratio, most likely suggesting the Solar System ratio is peculiar, and perhaps also the $^{18}$O/$^{17}$O ratio. A comparison of local interstellar and Solar System $^{32}$S/$^{34}$S and $^{34}$S/$^{33}$S ratios suggests that the Solar System may have been formed from gas and dust with a peculiarly high $^{34}$S abundance. The data also indicate that $^{33}$S is not a clean primary or secondary product of nucleosynthesis, similarly
to $^{34}$S .
    \item For the first time, we report a $^{32}$S/$^{33}$S gradient in our Galaxy: $^{32}{\rm S}/^{33}{\rm S}$ = $(2.64 \pm 0.77)R_{\rm GC}+(70.80 \pm 5.57)$, with a correlation coefficient of 0.46. 
    \item We find first potential indications for a positive $^{34}$S/$^{36}$S gradient with galactocentric radius. Combined $^{34}$S/$^{36}$S ratios from \citet{1996A&A...313L...1M} and our new data with corrections of opacity in the $J$ = 2-1 transition and applying new up-to-date distances yield a linear fit of $^{34}{\rm S}/^{36}{\rm S}$ = $(10.34 \pm 2.74)R_{\rm GC}+(57.45 \pm 18.59)$, with a correlation coefficient of 0.71. Considering the uniform $^{34}$S/$^{33}$S ratios in our Galaxy, a $^{33}$S/$^{36}$S gradient of (2.38~$\pm$~0.67)$R_{\rm GC}$+(13.21~$\pm$~4.48) is also obtained. 
    \item For the first time, we report a tentative $^{32}$S/$^{36}$S gradient with galactocentric radius: $^{32}{\rm S}/^{36}{\rm S}$ = $(314 \pm 55)R_{\rm GC}+(659 \pm 374)$, with a correlation coefficient of 0.84. Our measurements are consistent with $^{36}$S being a purely secondary nucleus. However, observations of $^{34}$S/$^{36}$S and $^{32}$S/$^{36}$S isotope ratios are still relatively few, especially in the CMZ and the inner disk within $R_{\rm GC}$~=~5.0~kpc.
    \item The predicted $^{12}$C/$^{13}$C ratios from the latest Galactic chemical evolution models \citep[e.g.,][]{2020ApJ...900..179K,2021A&A...653A..72R,2022arXiv220910620C} are in good agreement with our results, while $^{32}$S/$^{34}$S and $^{32}$S/$^{36}$S ratios show significant differences at larger galactocentric distances. $^{32}$S/$^{33}$S ratios even show clear offsets along the entire inner 12 kpc of the Milky Way. Taken together, these findings provide useful guidelines for further refinements of models of the chemical evolution of the Galaxy.
\end{itemize}

\begin{acknowledgements}
We wish to thank the referee for useful comments. Y.T.Y. is a member of the International Max Planck Research School (IMPRS) for Astronomy and Astrophysics at the Universities of Bonn and Cologne. Y.T.Y. would like to thank the China Scholarship Council (CSC) and the Max-Planck-Institut f\"{u}r Radioastronomie (MPIfR) for the financial support. Y.T.Y. also thanks his fiancee, Siqi Guo, for her support during this pandemic period. C.K. acknowledges funding from the UK Science and Technology Facility Council through grant ST/R000905/1 and ST/V000632/1. We thank the IRAM staff for help provided during the observations.
\end{acknowledgements}

\bibliographystyle{aa}
\bibliography{cs}

\clearpage
\onecolumn

\setcounter{table}{2}

\begin{center}
\begin{longtable}{lc|cccc|cccc}
\caption{\label{table_3433}Corrected $^{34}$S/$^{33}$S isotope ratios.}\\
\hline\hline
Source &  $R_{GC}$  &   \multicolumn{4}{c}{$J$ = 2-1} &  \multicolumn{4}{c}{$J$ = 3-2} \\
       &          & FWHM  & C$^{34}$S/C$^{33}$S & $f_{21HFS}$ & $^{34}$S/$^{33}$S &  FWHM & C$^{34}$S/C$^{33}$S & $f_{32HFS}$ & $^{34}$S/$^{33}$S  \\
       &  (kpc)     &   (km s$^{-1}$)   &                     &                               &              &    (km s$^{-1}$)      &           &            &        \\
\hline
\endfirsthead
\caption{continued.}\\
\hline\hline
Source &  $R_{GC}$  &   \multicolumn{4}{c}{$J$ = 2-1} &  \multicolumn{4}{c}{$J$ = 3-2} \\
       &            & FWHM  & C$^{34}$S/C$^{33}$S & $f21_{HFS}$ & $^{34}$S/$^{33}$S  &  FWHM  & C$^{34}$S/C$^{33}$S & $f32_{HFS}$ & $^{34}$S/$^{33}$S  \\
       &  (kpc)     &   (km s$^{-1}$)   &                     &                               &                   &           (km s$^{-1}$)        &           &        &            \\
\hline
\endhead
\hline
\endfoot
WB89-380                & 14.19 $\pm$ 0.92 & 3.60  & 5.70 $\pm$ 0.58 & 0.70 & 4.28 $\pm$ 0.44  & 3.87  & 3.70 $\pm$ 0.37 & 0.91 & 3.36 $\pm$ 0.34  \\ 
WB89-391                & 14.28 $\pm$ 0.94 & 1.03  & 6.75 $\pm$ 1.71 & 0.69 & 5.21 $\pm$ 1.32  & 2.19  & 3.67 $\pm$ 0.42 & 0.85 & 3.12 $\pm$ 0.36  \\ 
W3OH                    & 9.64 $\pm$ 0.03  & 4.41  & 3.04 $\pm$ 0.12 & 1.00 & 3.54 $\pm$ 0.14  & 6.14  & 2.59 $\pm$ 0.04 & 0.96 & 2.49 $\pm$ 0.04  \\ 
Orion-KL                & 8.54 $\pm$ 0.00  & 4.25  & 2.77 $\pm$ 0.22 & 1.00 & 3.01 $\pm$ 0.24  & 8.33  & 2.74 $\pm$ 0.12 & 0.99 & 2.72 $\pm$ 0.11  \\ 
$+$20~km~s$^{-1}$~cloud & 0.03 $\pm$ 0.03  & 24.78 & 4.89 $\pm$ 0.14 & 1.00 & 5.64 $\pm$ 0.16  & 20.12 & 5.22 $\pm$ 0.30 & 1.00 & 5.22 $\pm$ 0.30  \\ 
G359.61$-$00.24         & 5.51 $\pm$ 0.15  & 3.17  & 3.63 $\pm$ 0.12 & 1.00 & 4.04 $\pm$ 0.13  & 10.44 & 3.33 $\pm$ 0.45 & 0.99 & 3.31 $\pm$ 0.44  \\ 
$+$50~km~s$^{-1}$~cloud & 0.02 $\pm$ 0.04  & 24.29 & 4.80 $\pm$ 0.06 & 1.00 & 5.29 $\pm$ 0.07  & 21.28 & 4.57 $\pm$ 0.15 & 1.00 & 4.57 $\pm$ 0.15  \\ 
G000.31$-$00.20         & 5.25 $\pm$ 0.36  & 3.58  & 4.65 $\pm$ 0.70 & 0.70 & 3.54 $\pm$ 0.53  & $\cdots$ & $\cdots$        & $\cdots$ & $\cdots$   \\ 
SgrB2                   & 0.44 $\pm$ 0.70  & 21.14 & 2.23 $\pm$ 0.10 & 0.99 & 2.46 $\pm$ 0.11  & 21.22 & 2.00 $\pm$ 0.03 & 1.00 & 2.00 $\pm$ 0.03  \\
SgrD                    & 0.45 $\pm$ 0.07  & 2.73  & 7.01 $\pm$ 0.47 & 0.70 & 4.96 $\pm$ 0.33  & 4.00  & 3.56 $\pm$ 0.24 & 0.91 & 3.24 $\pm$ 0.22  \\ 
G006.79$-$00.25         & 4.75 $\pm$ 0.25  & 2.73  & 3.79 $\pm$ 0.08 & 0.98 & 4.79 $\pm$ 0.10  & 3.32  & 3.68 $\pm$ 0.09 & 0.89 & 3.29 $\pm$ 0.08  \\ 
G007.47$+$00.05         & 12.35 $\pm$ 2.49 & 3.12  & 5.41 $\pm$ 0.97 & 0.70 & 4.07 $\pm$ 0.73  & $\cdots$ & $\cdots$        & $\cdots$ & $\cdots$   \\ 
G010.32$-$00.15         & 5.34 $\pm$ 0.29  & 2.66  & 4.16 $\pm$ 0.22 & 0.98 & 4.60 $\pm$ 0.25  & 3.74  & 4.07 $\pm$ 0.21 & 0.90 & 3.68 $\pm$ 0.19  \\ 
G010.62$-$00.33         & 5.49 $\pm$ 0.56  & 2.70  & 6.37 $\pm$ 1.45 & 0.98 & 6.80 $\pm$ 1.54  & $\cdots$ & $\cdots$        & $\cdots$ & $\cdots$   \\ 
G011.10$-$00.11         & 4.26 $\pm$ 0.21  & 1.45  & 5.36 $\pm$ 1.52 & 0.69 & 3.72 $\pm$ 1.05  & $\cdots$ & $\cdots$        & $\cdots$ & $\cdots$   \\ 
G016.86$-$02.15         & 5.97 $\pm$ 0.47  & 3.32  & 3.47 $\pm$ 0.12 & 0.98 & 3.83 $\pm$ 0.14  & 4.33  & 3.42 $\pm$ 0.14 & 0.92 & 3.15 $\pm$ 0.13  \\ 
G017.02$-$02.40         & 6.40 $\pm$ 0.36  & 3.74  & 4.49 $\pm$ 0.33 & 0.98 & 4.83 $\pm$ 0.35  & 4.03  & 4.11 $\pm$ 0.26 & 0.91 & 3.75 $\pm$ 0.24  \\ 
G017.63$+$00.15         & 6.77 $\pm$ 0.04  & 2.74  & 5.41 $\pm$ 1.27 & 0.70 & 4.02 $\pm$ 0.94  & $\cdots$ & $\cdots$        & $\cdots$ & $\cdots$   \\ 
G018.34$+$01.76         & 6.31 $\pm$ 0.07  & 2.30  & 4.79 $\pm$ 0.28 & 0.98 & 4.95 $\pm$ 0.28  & 3.00  & 4.59 $\pm$ 0.16 & 0.88 & 4.06 $\pm$ 0.15  \\ 
G019.36$-$00.03         & 5.58 $\pm$ 0.49  & 3.02  & 4.58 $\pm$ 0.20 & 0.98 & 5.40 $\pm$ 0.24  & 5.09  & 4.25 $\pm$ 0.18 & 0.94 & 3.99 $\pm$ 0.17  \\ 
G023.43$-$00.18         & 3.63 $\pm$ 0.49  & 3.81  & 3.75 $\pm$ 0.38 & 0.98 & 4.22 $\pm$ 0.43  & 5.06  & 4.65 $\pm$ 0.24 & 0.94 & 4.37 $\pm$ 0.23  \\ 
G024.78$+$00.08         & 3.51 $\pm$ 0.15  & 4.29  & 2.86 $\pm$ 0.08 & 1.00 & 4.05 $\pm$ 0.11  & 6.36  & 2.76 $\pm$ 0.06 & 0.96 & 2.67 $\pm$ 0.06  \\ 
G024.85$+$00.08         & 3.85 $\pm$ 0.23  & 1.27  & 9.43 $\pm$ 2.27 & 0.69 & 7.56 $\pm$ 1.82  & 3.20  & 3.03 $\pm$ 0.40 & 0.89 & 2.70 $\pm$ 0.36  \\ 
G028.30$-$00.38         & 4.71 $\pm$ 0.26  & $\cdots$  & $\cdots$        & $\cdots$ & $\cdots$ & 2.68  & 4.05 $\pm$ 0.63 & 0.88 & 3.55 $\pm$ 0.55  \\ 
G028.39$+$00.08         & 4.83 $\pm$ 0.17  & 3.04  & 3.33 $\pm$ 0.15 & 0.98 & 3.82 $\pm$ 0.18  & 4.39  & 3.84 $\pm$ 0.12 & 0.92 & 3.54 $\pm$ 0.11  \\ 
G028.83$-$00.25         & 4.50 $\pm$ 0.48  & 2.27  & 3.80 $\pm$ 0.19 & 0.98 & 4.56 $\pm$ 0.23  & 3.94  & 4.65 $\pm$ 0.10 & 0.91 & 4.23 $\pm$ 0.09  \\ 
G030.70$-$00.06         & 4.20 $\pm$ 0.10  & 4.85  & 2.80 $\pm$ 0.07 & 1.00 & 3.55 $\pm$ 0.09  & 6.29  & 2.84 $\pm$ 0.12 & 0.96 & 2.73 $\pm$ 0.11  \\ 
G030.74$-$00.04         & 5.76 $\pm$ 0.36  & 3.55  & 3.61 $\pm$ 0.15 & 0.98 & 4.36 $\pm$ 0.19  & 4.24  & 3.48 $\pm$ 0.08 & 0.92 & 3.19 $\pm$ 0.08  \\ 
G030.78$+$00.20         & 4.19 $\pm$ 0.05  & $\cdots$  & $\cdots$        & $\cdots$ & $\cdots$ & 4.17  & 4.20 $\pm$ 0.51 & 0.92 & 3.85 $\pm$ 0.47  \\ 
G030.81$-$00.05         & 5.73 $\pm$ 0.24  & 6.15  & 3.05 $\pm$ 0.09 & 1.00 & 3.59 $\pm$ 0.11  & 7.34  & 2.88 $\pm$ 0.15 & 0.98 & 2.81 $\pm$ 0.15  \\ 
G031.24$-$00.11         & 7.48 $\pm$ 2.00  & $\cdots$  & $\cdots$        & $\cdots$ & $\cdots$ & 4.15  & 4.32 $\pm$ 0.52 & 0.92 & 3.95 $\pm$ 0.48  \\ 
G032.74$-$00.07         & 4.55 $\pm$ 0.23  & 4.54  & 3.44 $\pm$ 0.09 & 0.98 & 3.82 $\pm$ 0.10  & 8.07  & 3.15 $\pm$ 0.16 & 0.98 & 3.10 $\pm$ 0.16  \\
G032.79$+$00.19         & 5.26 $\pm$ 1.57  & 4.78  & 3.75 $\pm$ 0.21 & 0.98 & 4.04 $\pm$ 0.22  & 8.17  & 3.56 $\pm$ 0.06 & 0.98 & 3.50 $\pm$ 0.06  \\
G034.41$+$00.23         & 5.99 $\pm$ 0.06  & 4.56  & 3.39 $\pm$ 0.13 & 1.00 & 3.76 $\pm$ 0.14  & 6.28  & 3.13 $\pm$ 0.18 & 0.96 & 3.01 $\pm$ 0.18  \\
G034.79$-$01.38         & 6.21 $\pm$ 0.09  & 2.40  & 4.21 $\pm$ 0.37 & 0.98 & 5.54 $\pm$ 0.50  & 3.34  & 4.14 $\pm$ 0.16 & 0.89 & 3.70 $\pm$ 0.14  \\
G036.11$+$00.55         & 5.45 $\pm$ 0.43  & 2.57  & 2.75 $\pm$ 0.63 & 0.70 & 2.35 $\pm$ 0.54  & $\cdots$ & $\cdots$        & $\cdots$ & $\cdots$   \\ 
G037.42$+$01.51         & 6.78 $\pm$ 0.05  & 3.18  & 5.02 $\pm$ 0.23 & 0.98 & 5.21 $\pm$ 0.24  & 4.78  & 3.50 $\pm$ 0.48 & 0.93 & 3.26 $\pm$ 0.44  \\ 
G040.28$-$00.21         & 6.02 $\pm$ 0.10  & 3.72  & 6.79 $\pm$ 1.33 & 0.70 & 5.65 $\pm$ 1.10  & 6.43  & 3.43 $\pm$ 0.47 & 0.97 & 3.31 $\pm$ 0.45  \\ 
G045.45$+$00.06         & 6.41 $\pm$ 0.50  & 6.08  & 4.07 $\pm$ 0.66 & 0.75 & 3.37 $\pm$ 0.54  & 5.28  & 3.20 $\pm$ 0.25 & 0.94 & 3.02 $\pm$ 0.24  \\ 
G048.99$-$00.29         & 6.18 $\pm$ 0.02  & 3.74  & 7.02 $\pm$ 1.10 & 0.70 & 5.20 $\pm$ 0.81  & 4.78  & 4.32 $\pm$ 0.54 & 0.93 & 4.03 $\pm$ 0.50  \\ 
W51-IRS2                & 6.22 $\pm$ 0.06  & 8.55  & 3.19 $\pm$ 0.09 & 1.00 & 3.42 $\pm$ 0.10  & 8.72  & 2.82 $\pm$ 0.01 & 0.99 & 2.79 $\pm$ 0.01  \\ 
G071.31$+$00.82         & 8.02 $\pm$ 0.16  & 2.25  & 7.18 $\pm$ 1.81 & 0.70 & 5.50 $\pm$ 1.39  & 2.42  & 5.43 $\pm$ 1.08 & 0.87 & 4.71 $\pm$ 0.94  \\ 
G073.65$+$00.19         & 13.54 $\pm$ 2.90 & 2.76  & 4.71 $\pm$ 1.02 & 0.70 & 3.53 $\pm$ 0.76  & 2.88  & 4.65 $\pm$ 0.62 & 0.88 & 4.10 $\pm$ 0.55  \\ 
G075.29$+$01.32         & 10.69 $\pm$ 0.58 & 2.86  & 6.27 $\pm$ 0.85 & 0.70 & 4.70 $\pm$ 0.63  & 2.83  & 4.47 $\pm$ 0.40 & 0.88 & 3.94 $\pm$ 0.35  \\ 
DR21                    & 8.10 $\pm$ 0.00  & 2.71  & 3.45 $\pm$ 0.05 & 1.00 & 3.96 $\pm$ 0.06  & 3.15  & 3.27 $\pm$ 0.11 & 0.99 & 3.24 $\pm$ 0.11  \\ 
G090.92$+$01.48         & 10.13 $\pm$ 0.63 & 3.40  & 5.31 $\pm$ 0.56 & 0.70 & 3.98 $\pm$ 0.42  & 3.79  & 4.02 $\pm$ 0.32 & 0.91 & 3.64 $\pm$ 0.29  \\ 
G097.53$+$03.18         & 11.81 $\pm$ 0.70 & 5.22  & 2.45 $\pm$ 0.13 & 1.00 & 2.68 $\pm$ 0.14  & 5.47  & 3.28 $\pm$ 0.12 & 0.95 & 3.11 $\pm$ 0.12  \\ 
G109.87$+$02.11         & 8.49 $\pm$ 0.01  & 3.19  & 4.28 $\pm$ 0.21 & 0.98 & 4.50 $\pm$ 0.21  & 3.90  & 4.88 $\pm$ 0.67 & 0.91 & 4.43 $\pm$ 0.61  \\ 
NGC7538                 & 9.47 $\pm$ 0.07  & 3.71 & 4.17 $\pm$ 0.07 & 0.98 & 4.37 $\pm$ 0.06  & $\cdots$ & $\cdots$        & $\cdots$ & $\cdots$   \\ 
\hline
 \end{longtable}
\tablefoot{  Full width at half maximum values were obtained from measurements of C$^{33}$S; see Table \ref{fitting_all}. The $^{34}$S/$^{33}$S isotope ratios from the $J$ = 2-1 transition are corrected for the optical depth effect and the ones in the 3-2 line without corrections for opacity.
}
\end{center}

\begin{appendix}
\section{Tables}
\label{appendix_table}

\begin{center}
\begin{longtable}{lcccccc}
\caption{Source list.\label{table_sources}}\\
\hline 
\hline
Sources &  R.A.(J2000)&  Dec.(J2000) &  $V_{\rm LSR}$   & Spiral & d & $R_{\rm GC}$  \\
  &  ($h \quad m \quad s$) & ($\degr \quad \arcmin \quad \arcsec$)  & (km s$^{-1}$) &  Arm &  (kpc) & (kpc) \\
\hline
\endfirsthead
\caption{continued.}\\
\hline
\hline
Sources &  R.A.(J2000)&  Dec.(J2000) &  $V_{\rm LSR}$   & Spiral & d & $R_{\rm GC}$  \\
  &  ($h \quad m \quad s$) & ($\degr \quad \arcmin \quad \arcsec$)  & (km s$^{-1}$) &  Arm &  (kpc) & (kpc) \\
\hline
\endhead
\hline
\endfoot
WB89-380\tablefootmark{*}         &  01:07:51.3000  &  $+$65:21:25.000  &      -86.62 $\pm$ 0.02         & $\cdots$ &  7.85 $\pm$ 1.04  &  14.19 $\pm$ 0.92 \\
WB89-391\tablefootmark{*}         &  01:19:25.4000  &  $+$65:45:50.000  &      -85.93 $\pm$ 0.01   & $\cdots$ &  7.83 $\pm$ 1.06  &  14.28 $\pm$ 0.94 \\
W3OH             &  02:27:03.9000  &  $+$61:52:24.000  & $  -47\pm 3$  &    Per   &  1.95 $\pm$ 0.04 & 9.64 $\pm$ 0.03   \\
Orion-KL         &  05:35:14.2000  &  $-$05:22:31.000  & $    3\pm 5$  &    Loc   &  0.41 $\pm$ 0.00 & 8.54 $\pm$ 0.00 \\
PointC1\tablefootmark{*}          &  17:32:33.2000  &  $-$31:40:31.400  &      -64.18 $\pm$ 0.98    & $\cdots$ &  6.80 $\pm$ 0.16  &  1.46 $\pm$ 0.15 \\
G359.13$+$00.03  & 17:43:25.6109  & $-$29:39:17.551  & $   -1\pm 10$  & GC   & 6.06 $\pm$ 1.14 & 2.12 $\pm$ 1.14 \\
SgrC\tablefootmark{*}             &  17:44:47.0000  &  $-$29:28:24.800  &      70.06 $\pm$ 0.07    & $\cdots$ &  8.05 $\pm$ 0.04  &  0.15 $\pm$ 0.03 \\
$+$20 km~s$^{-1}$ cloud\tablefootmark{*}          &  17:45:37.5700  &  $-$29:05:22.600  &      6.80 $\pm$ 0.12     & $\cdots$ &  8.20 $\pm$ 0.04  & 0.03 $\pm$ 0.03  \\
G359.61$-$00.24  & 17:45:39.0697  & $-$29:23:30.265  & $   21\pm 5$  & CtN  & 2.67 $\pm$ 0.15 & 5.51 $\pm$ 0.15 \\
SgrA             &  17:45:40.5400  &  $-$29:00:16.280  &  $ 53\pm 1$                       & GC &  8.18 $\pm$ 0.04 &  0.00 $\pm$ 0.04 \\
$+$50 km~s$^{-1}$ cloud\tablefootmark{*}       &  17:45:50.2000  &  $-$28:59:41.000  &     46.78 $\pm$ 0.07     & $\cdots$ &  8.20 $\pm$ 0.04 &  0.02 $\pm$ 0.04 \\
G359.93$-$00.14  & 17:46:01.9183  & $-$29:03:58.674  & $  -10\pm 10$  & GC   & 5.52 $\pm$ 0.89 & 2.65 $\pm$ 0.89 \\
G000.37$+$00.03  & 17:46:21.4012  & $-$28:35:39.821  & $   37\pm 10$  & 3kF  & 8.00 $\pm$ 3.01 & 0.19 $\pm$ 2.88 \\
G0.25\tablefootmark{*}            &  17:46:09.1000  &  $-$28:42:12.000  &     39.70 $\pm$ 0.19     & $\cdots$ & 8.55 $\pm$ 0.04  & 0.37 $\pm$ 0.04 \\
CloudD\tablefootmark{*}           &  17:46:24.4000  &  $-$28:33:20.000  &     23.32 $\pm$ 0.55     & $\cdots$ & 8.85 $\pm$ 0.17  & 0.67 $\pm$ 0.17 \\
G000.31$-$00.20  & 17:47:09.1092  & $-$28:46:16.278  & $   18\pm 3$  & ScN  & 2.92 $\pm$ 0.36 & 5.25 $\pm$ 0.36 \\
SgrB2            &  17:47:20.4000  &  $-$28:23:02.000  & $   62\pm 5$  &    GC    &  7.75 $\pm$ 0.72 & 0.44 $\pm$ 0.70 \\
G1.28$+$0.07\tablefootmark{*}     &  17:48:22.0000  &  $-$27:48:19.000  &      98.64 $\pm$ 0.61       & $\cdots$ &  7.94 $\pm$ 0.04  &  0.30 $\pm$ 0.03 \\
SgrD\tablefootmark{*}             &  17:48:42.2400  &  $-$28:01:27.770  &      63.37 $\pm$ 0.86    & $\cdots$ &  7.76 $\pm$ 0.07 &  0.45 $\pm$ 0.07 \\
G001.00$-$00.23  & 17:48:55.2845  & $-$28:11:48.240  & $    2\pm 5$  & $\cdots$  & 11.11 $\pm$ 7.04 & 2.94 $\pm$ 7.03 \\ 
G001.14$-$00.12  & 17:48:48.5410  & $-$28:01:11.350  & $  -16\pm 3$  & Nor  & 5.15 $\pm$ 4.28 & 3.03 $\pm$ 4.27 \\
Clump2\tablefootmark{*}           &  17:51:24.8000  &  $-$26:00:04.000  &     80.82 $\pm$ 0.61      & $\cdots$ &  7.21 $\pm$ 0.11 &  1.06 $\pm$ 0.10 \\
G002.70$+$00.04  & 17:51:45.9766  & $-$26:35:57.070  & $   93\pm 5$  & $\cdots$  & 9.90 $\pm$ 10.29 & 1.77 $\pm$ 10.05 \\
PointD1\tablefootmark{*}          &  17:59:17.8000  &  $-$24:24:37.900  &     117.60 $\pm$ 0.14     & $\cdots$ &  7.18 $\pm$ 0.09 &  1.24 $\pm$ 0.07 \\
G006.79$-$00.25  & 18:01:57.7525  & $-$23:12:34.245  & $   22\pm 5$  & Nor  & 3.47 $\pm$ 0.25 & 4.75 $\pm$ 0.25 \\
G007.47$+$00.05  & 18:02:13.1823  & $-$22:27:58.981  & $  -14\pm 10$  & OSC  & 20.41 $\pm$ 2.50 & 12.35 $\pm$ 2.49 \\
G008.34$-$01.00  & 18:08:04.0510  & $-$22:13:26.566  & $    5\pm 5$  & SgN  & 1.56 $\pm$ 0.10 & 6.64 $\pm$ 0.10 \\
G009.21$-$00.20  & 18:06:52.8421  & $-$21:04:27.878  & $  102\pm 5$  & ScN  & 3.30 $\pm$ 1.05 & 4.95 $\pm$ 1.01 \\
G010.32$-$00.15  & 18:09:01.4549  & $-$20:05:07.854  & $   10\pm 5$  & ScN  & 2.92 $\pm$ 0.30 & 5.34 $\pm$ 0.29 \\
G010.62$-$00.33  & 18:10:17.9849  & $-$19:54:04.646  & $   -6\pm 5$  & 3kN  & 2.76 $\pm$ 0.58 & 5.49 $\pm$ 0.56 \\
G011.10$-$00.11  & 18:10:28.2470  & $-$19:22:30.216  & $   29\pm 5$  & ScN  & 4.07 $\pm$ 0.23 & 4.26 $\pm$ 0.21 \\
W33              &  18:13:54.000  & $-$17:55:48.000  & $   34\pm 5$  &    ScN   &  2.92 $\pm$ 0.31 & 5.37 $\pm$ 0.30   \\
G013.71$-$00.08  & 18:15:36.9814  & $-$17:04:32.108  & $   44\pm 5$  & Nor  & 3.79 $\pm$ 0.20 & 4.59 $\pm$ 0.18 \\
G015.66$-$00.49  & 18:20:59.7470  & $-$15:33:09.800  & $   -4\pm 5$  & 3kN  & 4.55 $\pm$ 0.60 & 3.99 $\pm$ 0.50 \\
G016.86$-$02.15  & 18:29:24.4085  & $-$15:16:04.141  & $   17\pm 3$  & SgN  & 2.35 $\pm$ 0.51 & 5.97 $\pm$ 0.47 \\
G017.02$-$02.40  & 18:30:36.2931  & $-$15:14:28.384  & $   22\pm 3$  & SgN  & 1.88 $\pm$ 0.38 & 6.40 $\pm$ 0.36 \\
G017.55$-$00.12  & 18:23:17.9084  & $-$13:42:47.146  & $   44\pm 10$  & SgN  & 2.01 $\pm$ 0.15 & 6.29 $\pm$ 0.14 \\
G017.63$+$00.15  & 18:22:26.3821  & $-$13:30:11.951  & $   25\pm 10$  & SgN  & 1.49 $\pm$ 0.04 & 6.77 $\pm$ 0.04 \\
G018.34$+$01.76  & 18:17:58.1254  & $-$12:07:24.893  & $   28\pm 3$  & SgN  & 2.00 $\pm$ 0.08 & 6.31 $\pm$ 0.07 \\
G019.00$-$00.02  & 18:25:44.7778  & $-$12:22:45.886  & $   93\pm 5$  & Nor  & 4.05 $\pm$ 1.03 & 4.55 $\pm$ 0.84 \\
G019.36$-$00.03  & 18:26:25.7796  & $-$12:03:53.267  & $   27\pm 3$  & ScN  & 2.84 $\pm$ 0.56 & 5.58 $\pm$ 0.49 \\
G019.49$+$00.11  & 18:26:09.1691  & $-$11:52:51.354  & $  121\pm 5$  & $\cdots$  & 3.07 $\pm$ 0.94 & 5.38 $\pm$ 0.81 \\
G022.35$+$00.06  & 18:31:44.1199  & $-$09:22:12.336  & $   80\pm 5$  & Nor  & 4.33 $\pm$ 2.02 & 4.49 $\pm$ 1.46 \\
G023.20$-$00.37  & 18:34:55.1794  & $-$08:49:15.206  & $   82\pm 10$  & Nor  & 4.18 $\pm$ 0.60 & 4.64 $\pm$ 0.43 \\
G023.25$-$00.24  & 18:34:31.2397  & $-$08:42:47.306  & $   63\pm 3$  & ScN  & 5.92 $\pm$ 1.79 & 3.60 $\pm$ 0.79 \\
G023.38$+$00.18  & 18:33:14.3240  & $-$08:23:57.500  & $   75\pm 3$  & Nor  & 4.81 $\pm$ 0.58 & 4.22 $\pm$ 0.37 \\
G023.43$-$00.18  & 18:34:39.1870  & $-$08:31:25.405  & $   97\pm 3$  & Nor  & 5.88 $\pm$ 1.11 & 3.63 $\pm$ 0.49 \\
G024.63$-$00.32  & 18:37:22.7091  & $-$07:31:42.093  & $   43\pm 5$  & ScN  & 4.13 $\pm$ 0.77 & 4.75 $\pm$ 0.53 \\
G024.78$+$00.08  & 18:36:12.5614  & $-$07:12:10.840  & $  111\pm 3$  & Nor  & 6.67 $\pm$ 0.71 & 3.51 $\pm$ 0.15 \\
G024.85$+$00.08  & 18:36:18.3867  & $-$07:08:50.834  & $  111\pm 5$  & Nor  & 5.68 $\pm$ 0.52 & 3.85 $\pm$ 0.23 \\
G028.14$-$00.00  & 18:42:42.5896  & $-$04:15:35.128  & $  100\pm 5$  & ScF  & 6.33 $\pm$ 0.92 & 3.96 $\pm$ 0.21 \\
G028.30$-$00.38  & 18:44:21.9666  & $-$04:17:39.904  & $   87\pm 5$  & ScN  & 4.52 $\pm$ 0.45 & 4.71 $\pm$ 0.26 \\
G028.39$+$00.08  & 18:42:51.9822  & $-$03:59:54.494  & $   75\pm 5$  & ScN  & 4.33 $\pm$ 0.28 & 4.83 $\pm$ 0.17 \\
G028.83$-$00.25  & 18:44:51.0865  & $-$03:45:48.378  & $   87\pm 5$  & ScN  & 5.00 $\pm$ 1.00 & 4.50 $\pm$ 0.48 \\
G029.98$+$00.10  & 18:45:39.9622  & $-$02:34:32.581  & $  109\pm 10$  & ScF  & 6.41 $\pm$ 0.41 & 4.14 $\pm$ 0.07 \\
G030.19$-$00.16  & 18:47:03.0698  & $-$02:30:36.268  & $  108\pm 3$  & ScN  & 4.72 $\pm$ 0.22 & 4.74 $\pm$ 0.11 \\
G030.22$-$00.18  & 18:47:08.2979  & $-$02:29:29.330  & $  113\pm 3$  & ScN  & 3.52 $\pm$ 0.40 & 5.43 $\pm$ 0.26 \\
G030.41$-$00.23  & 18:47:40.7589  & $-$02:20:30.907  & $  103\pm 3$  & ScN  & 3.95 $\pm$ 0.33 & 5.17 $\pm$ 0.20 \\
G030.70$-$00.06  & 18:47:36.7983  & $-$02:00:54.341  & $   89\pm 3$  & ScF  & 6.54 $\pm$ 0.85 & 4.20 $\pm$ 0.10 \\
G030.74$-$00.04  & 18:47:39.7248  & $-$01:57:24.974  & $   88\pm 3$  & ScN  & 3.07 $\pm$ 0.52 & 5.76 $\pm$ 0.36 \\
G030.78$+$00.20  & 18:46:48.0864  & $-$01:48:53.946  & $   82\pm 5$  & ScN  & 7.14 $\pm$ 1.63 & 4.19 $\pm$ 0.05 \\
G030.81$-$00.05  & 18:47:46.9751  & $-$01:54:26.416  & $  105\pm 5$  & ScF  & 3.12 $\pm$ 0.36 & 5.73 $\pm$ 0.24 \\
G030.97$-$00.14  & 18:48:22.0433  & $-$01:48:30.750  & $   77\pm 3$  & ScN  & 3.40 $\pm$ 0.25 & 5.55 $\pm$ 0.17 \\
G031.24$-$00.11  & 18:48:45.0808  & $-$01:33:13.200  & $   24\pm 10$  & Per  & 13.16 $\pm$ 2.42 & 7.48 $\pm$ 2.00 \\
G032.74$-$00.07  & 18:51:21.8624  & $-$00:12:06.220  & $   56\pm 5$  & SgF  & 7.94 $\pm$ 1.01 & 4.55 $\pm$ 0.23 \\
G032.79$+$00.19  & 18:50:30.7330  & $-$00:01:59.280  & $   16\pm 10$  & Per  & 9.71 $\pm$ 2.92 & 5.26 $\pm$ 1.57 \\
G033.39$+$00.00  & 18:52:14.6412  & $+$00:24:54.374  & $   43\pm 5$  & ScF  & 8.85 $\pm$ 2.27 & 4.93 $\pm$ 0.93 \\
G034.41$+$00.23  & 18:53:18.0319  & $+$01:25:25.500  & $   60\pm 10$  & SgN  & 2.94 $\pm$ 0.10 & 5.99 $\pm$ 0.06 \\
G034.79$-$01.38  & 18:59:45.9838  & $+$01:01:18.947  & $   45\pm 5$  & SgN  & 2.62 $\pm$ 0.14 & 6.21 $\pm$ 0.09 \\
G035.79$-$00.17  & 18:57:16.8905  & $+$02:27:58.007  & $   61\pm 5$  & SgF  & 8.85 $\pm$ 1.02 & 5.27 $\pm$ 0.43 \\
G036.11$+$00.55  & 18:55:16.7927  & $+$03:05:05.392  & $   75\pm 5$  & AqS  & 4.07 $\pm$ 0.93 & 5.45 $\pm$ 0.43 \\
G037.42$+$01.51  & 18:54:14.3481  & $+$04:41:39.647  & $   41\pm 3$  & SgN  & 1.88 $\pm$ 0.07 & 6.78 $\pm$ 0.05 \\
G037.47$-$00.10  & 19:00:07.1430  & $+$03:59:52.975  & $   58\pm 3$  & SgF  & 11.36 $\pm$ 3.87 & 6.97 $\pm$ 2.71 \\
G038.03$-$00.30  & 19:01:50.4676  & $+$04:24:18.900  & $   30\pm 7$  & SgF  & 10.53 $\pm$ 2.44 & 6.49 $\pm$ 1.54 \\
G038.11$-$00.22  & 19:01:44.1513  & $+$04:30:37.400  & $   76\pm 5$  & AqS  & 4.13 $\pm$ 0.60 & 5.55 $\pm$ 0.25 \\
G040.28$-$00.21  & 19:05:41.2146  & $+$06:26:12.698  & $   74\pm 5$  & AqS  & 3.37 $\pm$ 0.22 & 6.02 $\pm$ 0.10 \\
G040.42$+$00.70  & 19:02:39.6192  & $+$06:59:09.052  & $   -7\pm 5$  & Per  & 12.82 $\pm$ 2.14 & 8.46 $\pm$ 1.67 \\
G040.62$-$00.13  & 19:06:01.6288  & $+$06:46:36.140  & $   31\pm 3$  & Per  & 12.50 $\pm$ 3.28 & 8.24 $\pm$ 2.50 \\
G041.15$-$00.20  & 19:07:14.3676  & $+$07:13:18.025  & $   60\pm 3$  & SgF  & 8.00 $\pm$ 1.15 & 5.69 $\pm$ 0.37 \\
G041.22$-$00.19  & 19:07:21.3772  & $+$07:17:08.115  & $   59\pm 5$  & SgF  & 8.85 $\pm$ 1.72 & 6.03 $\pm$ 0.77 \\
G042.03$+$00.19  & 19:07:28.1834  & $+$08:10:53.433  & $  -36\pm 5$  & Per  & 14.08 $\pm$ 2.38 & 9.70 $\pm$ 1.97 \\
G043.03$-$00.45  & 19:11:38.9819  & $+$08:46:30.665  & $   56\pm 5$  & SgF  & 7.69 $\pm$ 1.12 & 5.84 $\pm$ 0.33 \\
G045.45$+$00.06  & 19:14:21.2658  & $+$11:09:15.872  & $   19\pm 5$  & SgF  & 8.40 $\pm$ 1.20 & 6.41 $\pm$ 0.50 \\
G045.49$+$00.12  & 19:14:11.3553  & $+$11:13:06.370  & $   58\pm 3$  & SgF  & 6.94 $\pm$ 1.16 & 5.96 $\pm$ 0.24 \\
G045.80$-$00.35  & 19:16:31.0795  & $+$11:16:11.985  & $   64\pm 5$  & SgF  & 7.30 $\pm$ 1.23 & 6.08 $\pm$ 0.32 \\
G048.99$-$00.29  & 19:22:26.1348  & $+$14:06:39.133  & $   67\pm 10$  & SgF  & 5.62 $\pm$ 0.54 & 6.18 $\pm$ 0.02 \\
G049.04$-$01.07  & 19:25:22.2504  & $+$13:47:19.525  & $   56\pm 3$  & SgN  & 6.10 $\pm$ 0.82 & 6.22 $\pm$ 0.10 \\
G049.26$+$00.31  & 19:20:44.8571  & $+$14:38:26.864  & $    0\pm 5$  & Per  & 8.85 $\pm$ 1.25 & 7.12 $\pm$ 0.62 \\
G049.34$+$00.41  & 19:20:32.4472  & $+$14:45:45.390  & $   68\pm 5$  & SgF  & 4.15 $\pm$ 0.53 & 6.32 $\pm$ 0.10 \\
G049.41$+$00.32  & 19:20:59.2098  & $+$14:46:49.613  & $  -12\pm 5$  & Per  & 7.58 $\pm$ 1.78 & 6.61 $\pm$ 0.61 \\
G049.59$-$00.24  & 19:23:26.6068  & $+$14:40:16.955  & $   63\pm 5$  & SgF  & 4.59 $\pm$ 0.19 & 6.27 $\pm$ 0.02 \\
W51-IRS2         &  19:23:39.8000  &  $+$14:31:05.000  & $   38\pm 5$  & SgN  & 5.13 $\pm$ 1.87 & 6.22 $\pm$ 0.06 \\
G054.10$-$00.08  & 19:31:48.7978  & $+$18:42:57.096  & $   40\pm 5$  & LoS  & 4.33 $\pm$ 0.58 & 6.64 $\pm$ 0.04 \\
G058.77$+$00.64  & 19:38:49.1269  & $+$23:08:40.205  & $   33\pm 3$  & LoS  & 3.34 $\pm$ 0.45 & 7.05 $\pm$ 0.06 \\
G059.47$-$00.18  & 19:43:28.3504  & $+$23:20:42.522  & $   26\pm 3$  & LoS  & 1.87 $\pm$ 0.08 & 7.41 $\pm$ 0.03 \\
G059.83$+$00.67  & 19:40:59.2938  & $+$24:04:44.177  & $   34\pm 3$  & LoS  & 4.13 $\pm$ 0.24 & 7.07 $\pm$ 0.00 \\
G060.57$-$00.18  & 19:45:52.4949  & $+$24:17:43.237  & $    4\pm 5$  & Per  & 8.26 $\pm$ 1.02 & 8.29 $\pm$ 0.52 \\
G070.18$+$01.74  & 20:00:54.4874  & $+$33:31:28.224  & $  -23\pm 5$  & Per  & 6.41 $\pm$ 0.66 & 8.51 $\pm$ 0.28 \\
G071.31$+$00.82  & 20:07:31.2593  & $+$33:59:41.491  & $    9\pm 5$  & Loc  & 4.69 $\pm$ 0.62 & 8.02 $\pm$ 0.16 \\
G071.52$-$00.38  & 20:12:57.8943  & $+$33:30:27.083  & $   11\pm 3$  & Loc  & 3.61 $\pm$ 0.34 & 7.82 $\pm$ 0.04 \\
G073.65$+$00.19  & 20:16:21.9320  & $+$35:36:06.094  & $  -76\pm 10$  & Out  & 13.33 $\pm$ 3.56 & 13.54 $\pm$ 2.90 \\
G074.56$+$00.84  & 20:16:13.3617  & $+$36:43:33.920  & $   -1\pm 5$  & Loc  & 2.72 $\pm$ 0.62 & 7.90 $\pm$ 0.04 \\
G075.29$+$01.32  & 20:16:16.0120  & $+$37:35:45.810  & $  -58\pm 5$  & Out  & 9.26 $\pm$ 0.86 & 10.69 $\pm$ 0.58 \\
DR21             &  20:39:01.8000  &  $+$42:19:40.900  & $   -3\pm 3$  &    Loc   &  1.50 $\pm$ 0.08 & 8.10 $\pm$ 0.00 \\
G090.92$+$01.48  & 21:09:12.9685  & $+$50:01:03.664  & $  -70\pm 5$  & Out  & 5.85 $\pm$ 1.06 & 10.13 $\pm$ 0.63 \\
G097.53$+$03.18  & 21:32:12.4343  & $+$55:53:49.689  & $  -73\pm 5$  & Out  & 7.52 $\pm$ 0.96 & 11.81 $\pm$ 0.70 \\
G108.20$+$00.58  & 22:49:31.4775  & $+$59:55:42.006  & $  -49\pm 10$  & Per  & 4.41 $\pm$ 0.72 & 10.43 $\pm$ 0.48 \\
G108.42$+$00.89  & 22:49:58.8760  & $+$60:17:56.650  & $  -51\pm 5$  & Per  & 2.50 $\pm$ 0.31 & 9.28 $\pm$ 0.17 \\
G109.87$+$02.11  & 22:56:18.0559  & $+$62:01:49.562  & $   10\pm 5$  & Loc  & 0.81 $\pm$ 0.02 & 8.49 $\pm$ 0.01 \\
NGC7538          &  23:13:45.4000  &  $+$61:29:11.400  & $  -57\pm 5$  &    Per   &  2.65 $\pm$ 0.12 & 9.47 $\pm$ 0.07   \\
\hline
\end{longtable}
\tablefoot{Column (1): source name; Columns (2) and (3): Equatorial coordinates; Column (4): Local Standard of Rest velocity; Column (5): the spiral arm, GC, Con, 3kN/F, Nor, ScN/F, SgN/F, Loc, Per, Out, OSC, LoS, and AqS indicate Galactic center region, connecting arm, 3 kpc arm (near/far), Norma arm, Scutum-Centaurus arm (near/far), Sagittarius arm (near/far), Local arm, Perseus arm, Outer arm, Outer-Scutum-Centaurus arm, Local arm spur, and Aquarius spur, respectively. See details in \citet{2019ApJ...885..131R}; Column (6): the heliocentric distance, derived with parallaxes from the BeSSeL Survey \citep{2019ApJ...885..131R}; Column (7): galactocentric distance following equation (1). \tablefoottext{*}{For those 12 sources without parallax data and an asterisk in column 1, kinematic distances were estimated (see details in Section~\ref{section_distance}). }}
\end{center}

\begin{center}
\small
\begin{longtable}{lcc|ccc|cc|cc}
\caption{\label{table_doubleisotope} $^{32}$S/$^{34}$S isotope ratios derived through the double isotope method with the $J$ = 2-1 and $J$ = 3-2 transitions of C$^{34}$S and $^{13}$CS.}\\
\hline\hline
Source &  $V_{\rm LSR}$ & $R_{\rm GC}$  &   \multicolumn{3}{c}{Optical depth $J$ = 2-1} &  \multicolumn{2}{c}{Corrections for} &  \multicolumn{2}{c}{$^{32}$S/$^{34}$S}  \\
  &                &         &     &           &           &  \multicolumn{2}{c}{optical depth} &      &    \\
  &  (km s$^{-1}$) & (kpc)   &  CS & C$^{34}$S & $^{13}$CS &  $f_1$  &  $f_2$ &  2-1 & 3-2 \\
\hline
\endfirsthead
\caption{continued.}\\
\hline\hline
Source &  $V_{\rm LSR}$ & $R_{\rm GC}$  &   \multicolumn{3}{c}{Optical depth $J$ = 2-1} &  \multicolumn{2}{c}{Corrections for} &  \multicolumn{2}{c}{$^{32}$S/$^{34}$S}  \\
  &                &         &     &           &           &  \multicolumn{2}{c}{optical depth} &      &    \\
  &  (km s$^{-1}$) & (kpc)   &  CS & C$^{34}$S & $^{13}$CS &  $f_1$  &  $f_2$ &  2-1 & 3-2 \\
\hline
\endhead
\hline
\endfoot
WB89-380                & -86.62 & 14.19 $\pm$ 0.92 & 3.111 $\pm$ 0.598  & 0.135 $\pm$ 0.137 & 0.035 $\pm$ 0.000 & 1.1 & 1.0 &  21.20 $\pm$ 4.01  & 24.98 $\pm$ 4.69  \\ 
WB89-391                & -85.93 & 14.28 $\pm$ 0.94 & 5.189 $\pm$ 0.996  & 0.223 $\pm$ 0.226 & 0.058 $\pm$ 0.001 & 1.1 & 1.0 &  24.39 $\pm$ 4.64  & 20.98 $\pm$ 4.04  \\ 
W3OH                    & -46.96 & 9.64  $\pm$ 0.03 & 8.520 $\pm$ 1.671  & 0.310 $\pm$ 0.003 & 0.128 $\pm$ 0.001 & 1.2 & 1.1 &  26.15 $\pm$ 4.87  & 31.42 $\pm$ 5.85  \\ 
Orion-KL                & 8.24   & 8.54  $\pm$ 0.00 & 6.035 $\pm$ 1.189  & 0.166 $\pm$ 0.002 & 0.096 $\pm$ 0.001 & 1.1 & 1.0 &  24.75 $\pm$ 5.24  & 25.41 $\pm$ 4.86  \\ 
PointC1                 & -64.18 & 1.46  $\pm$ 0.15 & 0.588 $\pm$ 0.128  & 0.037 $\pm$ 0.000 & 0.021 $\pm$ 0.000 & 1.0 & 1.0 &  19.40 $\pm$ 4.55  &     $\cdots$      \\ 
G359.13$+$00.03         & -2.03  & 2.12  $\pm$ 1.14 & 5.526 $\pm$ 1.188  & 0.420 $\pm$ 0.004 & 0.179 $\pm$ 0.002 & 1.2 & 1.1 &  14.34 $\pm$ 4.11  &     $\cdots$      \\ 
SgrC                    & -44.31 & 0.24  $\pm$ 0.07 & 0.538 $\pm$ 0.123  & 0.048 $\pm$ 0.000 & 0.025 $\pm$ 0.000 & 1.0 & 1.0 &  12.81 $\pm$ 2.95  &     $\cdots$      \\ 
SgrC                    & 70.06  & 0.15  $\pm$ 0.03 & 0.406 $\pm$ 0.094  & 0.031 $\pm$ 0.000 & 0.019 $\pm$ 0.000 & 1.0 & 1.0 &  14.87 $\pm$ 3.38  &     $\cdots$      \\
\hline
 \end{longtable}
\tablefoot{ Velocities were obtained from measurements of C$^{34}$S, see Table \ref{fitting_all}. The $^{32}$S/$^{34}$S isotope ratios from the $J$ = 2-1 transition are corrected for the optical depth effect and the ones in the 3-2 line without corrections of opacity.  }
\end{center}

\begin{center}
\begin{longtable}{lccccccc}
\caption{\label{fitting_all}Line Parameters derived from Gaussian-fitting.}\\
\hline\hline
    &  transitions    &   Time  &   rms  &  $V_{LSR}$  &  FWHM  &  $\int T_{MB} dv$  &  T$_{MB}$ \\
 Source  &    &  (mins)   &  (mK)    &  (km s$^{-1}$) &  (km s$^{-1}$) &  (K km s$^{-1}$)  &  (K)  \\
\hline
\endfirsthead
\caption{continued.}\\
\hline\hline
    &  transitions    &   Time  &   rms  &  $V_{LSR}$  &  FWHM  &  $\int T_{MB} dv$  &  T$_{MB}$ \\
 Source  &    &  (mins)   &  (mK)    &  (km s$^{-1}$) &  (km s$^{-1}$) &  (K km s$^{-1}$)  &  (K)  \\
\hline
\endhead
\hline
\endfoot
WB89-380 & CS (2-1)  & 533 & 3.48 & -86.48 $\pm$ 0.00 & 3.66 $\pm$ 0.00 &  4.87 $\pm$ 0.00 &  1.250 \\ 
   & C$^{33}$S (2-1)  & 533 & 3.16 & -85.86 $\pm$ 0.12 & 3.60 $\pm$ 0.28 &  0.10 $\pm$ 0.01 &  0.026 \\ 
   & C$^{34}$S (2-1)  & 533 & 2.91 & -86.62 $\pm$ 0.02 & 3.24 $\pm$ 0.04 &  0.57 $\pm$ 0.01 &  0.166 \\ 
  & C$^{36}$S (2-1) & 533  & 3.38 & $\cdots$ & $\cdots$ &  $\cdots$ &  $\cdots$ \\ 
   & $^{13}$CS (2-1)  & 533 & 3.57 & -86.49 $\pm$ 0.08 & 2.99 $\pm$ 0.16 &  0.14 $\pm$ 0.01 &  0.045 \\ 
  & $^{13}$C$^{33}$S (2-1) & 533  & 3.74 & $\cdots$ & $\cdots$ &  $\cdots$ &  $\cdots$ \\ 
  & $^{13}$C$^{34}$S (2-1) & 533  & 3.57 & $\cdots$ & $\cdots$ &  $\cdots$ &  $\cdots$ \\ 
   & C$^{33}$S (3-2)  & 533 & 7.82 & -86.33 $\pm$ 0.15 & 3.87 $\pm$ 0.43 &  0.20 $\pm$ 0.02 &  0.049 \\
\end{longtable}
\tablefoot{Column (1): source name; Column (2): transition; Column (3): on$+$off source integration time; Column (4): rms noise obtained from Gaussian-fitting; Columns (5)--(8): LSR velocity, line width (FWHM), integrated line intensity and peak main beam brightness temperature, respectively. 
}
\end{center}

\section{Figures}
\label{appendix_spectra}

\begin{figure*}
\centering
\includegraphics[width=90pt,height=300pt]{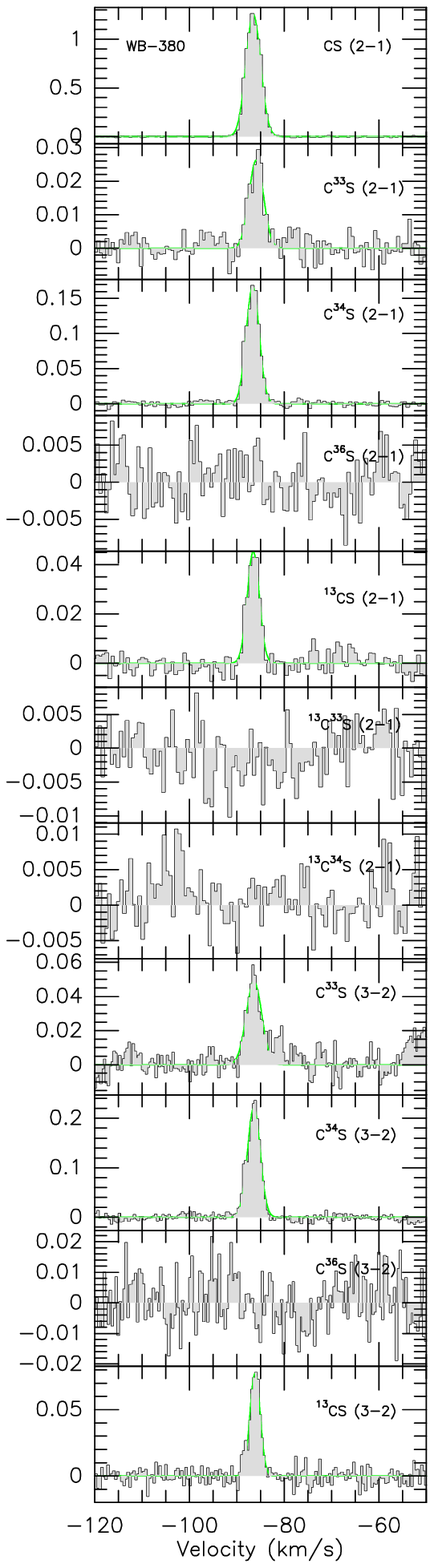}
\includegraphics[width=90pt,height=300pt]{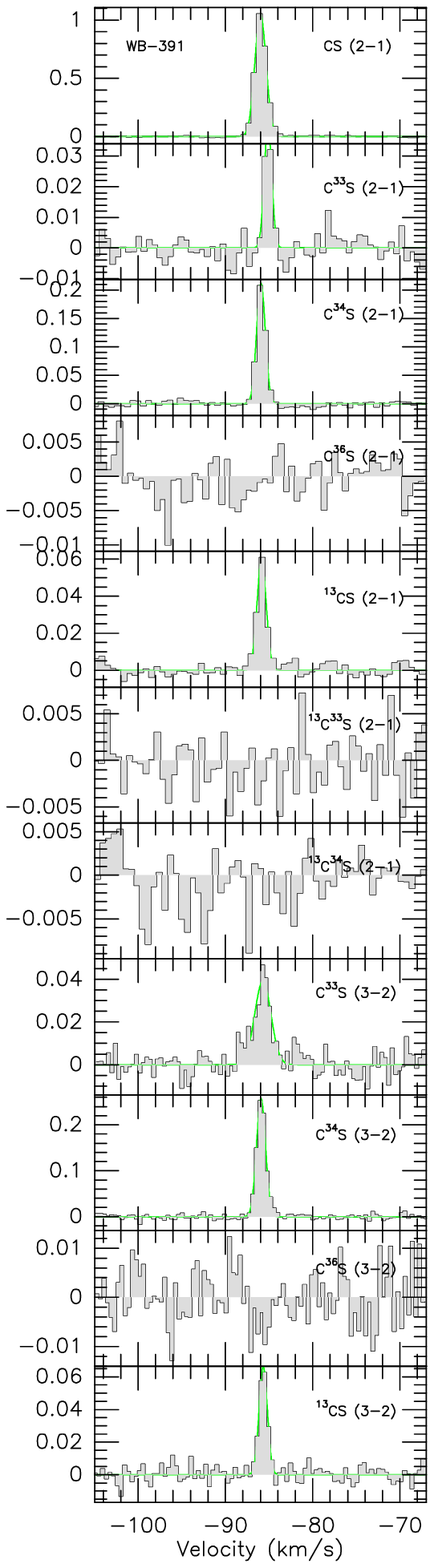}
\includegraphics[width=90pt,height=300pt]{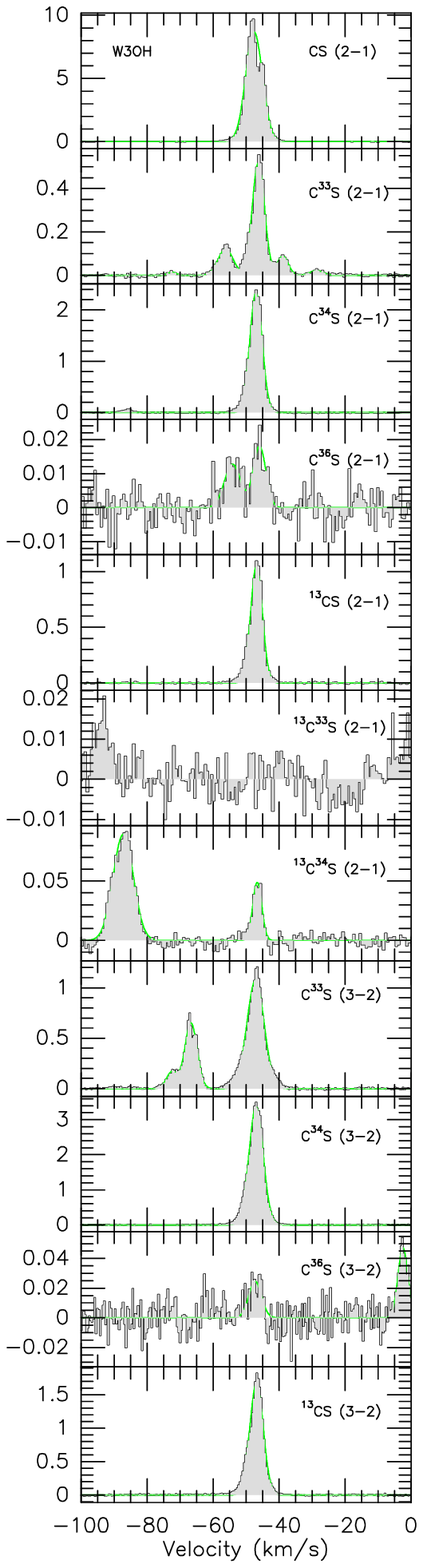}
\includegraphics[width=90pt,height=300pt]{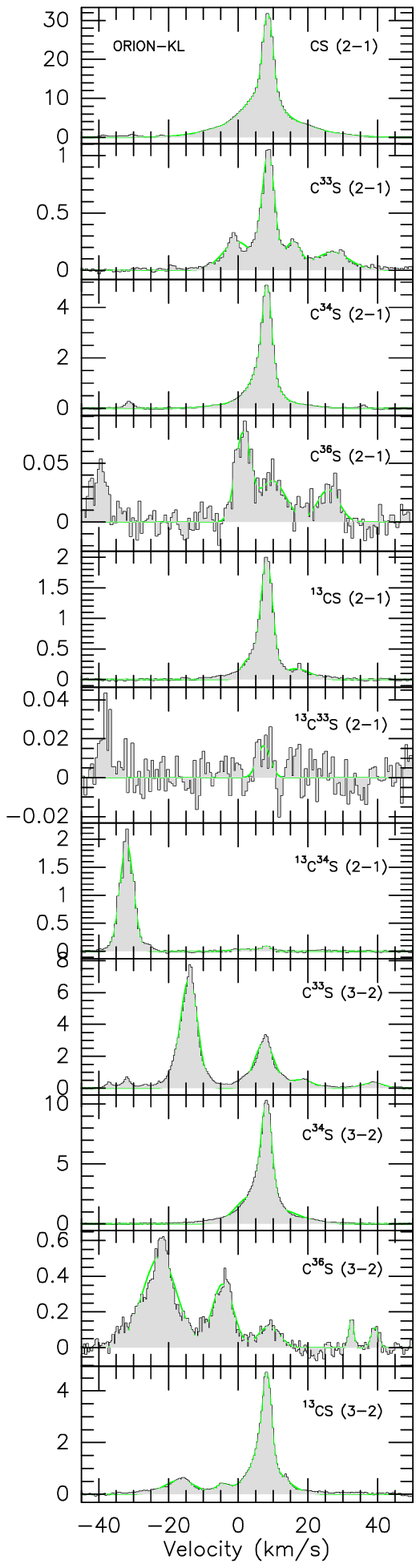}
\includegraphics[width=90pt,height=300pt]{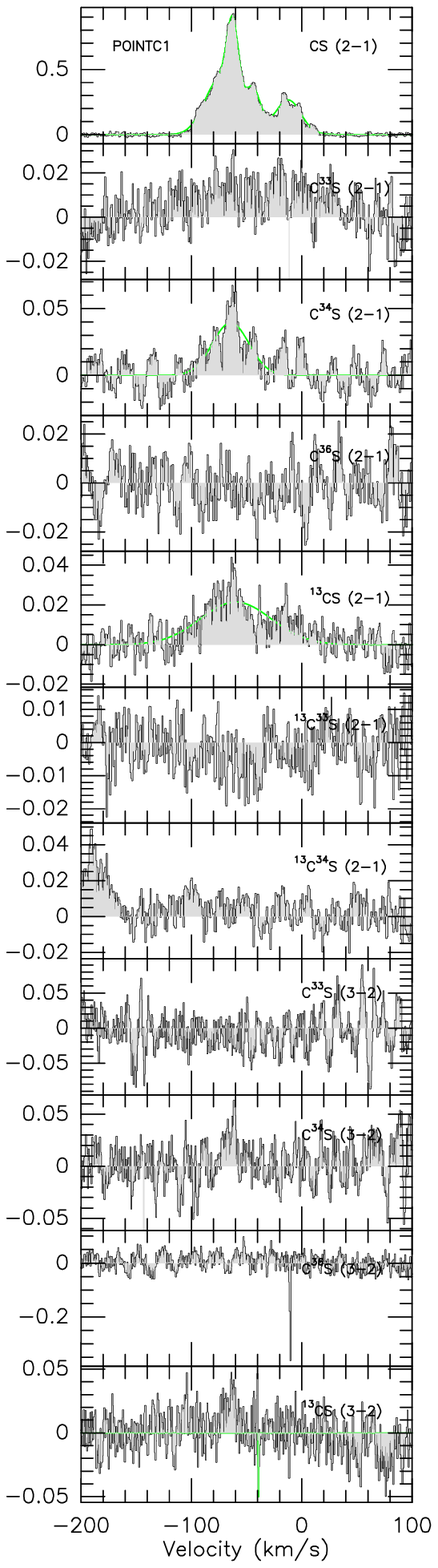}
\includegraphics[width=90pt,height=300pt]{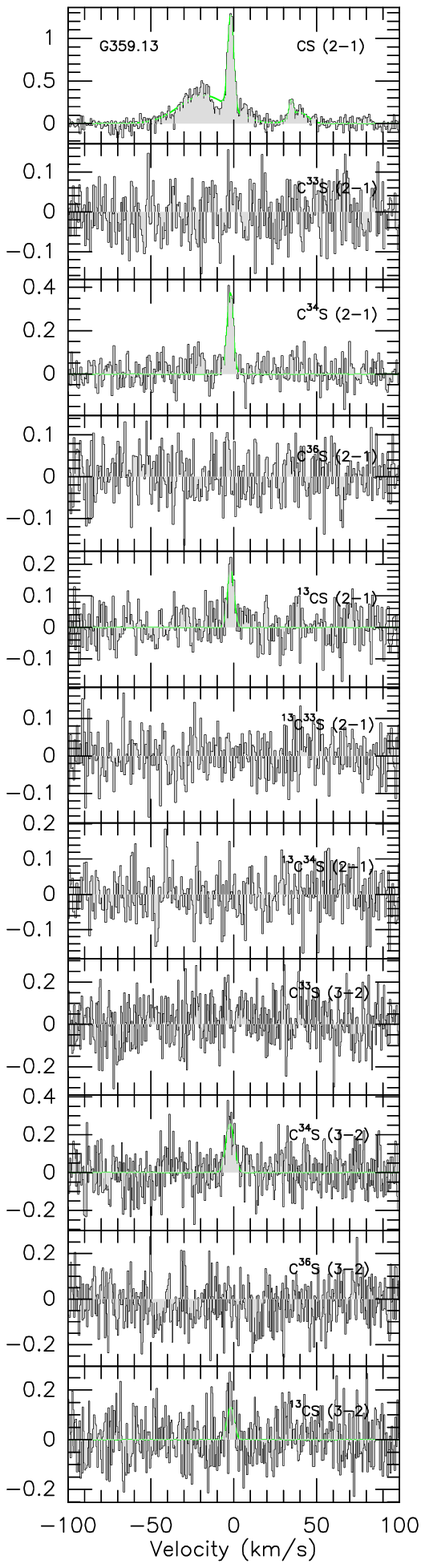}
\includegraphics[width=90pt,height=300pt]{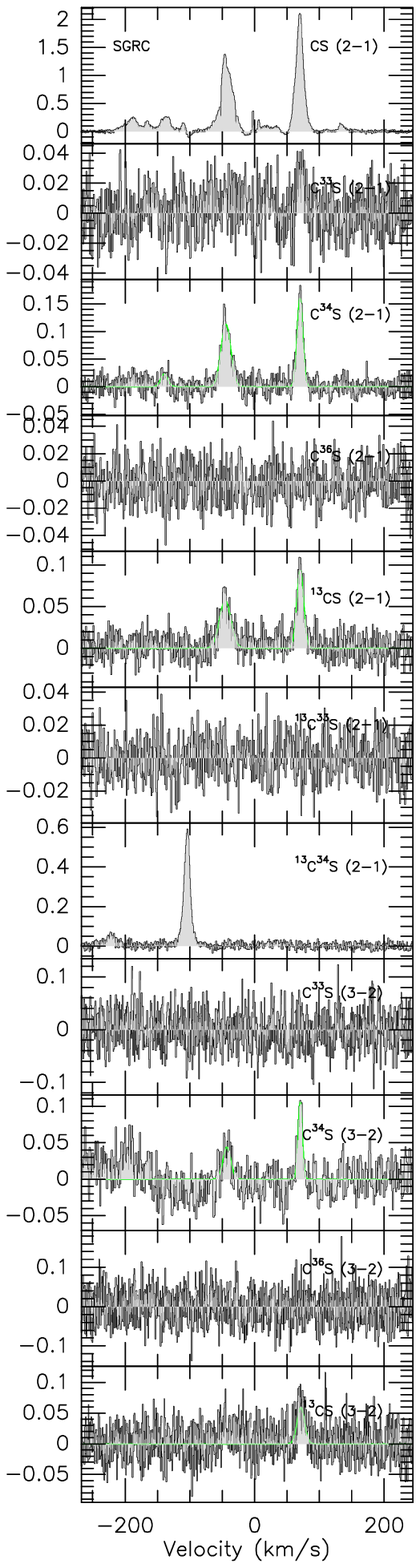}
\includegraphics[width=90pt,height=300pt]{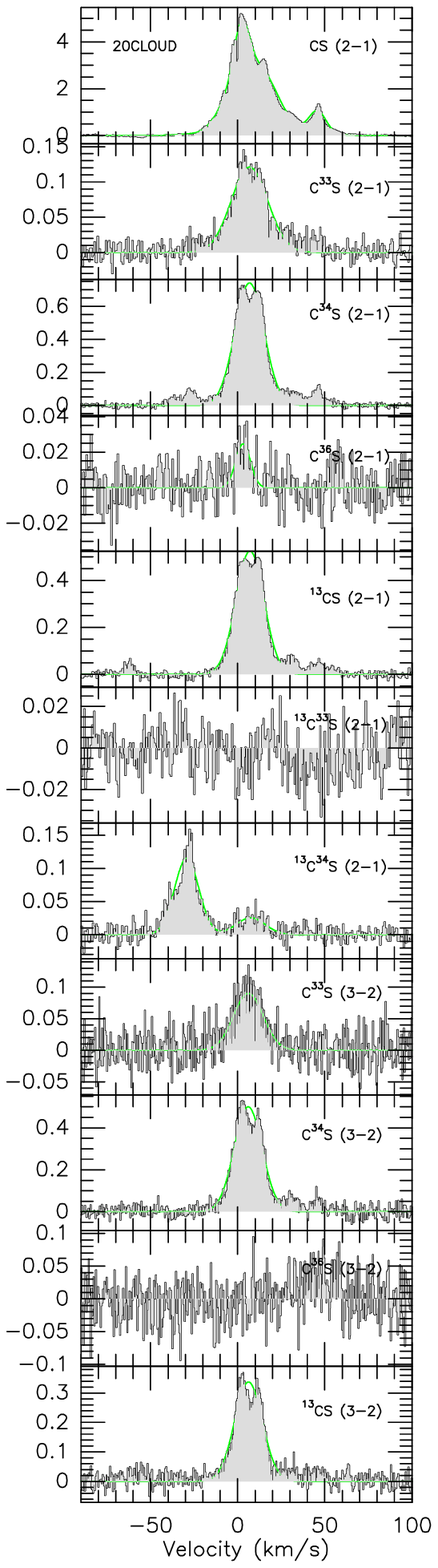}
\includegraphics[width=90pt,height=300pt]{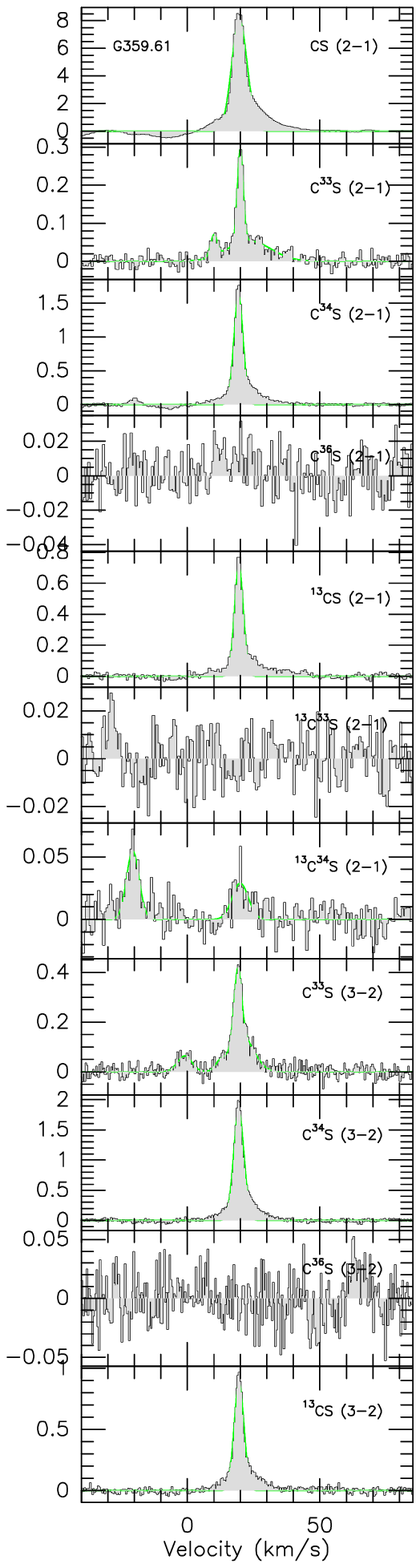}
\includegraphics[width=90pt,height=300pt]{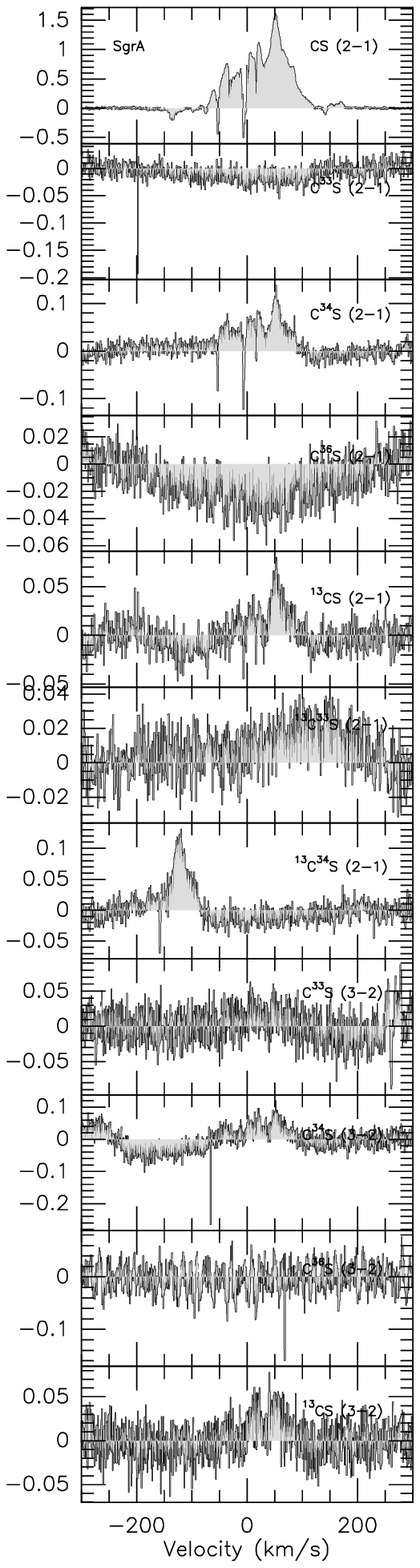}
\caption{\label{spectra_all} Line profiles of the $J$ = 2-1 transitions of CS, C$^{33}$S, C$^{34}$S, C$^{36}$S, $^{13}$CS, $^{13}$C$^{33}$S and $^{13}$C$^{34}$S as well as the $J$ = 3-2 transitions of C$^{33}$S, C$^{34}$S, C$^{36}$S, and $^{13}$CS toward 110 targets of our sample, after subtracting first-order polynomial baselines. The main beam temperature scales are presented on the left hand side of the profiles.}
\end{figure*}

\begin{figure*}
\centering
\includegraphics[width=90pt,height=300pt]{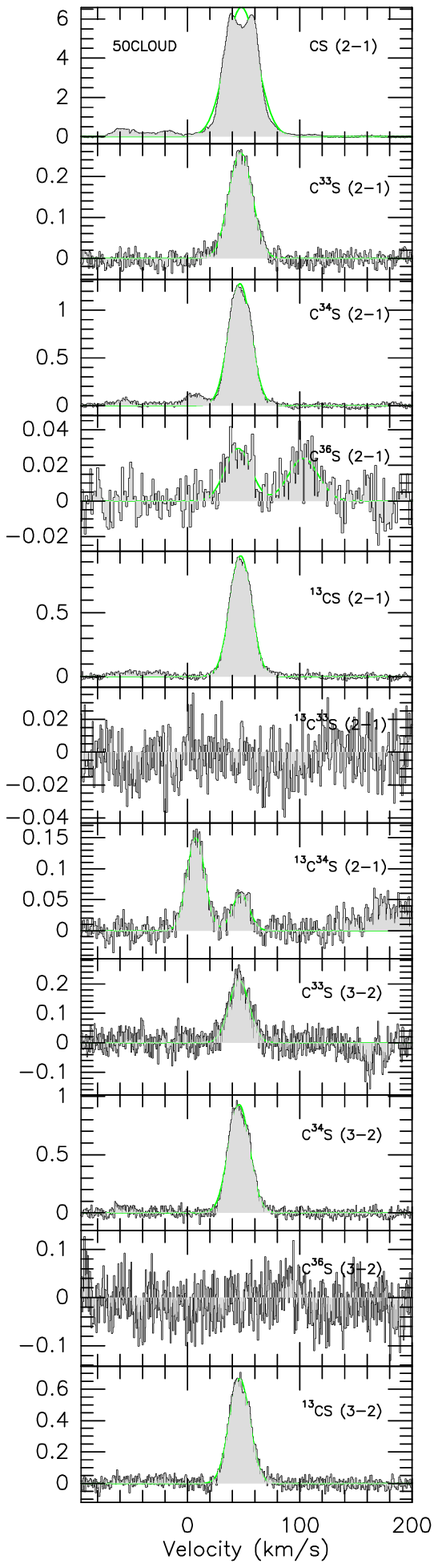}
\includegraphics[width=90pt,height=300pt]{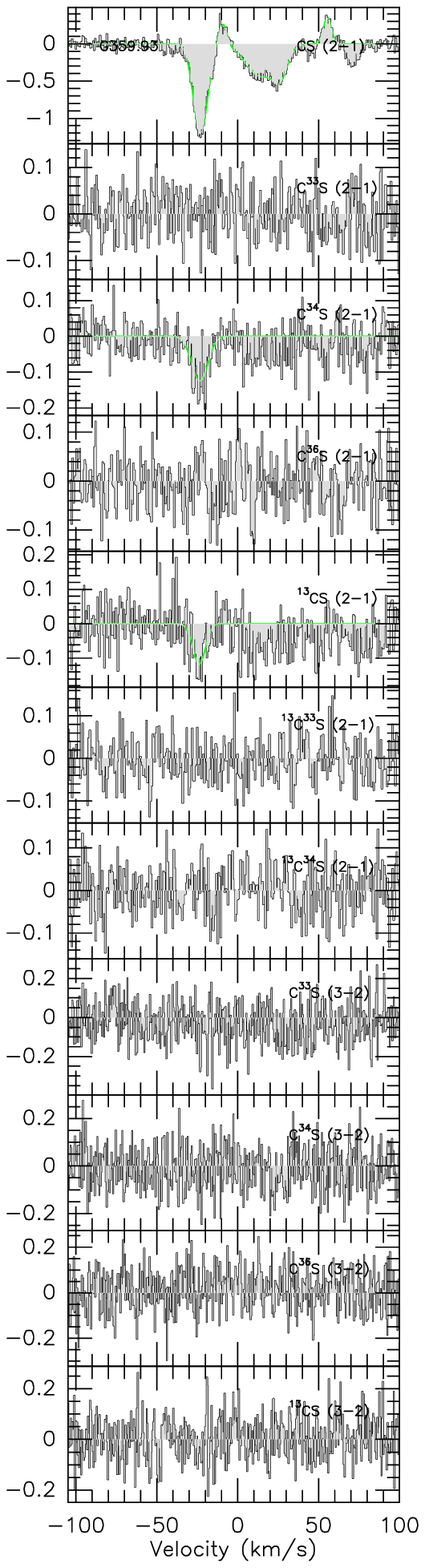}
\includegraphics[width=90pt,height=300pt]{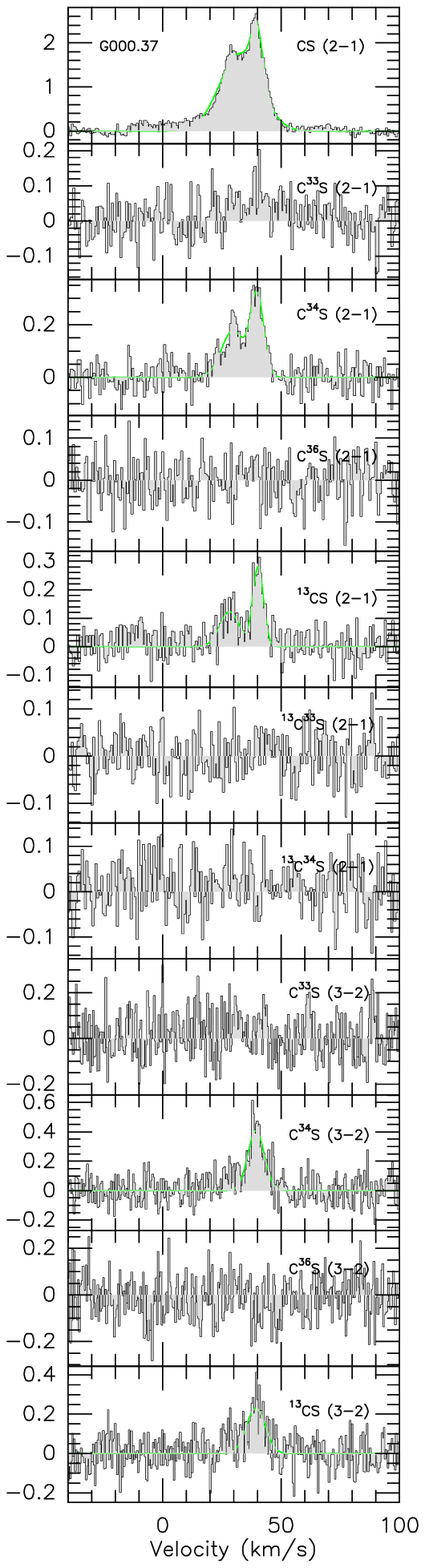}
\includegraphics[width=90pt,height=300pt]{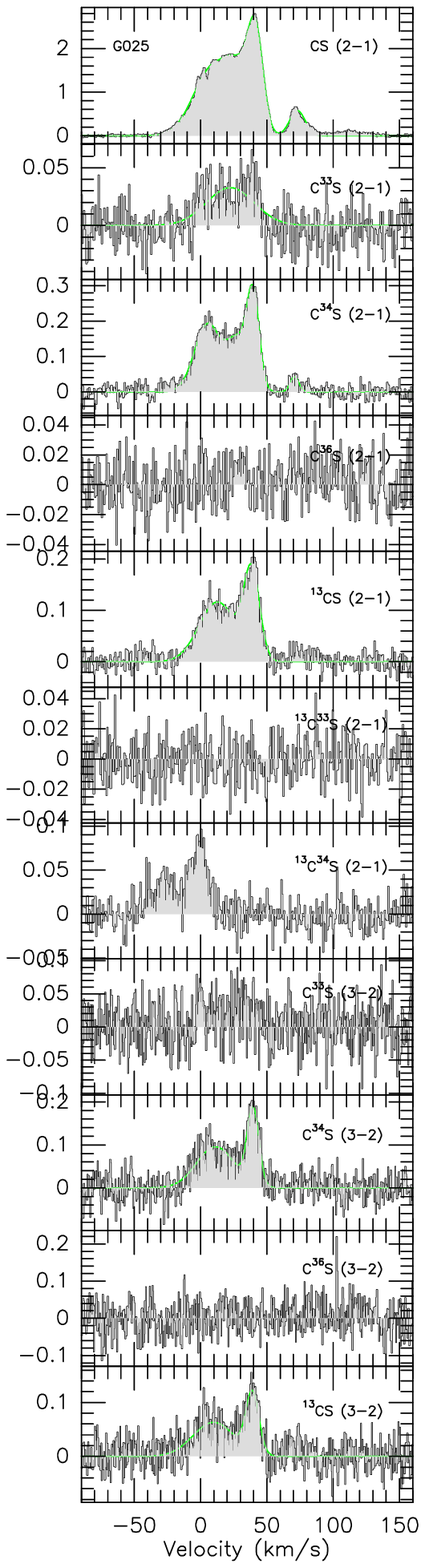}
\includegraphics[width=90pt,height=300pt]{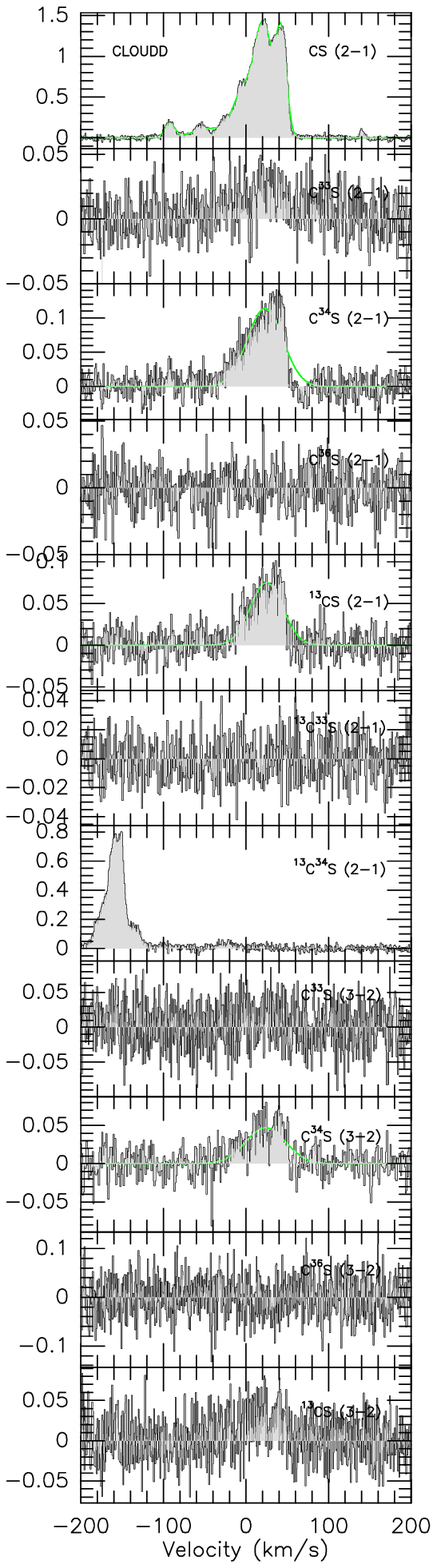}
\includegraphics[width=90pt,height=300pt]{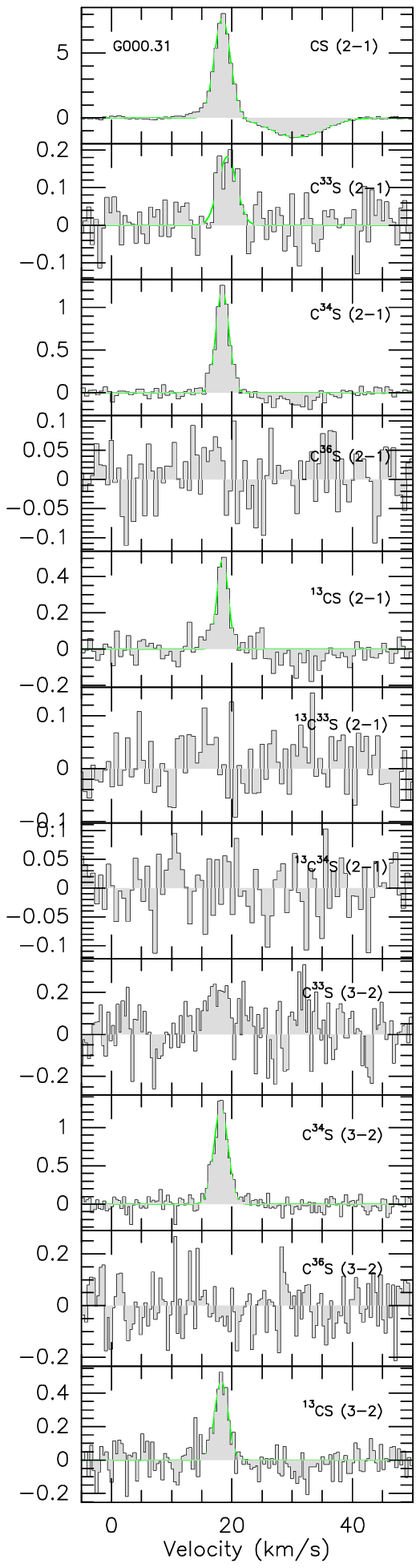}
\includegraphics[width=90pt,height=300pt]{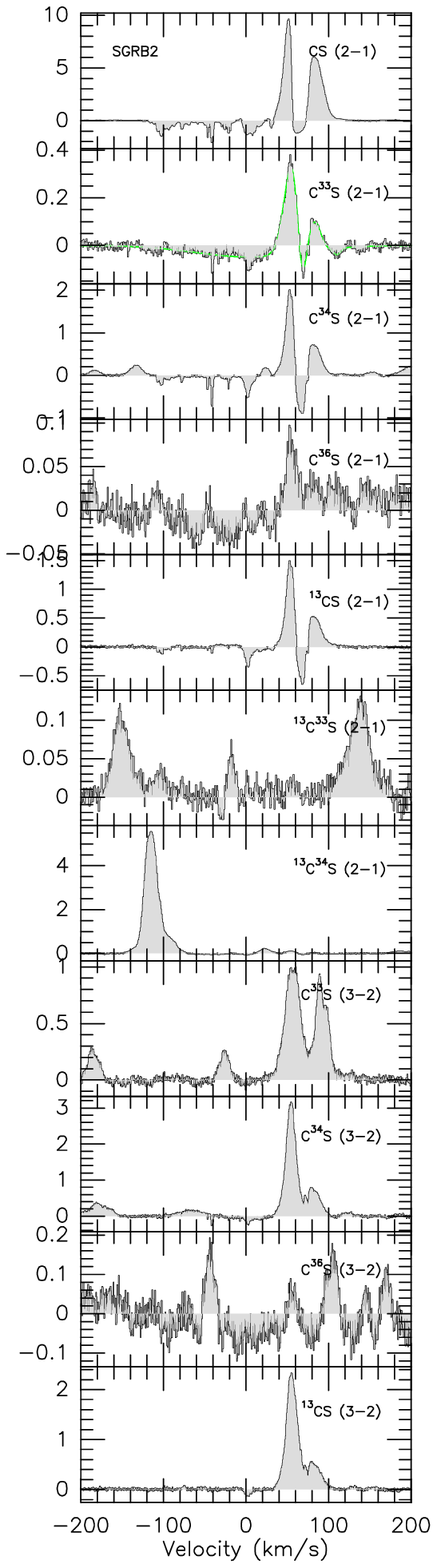}
\includegraphics[width=90pt,height=300pt]{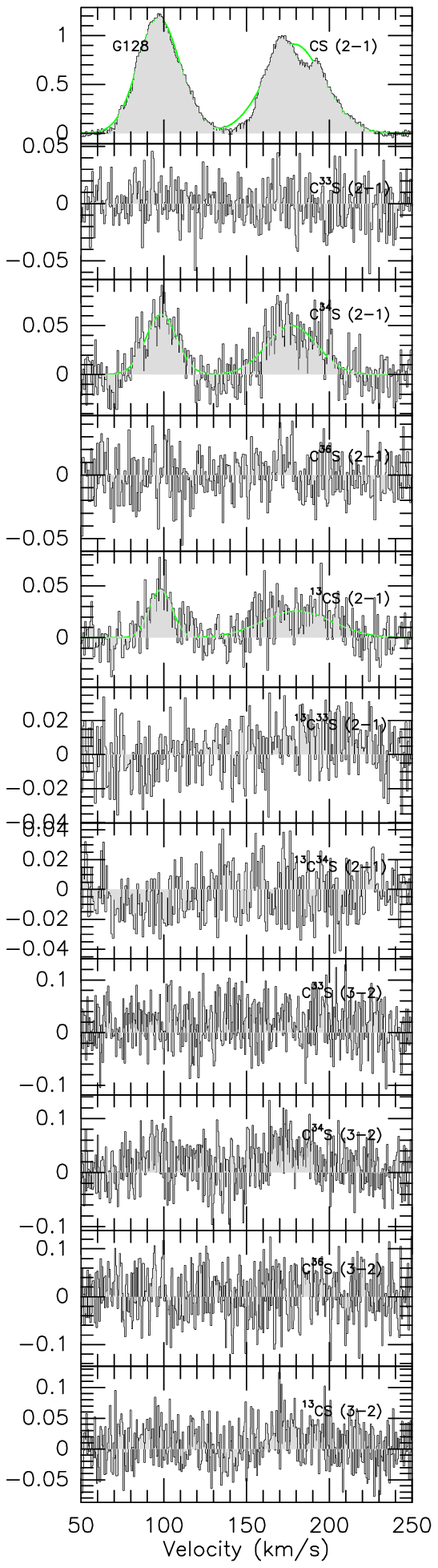}
\includegraphics[width=90pt,height=300pt]{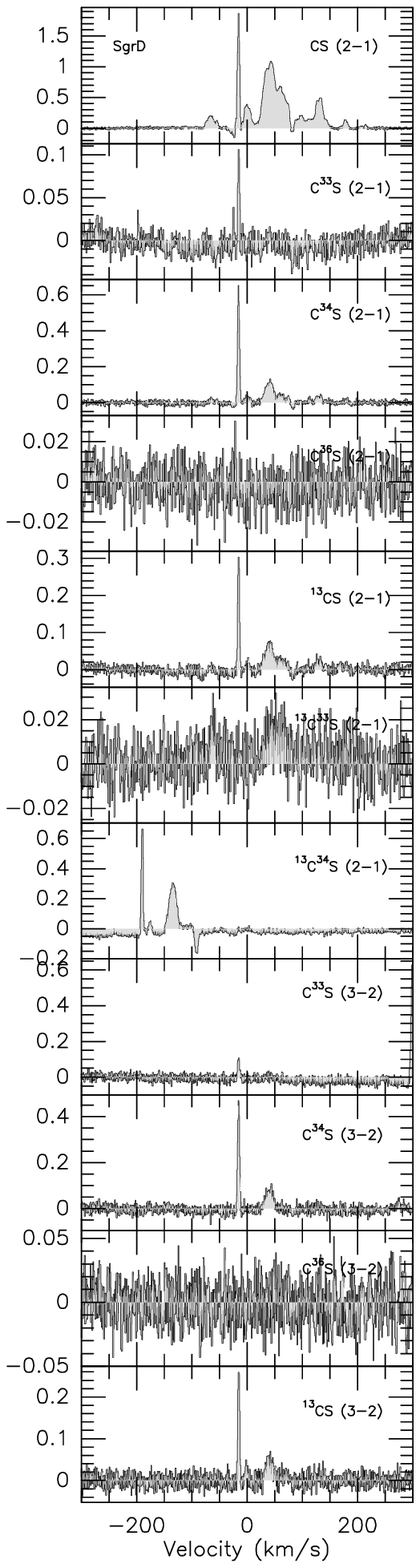}
\includegraphics[width=90pt,height=300pt]{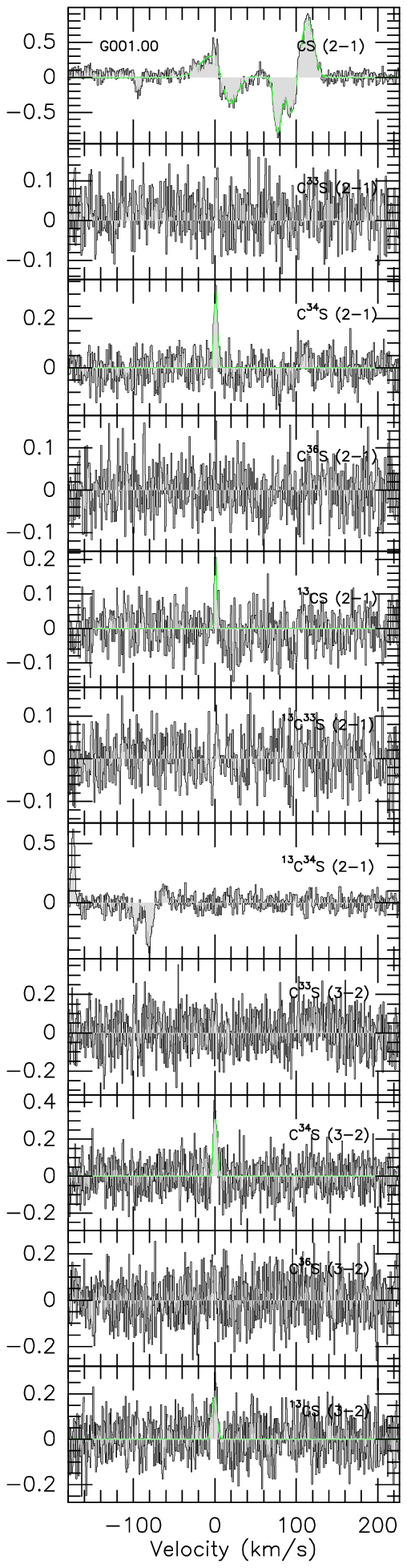}
\end{figure*}

\begin{figure*}
\centering
\includegraphics[width=90pt,height=300pt]{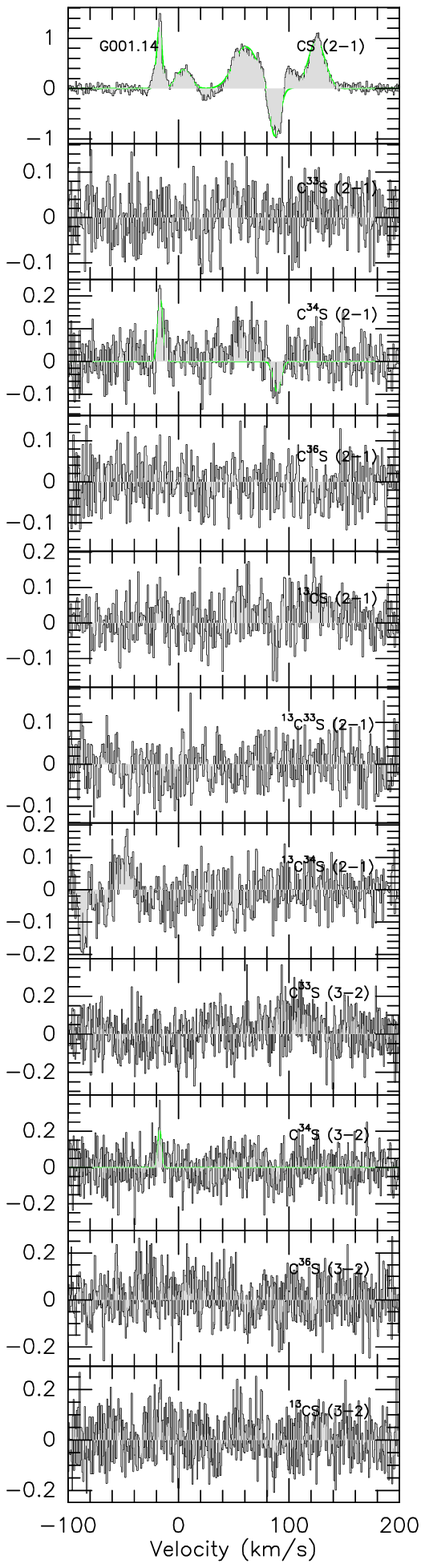}
\includegraphics[width=90pt,height=300pt]{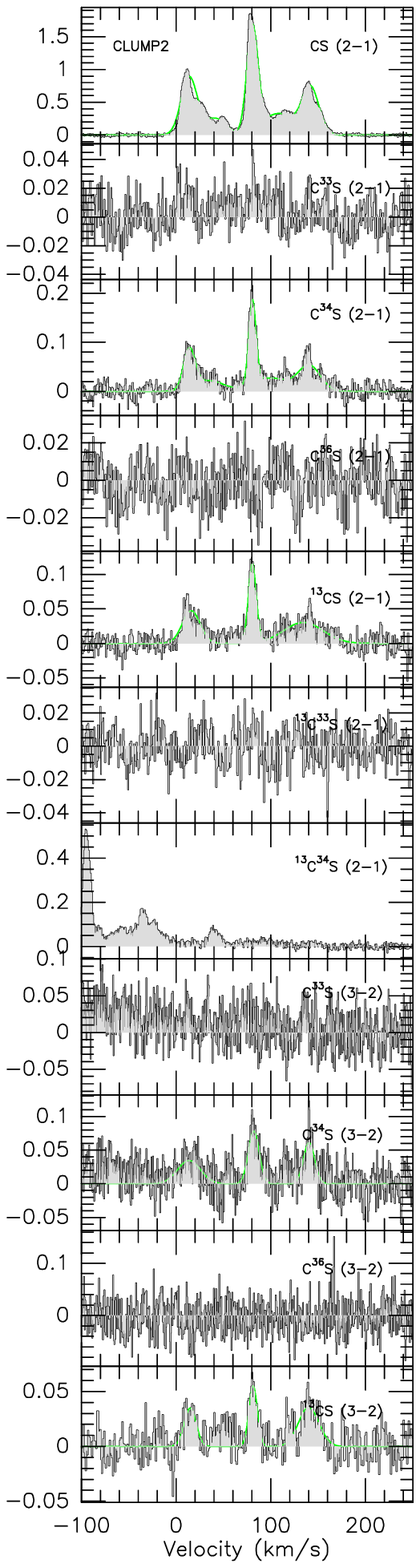}
\includegraphics[width=90pt,height=300pt]{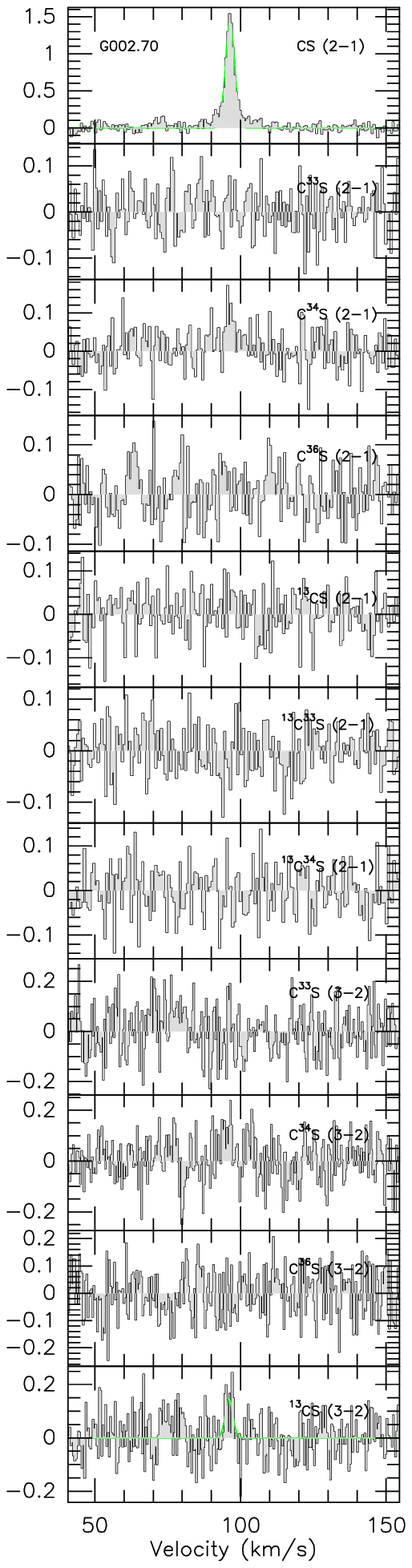}
\includegraphics[width=90pt,height=300pt]{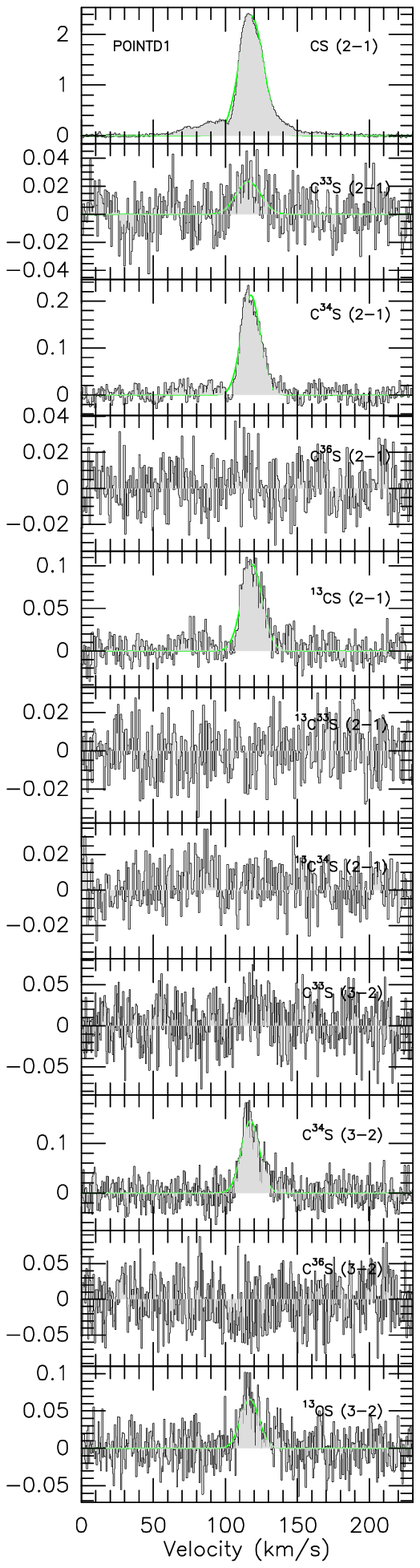}
\includegraphics[width=90pt,height=300pt]{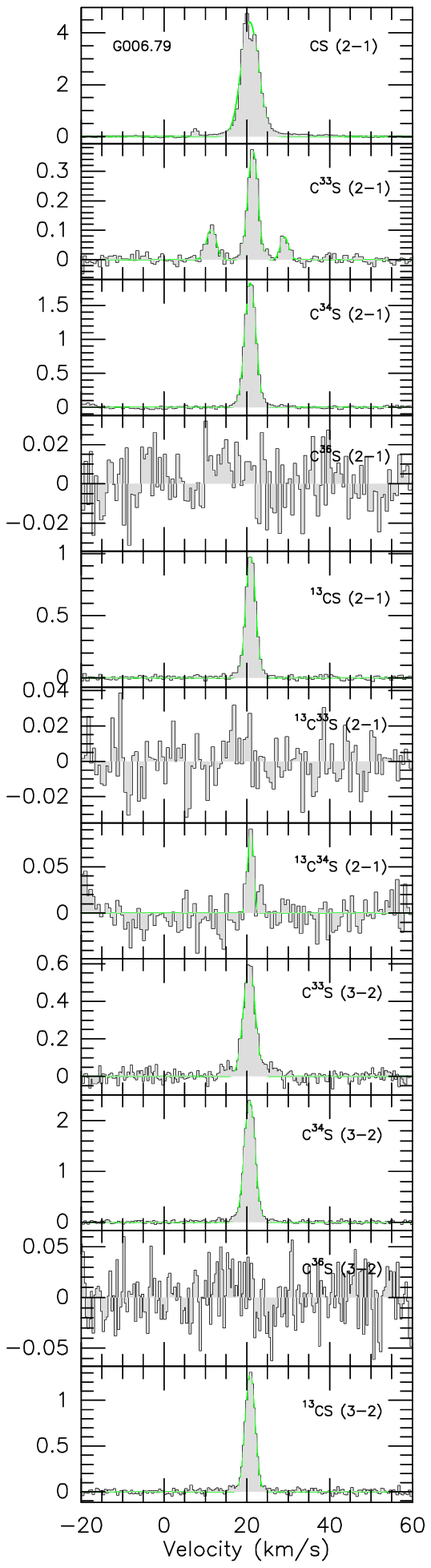}
\includegraphics[width=90pt,height=300pt]{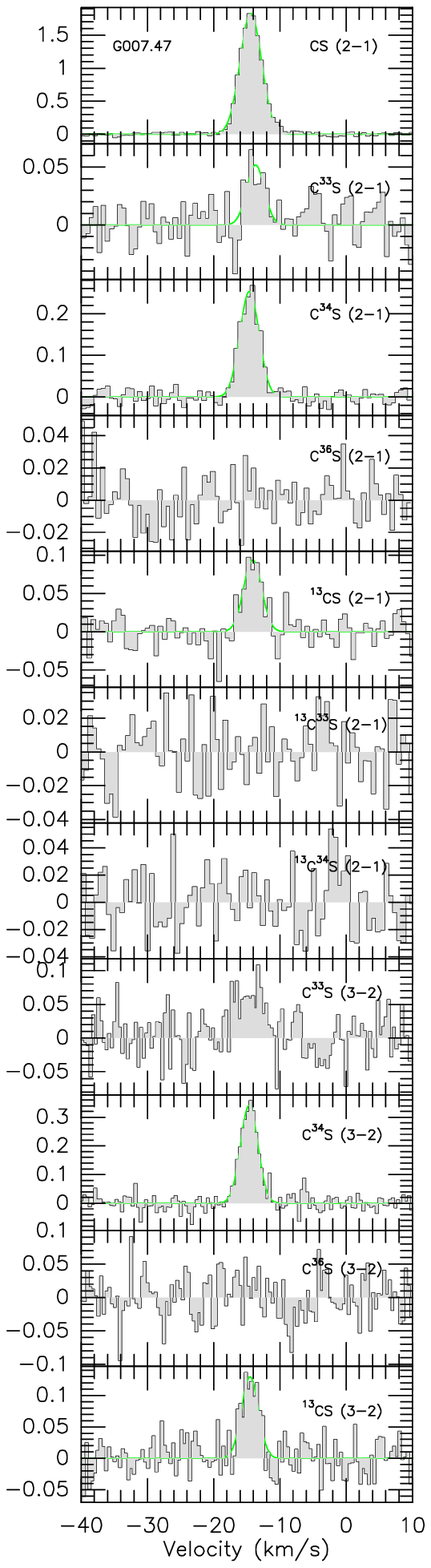}
\includegraphics[width=90pt,height=300pt]{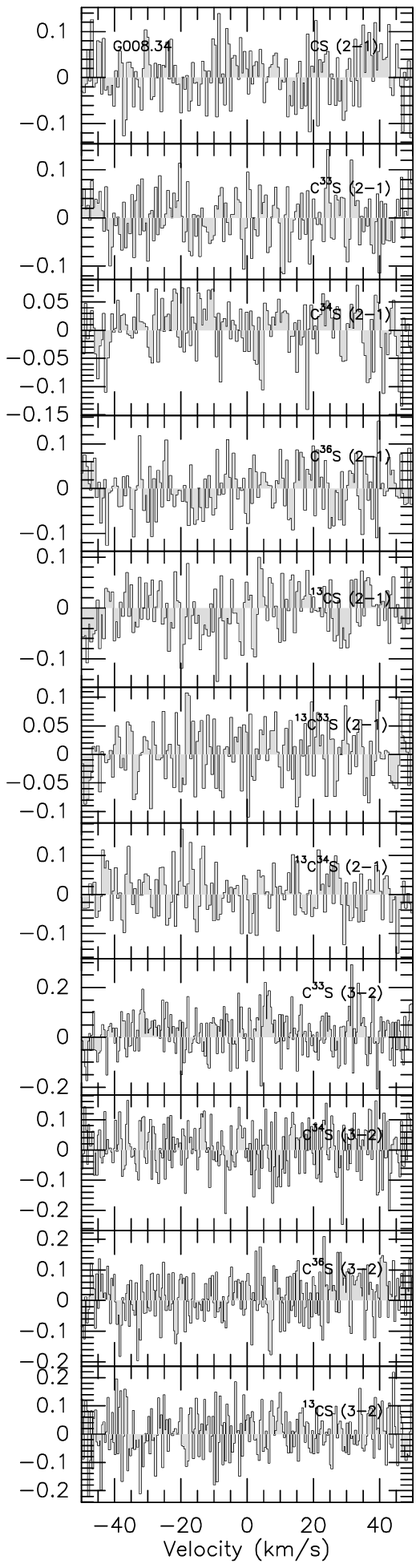}
\includegraphics[width=90pt,height=300pt]{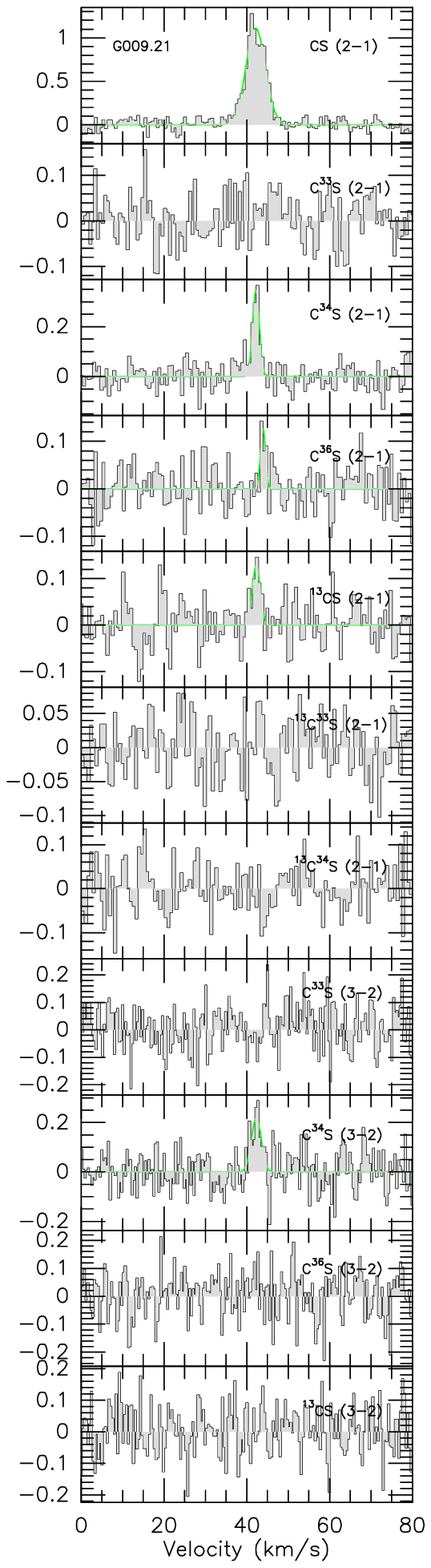}
\includegraphics[width=90pt,height=300pt]{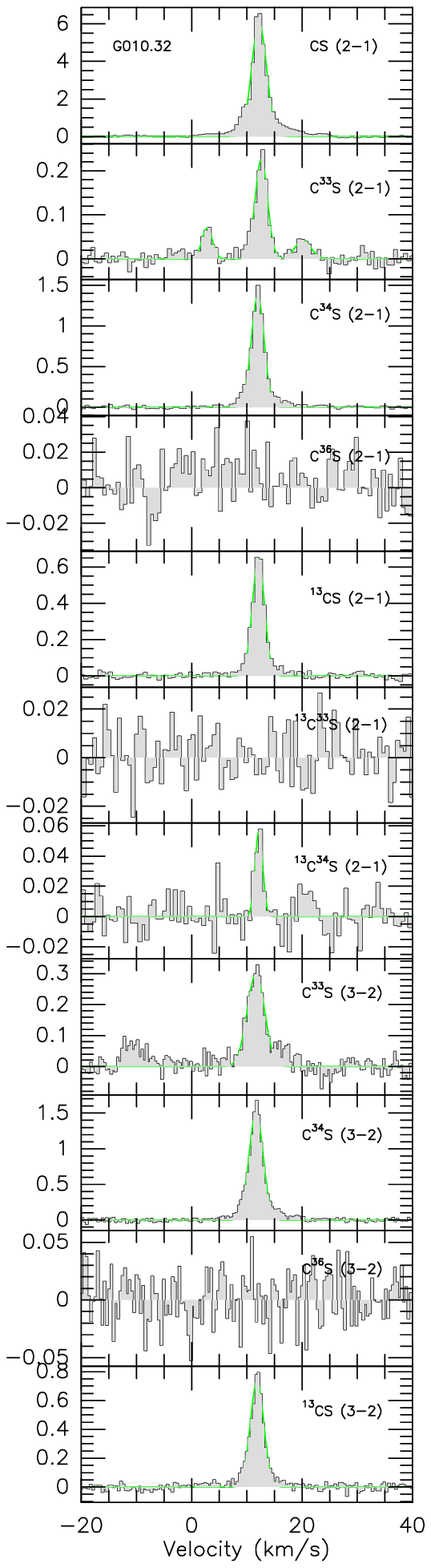}
\includegraphics[width=90pt,height=300pt]{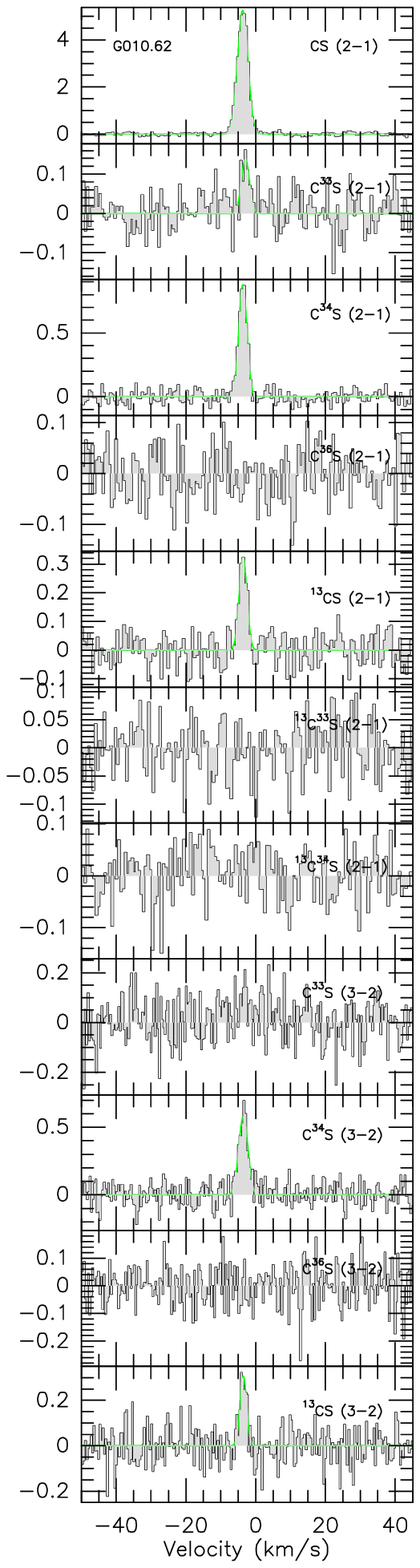}
\end{figure*}
\begin{figure*}
\centering
\includegraphics[width=90pt,height=300pt]{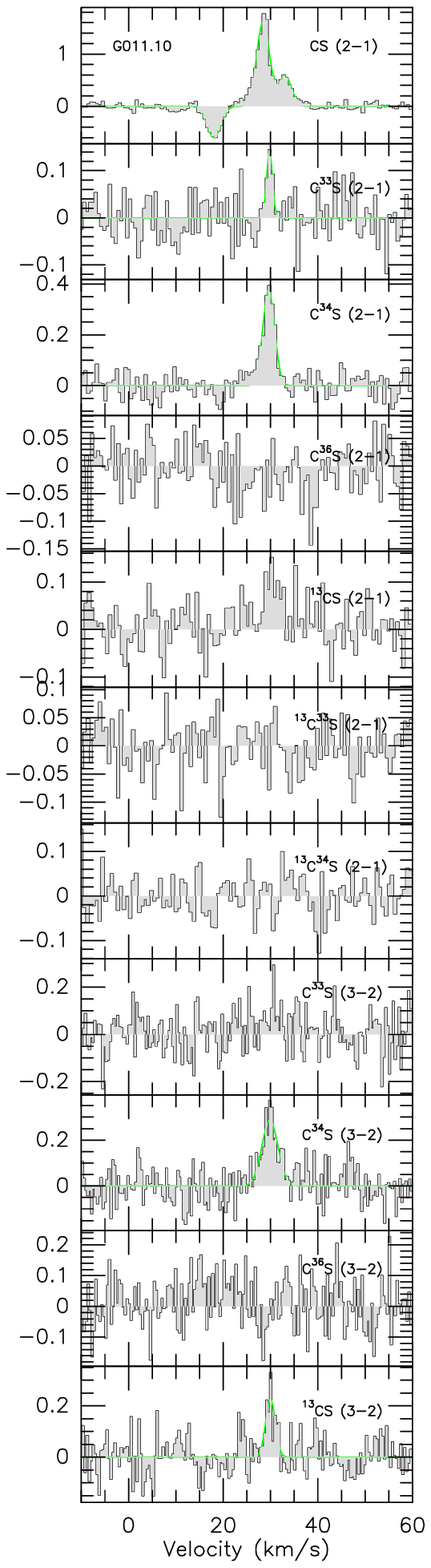}
\includegraphics[width=90pt,height=300pt]{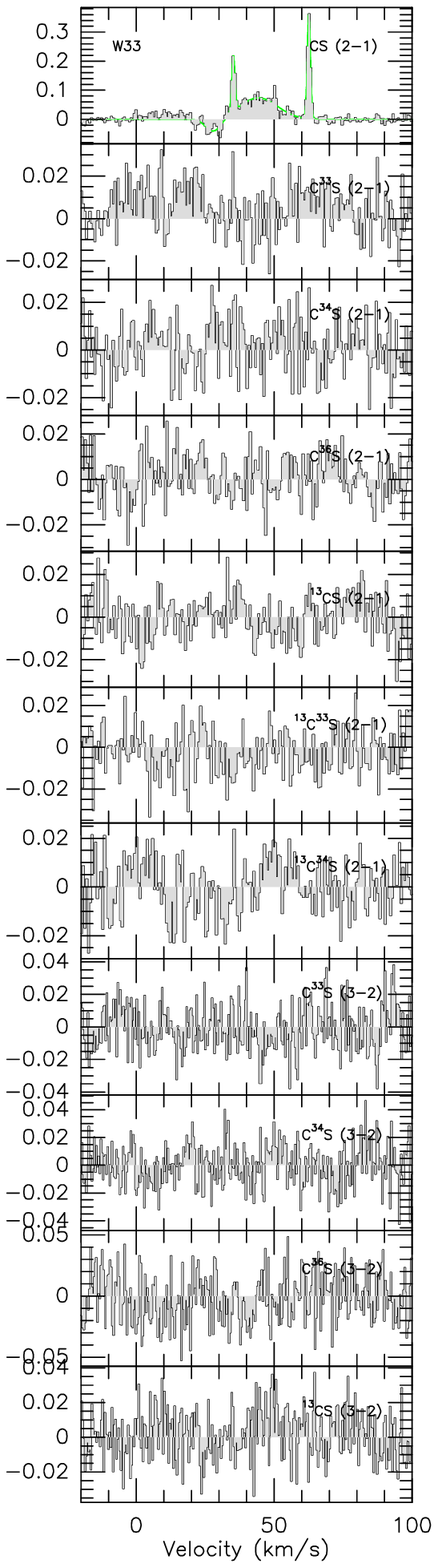}
\includegraphics[width=90pt,height=300pt]{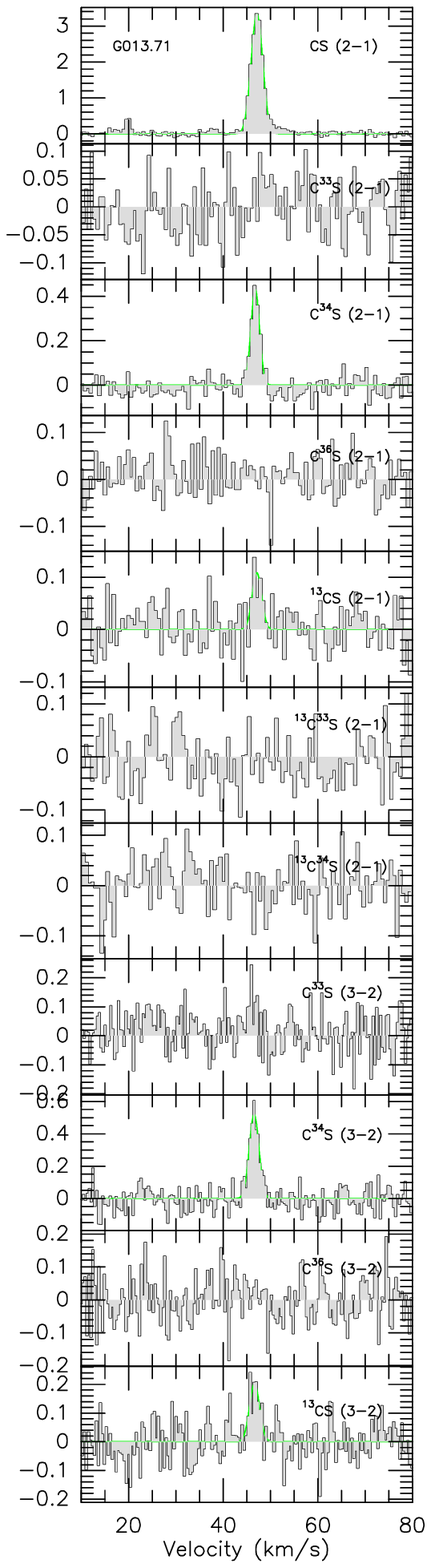}
\includegraphics[width=90pt,height=300pt]{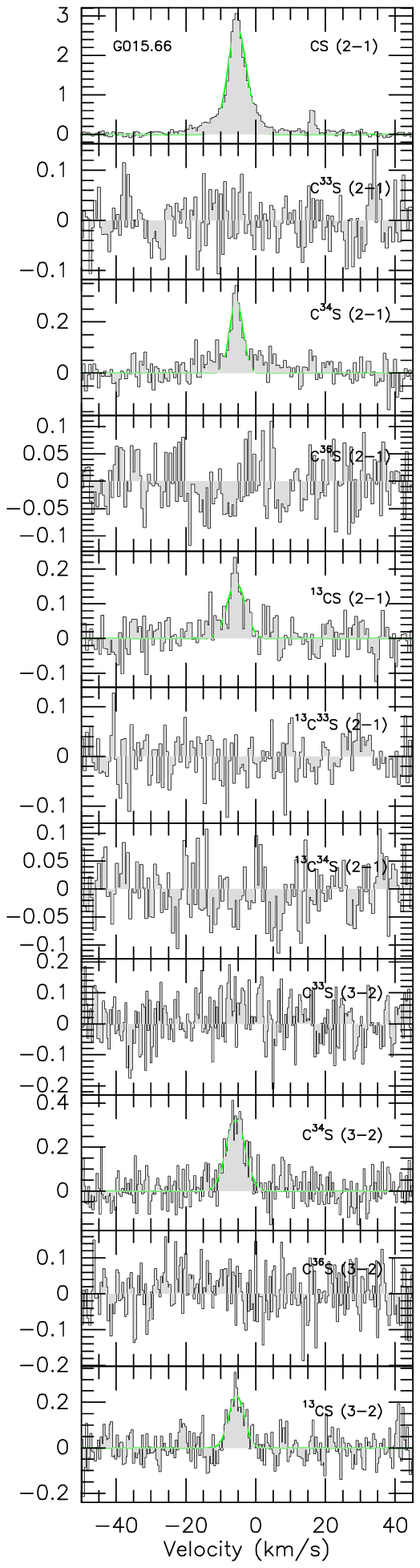}
\includegraphics[width=90pt,height=300pt]{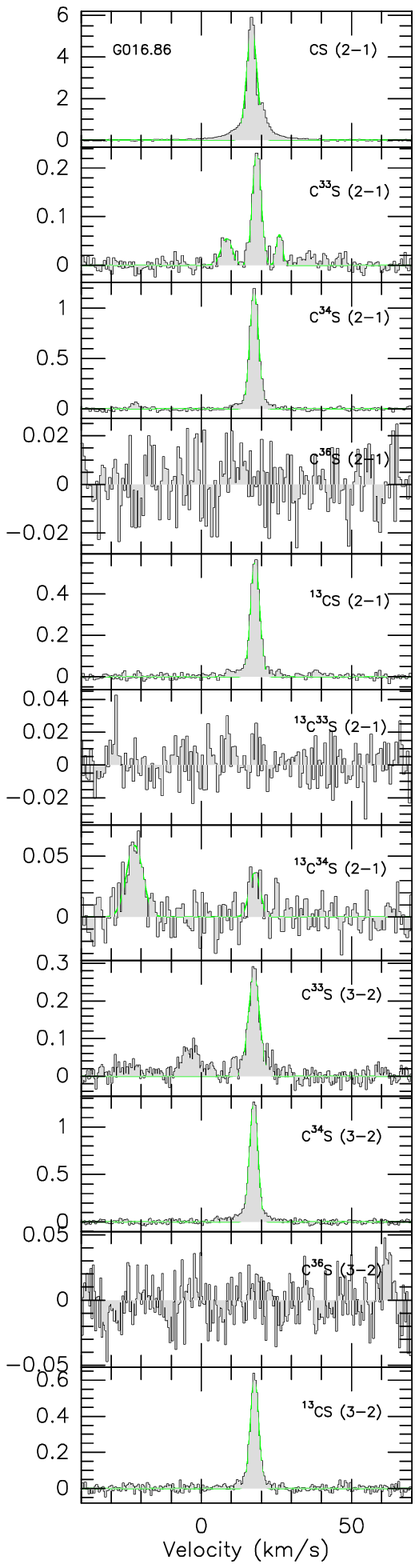}
\includegraphics[width=90pt,height=300pt]{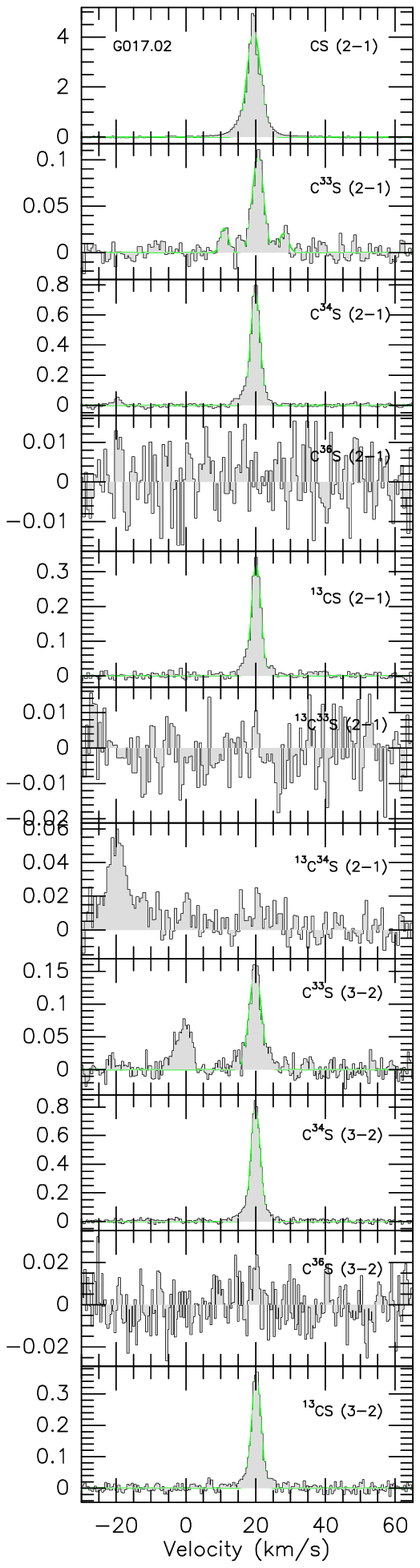}
\includegraphics[width=90pt,height=300pt]{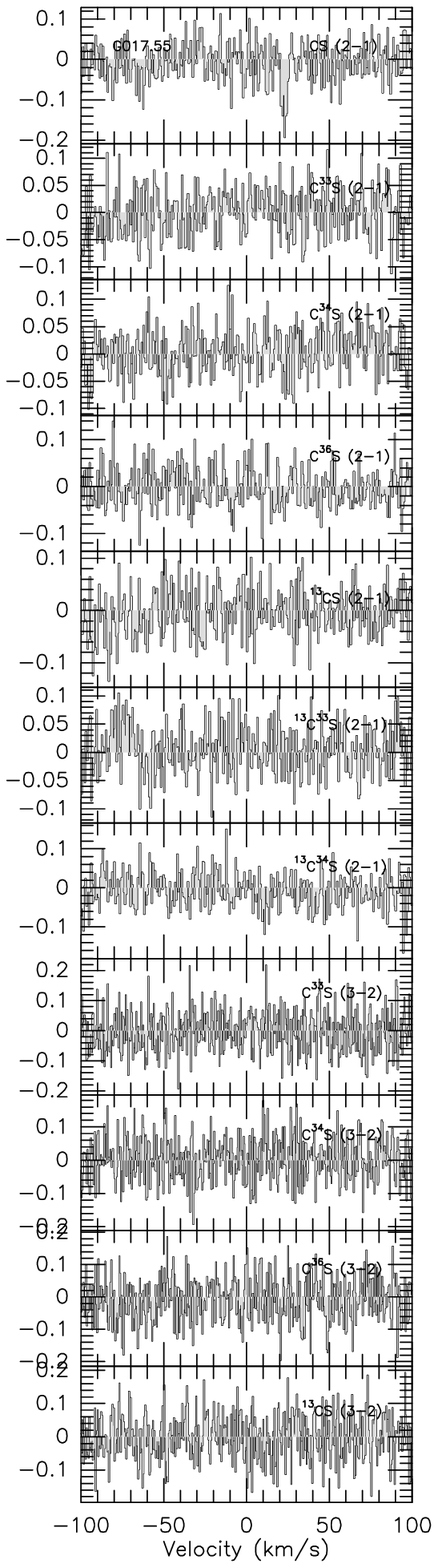}
\includegraphics[width=90pt,height=300pt]{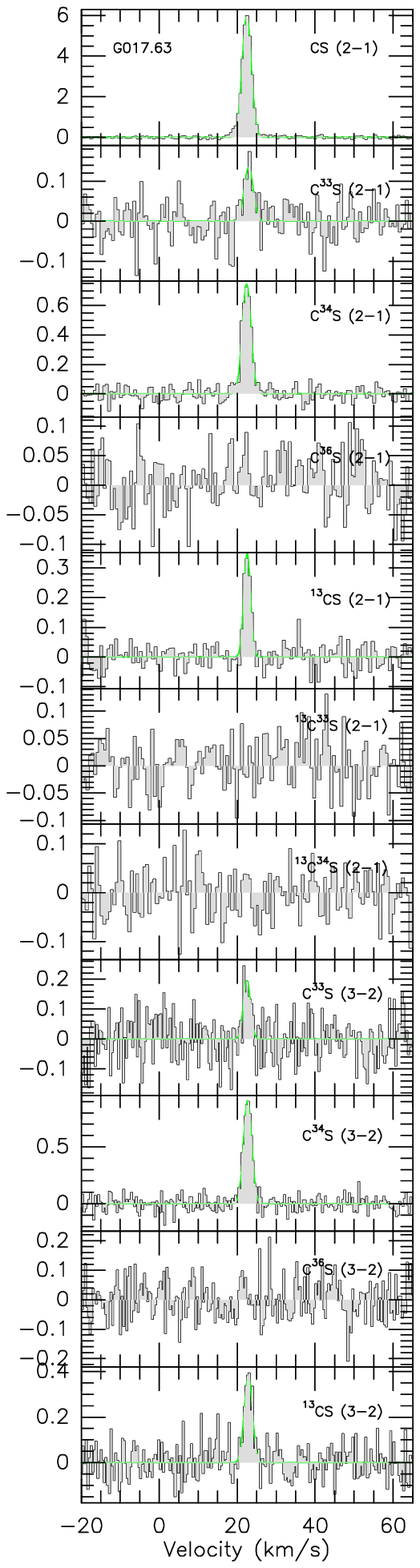}
\includegraphics[width=90pt,height=300pt]{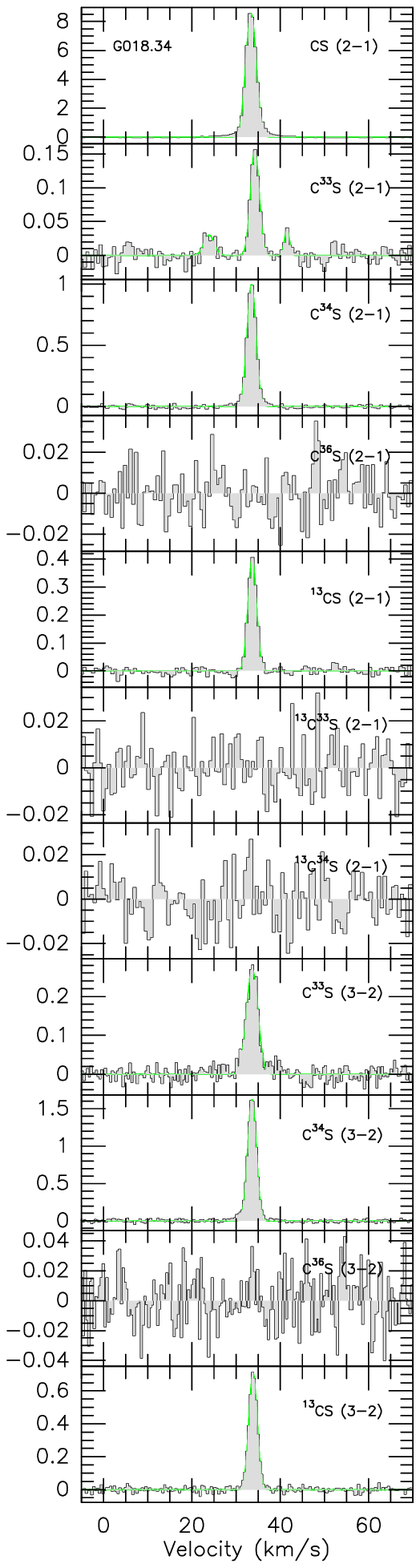}
\includegraphics[width=90pt,height=300pt]{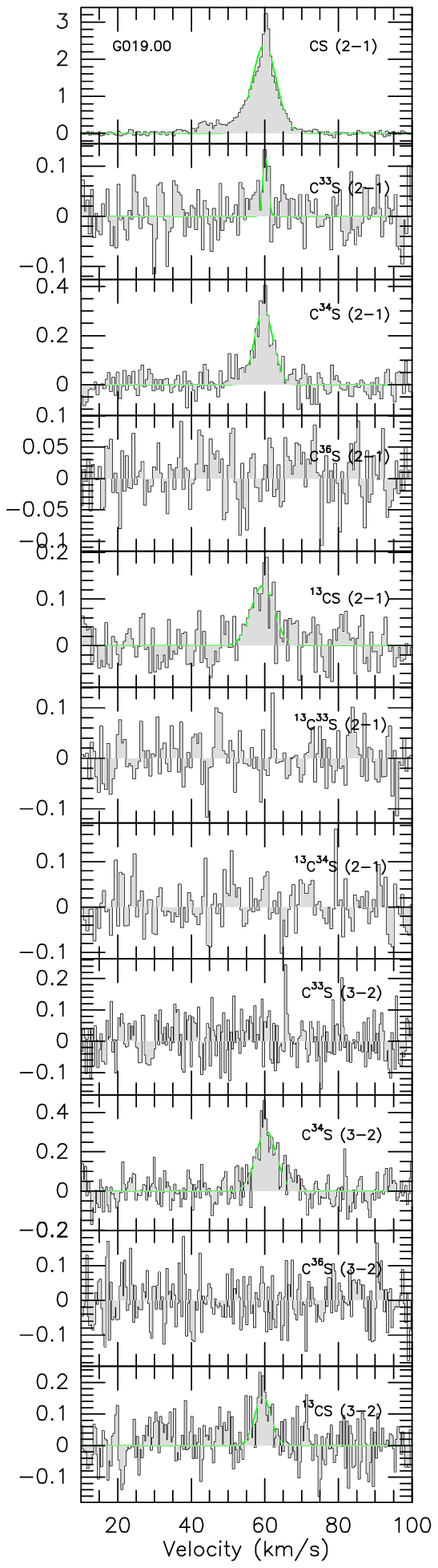}
\end{figure*}

\begin{figure*}
\centering
\includegraphics[width=90pt,height=300pt]{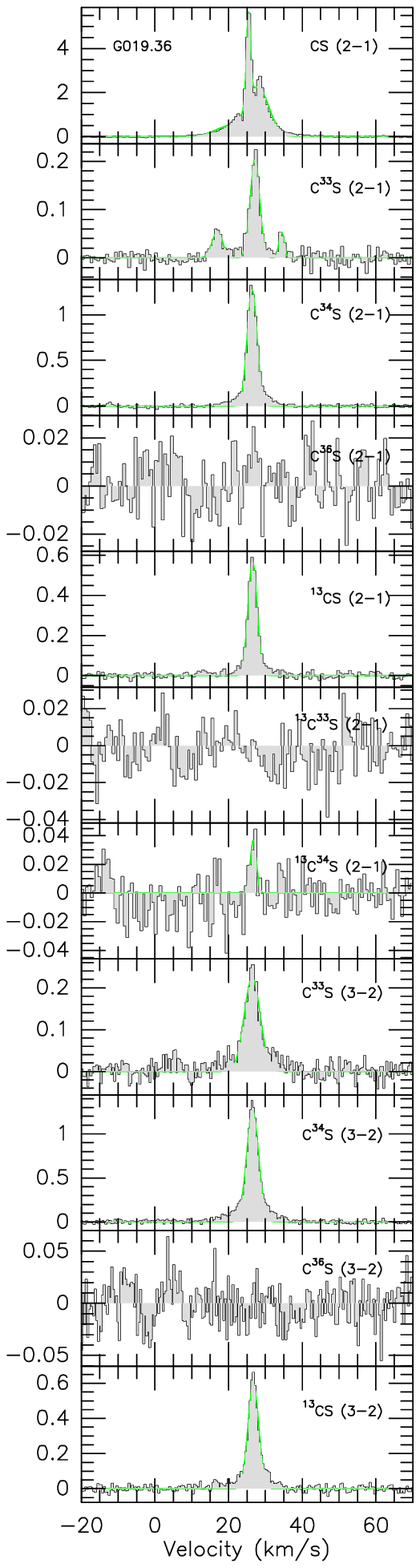}
\includegraphics[width=90pt,height=300pt]{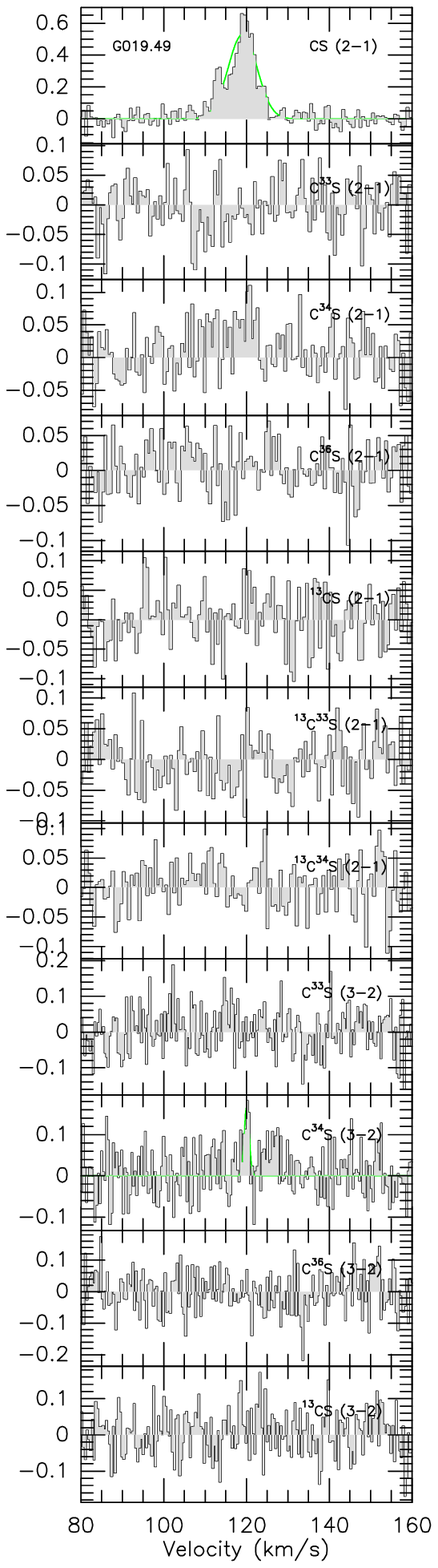}
\includegraphics[width=90pt,height=300pt]{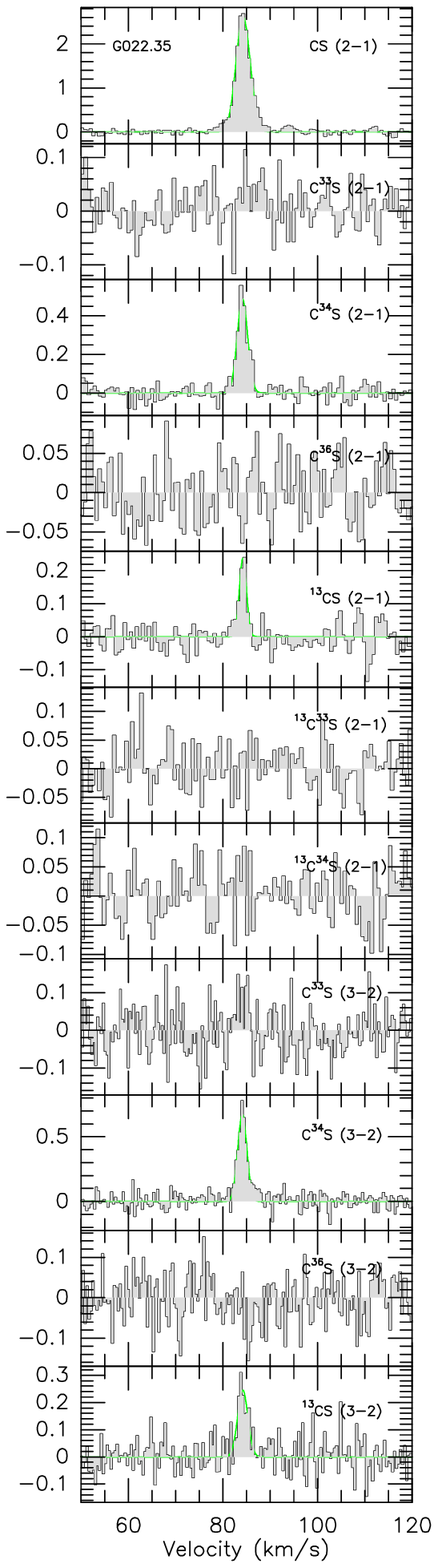}
\includegraphics[width=90pt,height=300pt]{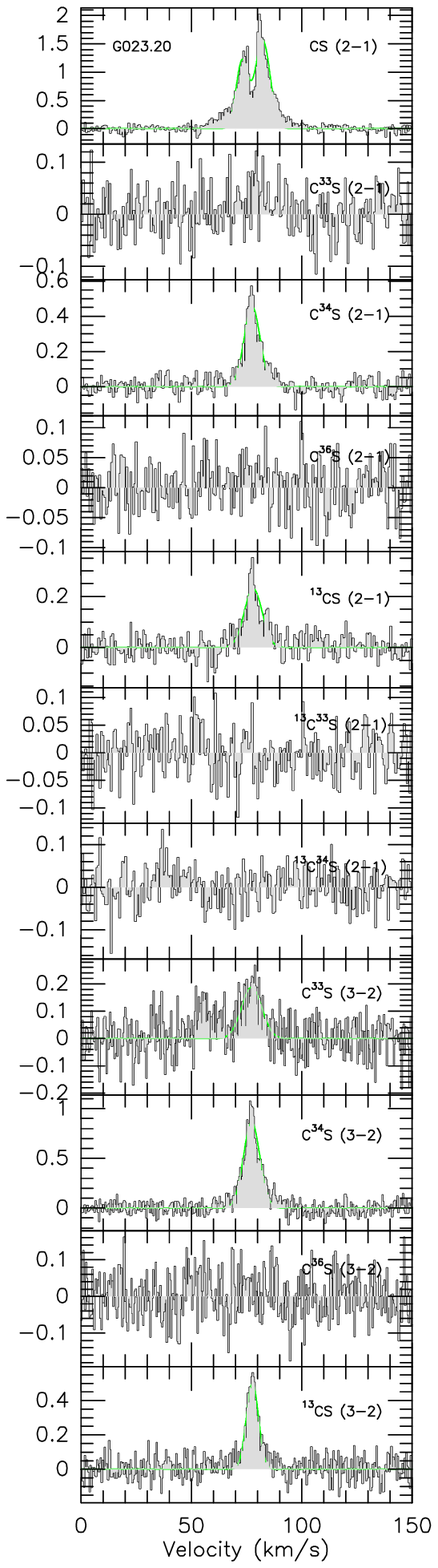}
\includegraphics[width=90pt,height=300pt]{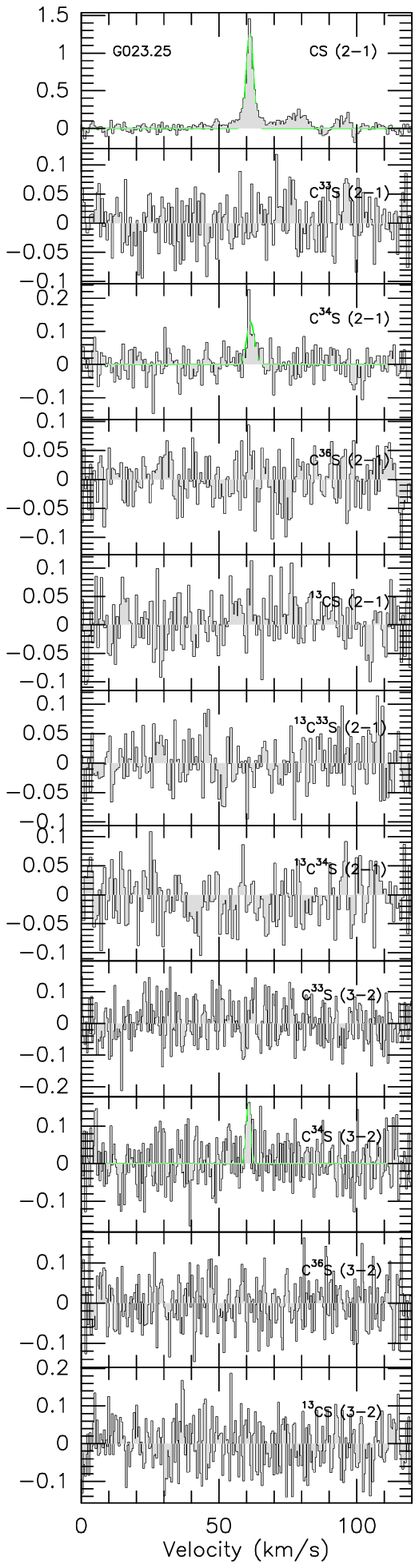}
\includegraphics[width=90pt,height=300pt]{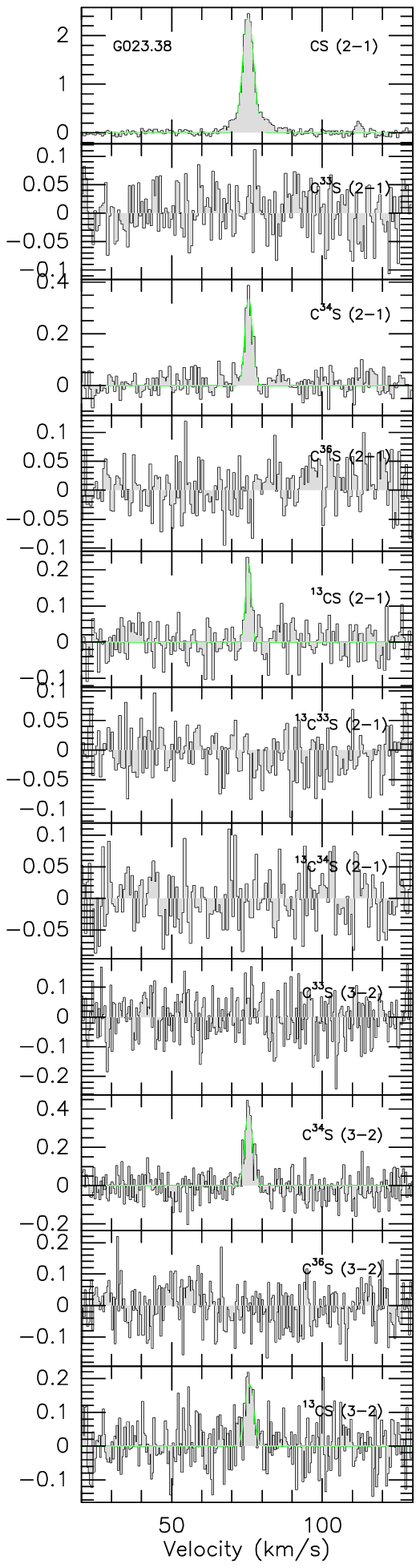}
\includegraphics[width=90pt,height=300pt]{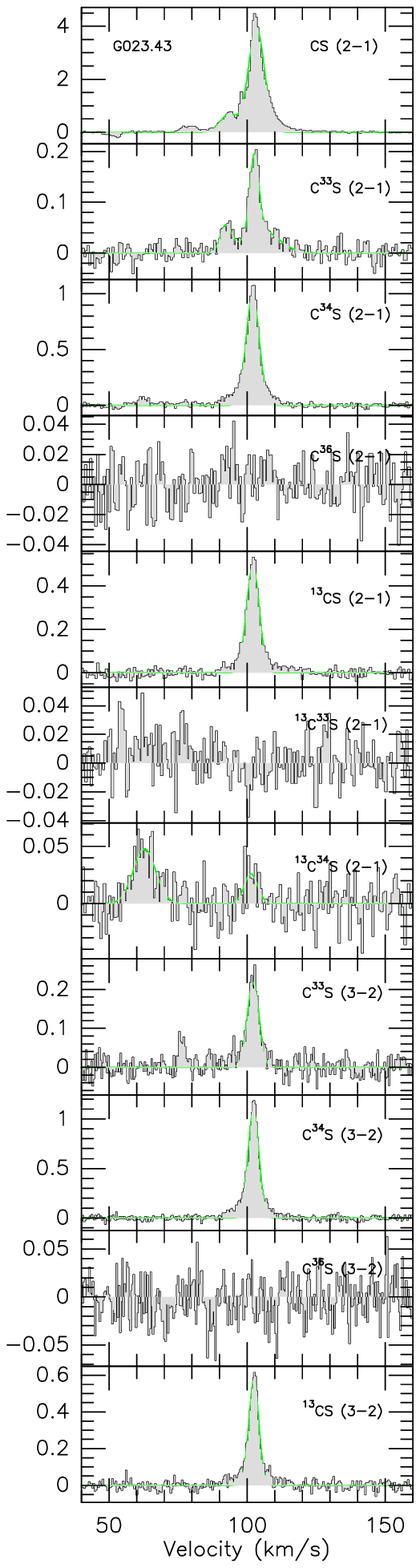}
\includegraphics[width=90pt,height=300pt]{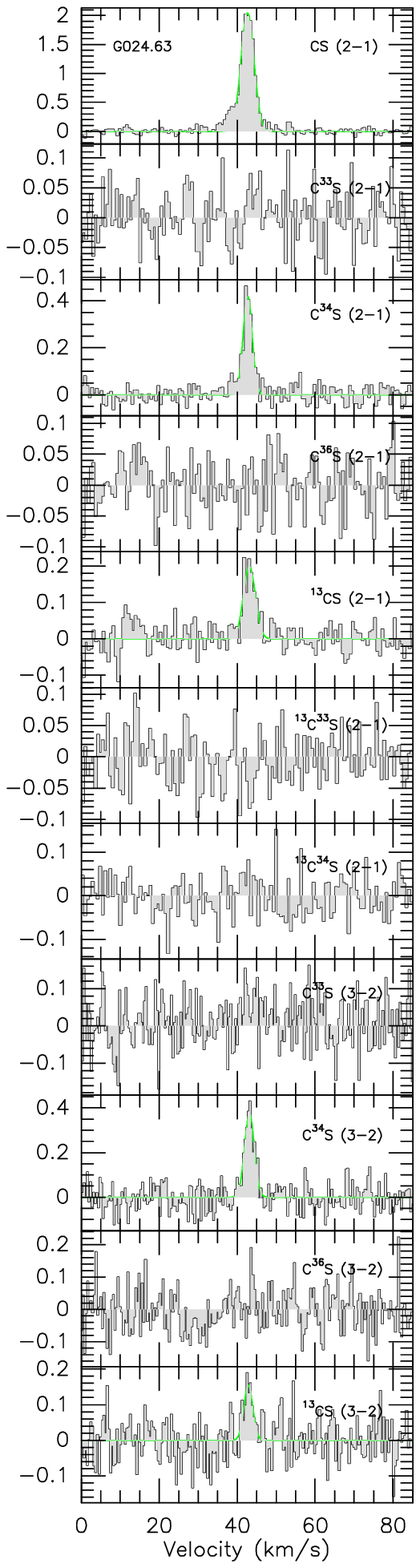}
\includegraphics[width=90pt,height=300pt]{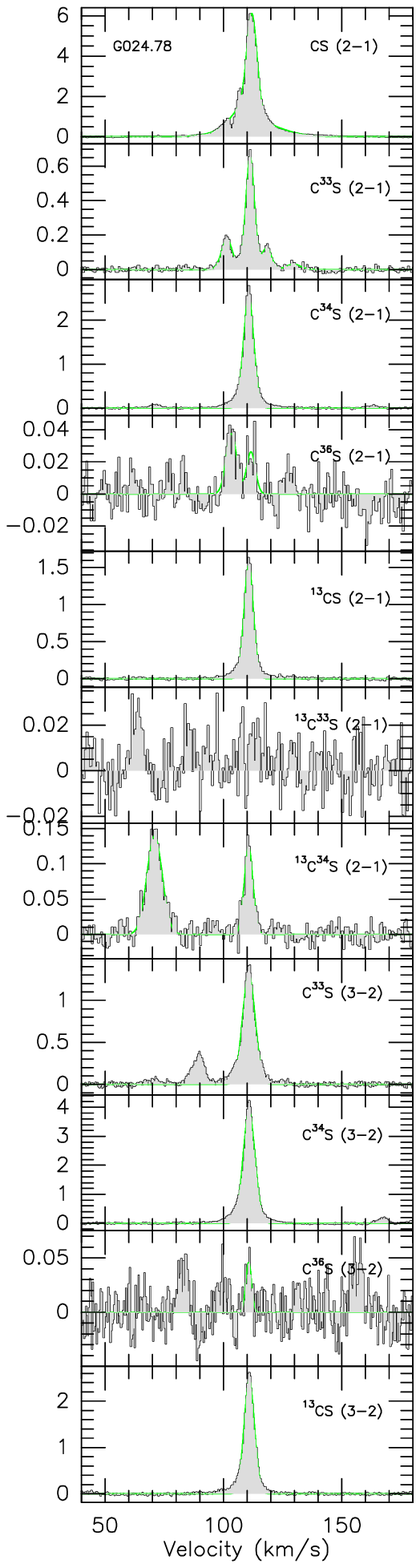}
\includegraphics[width=90pt,height=300pt]{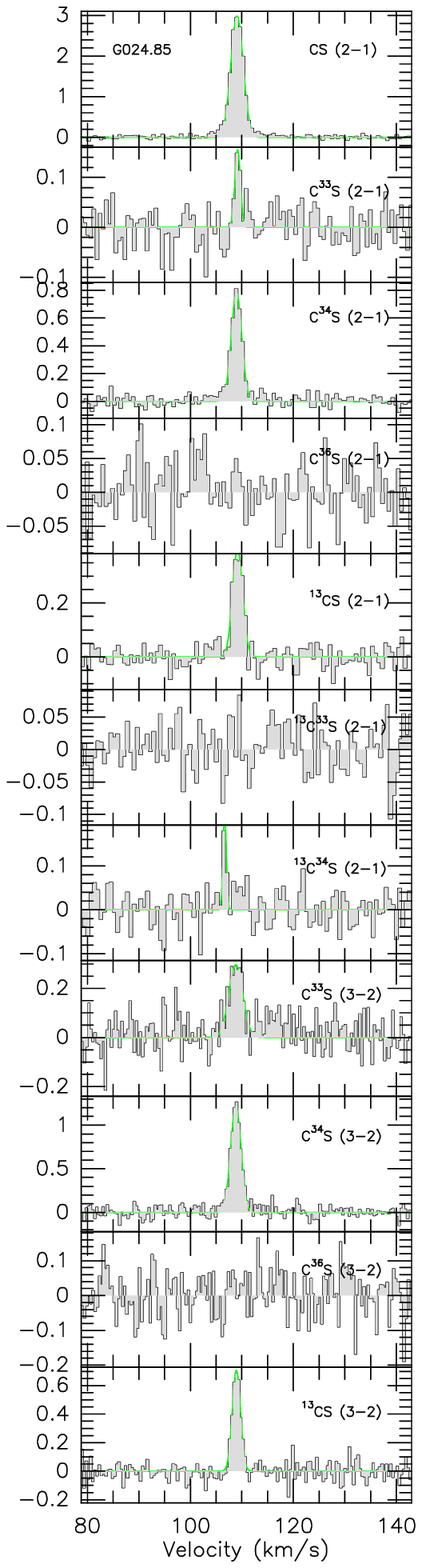}
\end{figure*}
\begin{figure*}
\centering
\includegraphics[width=90pt,height=300pt]{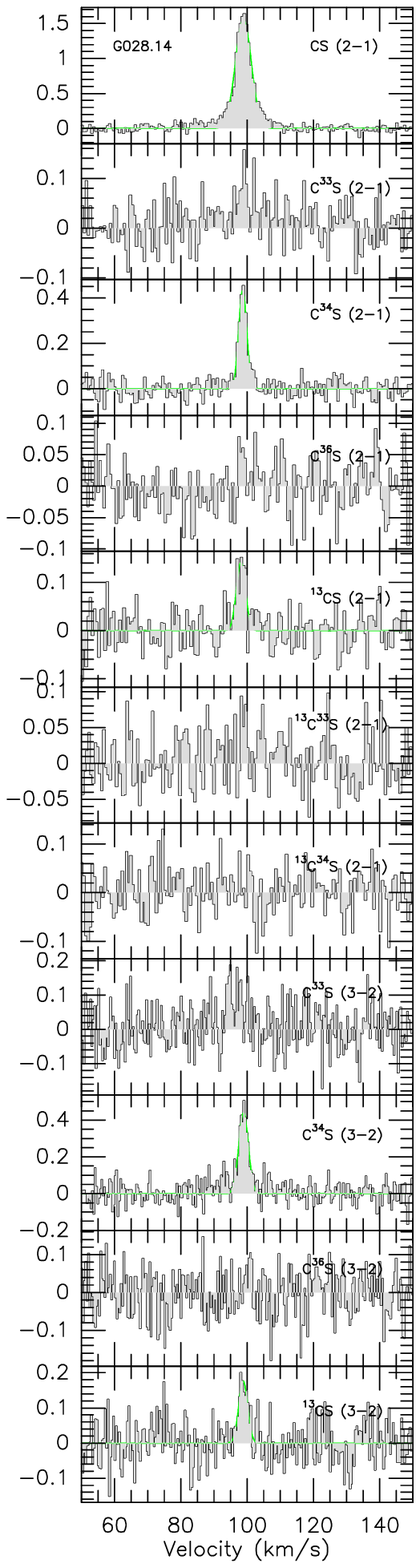}
\includegraphics[width=90pt,height=300pt]{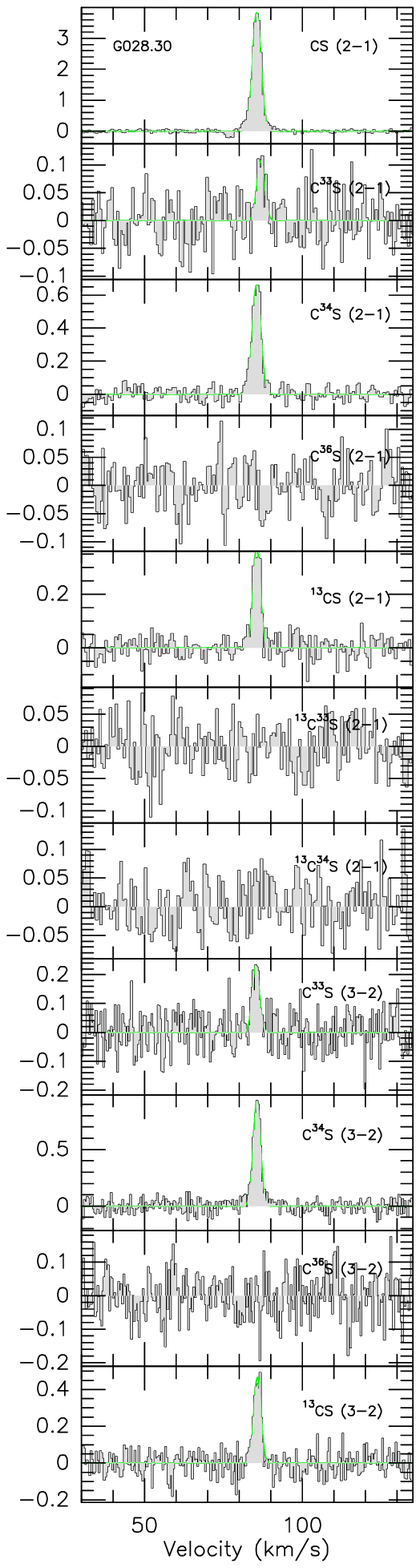}
\includegraphics[width=90pt,height=300pt]{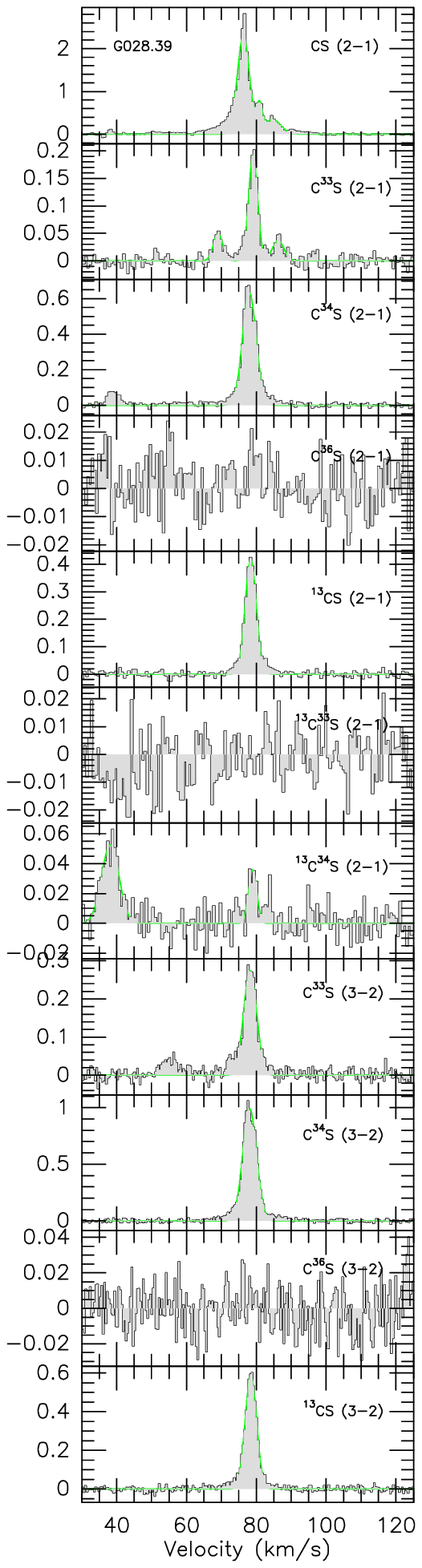}
\includegraphics[width=90pt,height=300pt]{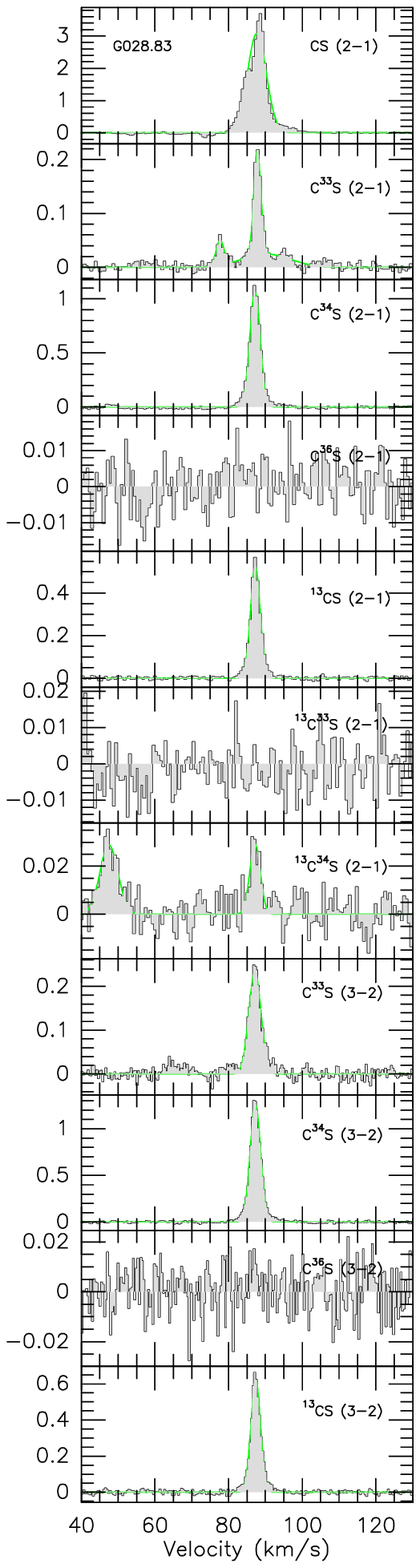}
\includegraphics[width=90pt,height=300pt]{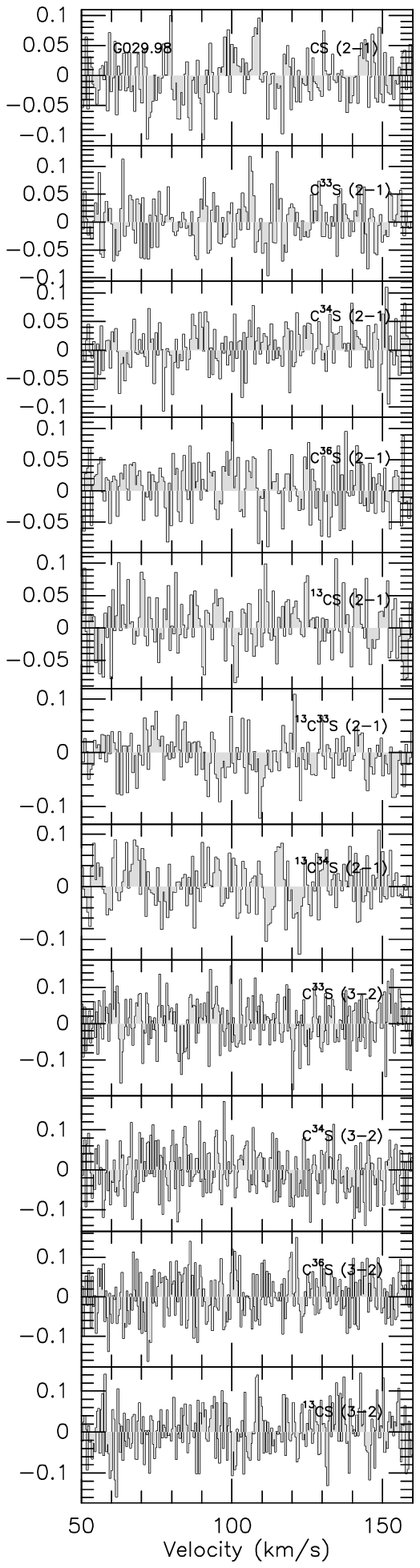}
\includegraphics[width=90pt,height=300pt]{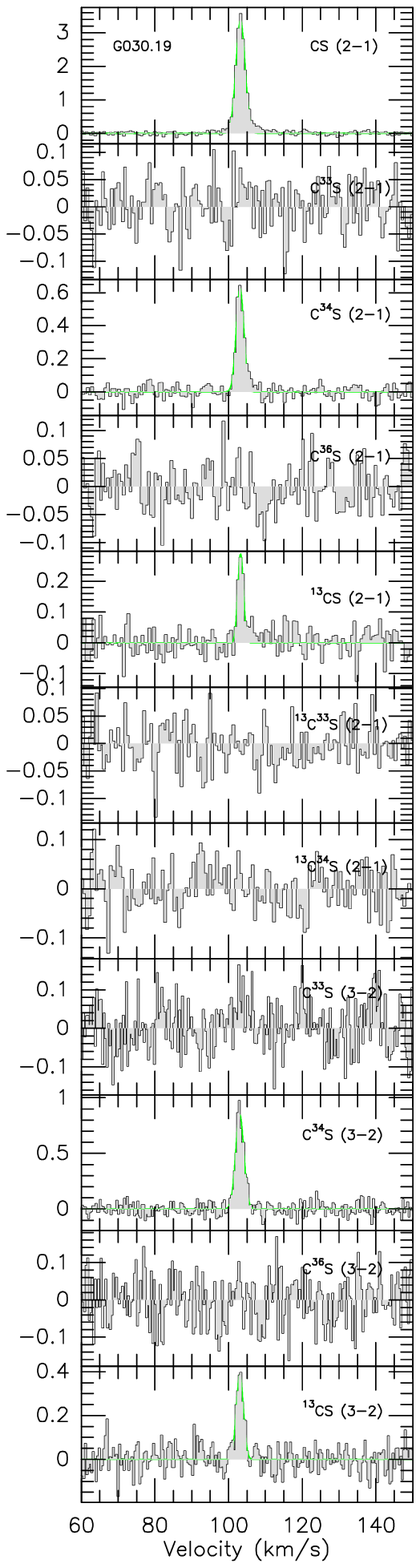}
\includegraphics[width=90pt,height=300pt]{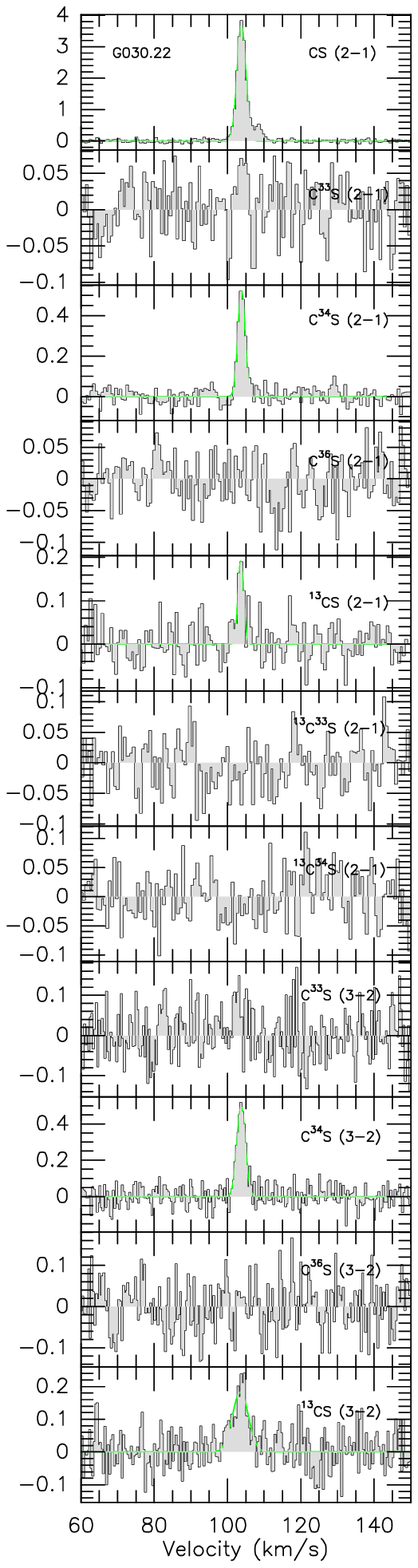}
\includegraphics[width=90pt,height=300pt]{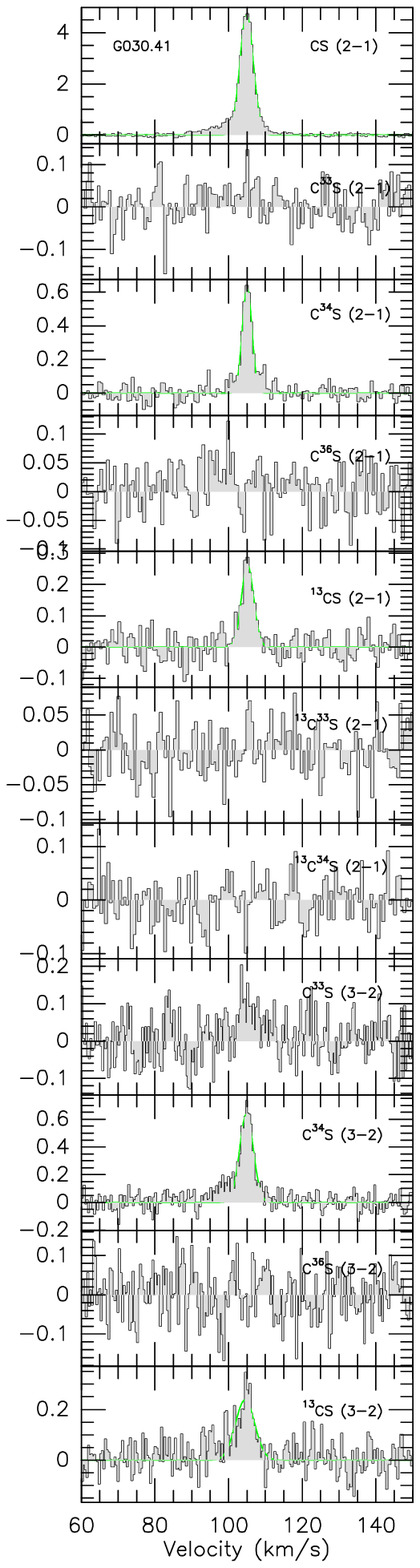}
\includegraphics[width=90pt,height=300pt]{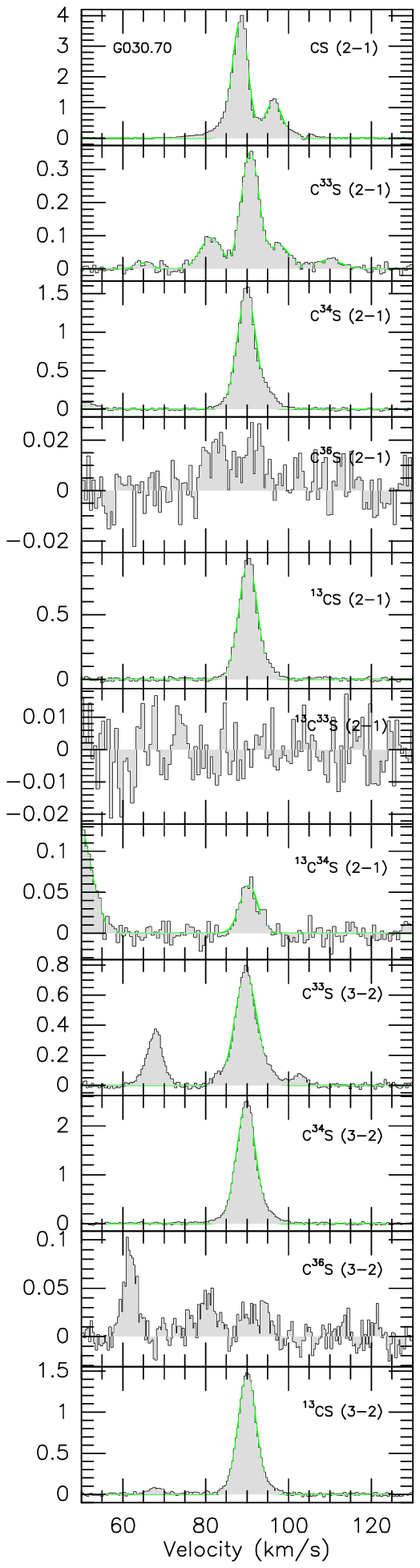}
\includegraphics[width=90pt,height=300pt]{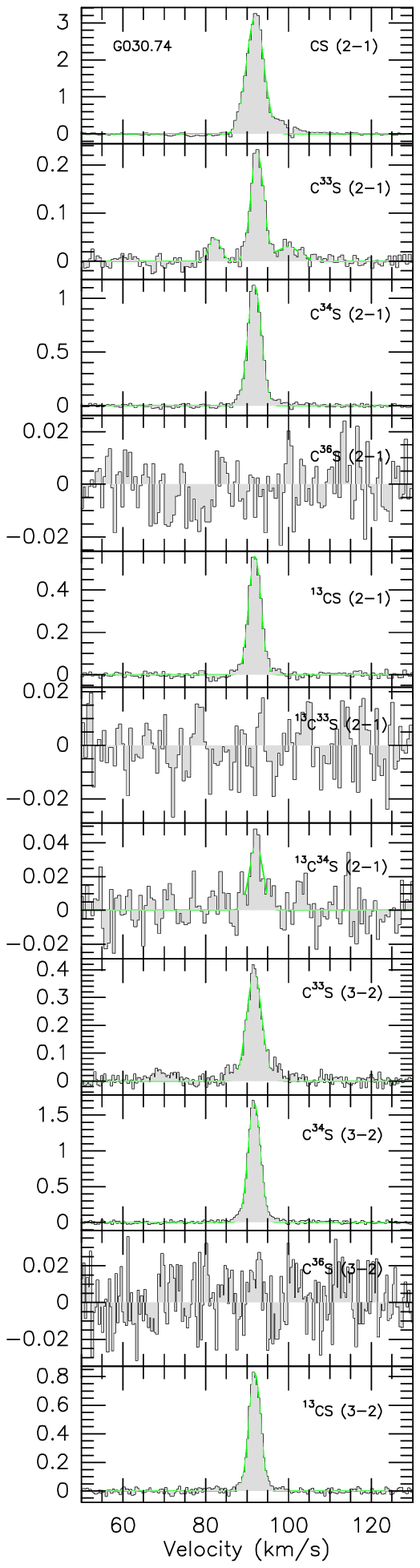}
\end{figure*}

\begin{figure*}
\centering
\includegraphics[width=90pt,height=300pt]{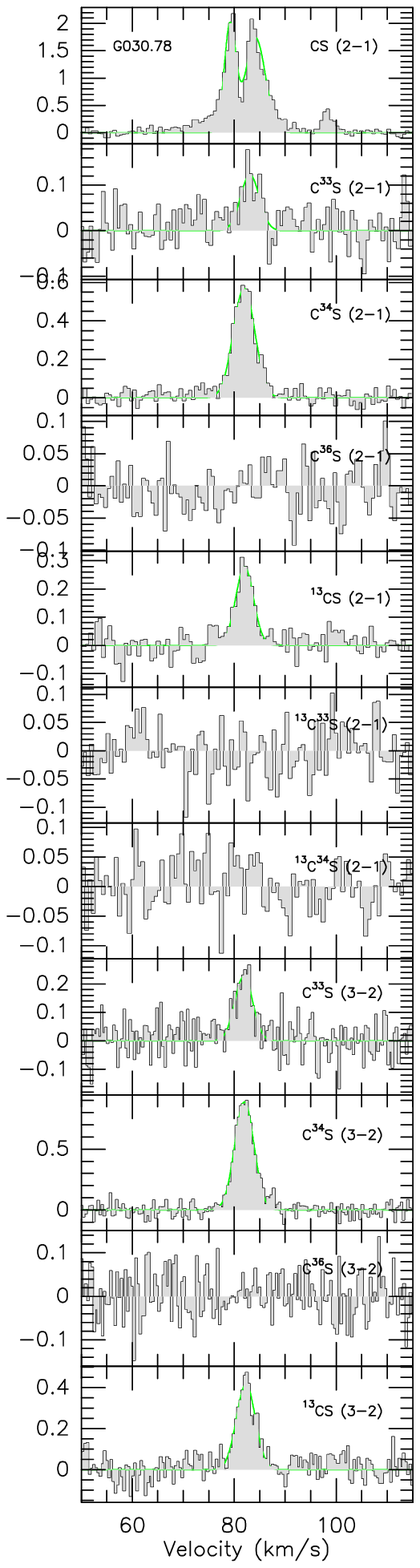}
\includegraphics[width=90pt,height=300pt]{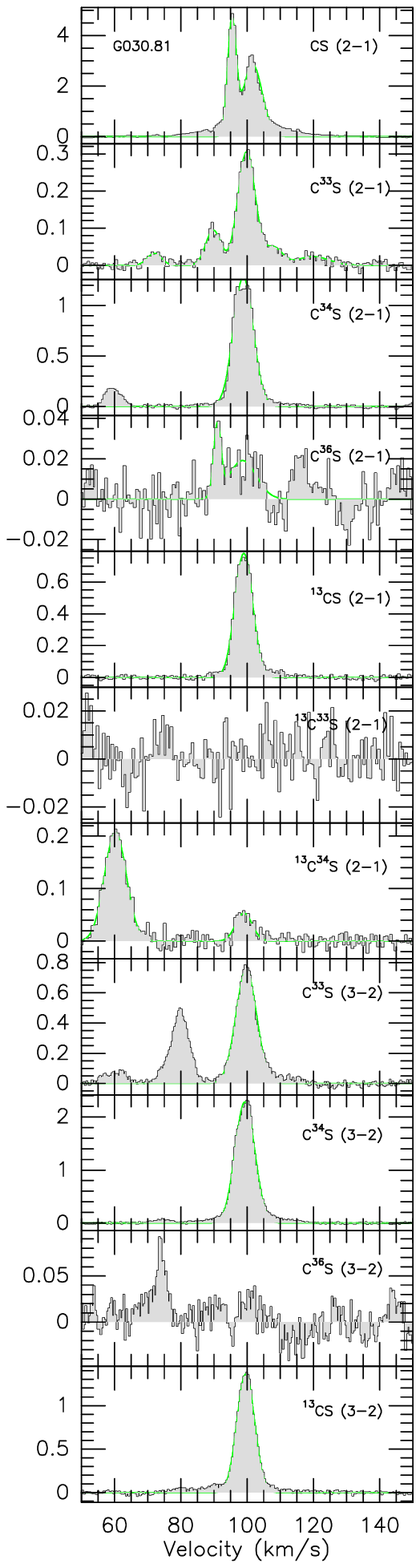}
\includegraphics[width=90pt,height=300pt]{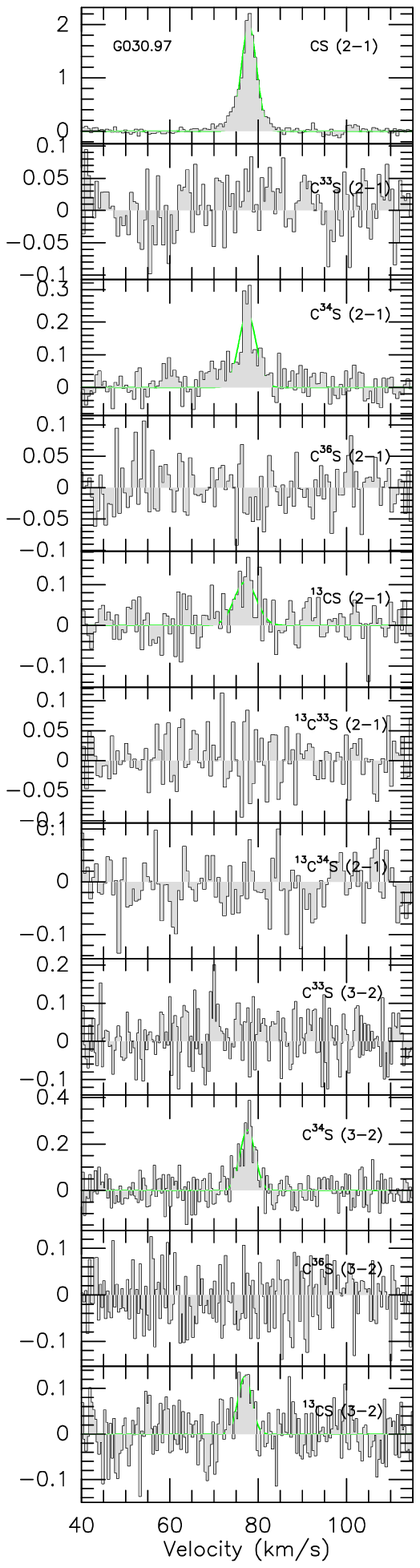}
\includegraphics[width=90pt,height=300pt]{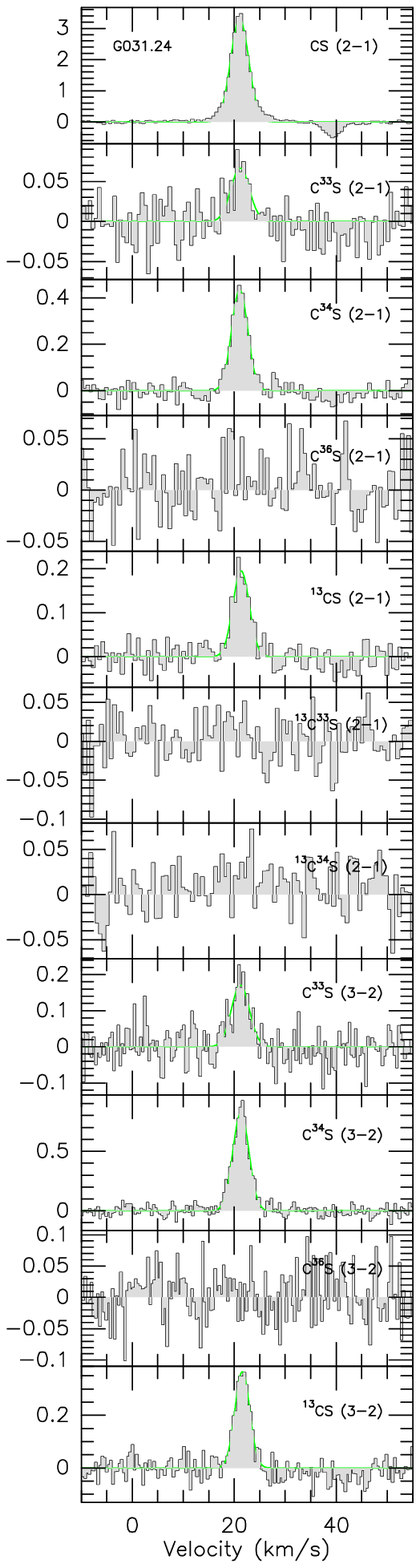}
\includegraphics[width=90pt,height=300pt]{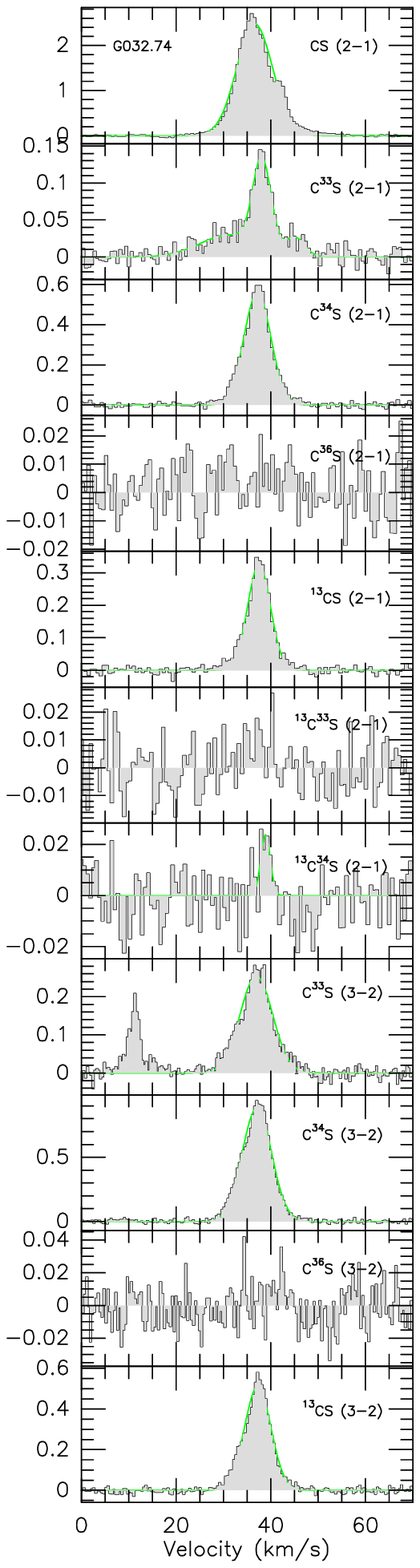}
\includegraphics[width=90pt,height=300pt]{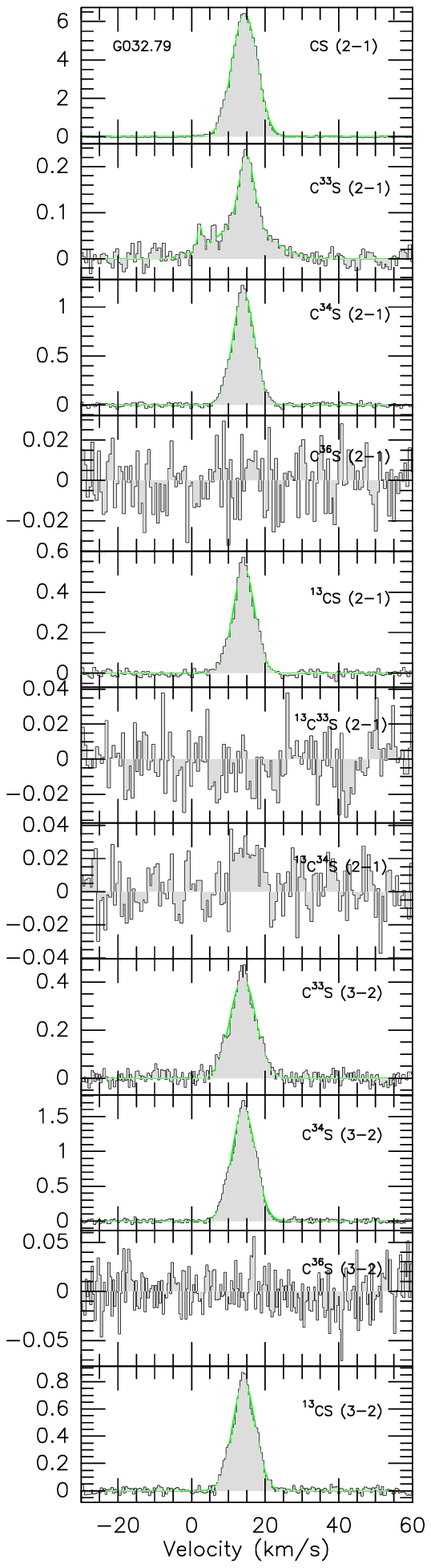}
\includegraphics[width=90pt,height=300pt]{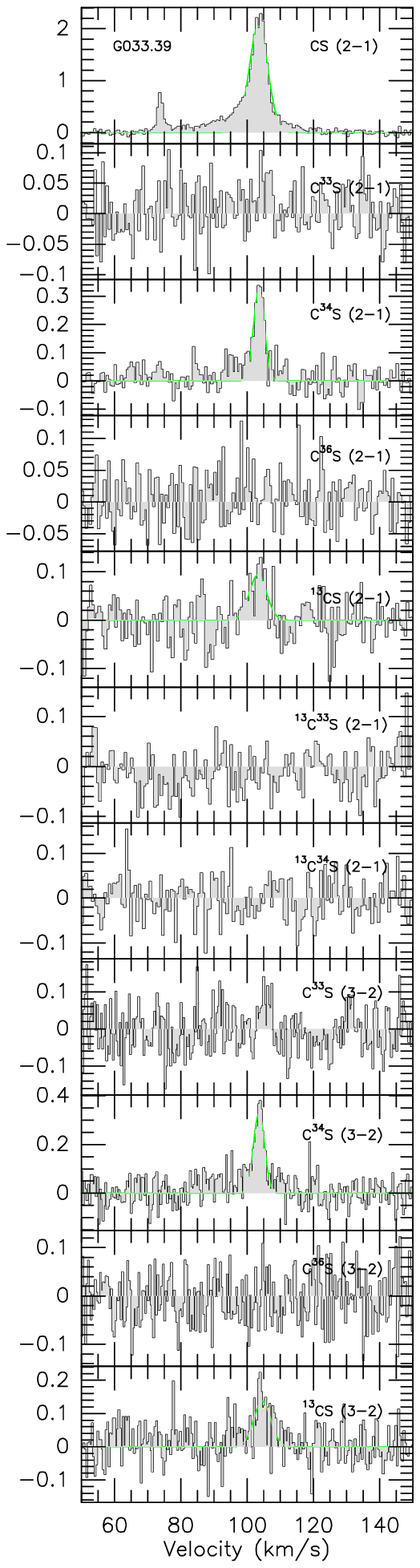}
\includegraphics[width=90pt,height=300pt]{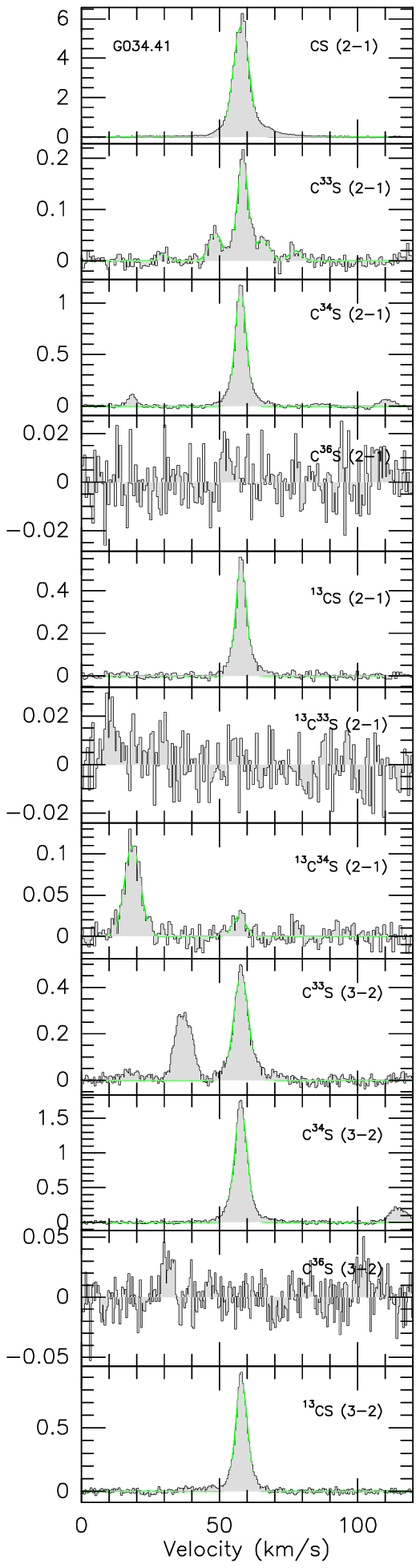}
\includegraphics[width=90pt,height=300pt]{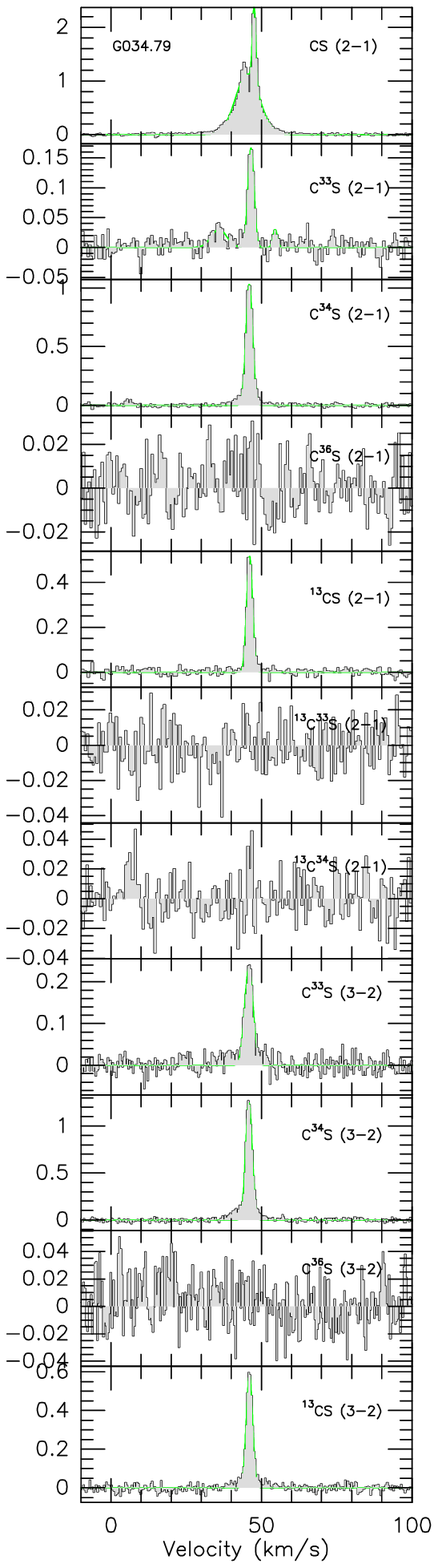}
\includegraphics[width=90pt,height=300pt]{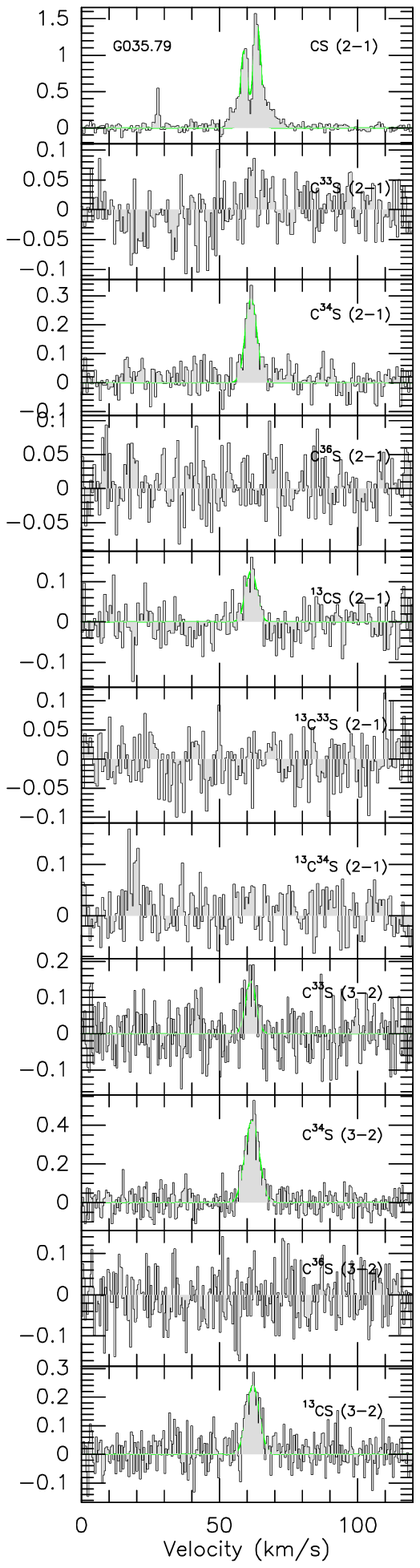}
\end{figure*}

\begin{figure*}
\centering
\includegraphics[width=90pt,height=300pt]{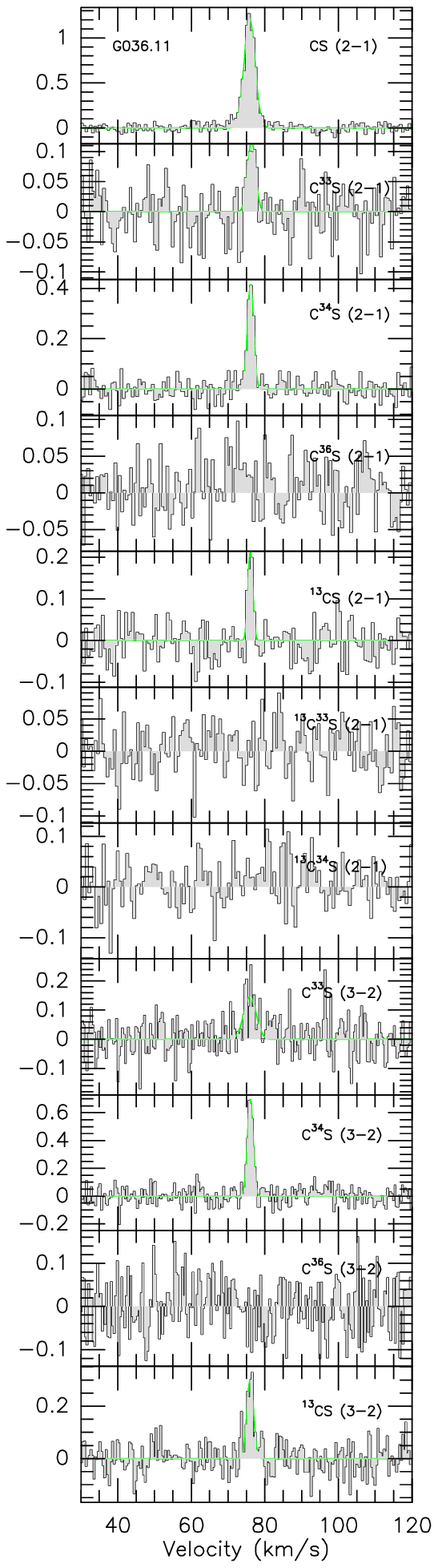}
\includegraphics[width=90pt,height=300pt]{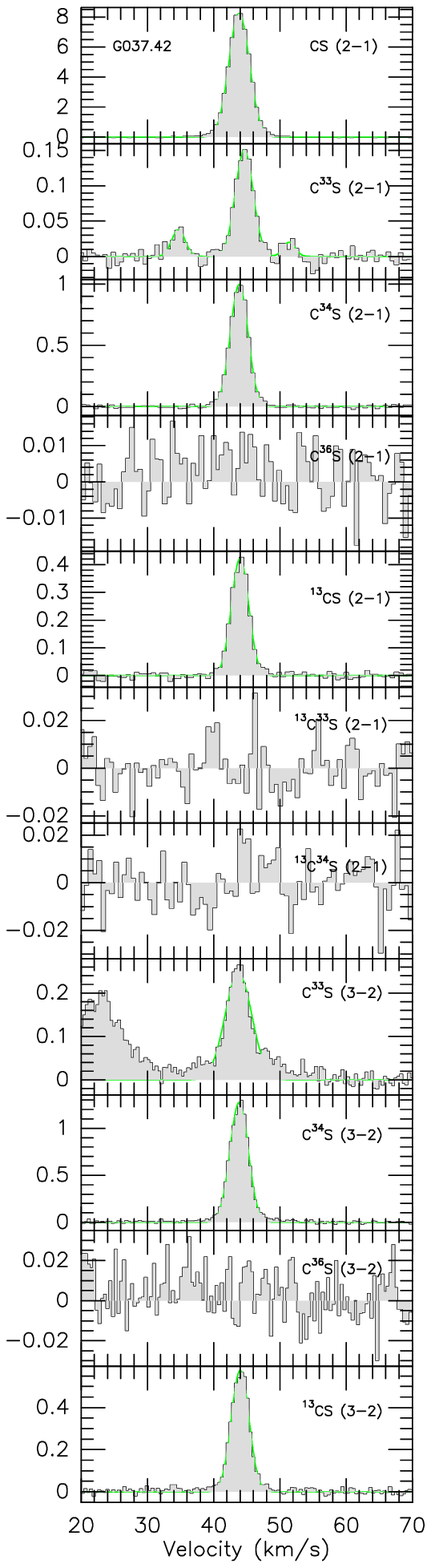}
\includegraphics[width=90pt,height=300pt]{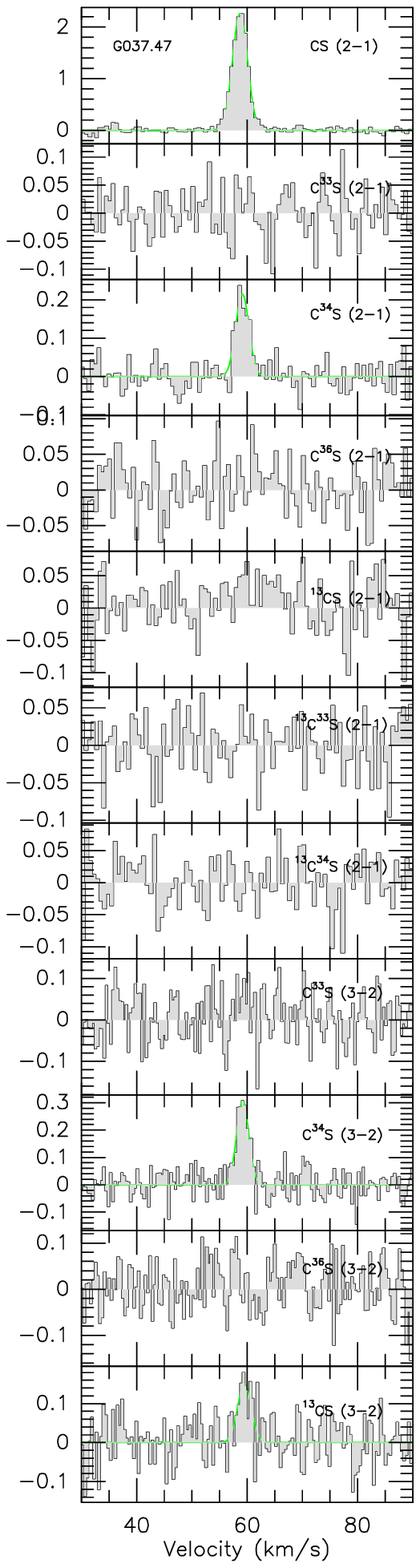}
\includegraphics[width=90pt,height=300pt]{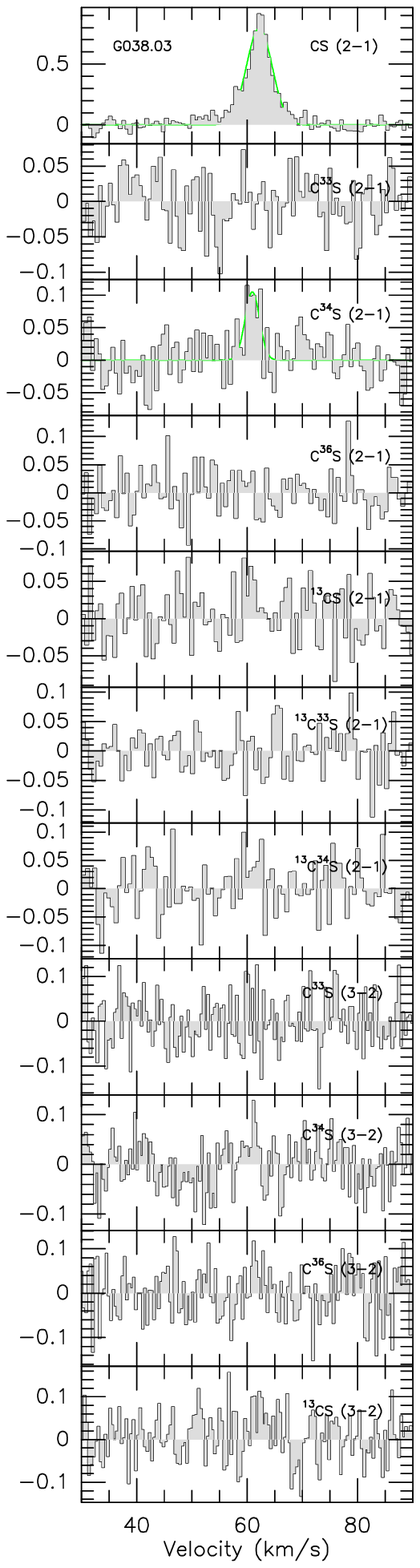}
\includegraphics[width=90pt,height=300pt]{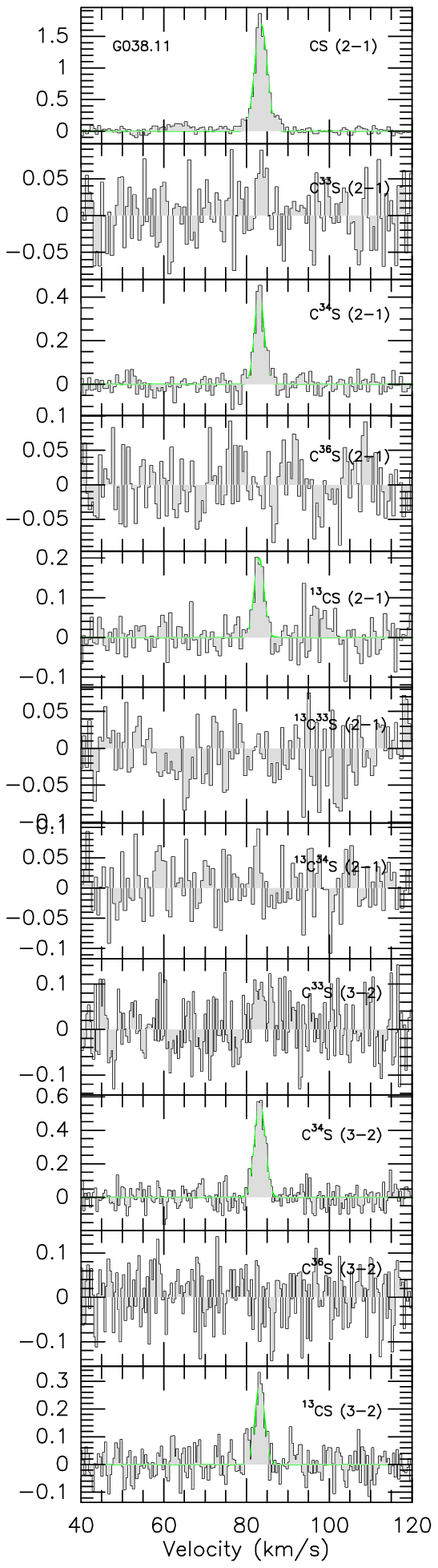}
\includegraphics[width=90pt,height=300pt]{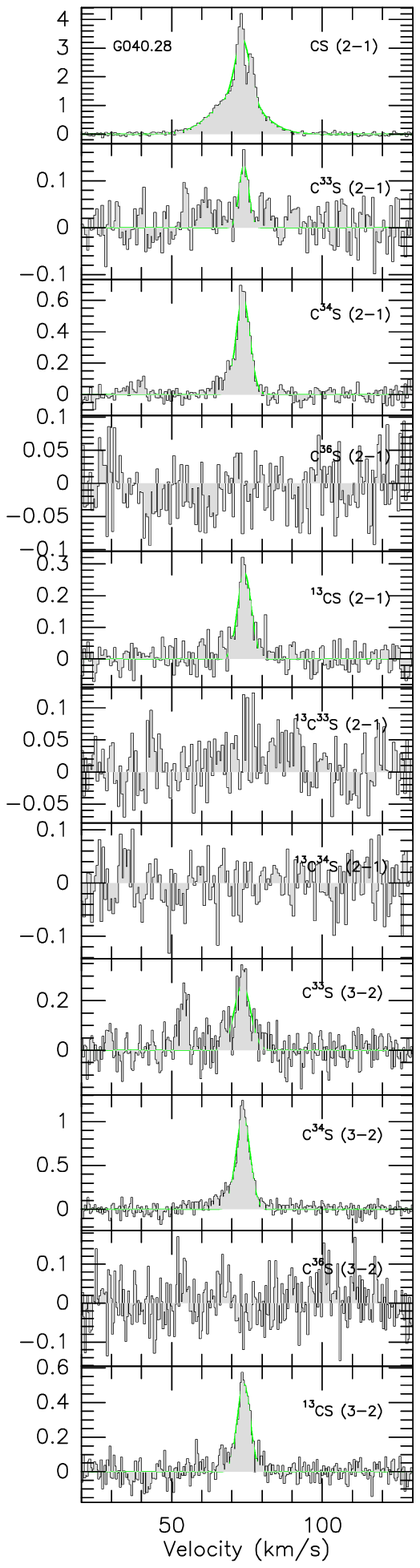}
\includegraphics[width=90pt,height=300pt]{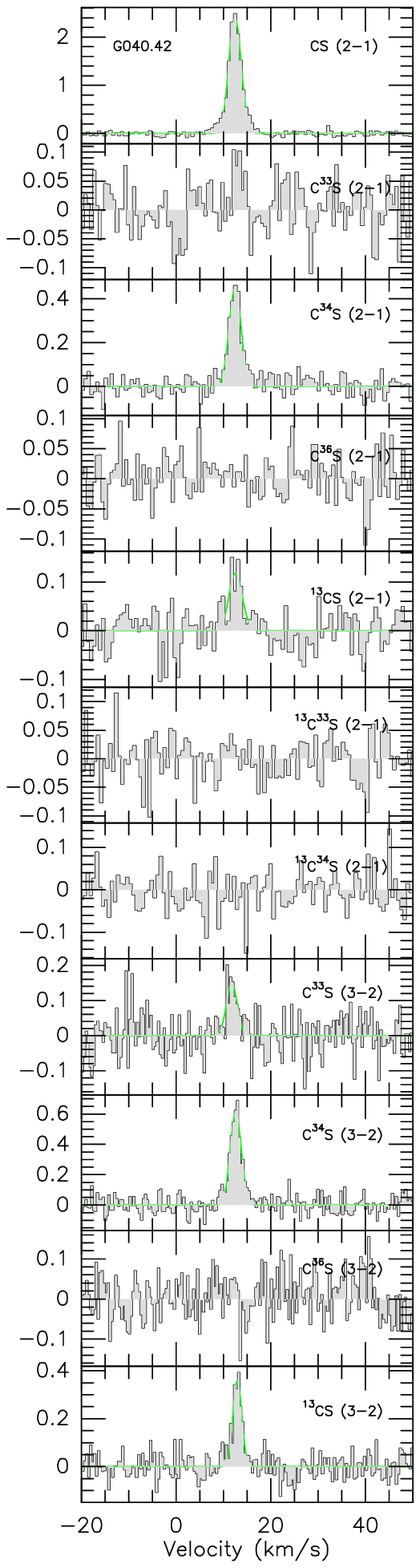}
\includegraphics[width=90pt,height=300pt]{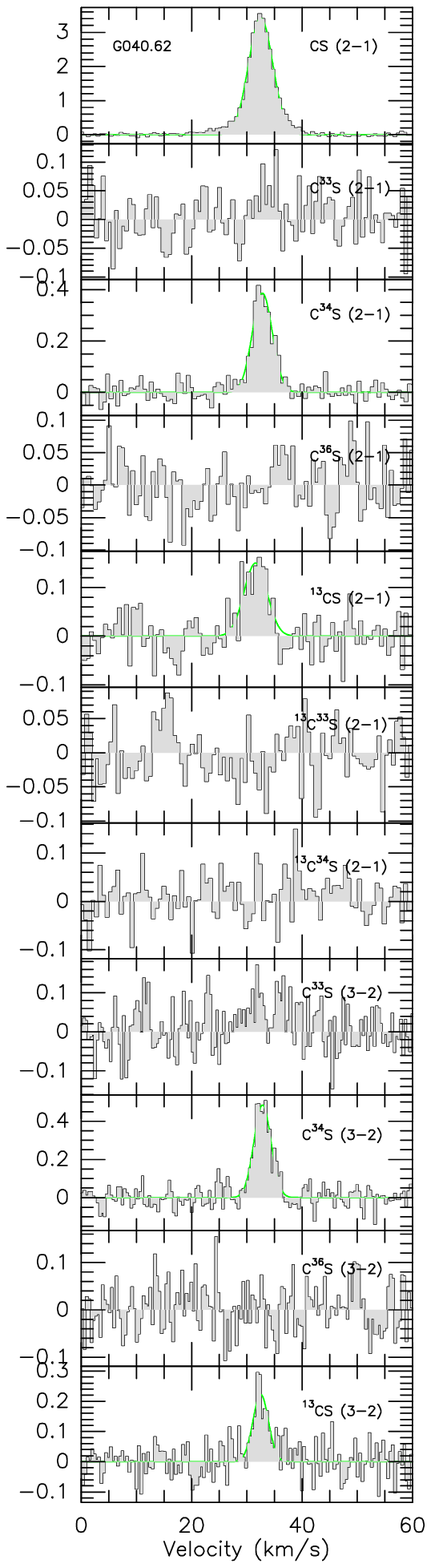}
\includegraphics[width=90pt,height=300pt]{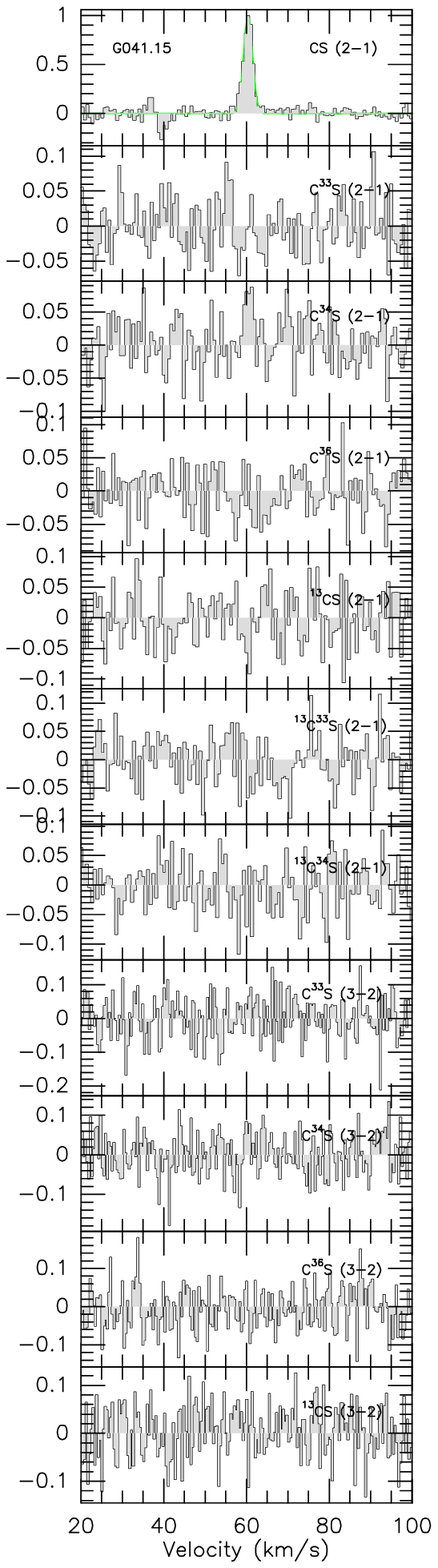}
\includegraphics[width=90pt,height=300pt]{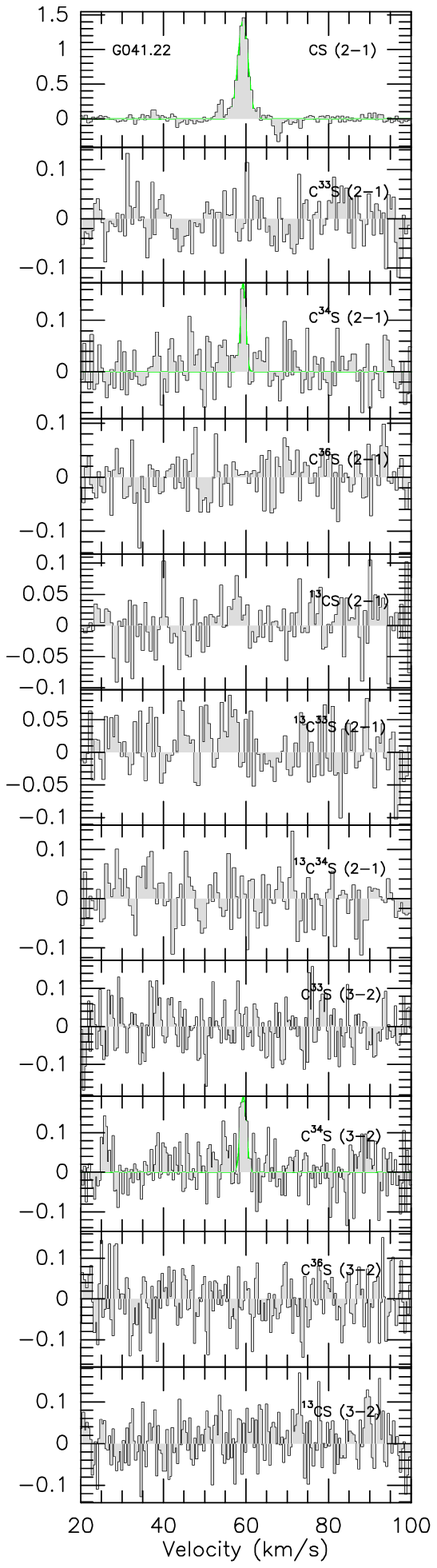}
\end{figure*}

\begin{figure*}
\centering
\includegraphics[width=90pt,height=300pt]{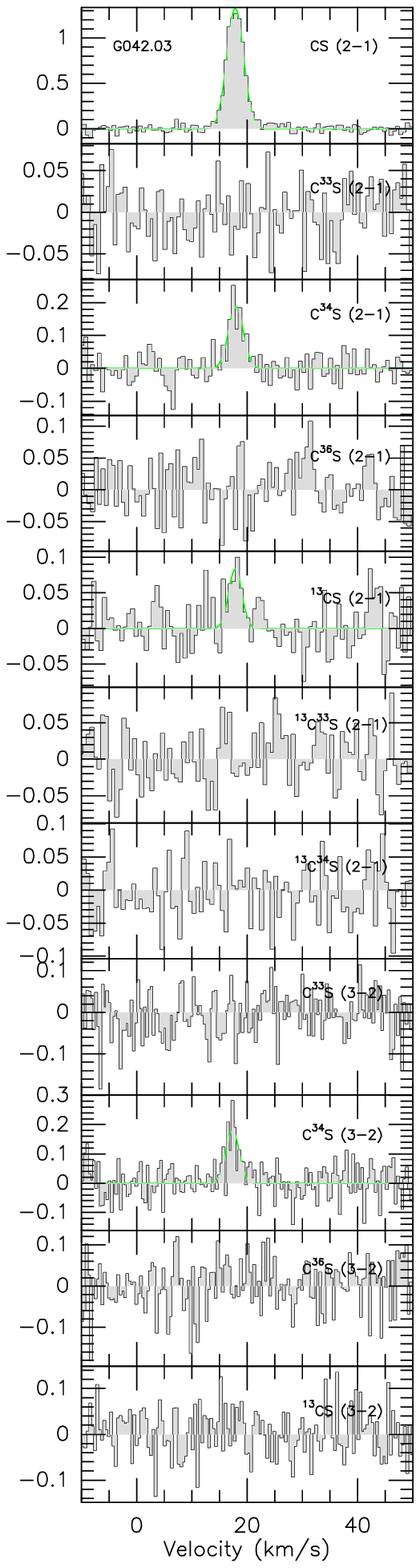}
\includegraphics[width=90pt,height=300pt]{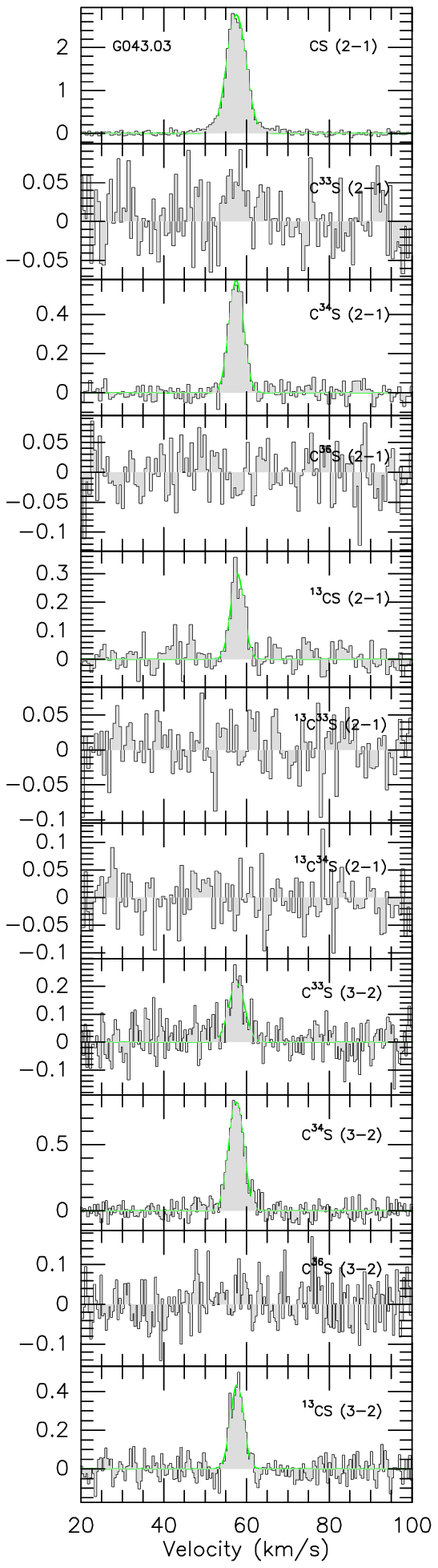}
\includegraphics[width=90pt,height=300pt]{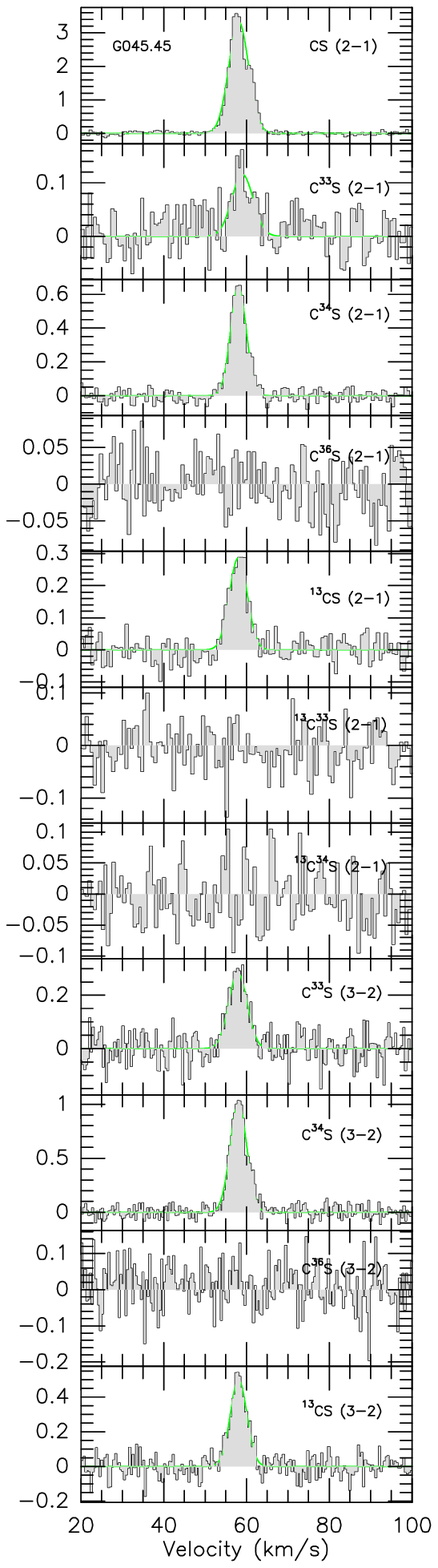}
\includegraphics[width=90pt,height=300pt]{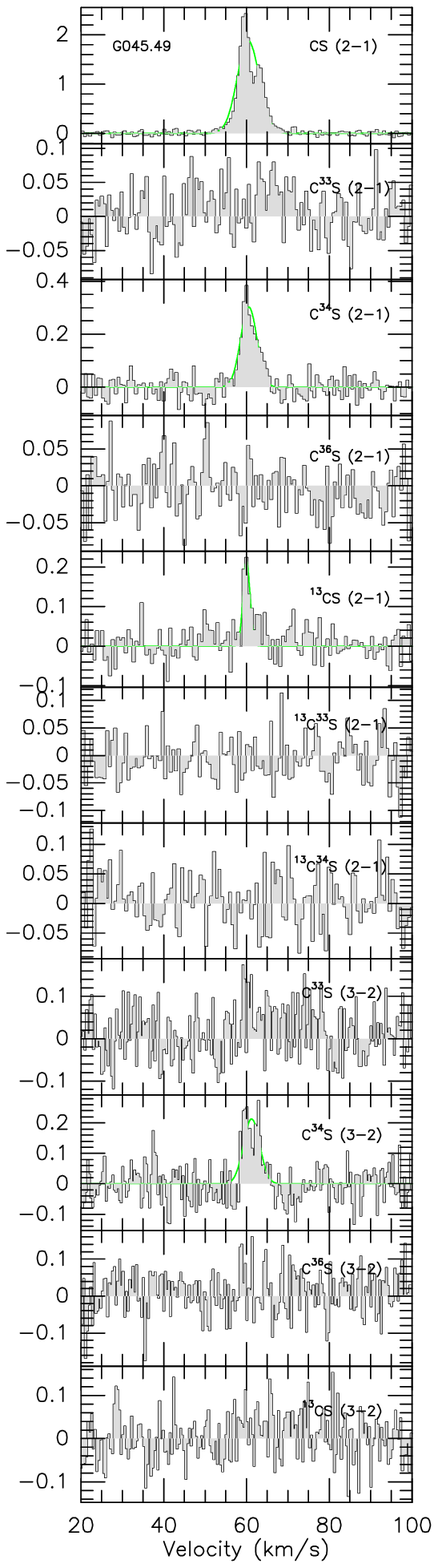}
\includegraphics[width=90pt,height=300pt]{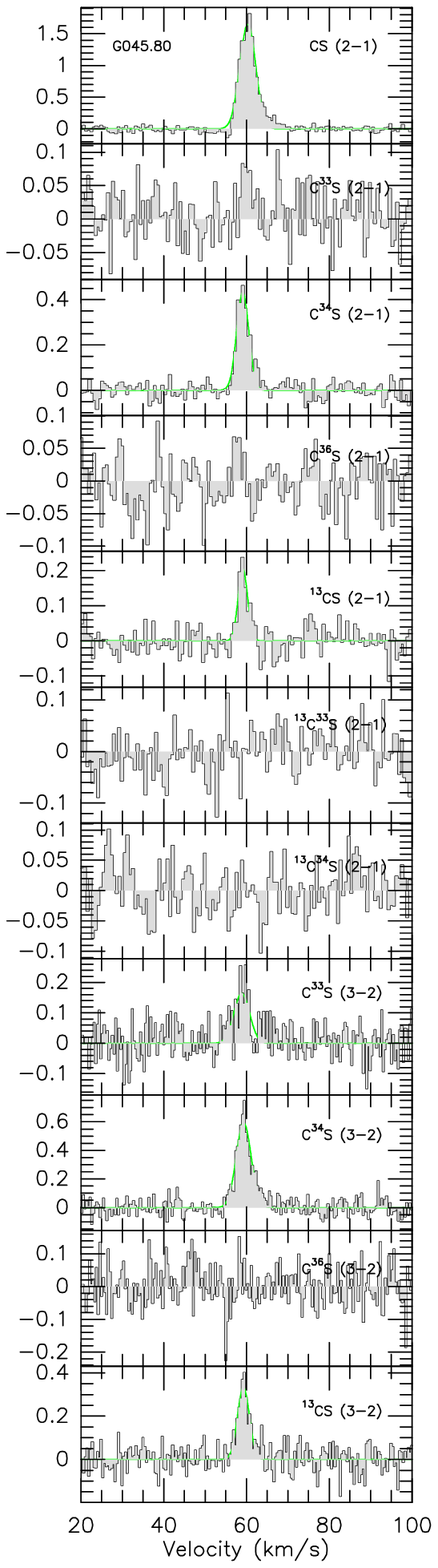}
\includegraphics[width=90pt,height=300pt]{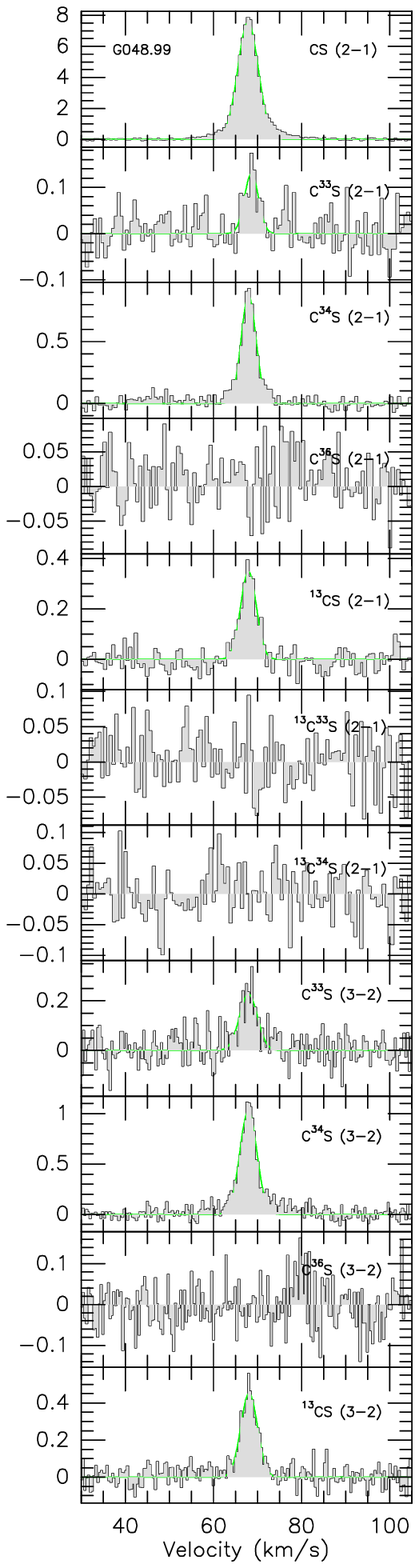}
\includegraphics[width=90pt,height=300pt]{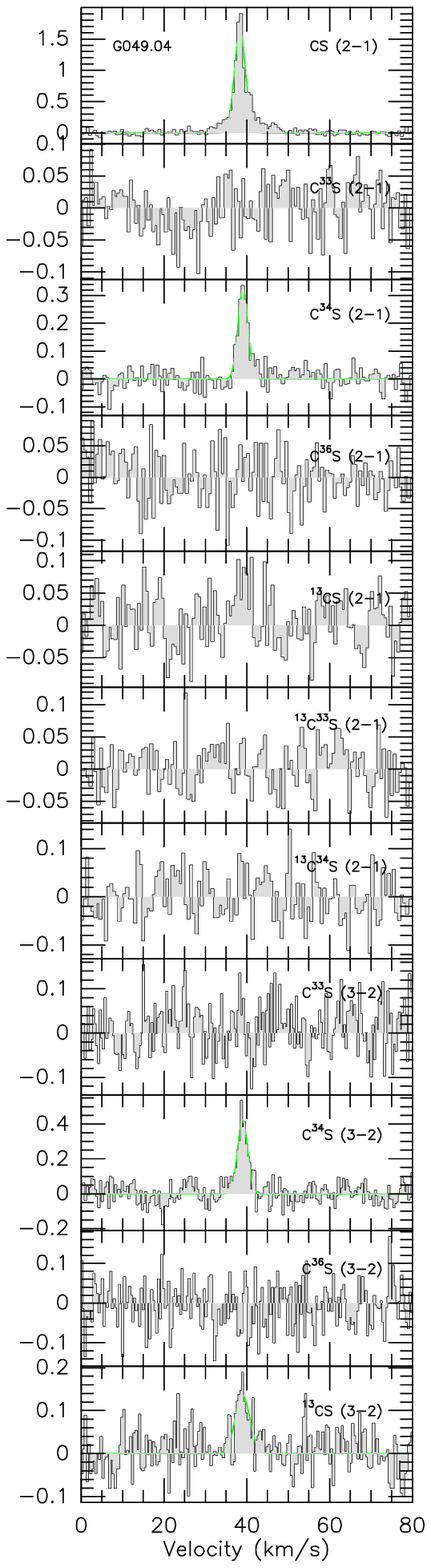}
\includegraphics[width=90pt,height=300pt]{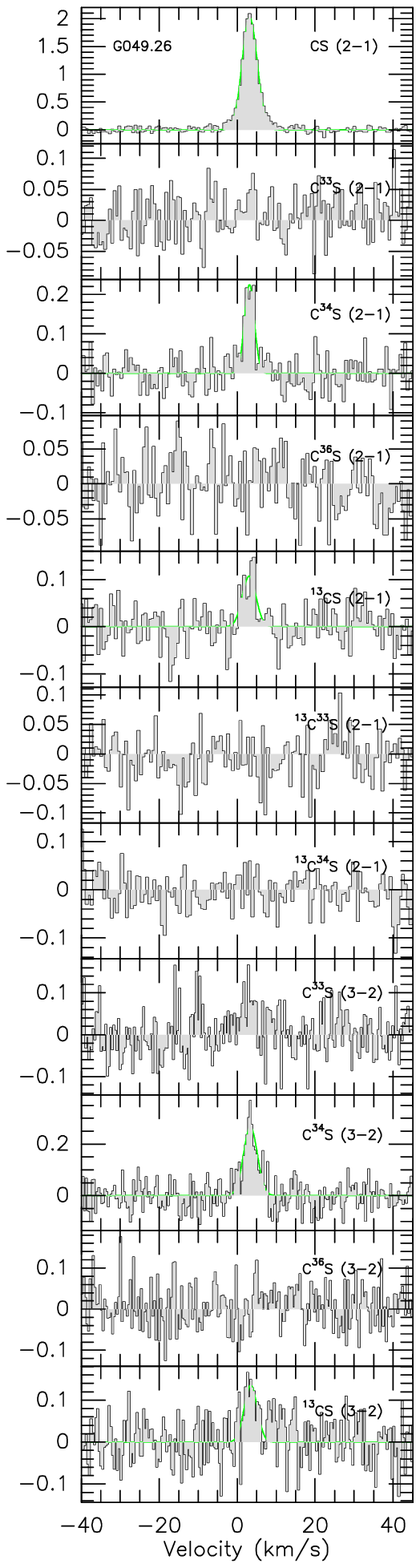}
\includegraphics[width=90pt,height=300pt]{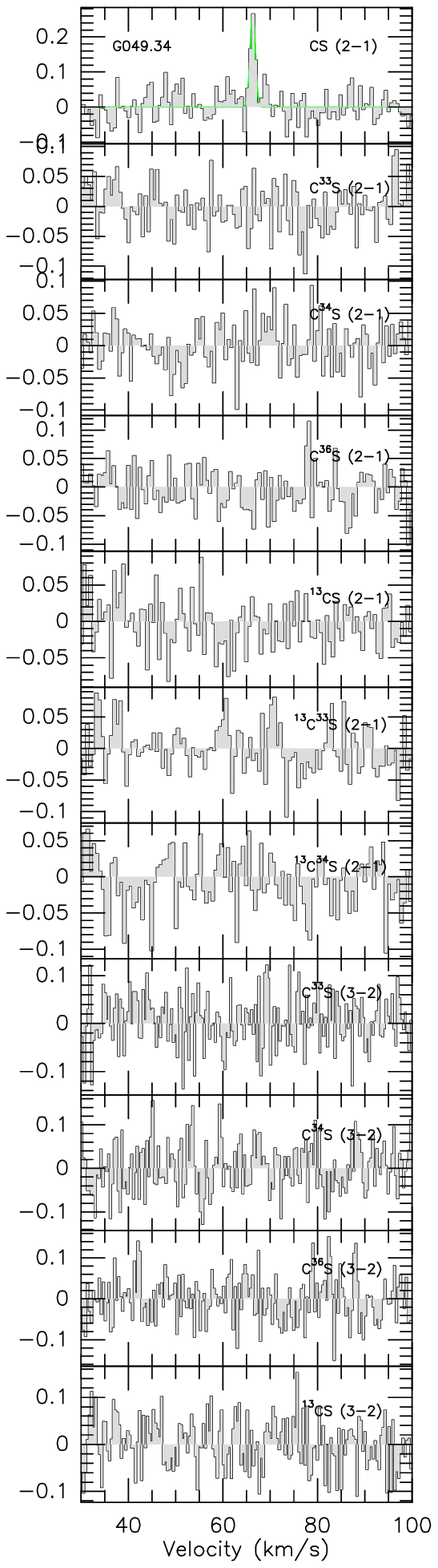}
\includegraphics[width=90pt,height=300pt]{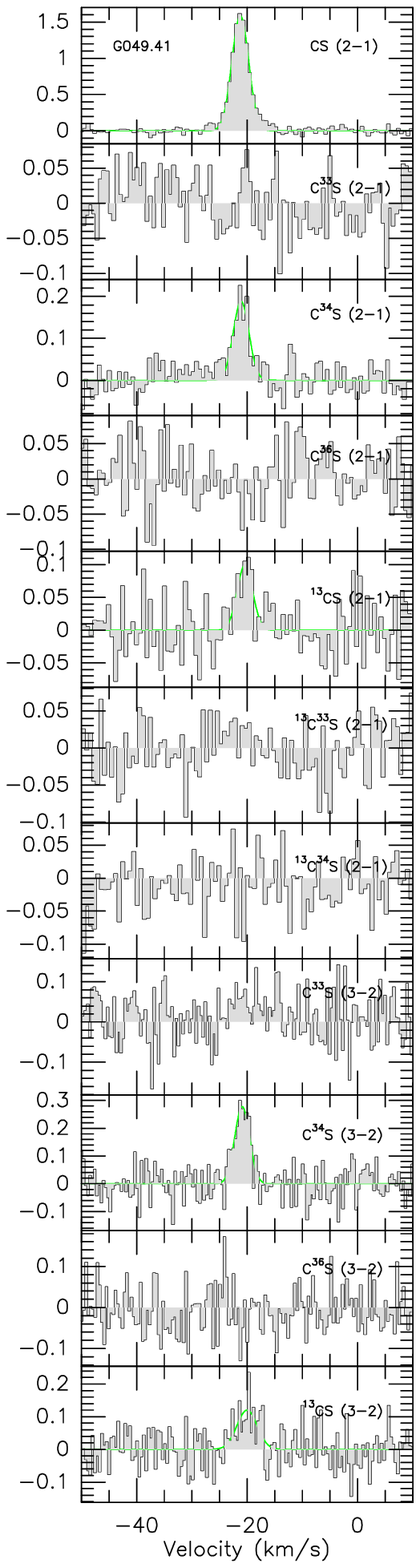}
\end{figure*}

\begin{figure*}
\centering
\includegraphics[width=90pt,height=300pt]{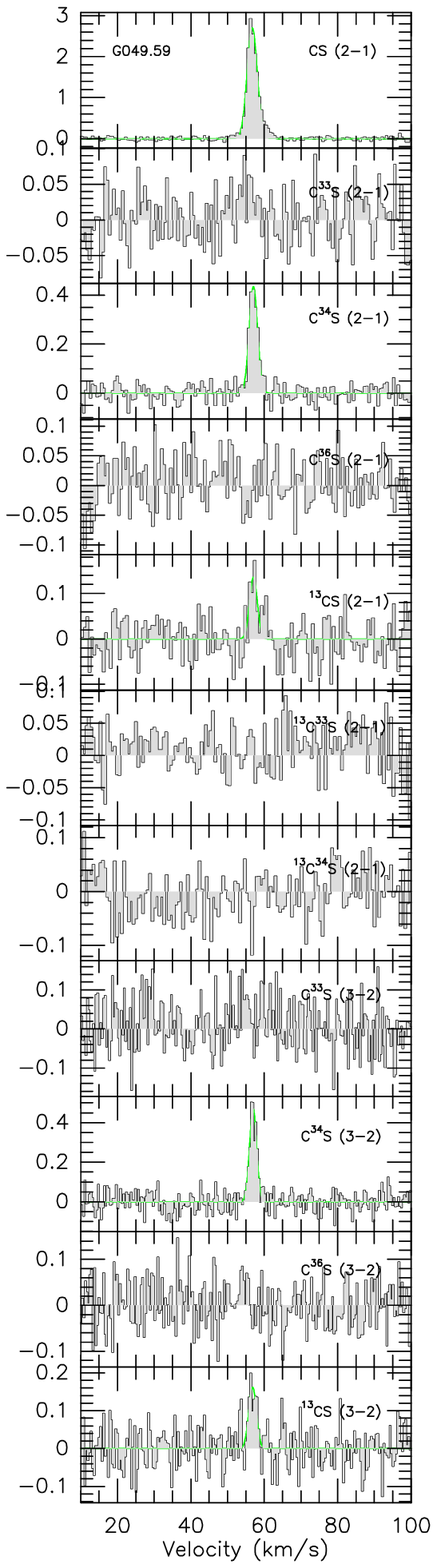}
\includegraphics[width=90pt,height=300pt]{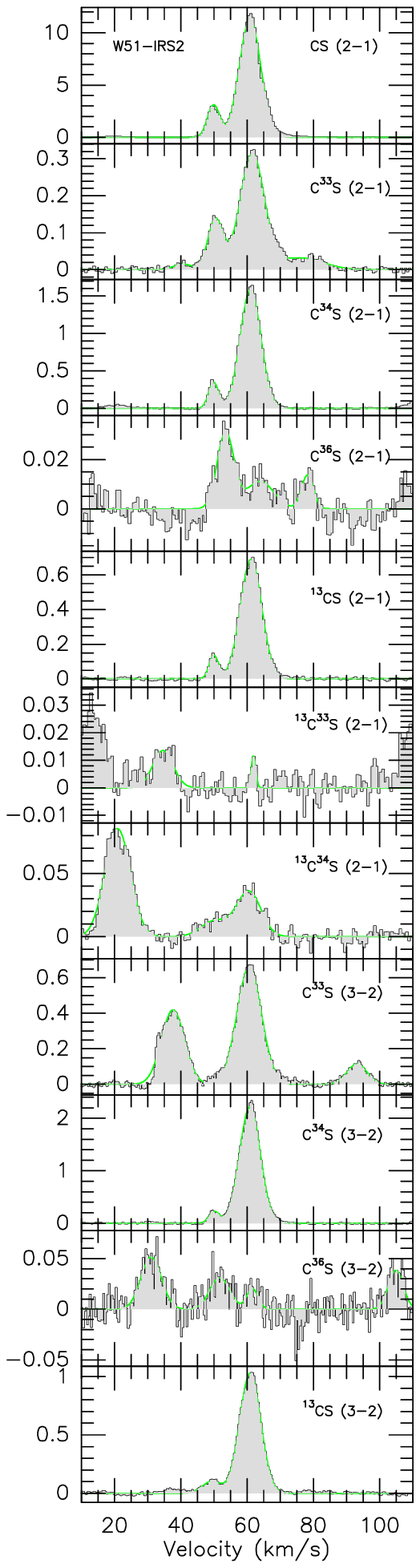}
\includegraphics[width=90pt,height=300pt]{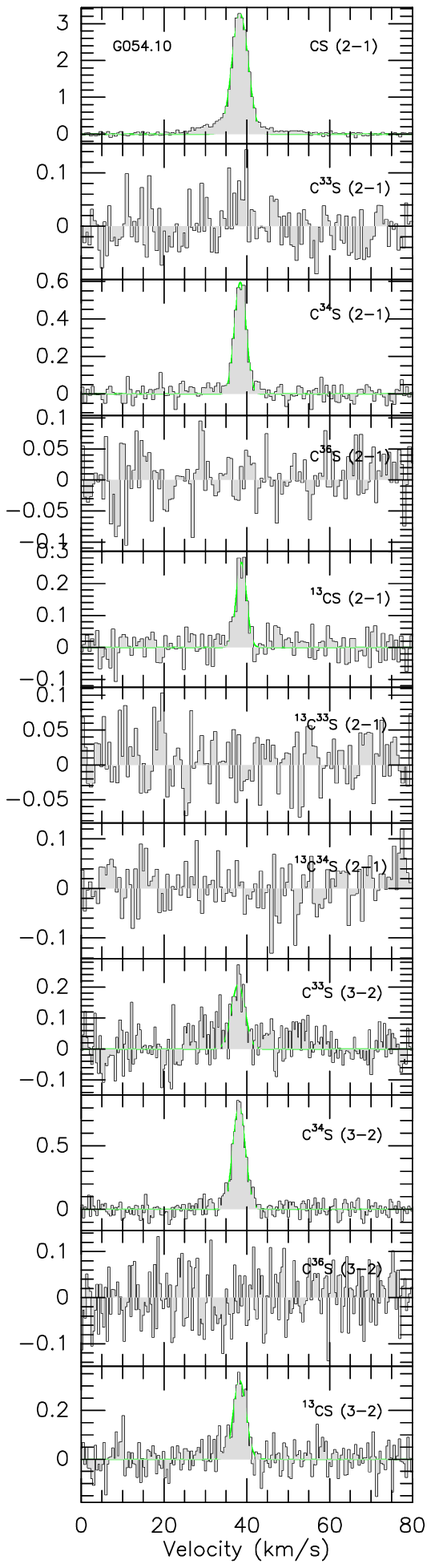}
\includegraphics[width=90pt,height=300pt]{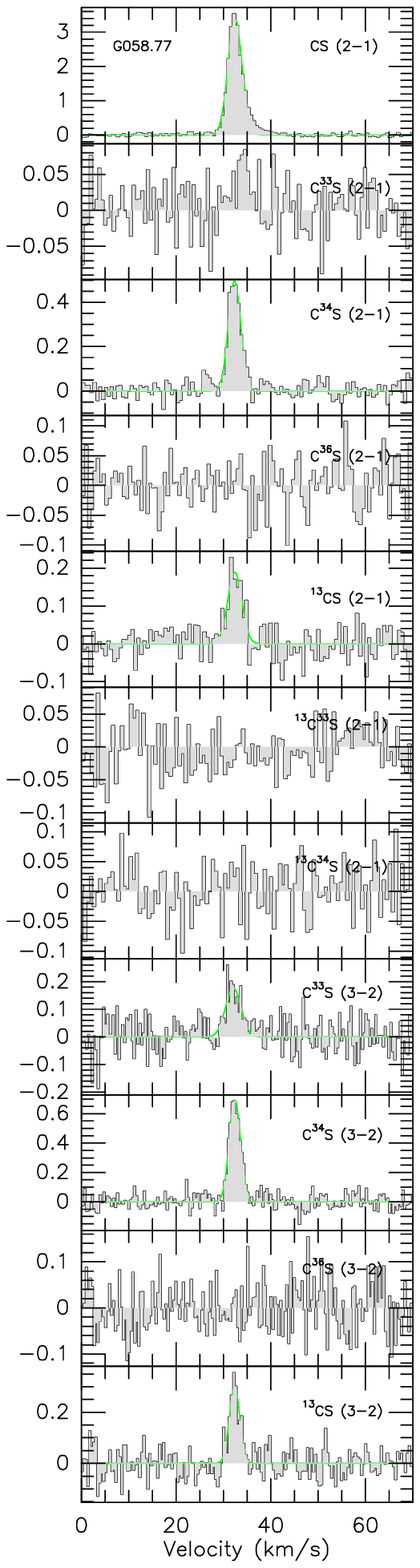}
\includegraphics[width=90pt,height=300pt]{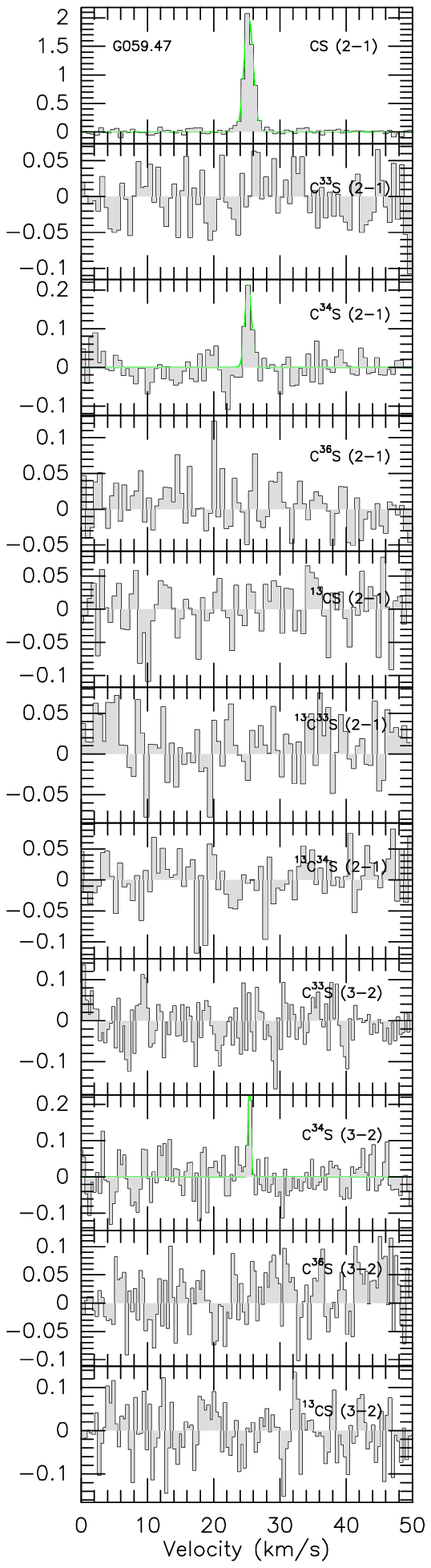}
\includegraphics[width=90pt,height=300pt]{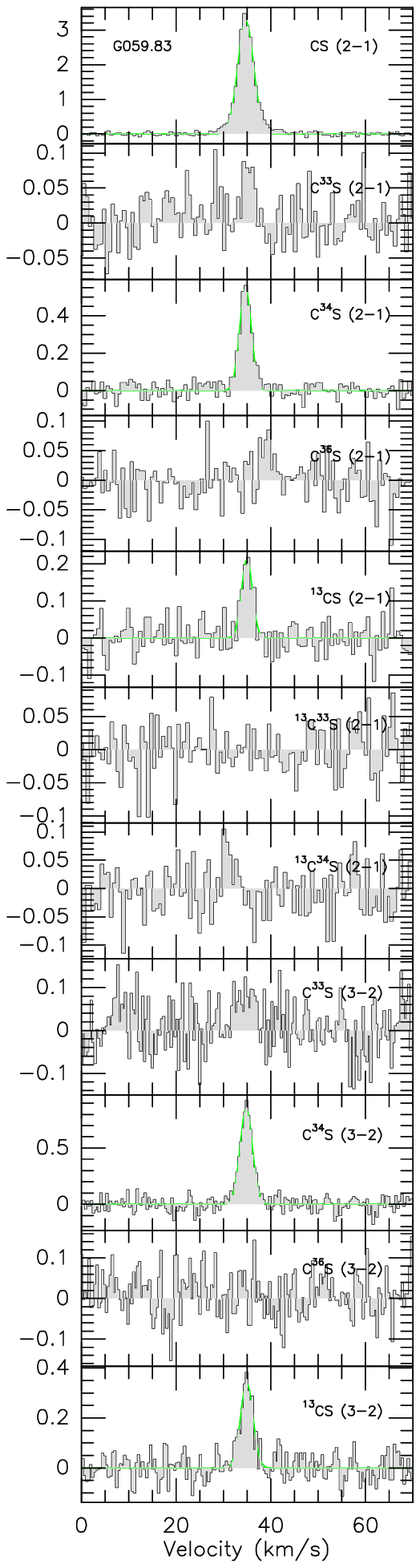}
\includegraphics[width=90pt,height=300pt]{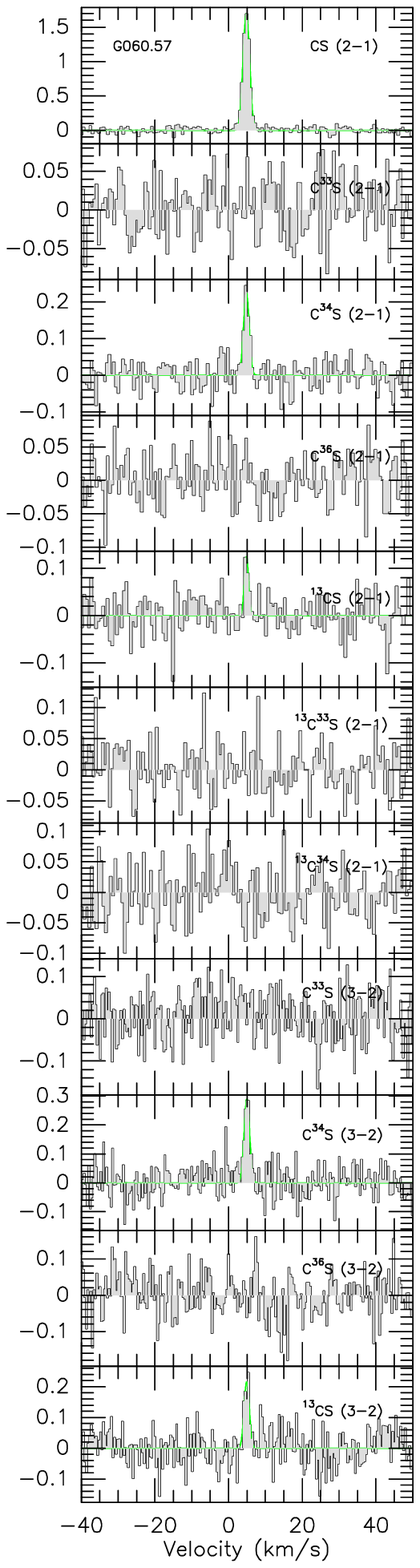}
\includegraphics[width=90pt,height=300pt]{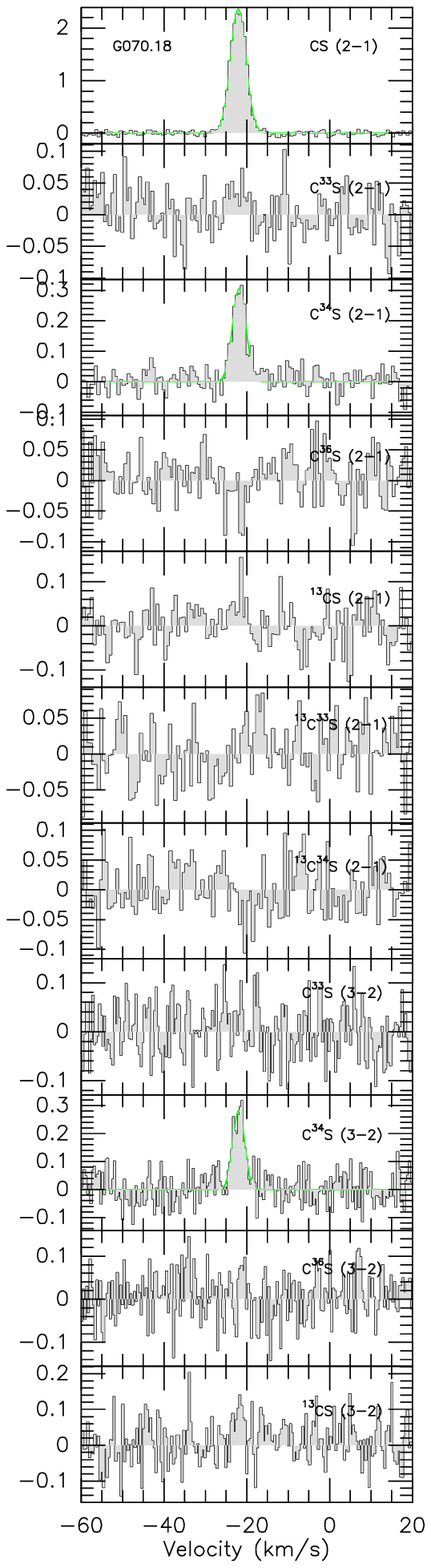}
\includegraphics[width=90pt,height=300pt]{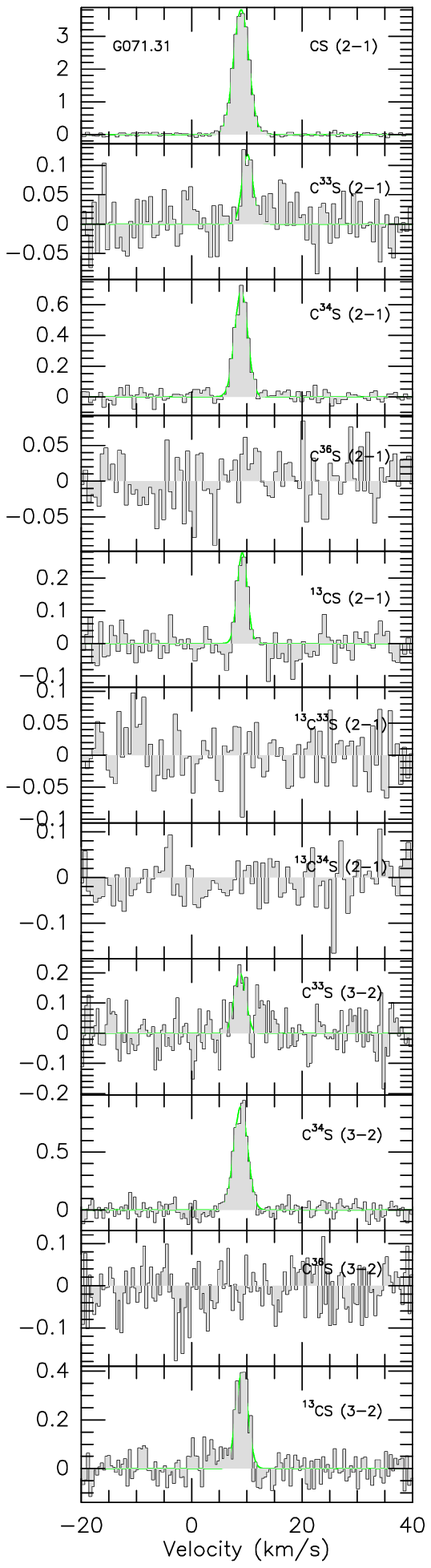}
\includegraphics[width=90pt,height=300pt]{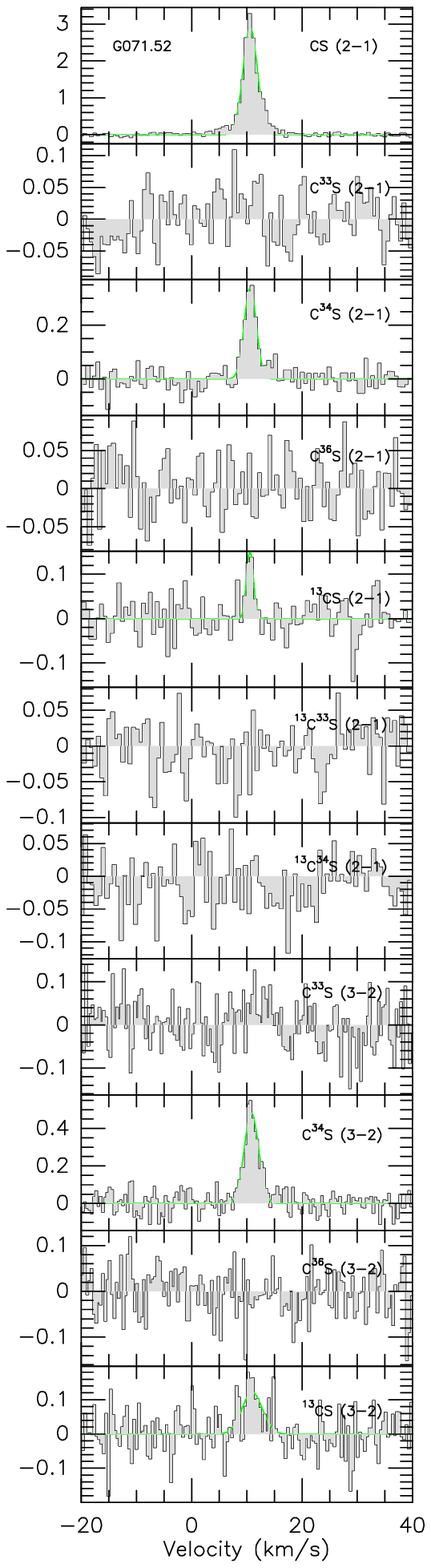}
\end{figure*}

\begin{figure*}
\centering
\includegraphics[width=90pt,height=300pt]{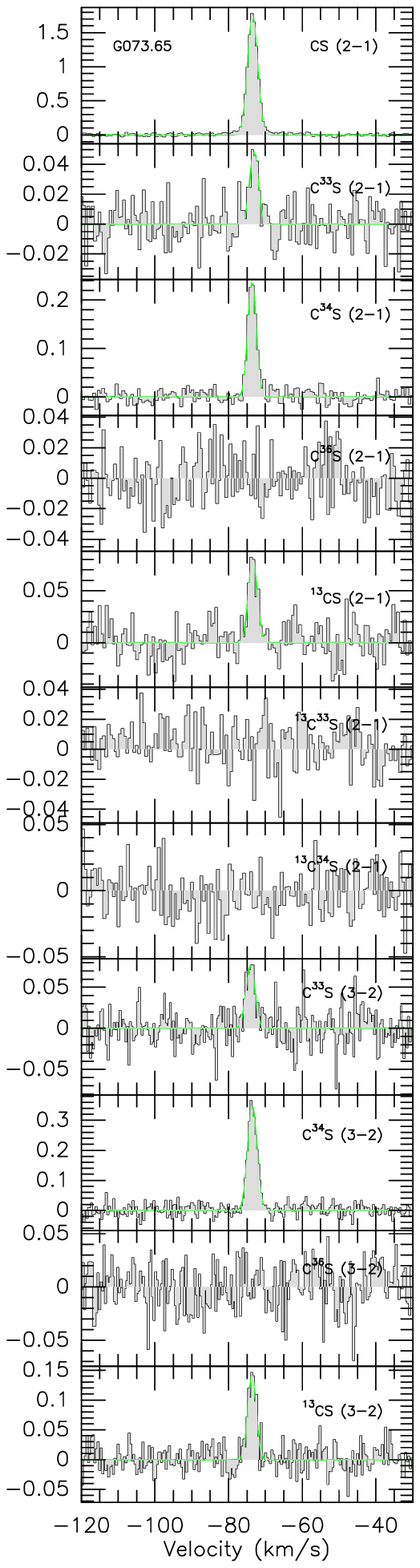}
\includegraphics[width=90pt,height=300pt]{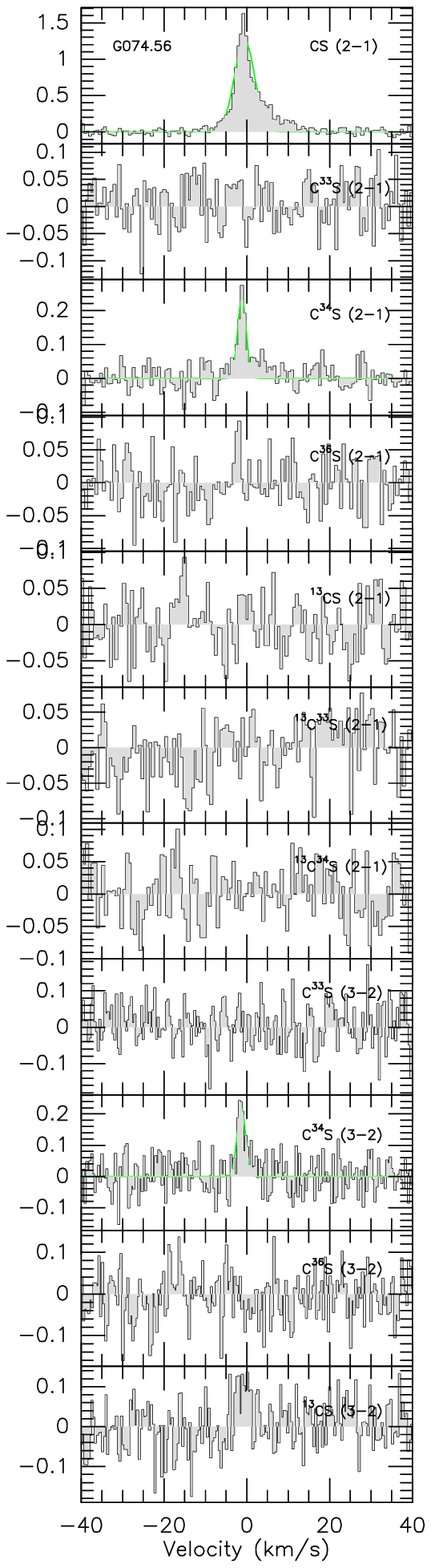}
\includegraphics[width=90pt,height=300pt]{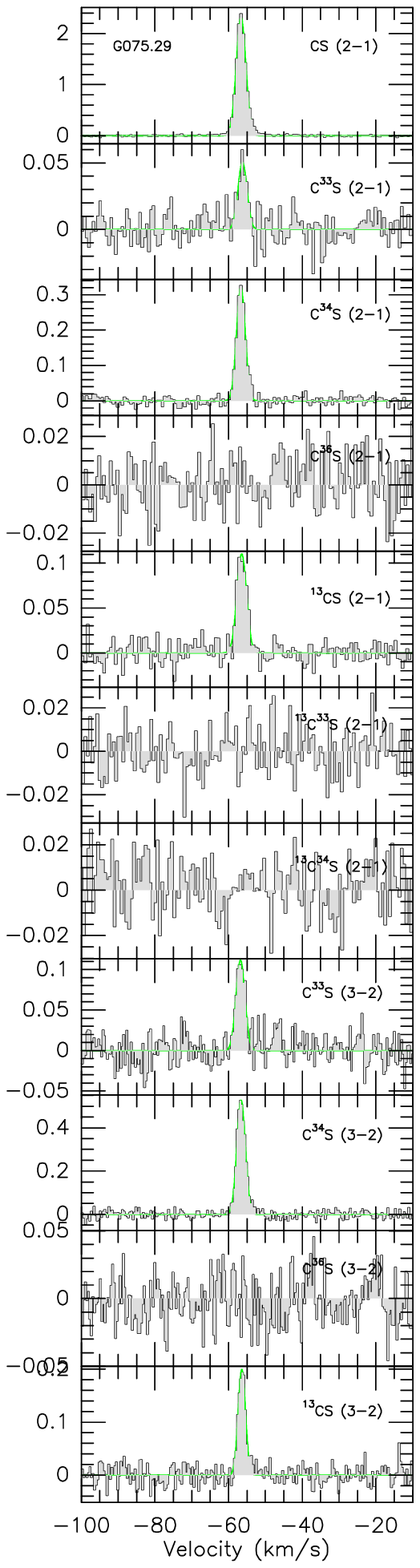}
\includegraphics[width=90pt,height=300pt]{spectra/DR21.eps}
\includegraphics[width=90pt,height=300pt]{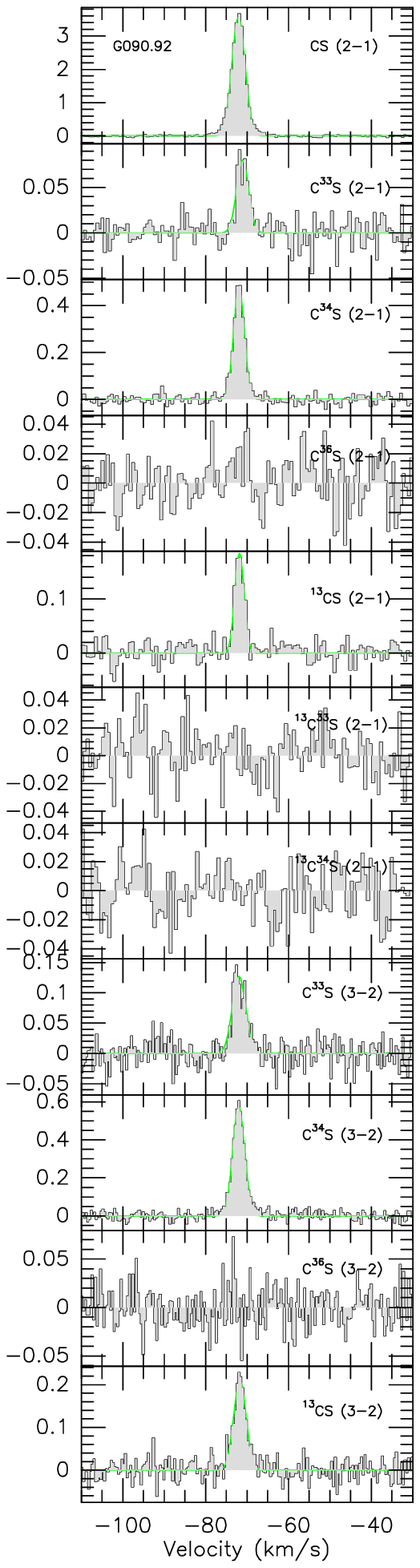}
\includegraphics[width=90pt,height=300pt]{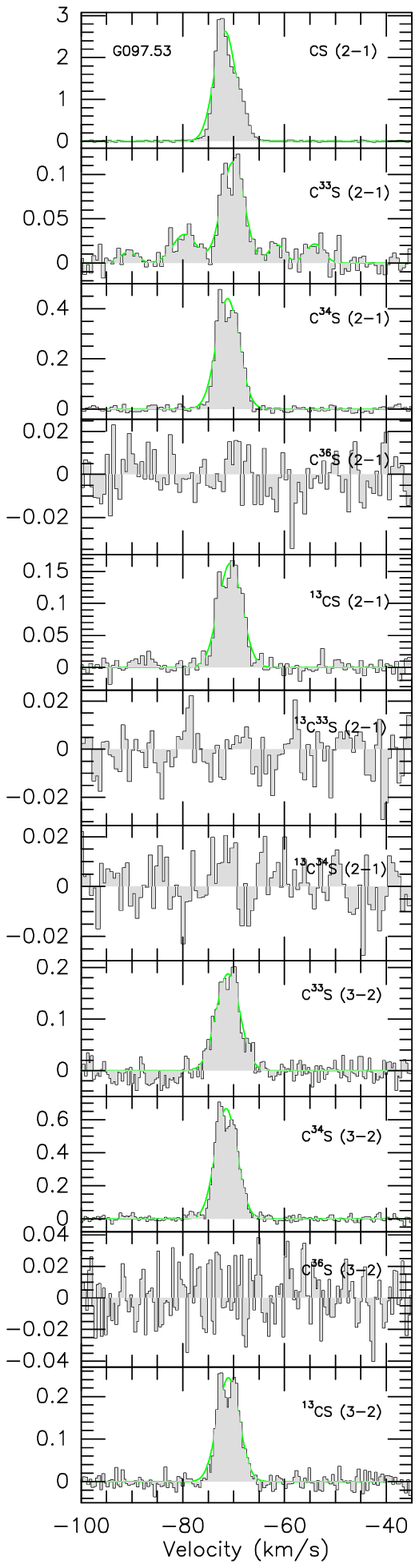}
\includegraphics[width=90pt,height=300pt]{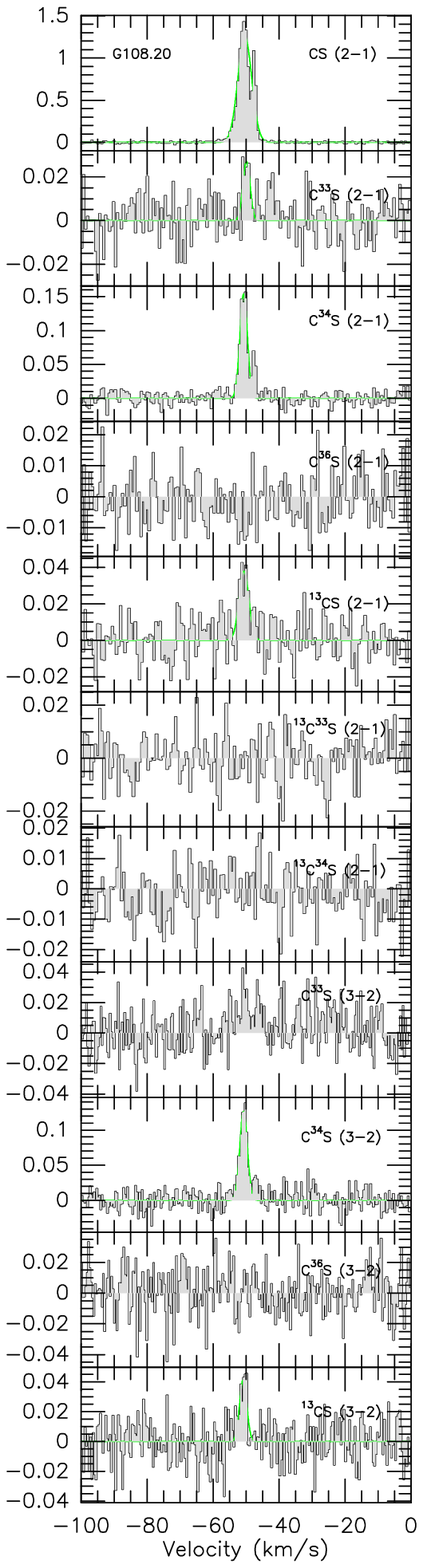}
\includegraphics[width=90pt,height=300pt]{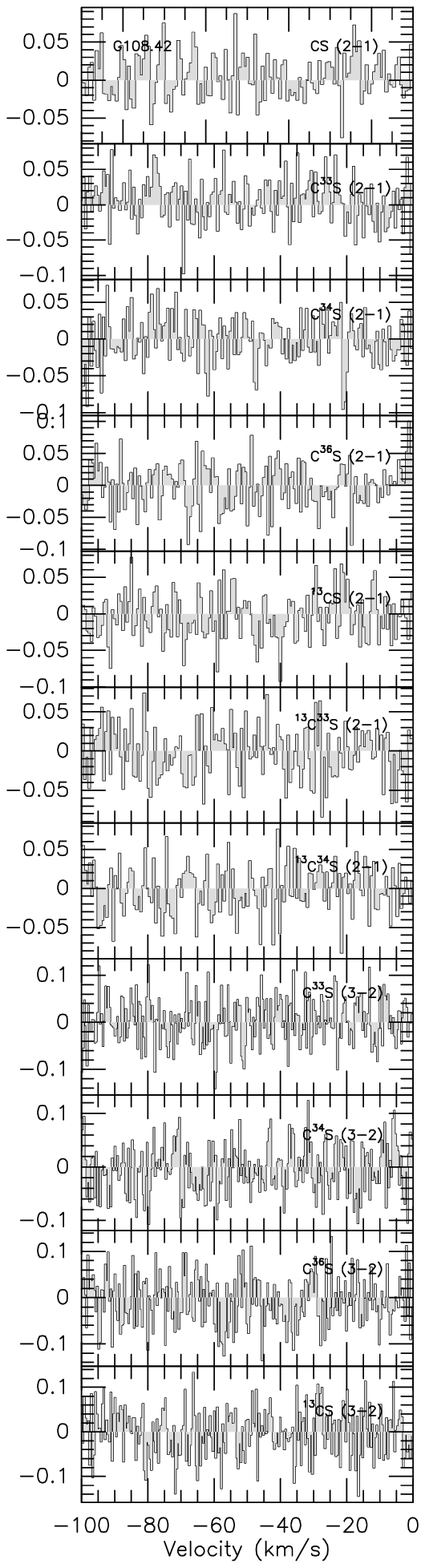}
\includegraphics[width=90pt,height=300pt]{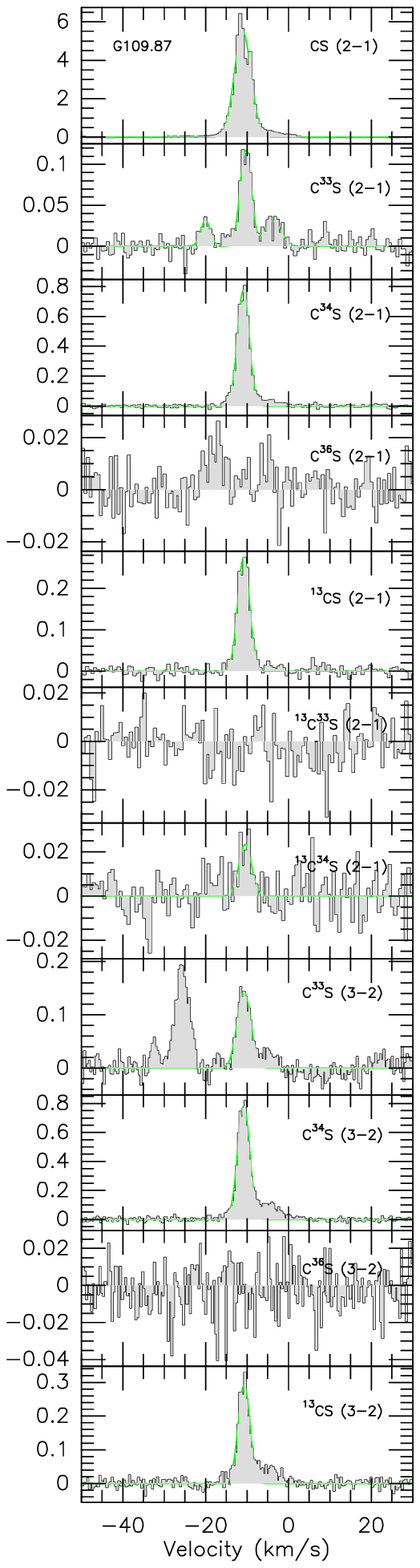}
\includegraphics[width=90pt,height=300pt]{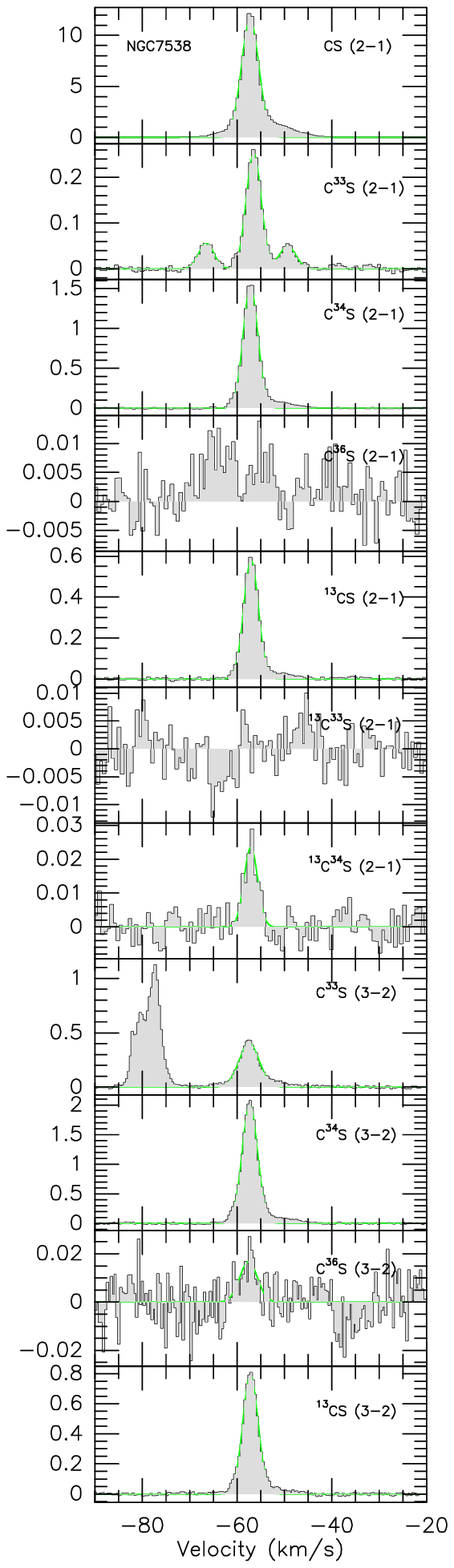}
\end{figure*}

\begin{figure*}[h]
\centering
   \includegraphics[width=460pt]{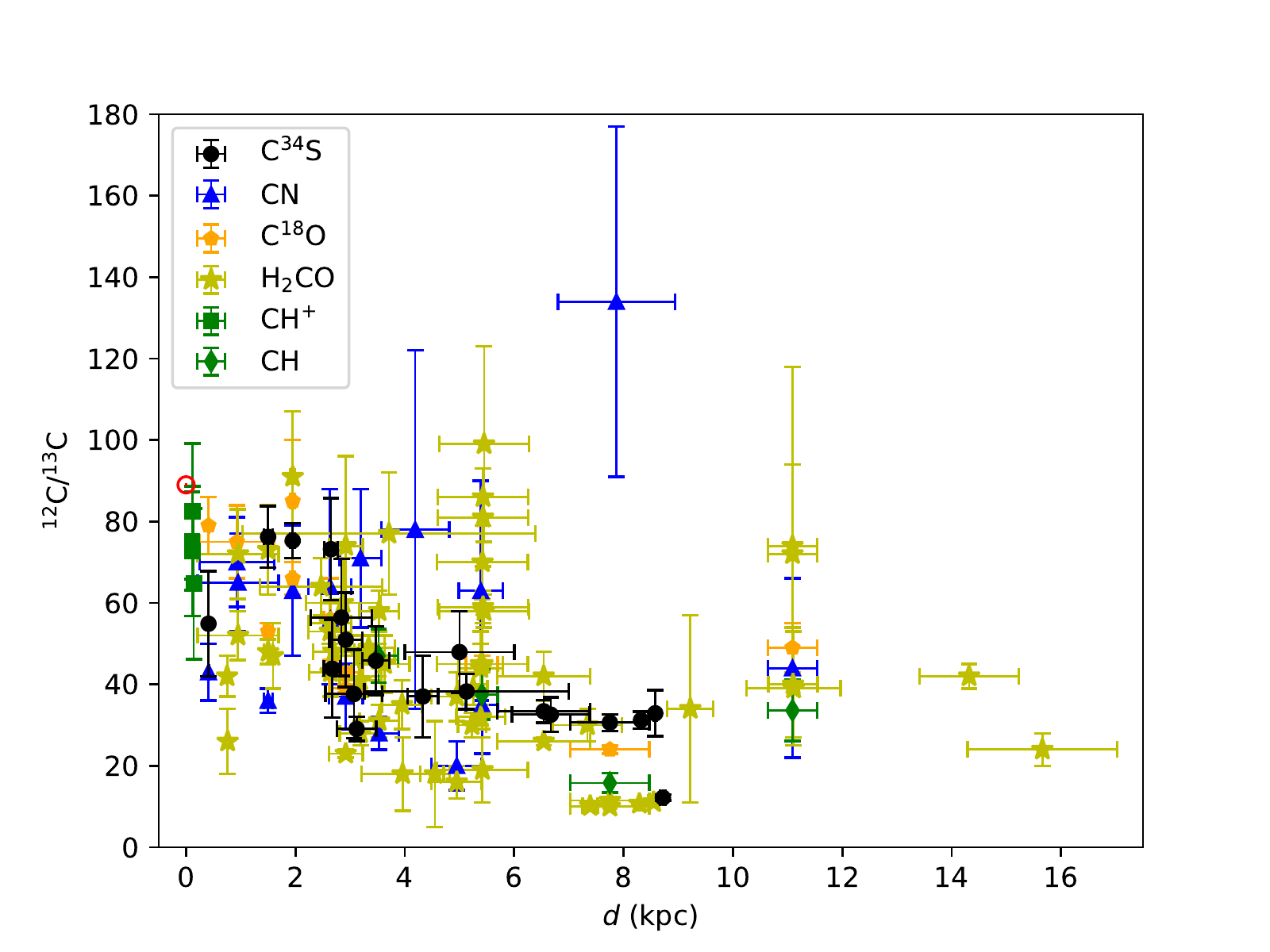}
     \caption{$^{12}$C/$^{13}$C isotope ratios from C$^{34}$S/$^{13}$C$^{34}$S, CN/$^{13}$CN, C$^{18}$O/$^{13}$C$^{18}$O, H$_2$CO/H$_2^{13}$CO, CH$^+$/$^{13}$CH$^+$, and CH/$^{13}$CH plotted as a function of the distance from the Sun. The red symbol $\odot$ indicates the $^{12}$C/$^{13}$C isotope ratio of the Sun. The $^{12}$C/$^{13}$C ratios directly derived from C$^{34}$S/$^{13}$C$^{34}$S in the $J$ = 2-1 transition with minor opacity corrections in the current work are plotted as black filled circles. The blue triangles, orange pentagons, yellow stars, green squares, and green diamonds are values determined from CN \citep{2002ApJ...578..211S,2005ApJ...634.1126M}, C$^{18}$O \citep{1990ApJ...357..477L,1996A&AS..119..439W,1998ApJ...494L.107K}, H$_2$CO \citep{1980A&A....82...41H,1982A&A...109..344H,1983A&A...127..388H,1985A&A...143..148H,2019ApJ...877..154Y}, CH$^+$ \citep{2011ApJ...728...36R}, and CH \citep{2020A&A...640A.125J}, respectively, using state of the art distances.}
  \label{fig_12C13C_2Sun}
\end{figure*}

\begin{figure*}[h]
\centering
   \includegraphics[width=460pt]{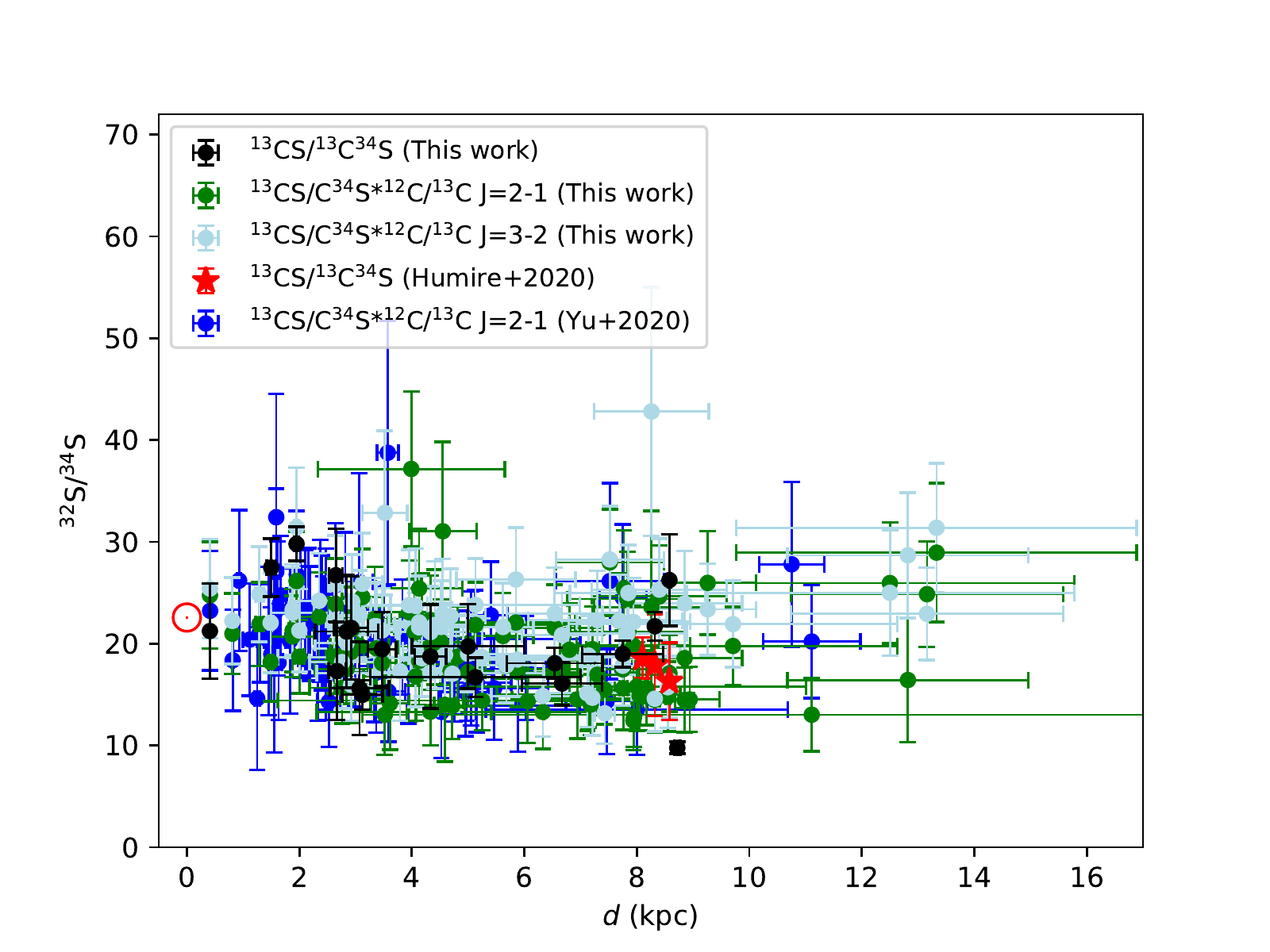}
     \caption{$^{32}$S/$^{34}$S isotope ratios plotted as a function of the distance from the Sun. The symbol $\odot$ indicates the $^{32}$S/$^{34}$S isotope ratio in the Solar System. The $^{32}$S/$^{34}$S ratios with corrections of opacity in the $J$ = 2-1 transition derived from $^{13}$CS/$^{13}$C$^{34}$S and obtained from the double isotope method are plotted as black and green dots, respectively. The $^{32}$S/$^{34}$S ratios without corrections for optical depth in the $J$ = 3-2 transition derived from the double isotope method are shown as light blue dots. The $^{32}$S/$^{34}$S ratios in \citet{2020ApJ...899..145Y} derived from the double isotope method in the $J$ = 2-1 transitions are shown as blue dots. The $^{32}$S/$^{34}$S values in the CMZ obtained from $^{13}$CS/$^{13}$C$^{34}$S in \citet{2020A&A...642A.222H} are plotted as red stars. }
  \label{fig_32S34S_2Sun}
\end{figure*}

\begin{figure*}[h]
\centering
   \includegraphics[height=620pt]{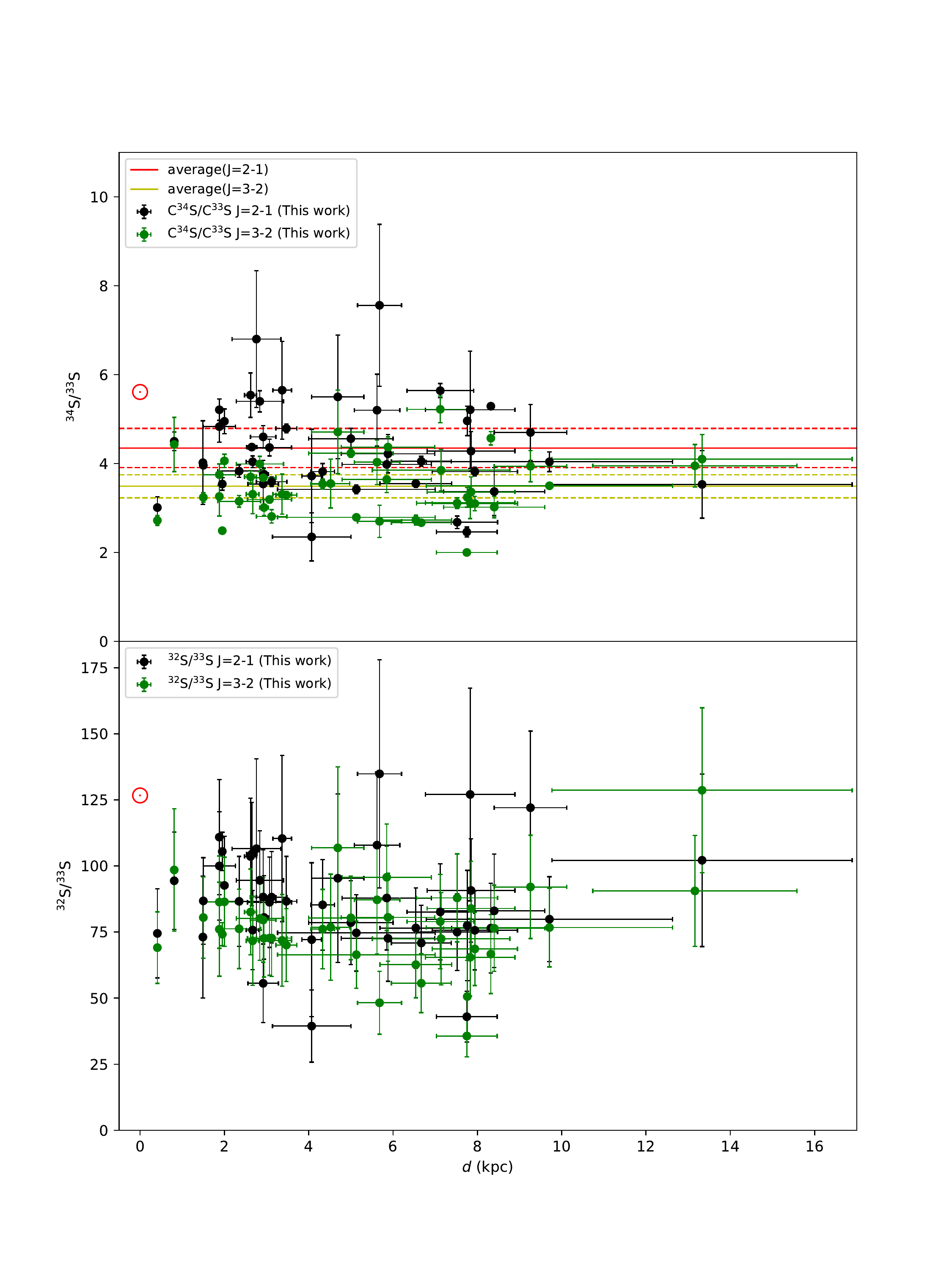}
     \caption{ $^{34}$S/$^{33}$S and $^{32}$S/$^{33}$S isotope ratios plotted as functions of the distance from the Sun. \textbf{Top:}  $^{34}$S/$^{33}$S ratios derived from C$^{34}$S/C$^{33}$S in the $J$ = 2-1 and $J$ = 3-2 transitions are plotted as black and green dots, respectively. The red solid and the two dashed lines show the average value and its standard deviation, 4.35~$\pm$~0.44, for $^{34}$S/$^{33}$S with corrections of optical depth toward our sample in the $J$ = 2-1 transition. The yellow solid and the two dashed lines show the average value and its standard deviation, 3.49~$\pm$~0.26, of $^{34}$S/$^{33}$S toward our sample without opacity corrections for the $J$ = 3-2 transition. The red symbol $\odot$ indicates the $^{34}$S/$^{33}$S isotope ratio in the Solar System. \textbf{Bottom:} Black and green dots show the $^{32}$S/$^{33}$S ratios in the $J$ = 2-1 and $J$ = 3-2 transitions, respectively. The red symbol $\odot$ indicates the $^{32}$S/$^{33}$S value in the Solar System. }
  \label{fig_all33S_2Sun}
\end{figure*}

\begin{figure*}[h]
\centering
   \includegraphics[height=600pt]{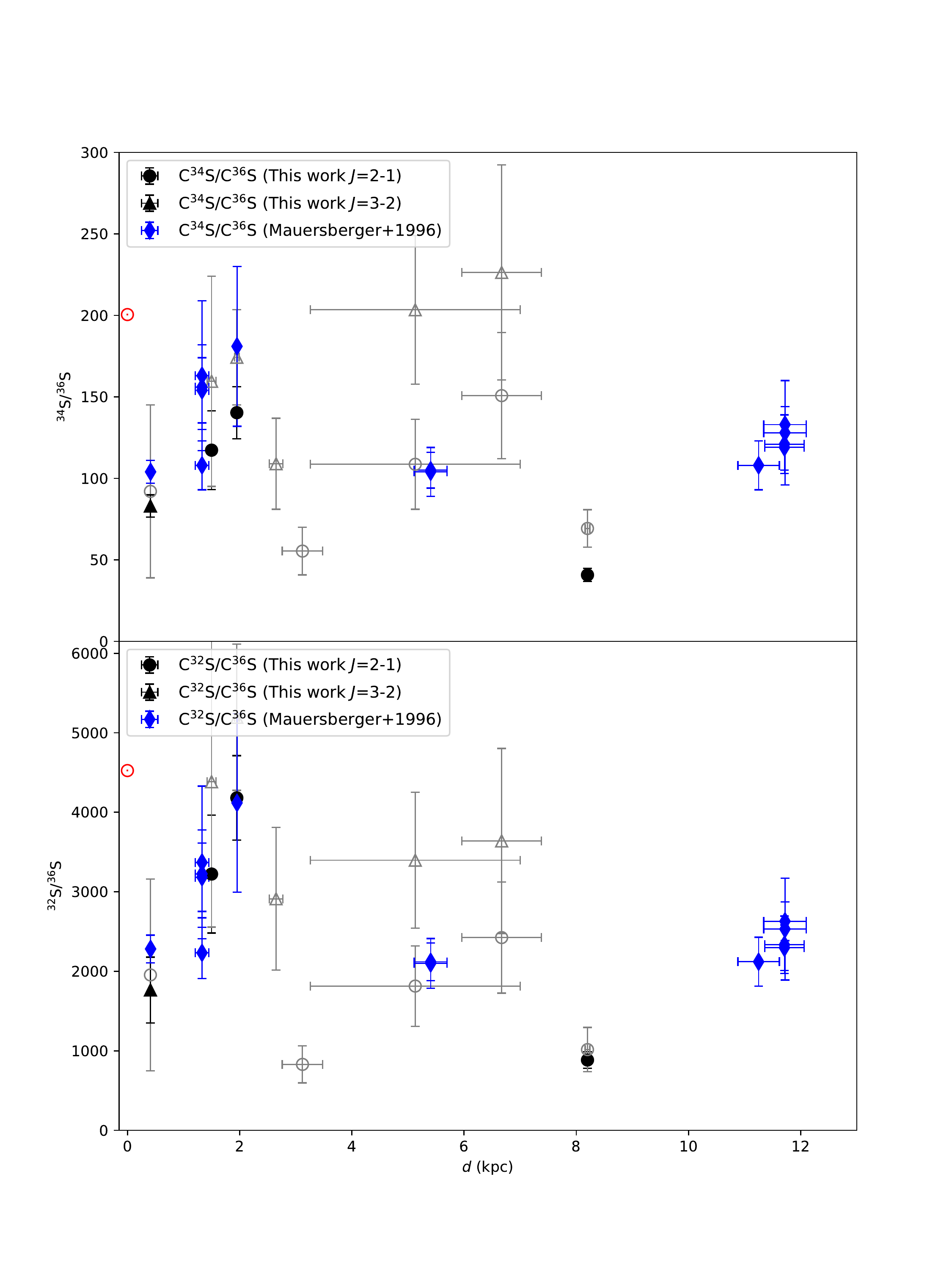}
     \caption{ $^{34}$S/$^{36}$S and $^{32}$S/$^{36}$S isotope ratios plotted as functions of the distance from the Sun. \textbf{Top:} Filled black circles and filled black triangle present the $^{34}$S/$^{36}$S ratios in the $J$ = 2-1 and $J$ = 3-2 transitions derived from C$^{34}$S/C$^{36}$S in this work with detections of C$^{36}$S, respectively. The open gray circles and open gray triangles present the $^{34}$S/$^{36}$S ratios in the $J$ = 2-1 and $J$ = 3-2 transitions derived from C$^{34}$S/C$^{36}$S in this work with tentative detections of C$^{36}$S, respectively. The blue diamonds show the $^{34}$S/$^{36}$S ratios in \citet{1996A&A...313L...1M}. The red symbol $\odot$ indicates the $^{34}$S/$^{36}$S isotope ratio in the Solar System. \textbf{Bottom:}  $^{32}$S/$^{36}$S ratios obtained from $^{34}$S/$^{36}$S ratios combined with the $^{32}$S/$^{34}$S ratios derived in this work. The filled black circles and filled black triangle present the values in the $J$ = 2-1 and $J$ = 3-2 transitions from this work with detections of C$^{36}$S, respectively. The open gray circles and open gray triangles present the ratios in the $J$ = 2-1 and $J$ = 3-2 transitions derived from this work with tentative detections of C$^{36}$S, respectively. The $^{32}$S/$^{36}$S ratios, derived with $^{34}$S/$^{36}$S values in \citet{1996A&A...313L...1M} and the $^{32}$S/$^{34}$S gradient in this work, are plotted as blue diamonds. The red symbol $\odot$ indicates the $^{32}$S/$^{36}$S isotope ratio in the Solar System. }
  \label{fig_all36S_2Sun}
\end{figure*}

\end{appendix}
\end{document}